\documentclass[aps,pra,superscriptaddress,showpacs,showkeys,reprint]{revtex4-1}
\usepackage{amsmath}
\usepackage{amsfonts}
\usepackage{amsthm}
\usepackage{graphics}
\usepackage{graphicx}
\usepackage{subfigure}
\usepackage{amssymb}
\usepackage{color}
\usepackage{url}
\usepackage[abs]{overpic}
\usepackage{tikz}
\usepackage{bm}
\usepackage[normalem]{ulem}
\usepackage{lipsum}  
\usepackage{morefloats}

\definecolor{darktangerine}{rgb}{1.0, 0.66, 0.07}

\begin{document}

\title{Particles and Fields in Superfluids: Insights from the Two-dimensional
Gross-Pitaevskii Equation}

\author{Vishwanath Shukla}
\email{research.vishwanath@gmail.com}
\affiliation{Service de Physique de l'\'Etat Condens\'e, Universit\'e Paris-Saclay, CEA Saclay, 91191 Gif-sur-Yvette, France}
\author{Rahul Pandit}
\email{rahul@physics.iisc.ernet.in}
\altaffiliation[\\ also at~]{Jawaharlal Nehru Centre For Advanced
Scientific Research, Jakkur, Bangalore, India.}
\affiliation{Centre for Condensed Matter Theory, Department of Physics, 
Indian Institute of Science, Bangalore 560012, India.} 
\author{Marc Brachet}
\email{brachet@physique.ens.fr}
\affiliation{Laboratoire de Physique Statistique de l'Ecole Normale 
Sup{\'e}rieure, \\
associ{\'e} au CNRS et aux Universit{\'e}s Paris VI et VII,
24 Rue Lhomond, 75231 Paris, France}

\date{\today}
\begin{abstract}
We carry out extensive direct numerical simulations (DNSs) to investigate the
interaction of active particles and fields in the two-dimensional (2D)
Gross-Pitaevskii (GP) superfluid, in both simple and turbulent flows. The
particles are active in the sense that they affect the superfluid even as they
are affected by it. We tune the mass of the particles, which is an important
control parameter. At the one-particle level, we show how light, neutral, and
heavy particles move in the superfluid, when a constant external force acts on
them; in particular, beyond a critical velocity, at which a vortex-antivortex
pair is emitted, particle motion can be periodic or chaotic.  We demonstrate
that the interaction of a particle with vortices leads to dynamics that depends
sensitively on the particle characteristics.  We also demonstrate that
assemblies of particles and vortices can have rich, and often turbulent
spatiotemporal evolution.  In particular, we consider the dynamics of the
following illustrative initial configurations: (a) one particle placed in front
of a translating vortex-antivortex pair; (b) two particles placed in front of a
translating vortex-antivortex pair; (c) a single particle moving in the
presence of counter-rotating vortex clusters; and (d) four particles in the
presence of counter-rotating vortex clusters. We compare our work with earlier
studies and examine its implications for recent experimental studies in
superfluid Helium and Bose-Einstein condensates.
\end{abstract}
\maketitle

\section{Introduction}

The study of the dynamics of particles immersed in a superfluid has a long
history~\cite{donnellybook}. This venerable subject has experienced a
renaissance because of recent experiments that (a) use particles in superfluid
helium~\cite{bewley2006superfluid,bewley2008characterization,bewleyparticlesdetails,
He2excimerT0tracer,mantiaprbaccpdf},
to track vortex lines or (b) to study impurities in cold-atom, Bose-Einstein
condensates (BECs)~\cite{Spethmann2012}.

Unlike particles moving through a viscous fluid, particles that move with a
constant speed through a superfluid move without resistance, so long as this
speed lies below a critical
threshold~\cite{frisch1992vcrit,NoreHB2000,huepe2000scaling,
Pham-Brachet,winiecki2}.
Above this threshold, superfluidity breaks down. This critical speed has been
related, in the Gross-Pitaevskii Equation (GPE), to vortex generation, which is
caused by a saddle-node bifurcation of steady
states~\cite{frisch1992vcrit,NoreHB2000,Pham-Brachet,winiecki2,winiecki1}.  At
supercritical speeds, vortex generation is associated with forces that act on
particles. The interaction of such vortices with single particles or assemblies
of particles are problems of central importance in this rapidly developing
field. 
The motion of a
single particle, which is affected by the superflow and acts on it too, has
been studied in Refs.~\cite{winiecki2,winiecki1,pitapart04,Varga2015} in a
Gross-Pitaevskii (GP) superfluid. We refer to this as an \textit{active
particle}.

To go beyond earlier theoretical and numerical
studies~\cite{winiecki1,pitapart04,Varga2015} of this problem we have developed
a minimal model~\cite{shuklaPRA16} recently. In this model, the equations of
motion of particles are coupled with the GP field $\psi$, which is used often
to describe a weakly interacting superfluid at low temperatures. Furthermore,
our model includes a particle-particle, short-range repulsive force. This model
generates naturally an effective superfluid-mediated attractive interaction
between the particles; we have shown~\cite{shuklaPRA16} that the interplay
between the short-range particle repulsion and the superfluid-mediated
attraction leads to a sticking transition at which the coefficient of
restitution, for two-particle collisions, vanishes.

We build on our study in Ref.~\cite{shuklaPRA16} and consider the interaction
of particles with vortices.  We begin with a brief, qualitative overview of our
principal results, which we obtain from extensive direct numerical simulations
(DNSs) of the interaction of particles and fields in the two-dimensional (2D)
GPE, in both simple and turbulent flows. At the one-particle level, we explore,
for light, neutral, and heavy particles, the nature of their dynamics in the
superfluid, when a constant external force acts on them; in particular, we
show, by a careful consideration of the effects of the particle mass, how the
motion of such particles can be temporally periodic or chaotic.  We demonstrate
that the interaction of a particle with vortices leads to dynamics that depends
sensitively on the particle characteristics. 
For assemblies of particles and
vortices we demonstrate that their dynamics show rich, turbulent spatiotemporal
evolution.  In particular, we systematize the spatiotemporal evolution of an
initial configuration in which one particle is placed in front of a translating
vortex-antivortex pair. We then examine the complicated motions of two particles
placed in front of a translating vortex-antivortex pair and show that, once a
particle traps a vortex, we can use this particle as a tracer that can track
vortex motion.  
As an illustrative example, we first examine the motion of a single particle
in the presence of many vortices in a decaying turbulent flow. We then study how 
the interactions of many particles with fields in the presence of a group of
counter-rotating vortices can lead to turbulent and spatiotemporally chaotic evolution.
We compare our work with earlier studies and discuss  the experimental implications of our
work.

The remainder of this paper is organized as follows. In the first part of Sec.
II we recapitulate the essentials of our model~\cite{shuklaPRA16} for particles
coupled to the GP field $\psi$; in a subsection, we present the numerical
methods we use in our studies of this model. Section III is devoted to our
results for different classes of initial conditions; the first subsection here
discusses vortex generation by the motion of a single particle (light, neutral,
or heavy) and the spatiotemporal evolution of $\psi$ and the particle position;
the second subsection explores the interaction between particles and a
vortex-antivortex pair; the third subsection considers illustrative
multi-particle and multi-vortex assemblies and studies their rich and turbulent
spatiotemporal evolution. In Section IV we give our conclusions and discuss the
significance of our results. 

\section{Model and Numerical Methods}

\subsection{Model}

We begin with a brief recapitulation of our minimal 
model~\cite{shuklaPRA16} of active and interacting particles in a weakly
interacting Bose superfluid at zero temperature. The dynamics of 
these particles is governed by the Lagrangian 
\begin{equation}\label{eq:Lagrangianfull}
\begin{split}
\mathcal{L} &= \int_{\mathcal{A}}\Bigl[\frac{i\hbar}{2}\Bigl(\psi^*(\mathbf{r},t)\frac{\partial \psi(\mathbf{r},t)}{\partial t}
-\psi(\mathbf{r},t)\frac{\partial \psi^*(\mathbf{r},t)}{\partial t}\Bigr) \\
&- \frac{\hbar^2}{2m}\nabla\psi(\mathbf{r},t)\cdot\nabla\psi^*(\mathbf{r},t)
+\mu|\psi(\mathbf{r},t)|^2 \\ 
&- \frac{g}{2}|\psi(\mathbf{r},t)|^4
-\sum^{\mathcal{N}_0}_{i=1}V_{\mathcal{P}}(\mathbf{r}-\mathbf{q}_i)|\psi(\mathbf{r},t)|^2\Bigr]\,d\mathbf{r} \\
&+\frac{m_{o}}{2}\sum^{\mathcal{N}_0}_{i=1}\dot{q}^2_{i}
-\sum^{\mathcal{N}_0,\mathcal{N}_0}_{i,j,i\neq j}\frac{\Delta_E r^{12}_{SR}}{|\mathbf{q}_i-\mathbf{q}_j|^{12}},
\end{split}
\end{equation}
where $\psi$ is the complex, condensate wave function, $\psi^*$ is its complex
conjugate, $m$ is the mass of the bosons, ${\mathcal{A}}$ is the simulation
domain, $g$ is the effective interaction strength, $\mu$ is the chemical
potential, $V_{\mathcal{P}}$ is the potential that accounts for the particles,
$q_i$ the position of particle $i$, and $\mathcal{N}_0$ the total number of
particles, each with mass $m_{o}$.  The last term in
Eq.~(\ref{eq:Lagrangianfull}) is the short-range (SR) repulsive, two-particle
potential, which is characterized by the parameters $\Delta_E$ and $r_{SR}$.

From the Lagrangian~(\ref{eq:Lagrangianfull}) we obtain the following
GP equation for the field $\psi$:
\begin{equation}\label{eq:GPEfull}
i\hbar\frac{\partial \psi}{\partial t} = -\frac{\hbar^2}{2m}\nabla^2\psi -\mu\psi + g|\psi|^2\psi
+ \sum^{\mathcal{N}_0}_{i=1}V_{\mathcal{P}}(\mathbf{r}-\mathbf{q}_i)\psi;
\end{equation}
and, for the particle $i$, we get 
\begin{equation}\label{eq:eqmpartfull}
	m_{o}\ddot{\mathbf{q}}_i = \mathbf{f}_{o,i} + \mathbf{f}_{SR,i},
\end{equation}
where
\begin{equation}\label{eq:frcbyfluid}
	\mathbf{f}_{o,i} = \int_{\mathcal{A}}|\psi|^2\nabla V_{\mathcal{P}}\,d\mathbf{r},
\end{equation}
and $\mathbf{f}_{SR,i}$ arises because of the SR potential.  If there is no
external force, the total energy of this system
\begin{equation}\label{eq:Etotal}
	E = E_{\rm F} + E^{\rm T}_{\rm o} + E_{\rm SR}
\end{equation}
is conserved; here, $E_{\rm F}$, the energy of the superfluid field, 
$E^{\rm T}_{\rm o}$, the total kinetic energy of the particles, and 
$E_{\rm SR}$, the energy from the SR repulsion between the particles,
are defined, respectively, as follows:
\begin{subequations}\label{eq:Ecomponents}
\begin{align}
	E_{\rm F} &= \frac{1}{\mathcal{A}}\int_{\mathcal{A}}\biggl[\frac{\hbar^2}{2m}|\nabla\psi|^2 
	+ \frac{1}{2}g\Bigl(|\psi|^2-\frac{\mu}{g}\Bigr)^2\\
&+\sum\nolimits_{\rm i=1}^{\mathcal{N}_{\rm o}}
V_{\mathcal{P}}(\mathbf{r}-\mathbf{q}_{i})|\psi|^2\biggr]\,d\mathbf{r}; \\
E^{\rm T}_{\rm o} &= \frac{1}{\mathcal{A}}\sum\nolimits_{\rm i=1}\frac{1}{2}m_{\rm o}\dot{\mathbf{q}}_{i}^2;\\
E_{\rm SR} &= \frac{1}{\mathcal{A}}\sum\nolimits_{\rm i,j, i\neq j}^
{\mathcal{N}_{\rm o},\mathcal{N}_{\rm o}} \frac{\Delta_E r^{12}_{\rm SR}}{|q_{i}-q_{j}|^{12}}.
\end{align}
\end{subequations}
The dynamical evolution of Eqs.~(\ref{eq:GPEfull})-(\ref{eq:eqmpartfull}) 
conserves the total momentum 
\begin{equation}
\mathbf{P}(t) 
=  \int_{\mathcal{A}}\frac{i\hbar}{2}(\psi^*\nabla\psi-\psi\nabla\psi^*)\,d\mathbf{r}
+ \sum\nolimits_{\rm i=1}^{\mathcal{N}_{\rm o}} m_{\rm o}\dot{\mathbf{q}}_{i}; 
\end{equation}
and the number of bosons
\begin{equation}
	N =\int_{\mathcal{A}}|\psi|^2\,d\mathbf{r}.
\end{equation}
We use the Madelung transformation
\begin{equation}
\psi(\mathbf{r},t)=\sqrt{\rho(\mathbf{r},t)/m}\exp(i\phi(\mathbf{r},t))
\end{equation}
to express the GP equation in terms of hydrodynamical variables; here, 
$\rho(\mathbf{r},t)$ and $\phi(\mathbf{r},t)$ are, respectively, the
density and phase fields, and the superfluid velocity is 
\begin{equation}
\mathbf{v}(\mathbf{r},t)=(\hbar/m)\nabla\phi(\mathbf{r},t),
\end{equation}
whence we note that the flow is irrotational in the absence of any
quantum vortices.

Linearization of the GP equation for a uniform system around the equilibrium 
state with constant density $\rho_0$ yields the Bogoliubov dispersion relation
\begin{equation}
	\omega(k) = c\,k\left(1+\frac{1}{2}\xi^2k^2\right)^{1/2},
\end{equation}
where $k$ is the wave number; for small wave numbers $k$ the spectrum
is sound-like, with sound velocity $c=\sqrt{g\rho_0/m^2}$; for 
length scales smaller than $\xi$, i.e., $k\gtrsim 1/\xi$ 
dispersive effects are observed. The length scale 
$\xi=\sqrt{\hbar^2/2g\,\rho_0}$ is called the healing length scale
and is also associated with the vortex core size.

We model the particles by specifying the potential $V_{\mathcal{P}}$,
which allows us to choose the shape and size of the particles. In this
study, we use the Gaussian potential
\begin{equation}\label{eq:potparticle}
V_{\mathcal{P}} = V_o\exp\Bigl(-\frac{r^2}{2d^2_p}\Bigr);
\end{equation}
here $V_o$ is the strength of the potential and $d_{\rm p}$ is the measure of
its width.
The introduction of a particle displaces 
some superfluid, with area comparable to that occupied by the particle. 
The mass of the displaced fluid is given by 
\begin{equation}
m_{\rm f}=m\int_{\mathcal{A}}(|\psi_{\rm uniform}|^2-|\psi_{\rm particle}|^2)\,d\mathbf{r},
\end{equation}
where $\psi_{\rm uniform}$ and $\psi_{\rm particle}$ are the wave functions of
the minimum energy state of the uniform system in the absence of any particle and in the
presence of a single particle, respectively.
We use the mass $m_{\rm f}$ of the displaced superfluid to define the ratio
\begin{equation}
	\mathcal{M}\equiv \frac{m_o}{m_{\rm f}},
\end{equation}
which allows us to distinguish between (1) heavy ($\mathcal{M}>1$), (2) neutral
($\mathcal{M}=1$), and (3) light ($\mathcal{M}<1$) particles.

\subsection{Numerical methods}

To study the dynamics of particles in complex superfluid flows, we solve
Eqs.~(\ref{eq:GPEfull})-(\ref{eq:eqmpartfull}) numerically.  We perform direct
numerical simulations (DNSs) of the GPE by using the Fourier pseudospectral
method on a square, periodic simulation domain $\mathcal{A}$ of side $L=2\pi$
with $N^2_c$ collocation points~\cite{vmrnjp13}. In this method, 
we evaluate the linear terms in Eqs.~(\ref{eq:GPEfull})-(\ref{eq:eqmpartfull}) 
in Fourier space and the nonlinear term in real (physical) space, which we then 
transform to Fourier space. For the Fourier transform operations we use the 
FFTW library~\cite{fftwsite}. A fourth-order, Runge-Kutta scheme, with time step $\Delta\,t$
is used to evolve these equations in time.
Any peusodspectral DNS retains a finite number of Fourier modes; therefore, we
introduce the Galerkin projector $\mathcal{P}_G$
\begin{equation}
\mathcal{P}_G[\hat{\psi}(\mathbf{k})]= \theta(k_{\rm max}-|\mathbf{k}|)
\hat{\psi} (\mathbf{k}),
\end{equation}
where $\hat{\psi}$ is the spatial Fourier transform of $\psi$, $k_{\rm max}$ is
a suitably chosen ultraviolet cutoff, and $\theta(\cdot)$ the Heaviside
function. We use the standard $2/3$-dealiasing rule, with $k_{\rm
max}=2/3\times N_c/2$
and we follow Ref.~\cite{giorgio2011longPRE} in our treatment of the nonlinear term in
the GPE: We first apply $\mathcal{P}_G$ on $|\psi|^2$ and then again on
$\mathcal{P}_G[|\psi|^2]\psi$. This ensures global momentum conservation in our
DNS, which is essential for the study of collisions between
particles~\cite{shuklaPRA16} and their interactions with the field $\psi$.
Thus, our Galerkin-truncated GPE (TGPE) becomes
\begin{equation}\label{eq:tgpe} 
\begin{split}
i\hbar\frac{\partial
\psi(\mathbf{r},t)}{\partial t} &= \mathcal{P}_G\Biggl[
	\Bigl(-\frac{\hbar^2}{2m}\nabla^2 + g\mathcal{P}_G[|\psi|^2] \\
&-\mu
+ \sum\nolimits_{\rm i=1}^{\mathcal{N}_{\rm o}}
V_{\mathcal{P}}(\mathbf{r}-\mathbf{q}_{i})\Bigr)\psi(\mathbf{r},t)
\Biggr].
\end{split}
\end{equation}
Given our Galerkin-truncation scheme, we can write the force acting on the 
particle as
\begin{equation}\label{eq:frcbyfluidGT}
\begin{split}
\mathbf{f}_{{\rm o},i} &= -\int_{\mathcal{A}}\biggl[\psi^*\mathcal{P}_G
[V_{\mathcal{P}}(\mathbf{r}-\mathbf{q}_{i})\nabla\psi] \\
& + \psi\mathcal{P}_G[V_{\mathcal{P}}(\mathbf{r}-\mathbf{q}_{i})\nabla\psi^*]\biggr]d^2x.
\end{split}
\end{equation}

\subsection{Units and parameters}

We write the important parameters of the GPE system in the following form:
the quantum of circulation $\kappa\equiv h/m\equiv4\pi\alpha_{\rm 0}$, 
the speed of sound $c=\sqrt{2\alpha_{\rm 0}\tilde{g}\rho_{\rm 0}}$, 
the healing length $\xi=\sqrt{\alpha_{\rm 0}/(\tilde{g}\rho_{\rm 0})}$, 
where $\tilde{g}=g/\hbar\,m$, and the mean density 
$\rho_{\rm 0}=\int_{\mathcal{A}}m\,|\psi|^2\,d\mathbf{r}/\mathcal{A}$;
see Appendix~\ref{app:units} for more details.
In all our calculations, we set
$\rho_{\rm 0}=1$, $c=1$, and $\xi=1.44\, dx$, where $dx=L/N_c$, $N^2_c=128^2$, 
$\tilde{\mu}\equiv \mu/\hbar=\tilde{g}\rho_0$, 
$\tilde{V}_{\rm o}\equiv V_{\rm o}/\hbar=10\, \tilde{g}\rho_0$, 
$d_{\rm p}=1.5\, \xi$, and $\Delta_{\rm E}=0.062$; 
except in Sec.~\ref{subsec:manypartcrotN} where we set $c=2$.
Energies are in units of $E_{\xi}=2\alpha\rho^2_{\rm 0}\tilde{g}=c^2\rho_0$, which
has the dimension $[M\,T^{-2}]$, as the energies defined in
Eqs.~\eqref{eq:Etotal} and \eqref{eq:Ecomponents} are energies per unit area; 
to obtain the energy integrated over the whole computation
domain $\mathcal{E}_{\mathcal{A}}$, the dimensionless quantity $E$ must
be multiplied by $c^2\,\rho_0\,\mathcal{A}$ and an appropriate units for $\mathcal{E}_{\mathcal{A}}$
is then $c^2\,\xi^2\,\rho_0$.
The force acting on the particle is expressed in units of $c^2\xi\rho_0$.

\section{Results}

In order to understand how particles and the field $\psi$ interact in 
superfluids, we have considered the following illustrative 
initial configurations in our DNSs:

\begin{enumerate}

\item $\tt ICP1$: An initial configuration with one particle on which a 
constant, external force acts for the entire duration of the DNS.

\item 
\begin{itemize} 
\item $\tt ICP2A$: An initial configuration with one particle placed in front 
of a translating vortex-antivortex pair.

\item $\tt ICP2B$: An initial configuration with two
particles placed in front of a translating vortex-antivortex pair.
\end{itemize}

\item 
\begin{itemize}
\item $\tt ICP3A$: An initial configuration with a single particle moving 
in the presence of counter-rotating vortex clusters.

\item $\tt ICP3B$: An initial configuration with four particles in the presence
of counter-rotating vortex clusters. 
\end{itemize}

\end{enumerate}

\subsection{Single-particle dynamics: Constant external force on the particle}

\label{subsec:1partextforce}

We study the dynamics of a single particle in the superfluid by using the
initial configuration $\tt ICP1$, where an external force acts on the particle;
in particular, we examine the dependence of this dynamics on $\mathcal{M}$. We
achieve this by first preparing an initial state with a single particle at rest
by using the Advective-Ginzburg-Landau equation (ARGLE) (see Appendix~\ref{ARGLE}).
We use this initial state in the GPE and switch on the external force
$\mathbf{F}_{\rm ext}= F_0\hat{\mathbf{x}}$, where $F_0$ is constant in time.
We now describe the dynamics of the heavy ($\mathcal{M}>1$), neutral
($\mathcal{M}=1$), and light particles ($\mathcal{M}<1$) for the initial
configuration $\tt ICP1$.

\begin{figure*}
\centering
\begin{overpic}
[height=4.5cm,unit=1mm]{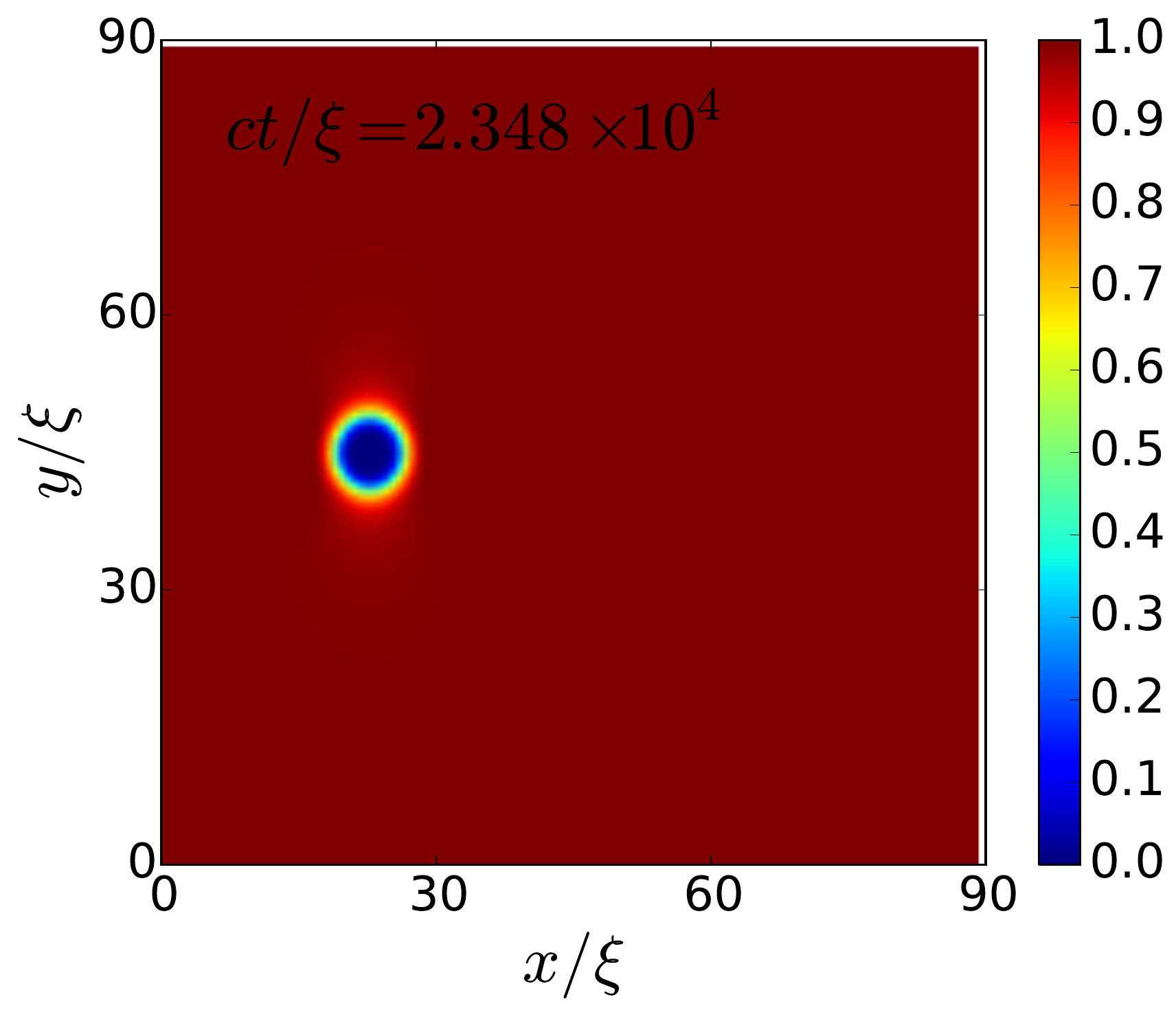}
\put(10.,10){\large{\bf (a)}}
\end{overpic}
\begin{overpic}
[height=4.5cm,unit=1mm]{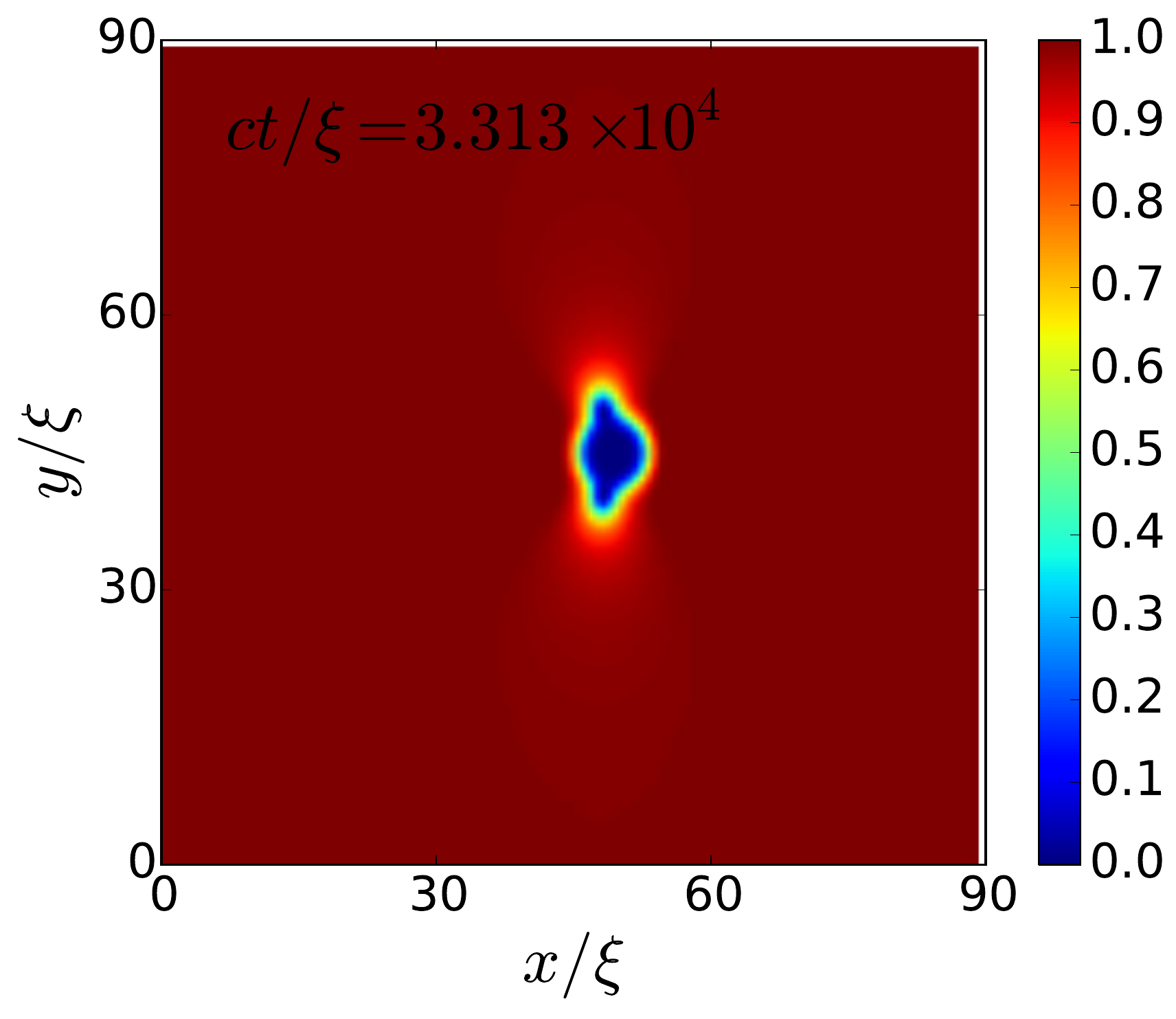}
\put(10,10){\large{\bf (b)}}
\end{overpic}
\begin{overpic}
[height=4.5cm,unit=1mm]{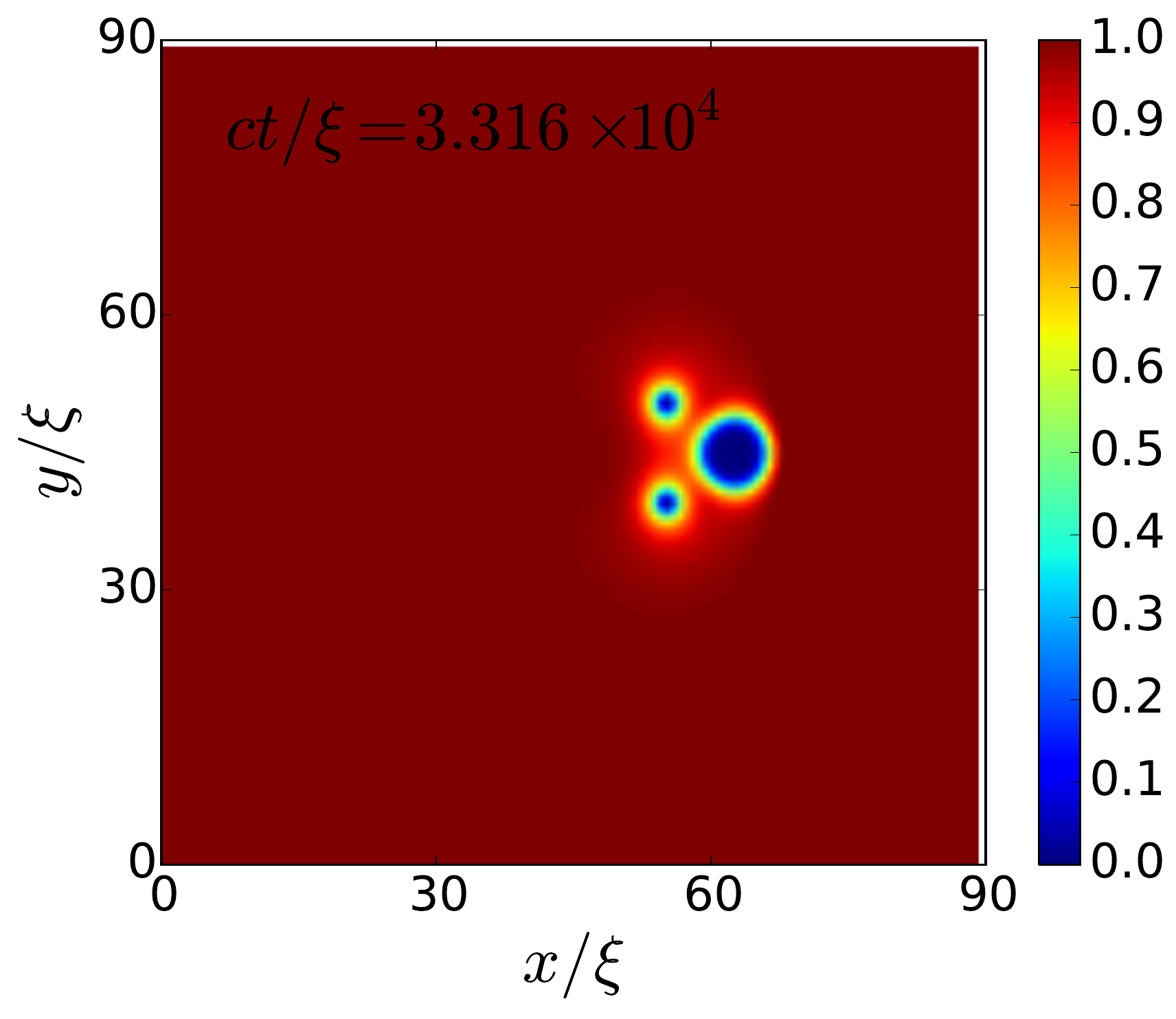}
\put(10,10){\large{\bf (c)}}
\end{overpic}
\\
\begin{overpic}
[height=4.5cm,unit=1mm]{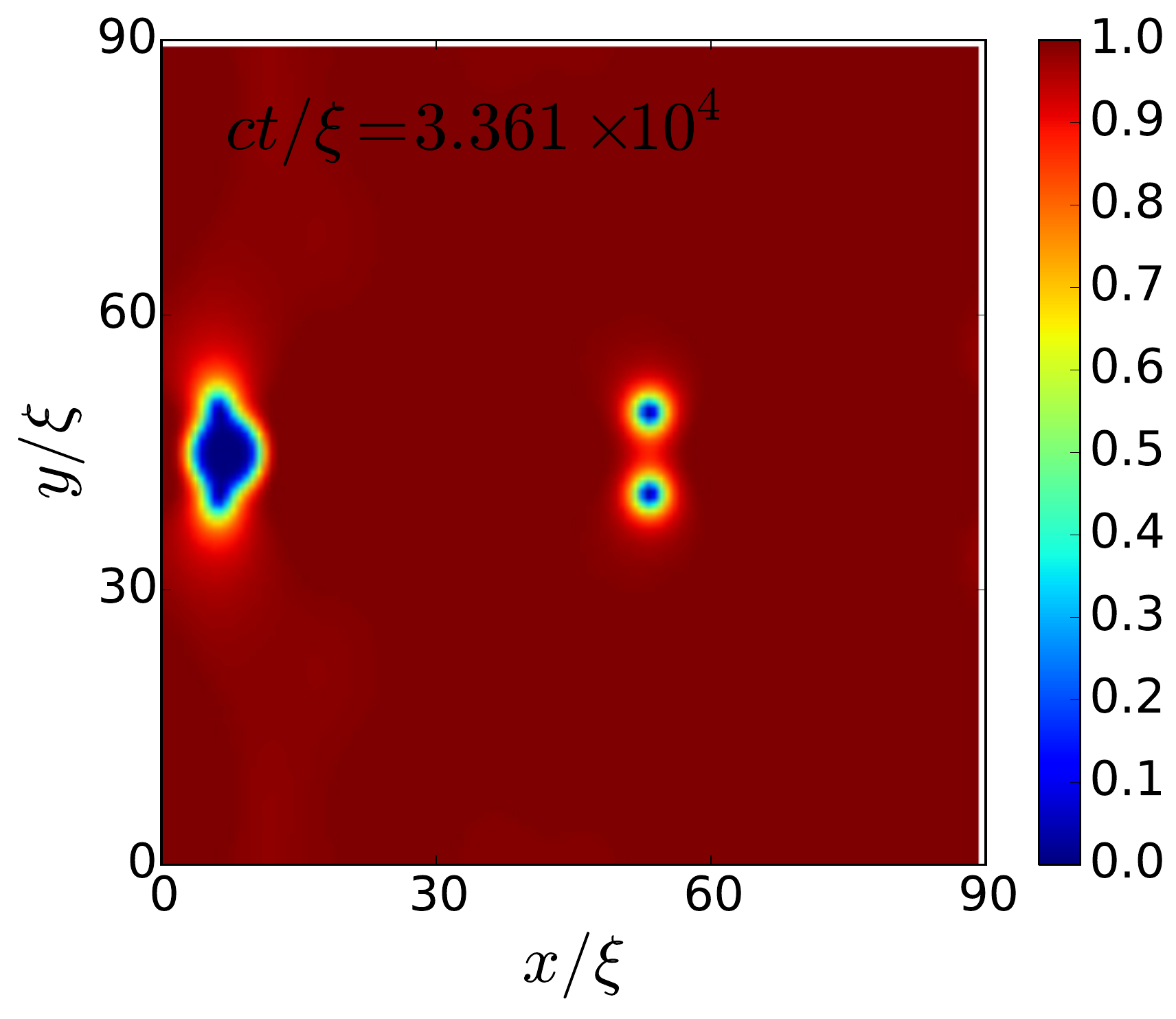}
\put(10.,10){\large{\bf (d)}}
\end{overpic}
\begin{overpic}
[height=4.5cm,unit=1mm]{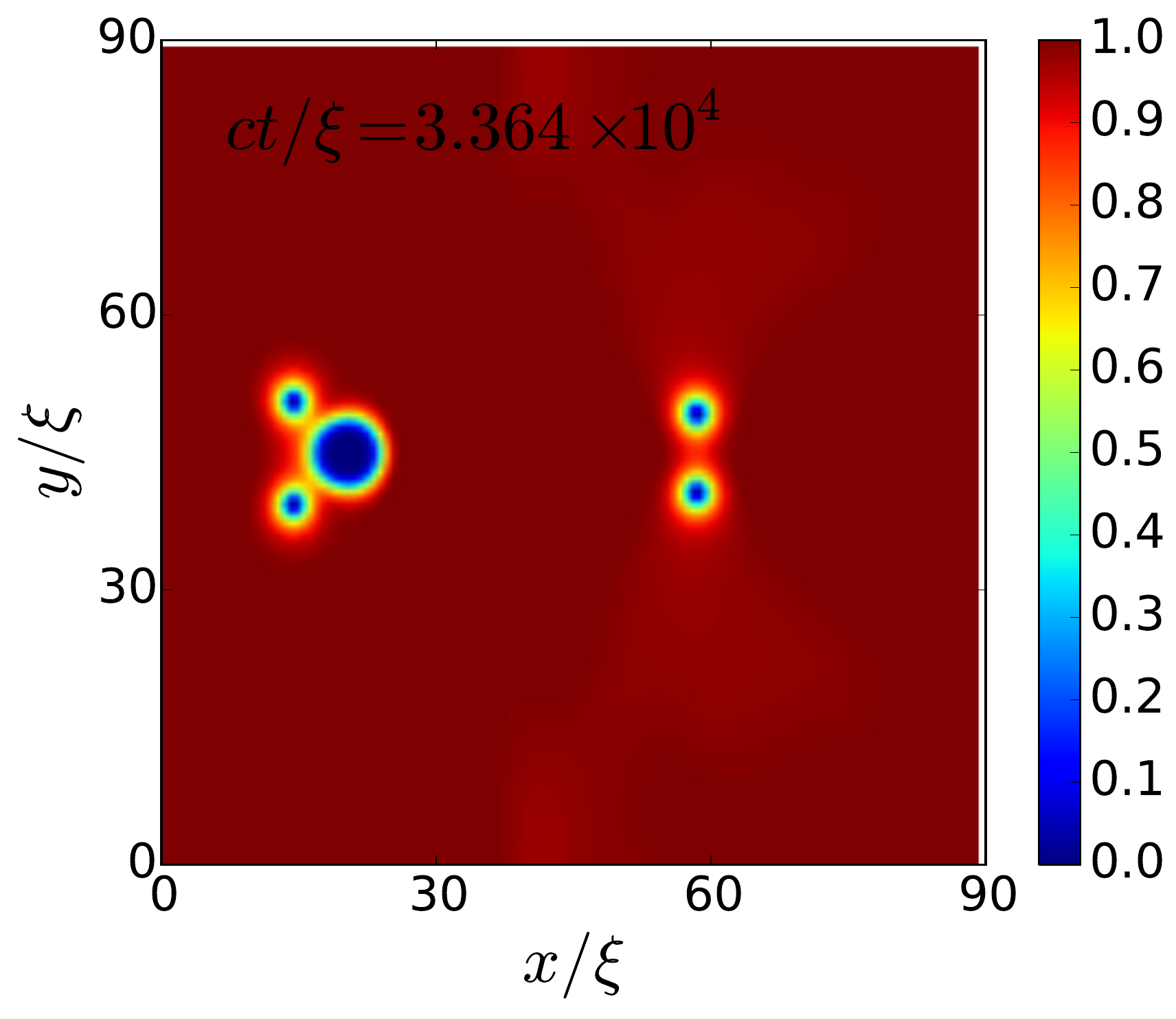}
\put(10,10){\large{\bf (e)}}
\end{overpic}
\begin{overpic}
[height=4.5cm,unit=1mm]{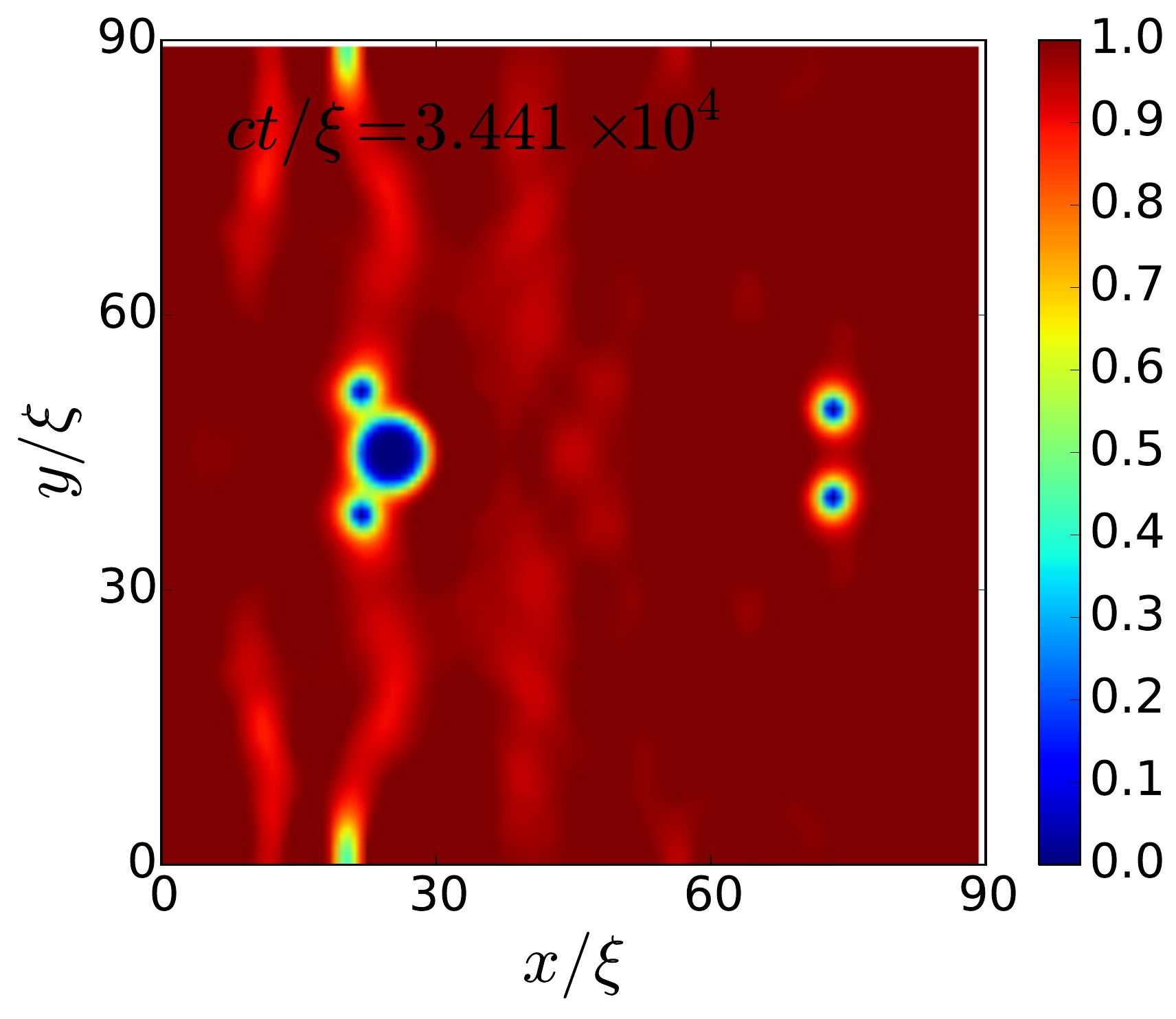}
\put(10,10){\large{\bf (f)}}
\end{overpic}
\\
\begin{overpic}
[height=4.5cm,unit=1mm]{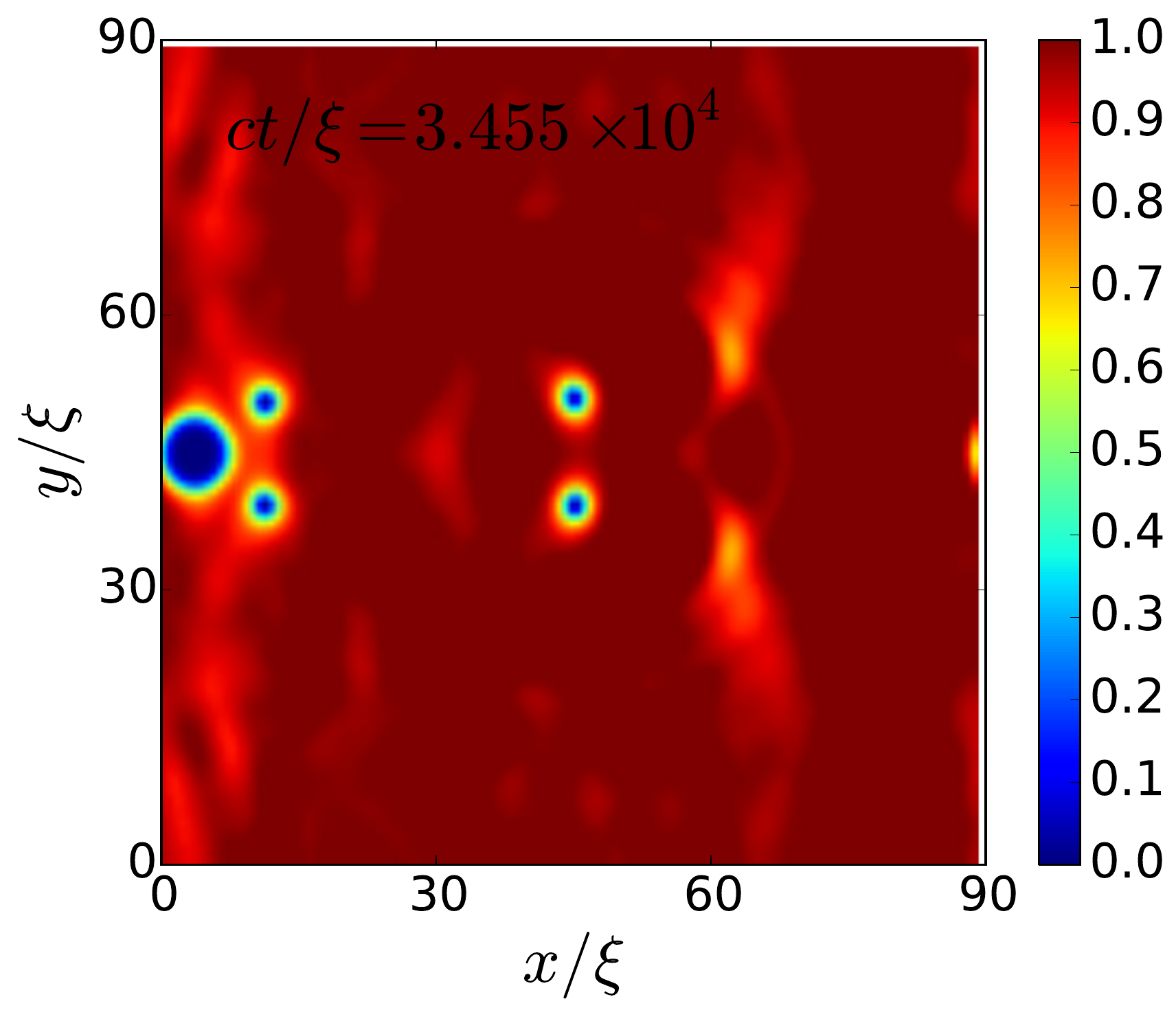}
\put(10.,10){\large{\bf (g)}}
\end{overpic}
\begin{overpic}
[height=4.5cm,unit=1mm]{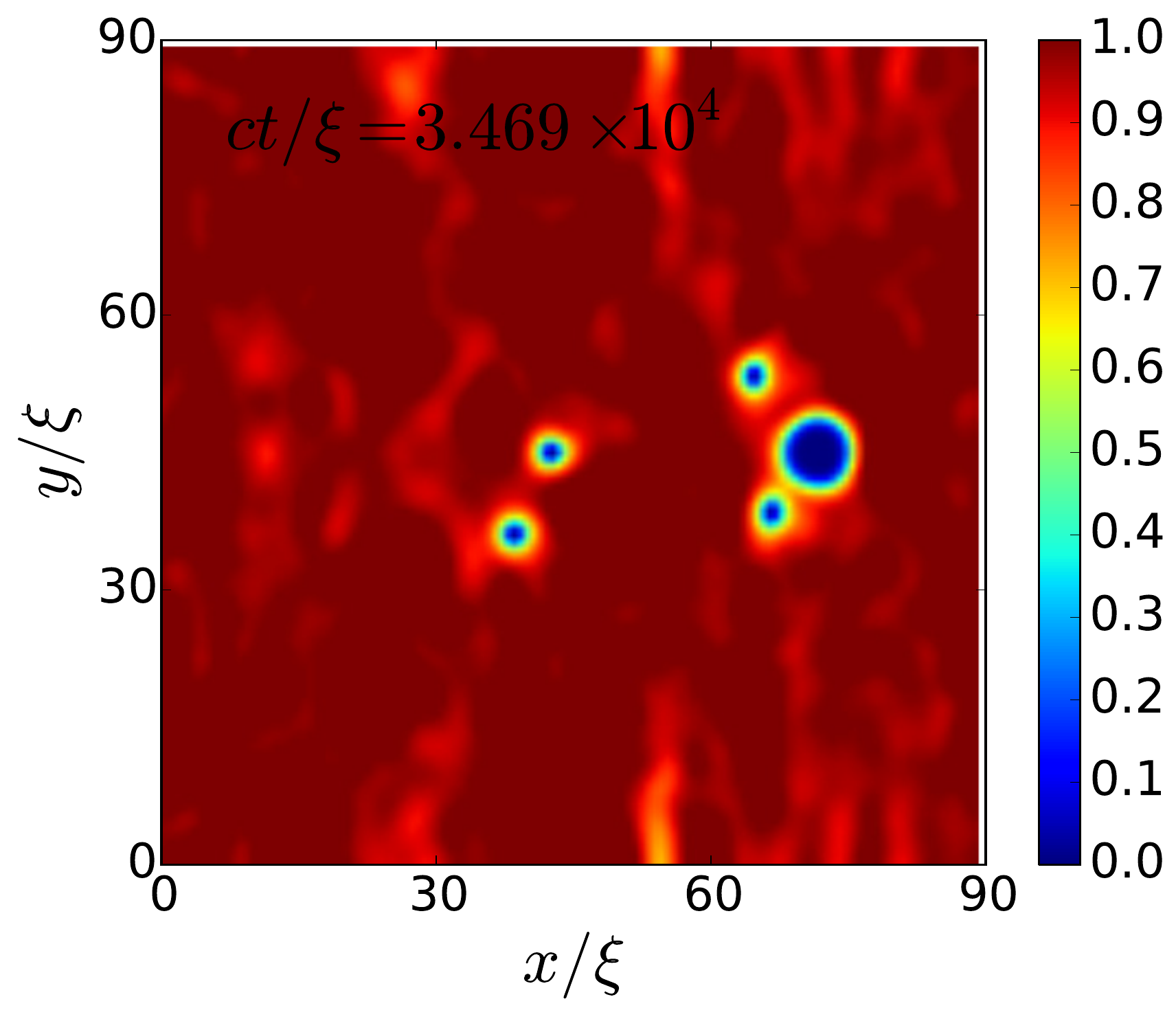}
\put(10,10){\large{\bf (h)}}
\end{overpic}
\begin{overpic}
[height=4.5cm,unit=1mm]{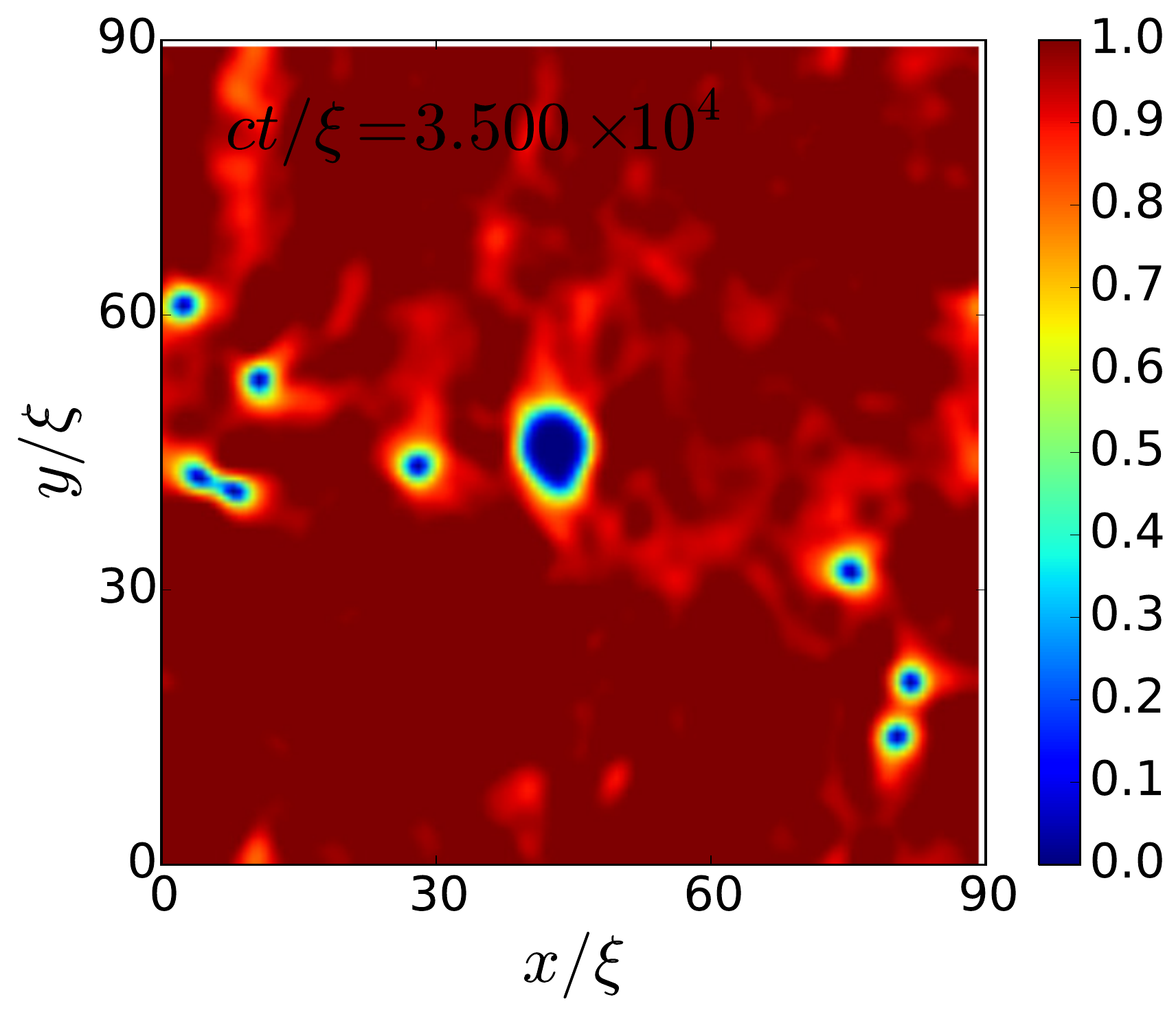}
\put(10,10){\large{\bf (i)}}
\end{overpic}
\\
\begin{overpic}
[height=4.5cm,unit=1mm]{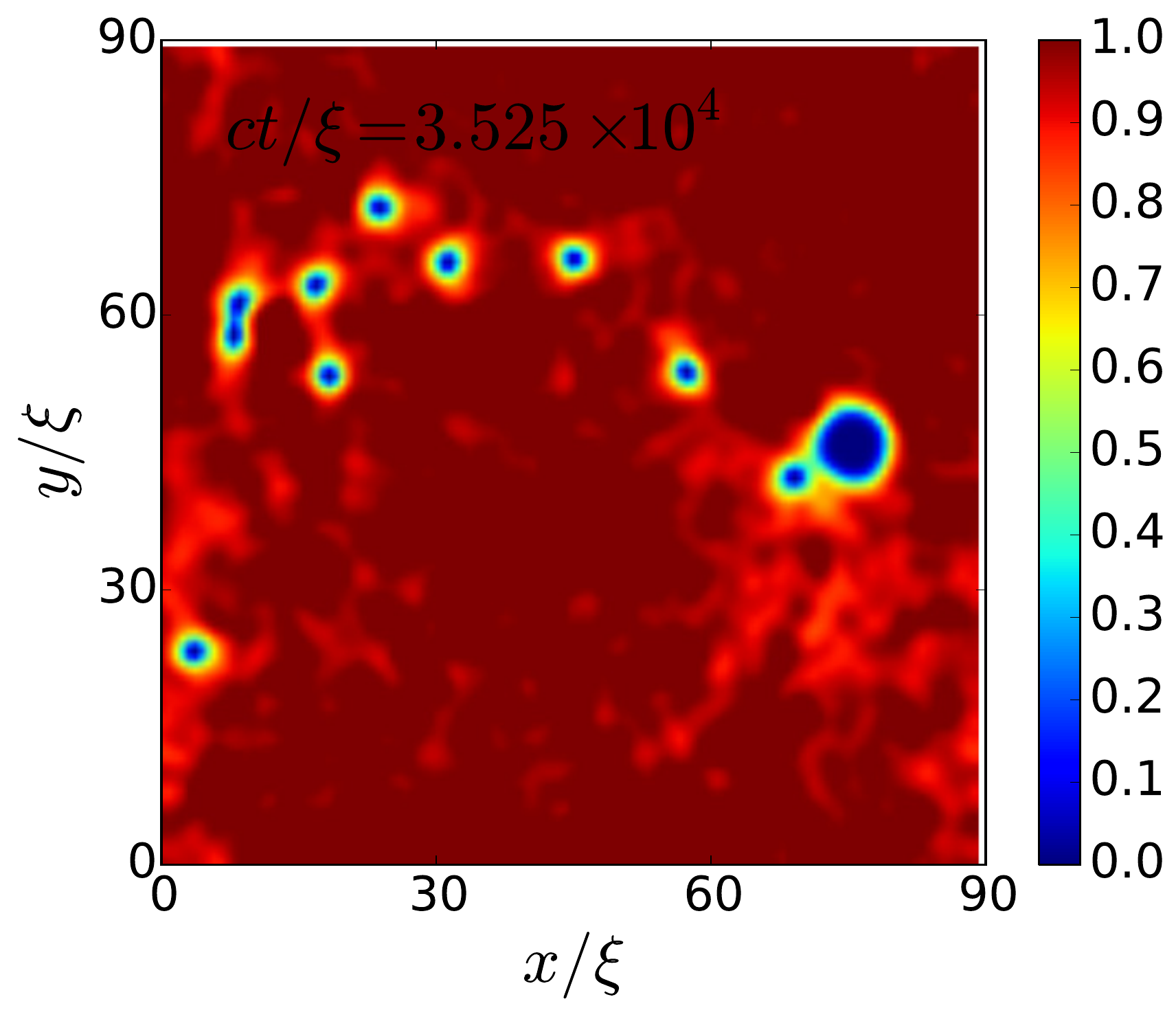}
\put(10.,10){\large{\bf (j)}}
\end{overpic}
\begin{overpic}
[height=4.5cm,unit=1mm]{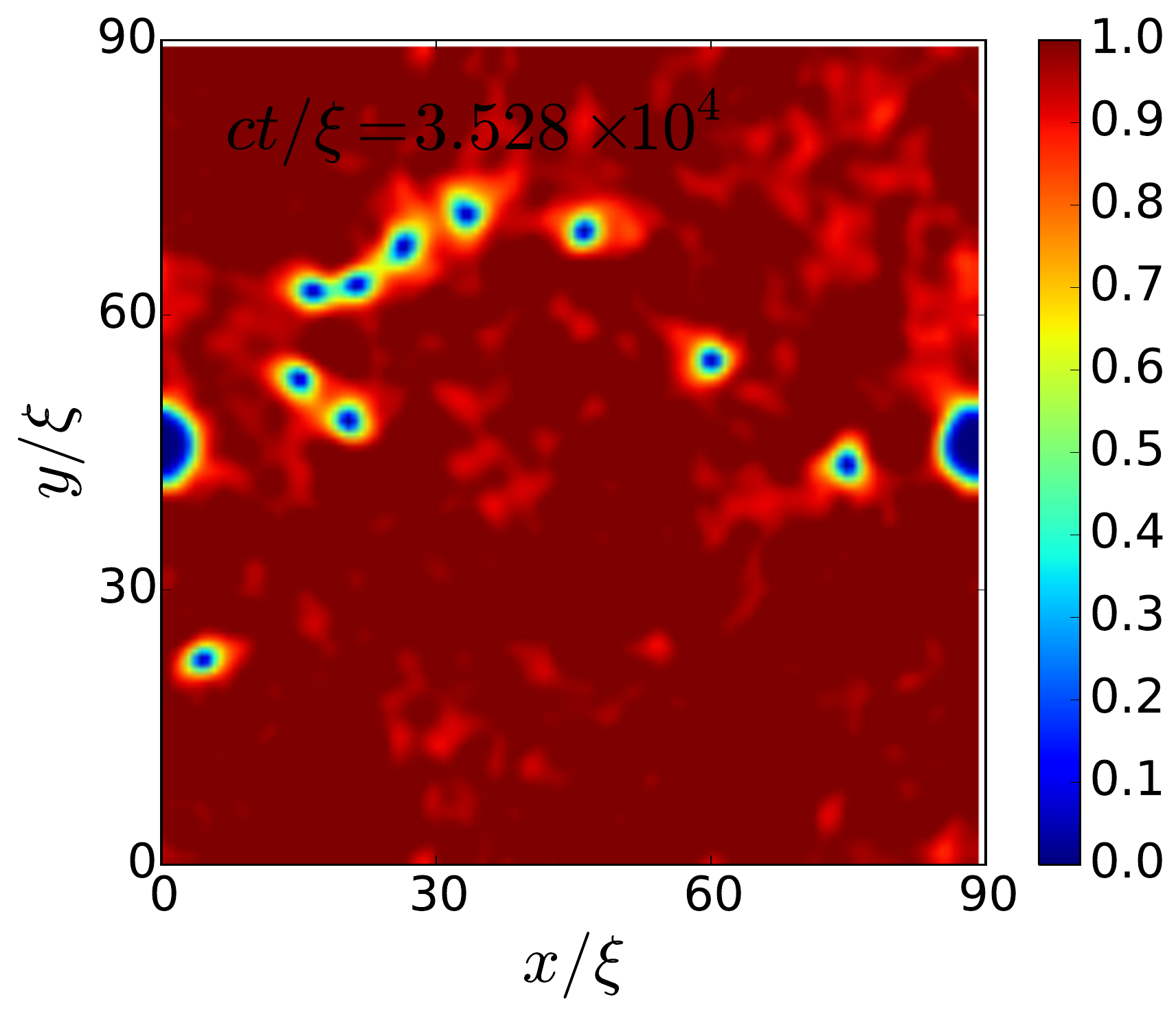}
\put(10,10){\large{\bf (k)}}
\end{overpic}
\begin{overpic}
[height=4.5cm,unit=1mm]{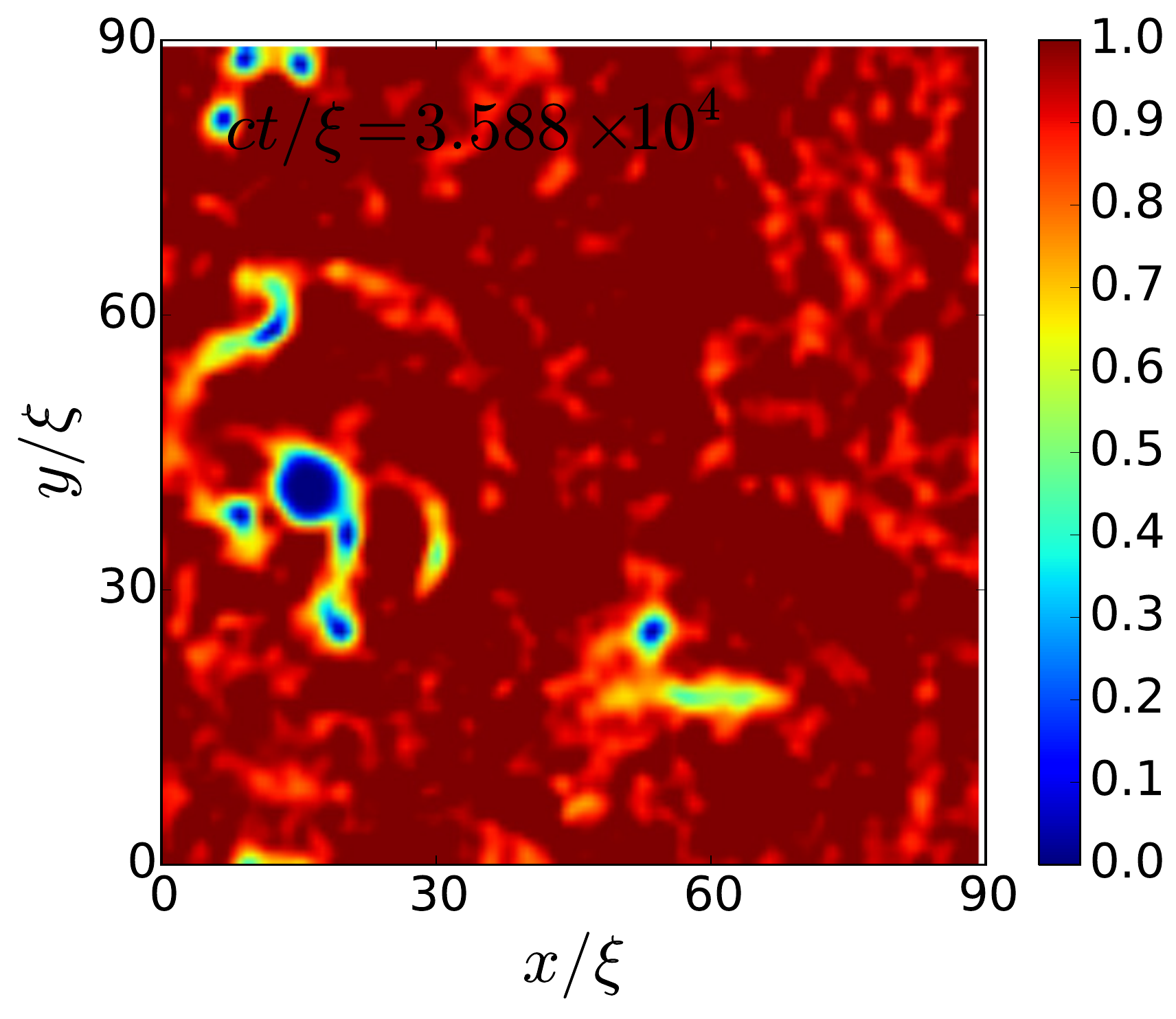}
\put(10,10){\large{\bf (l)}}
\end{overpic}
\caption{\small (Color online) Spatiotemporal evolution of the density field
$\rho(\mathbf{r},t)$ shown via pseudocolor plots, illustrating the dynamics
of a heavy particle, when a constant external force $\mathbf{F}_{\rm
ext}=0.28\,c^2\xi\rho_0\,\hat{\mathbf{x}}$ acts on it (initial configuration $\tt ICP1$).  The
particle appears as a large blue patch and the vortices as blue dots (for
details see text, subsection~\ref{subsec:1partextforce}).}
\label{fig:pdIC1f1H}
\end{figure*}

\begin{figure*}
\centering
\resizebox{\linewidth}{!}{
\includegraphics[height=4.cm,unit=1mm]{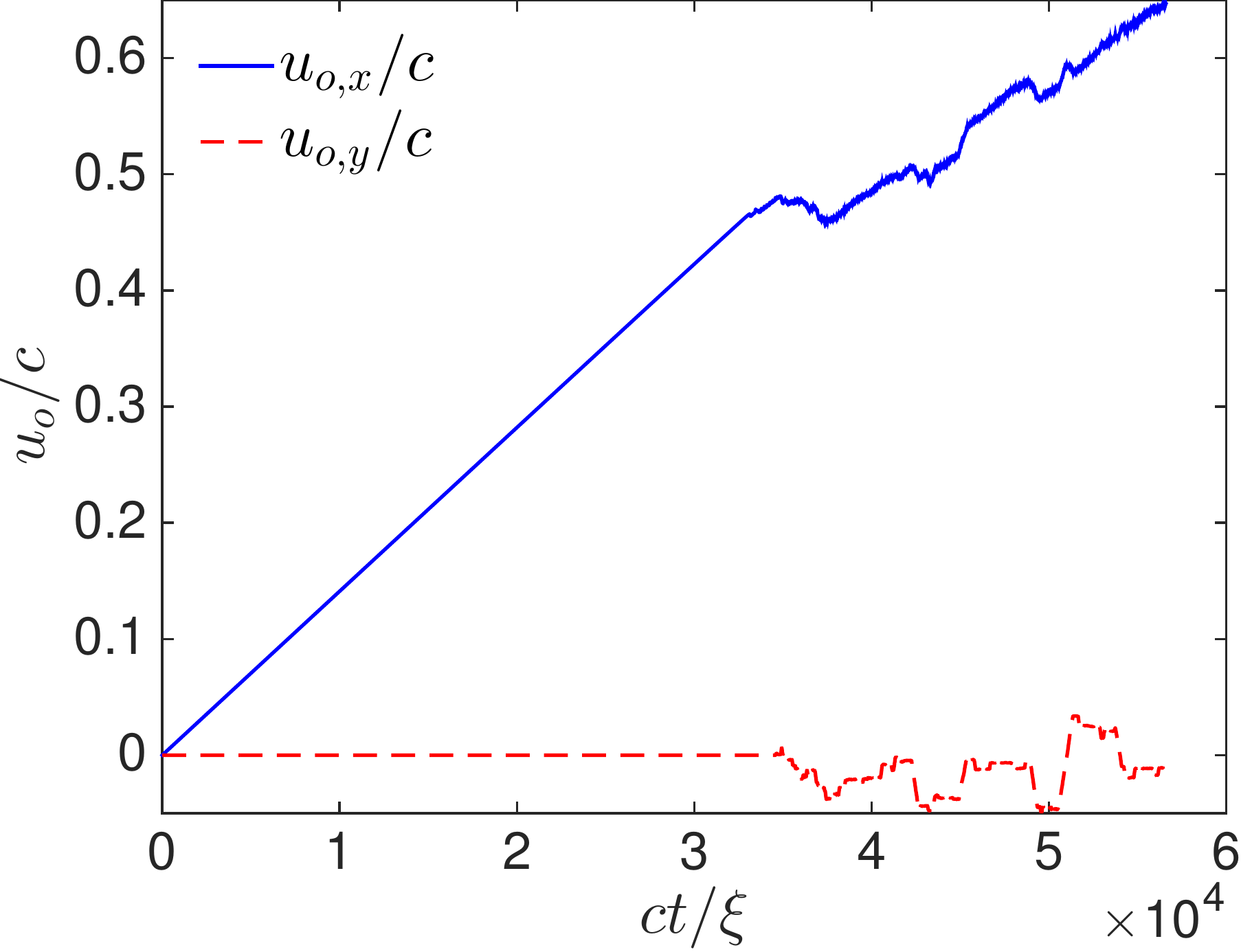}
\put(-75,30){\bf (a)}
\hspace{0.15 cm}
\includegraphics[height=4.cm,unit=1mm]{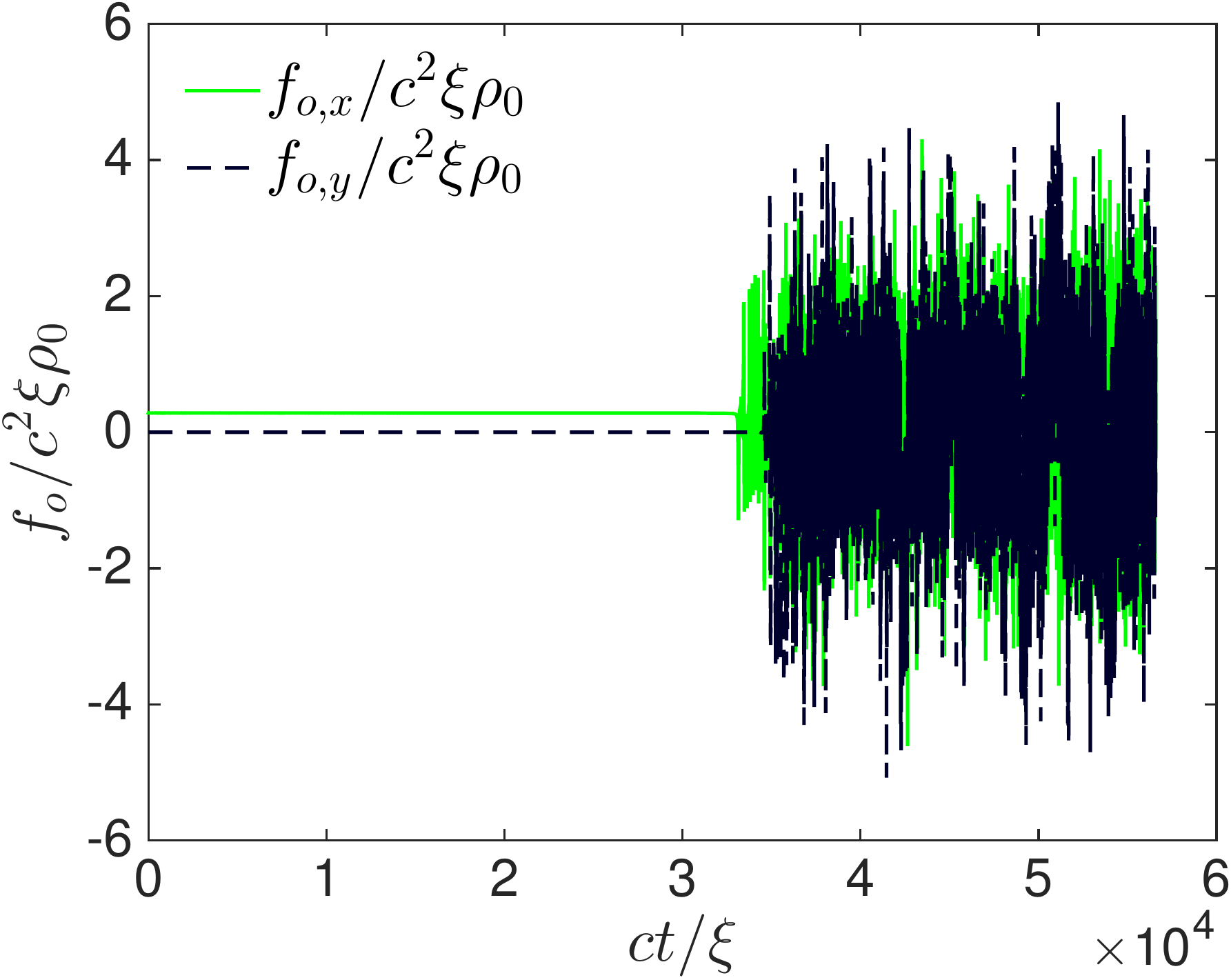}
\put(-75,30){\bf (b)}
\hspace{0.15 cm}
\includegraphics[height=4.cm,unit=1mm]{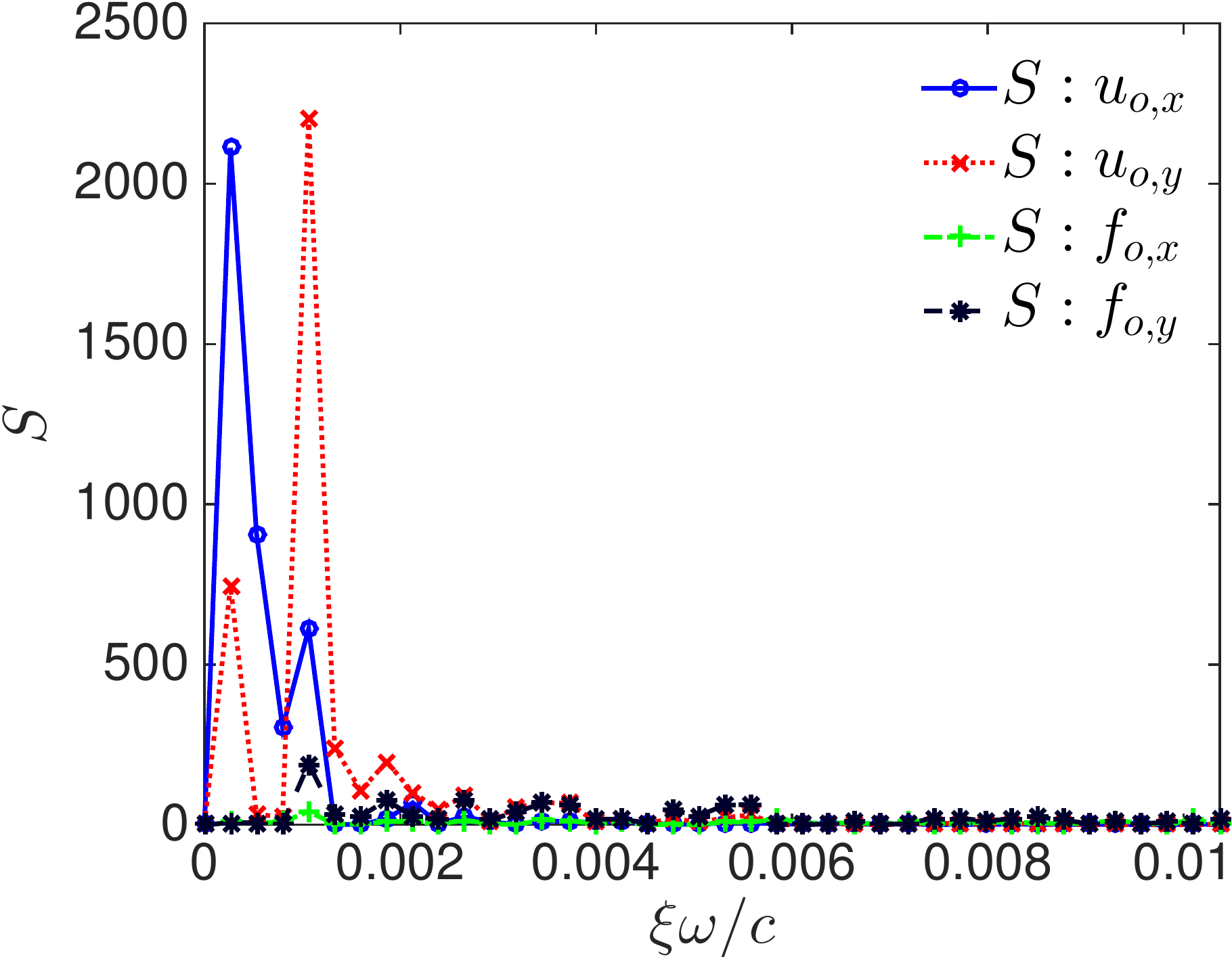}
\put(-75,30){\bf (c)}
}
\\
\vspace{0.25 cm}
\resizebox{\linewidth}{!}{
\includegraphics[height=4.cm,unit=1mm]{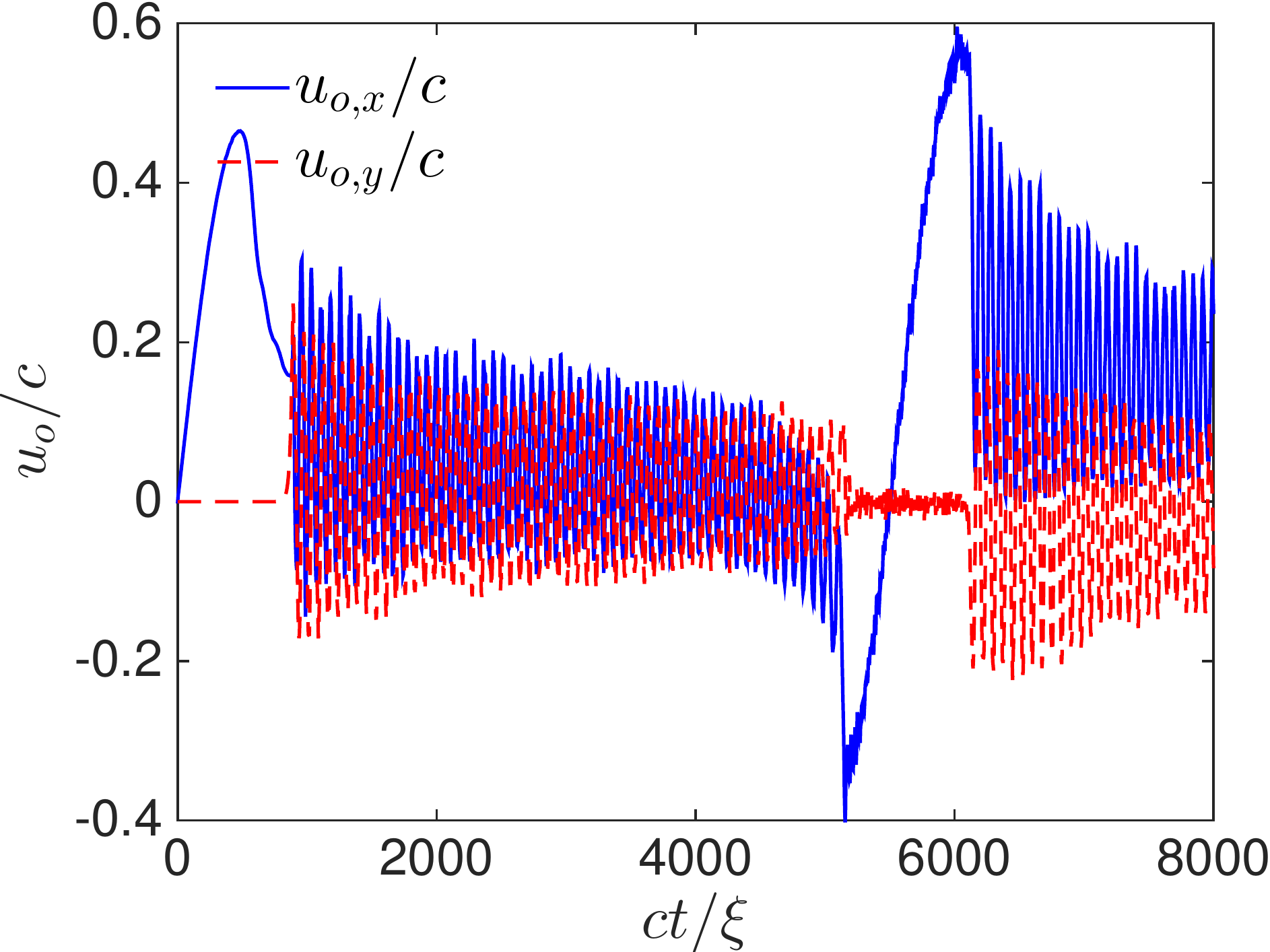}
\put(-75,30){\bf (d)}
\hspace{0.15 cm}
\includegraphics[height=4.cm,unit=1mm]{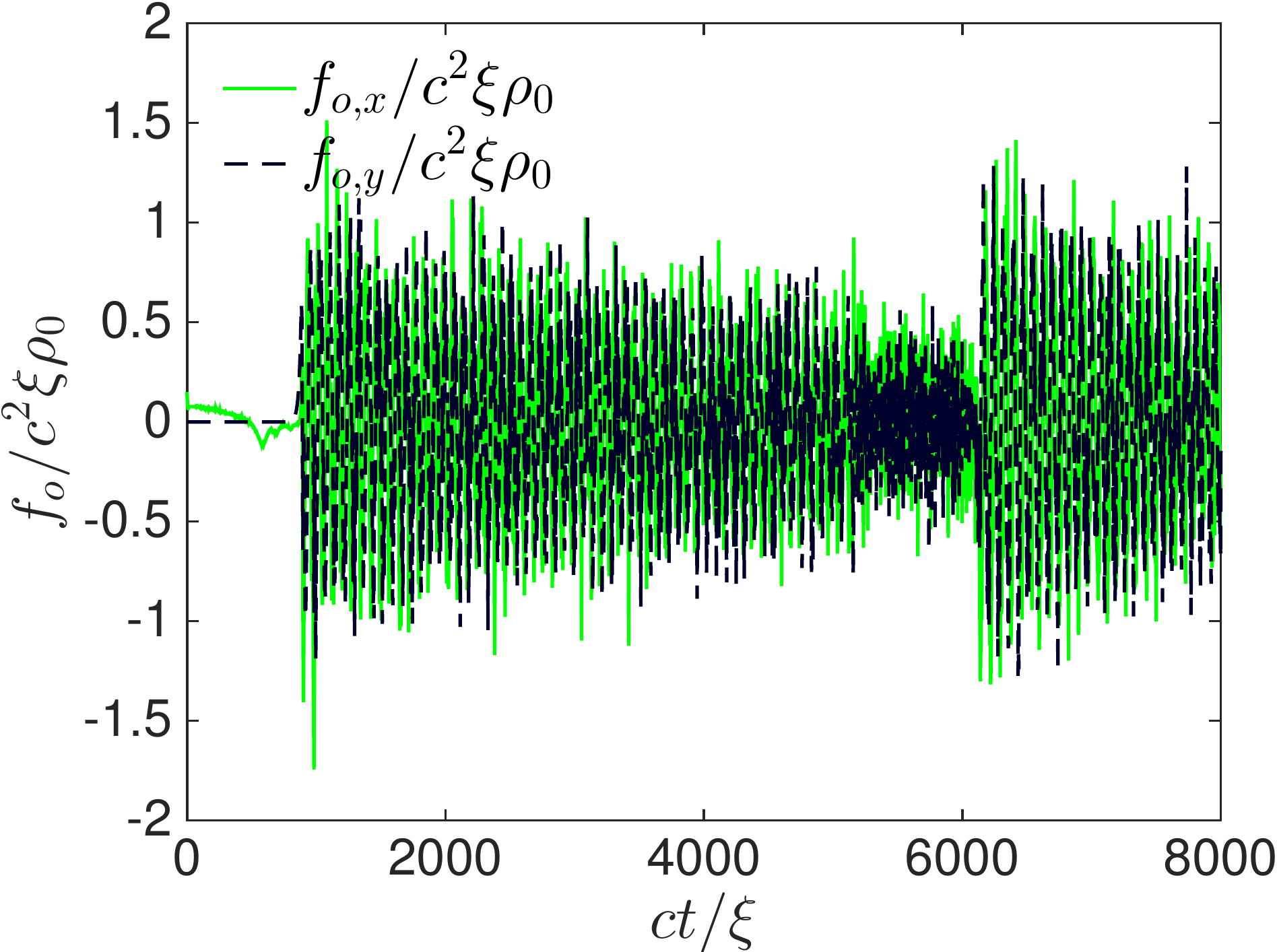}
\put(-75,25){\bf (e)}
\hspace{0.15 cm}
\includegraphics[height=4.cm,unit=1mm]{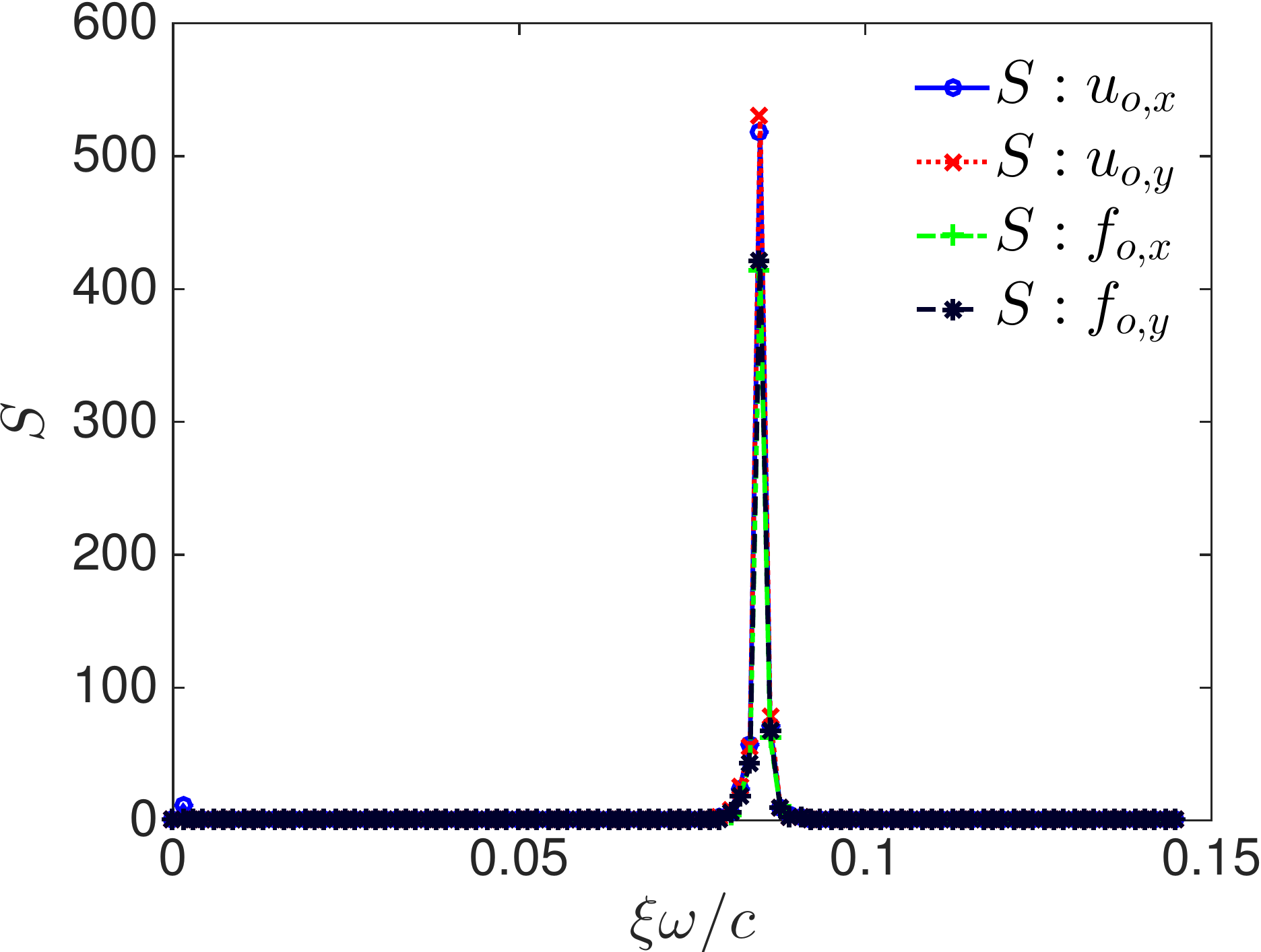}
\put(-75,25){\bf (f)}
}
\\
\vspace{0.25 cm}
\resizebox{\linewidth}{!}{
\includegraphics[height=4.cm,unit=1mm]{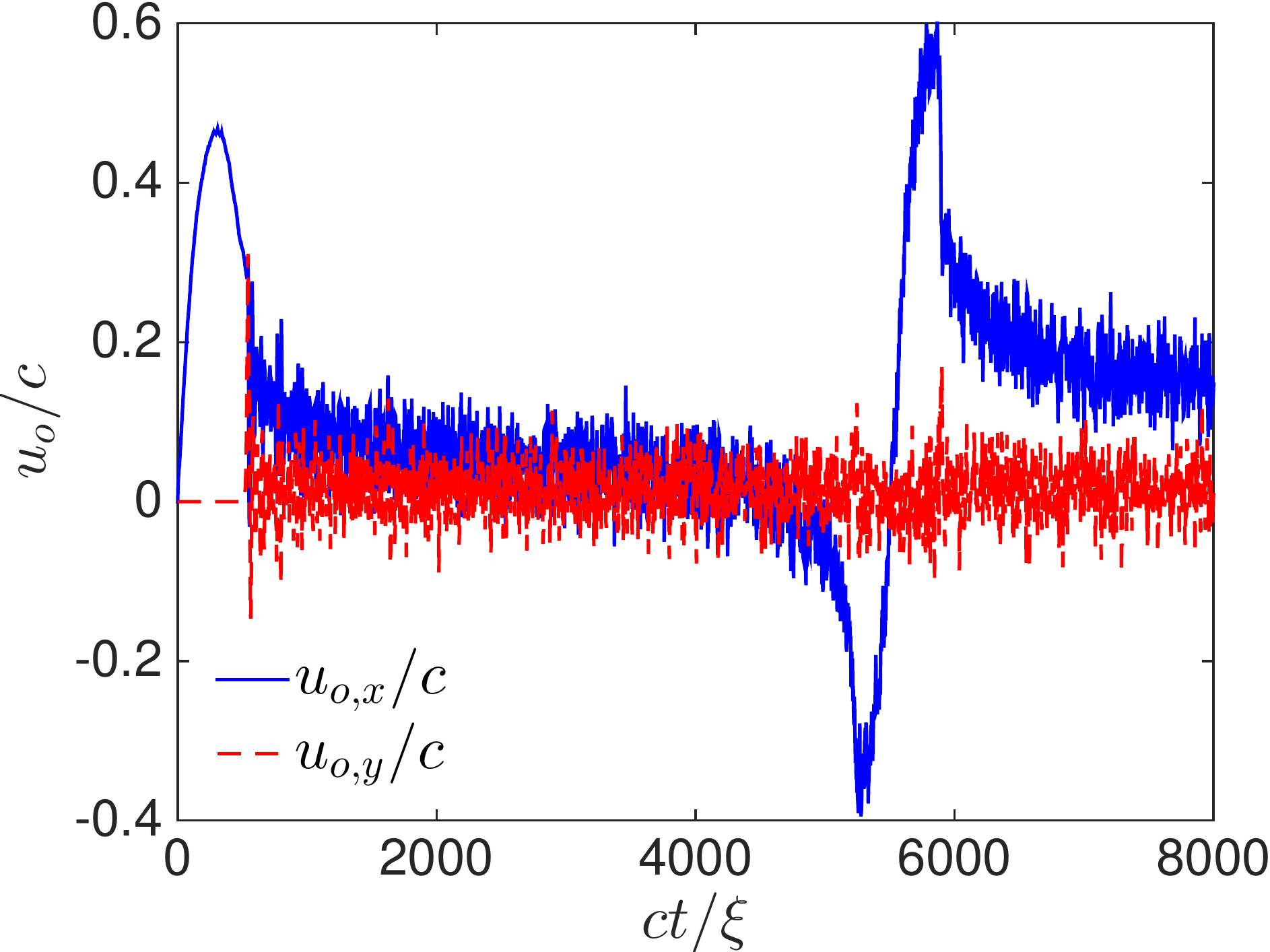}
\put(-75,30){\bf (g)}
\hspace{0.15 cm}
\includegraphics[height=4.cm,unit=1mm]{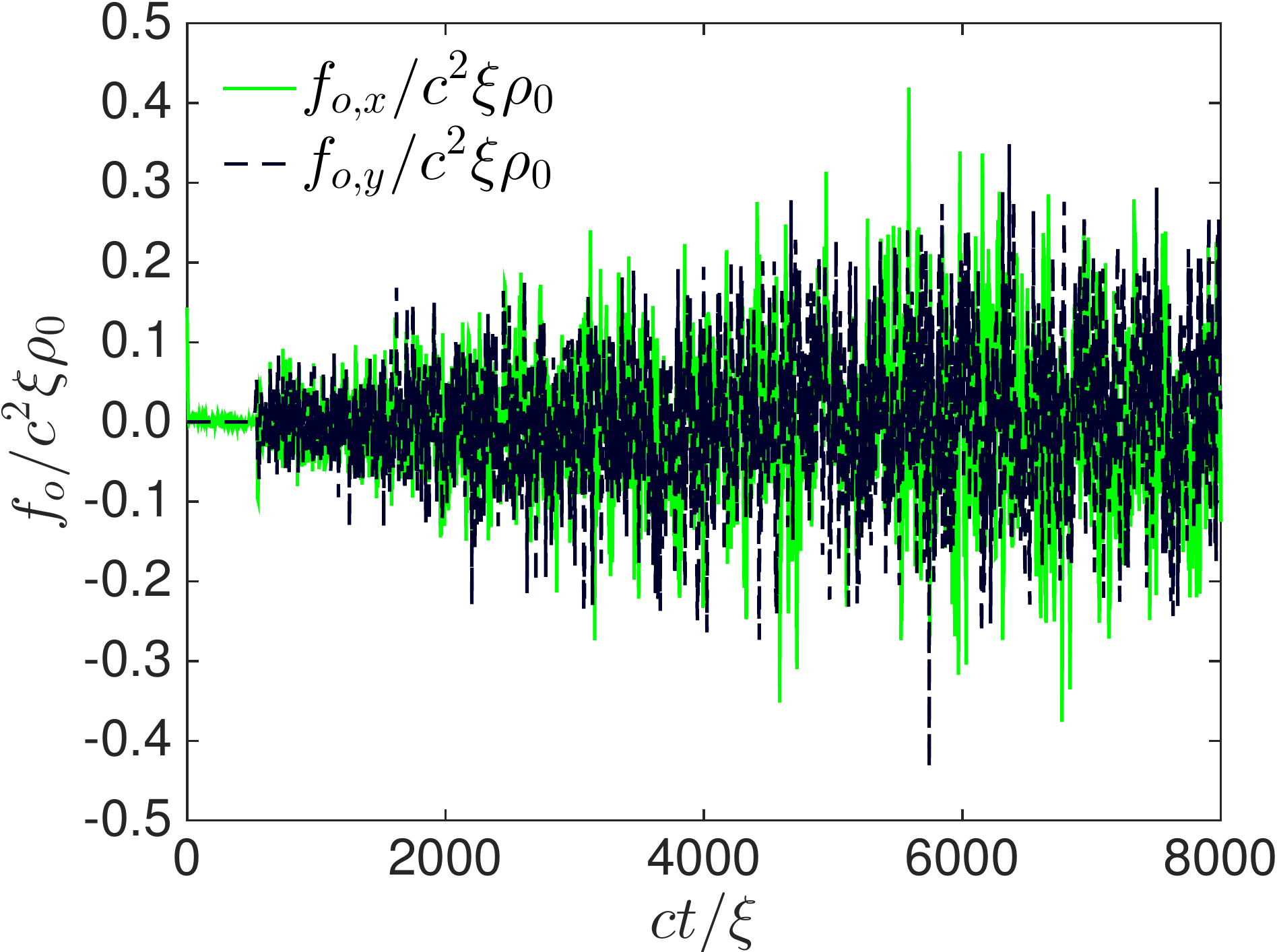}
\put(-75,30){\bf (h)}
\hspace{0.15 cm}
\includegraphics[height=4.cm,unit=1mm]{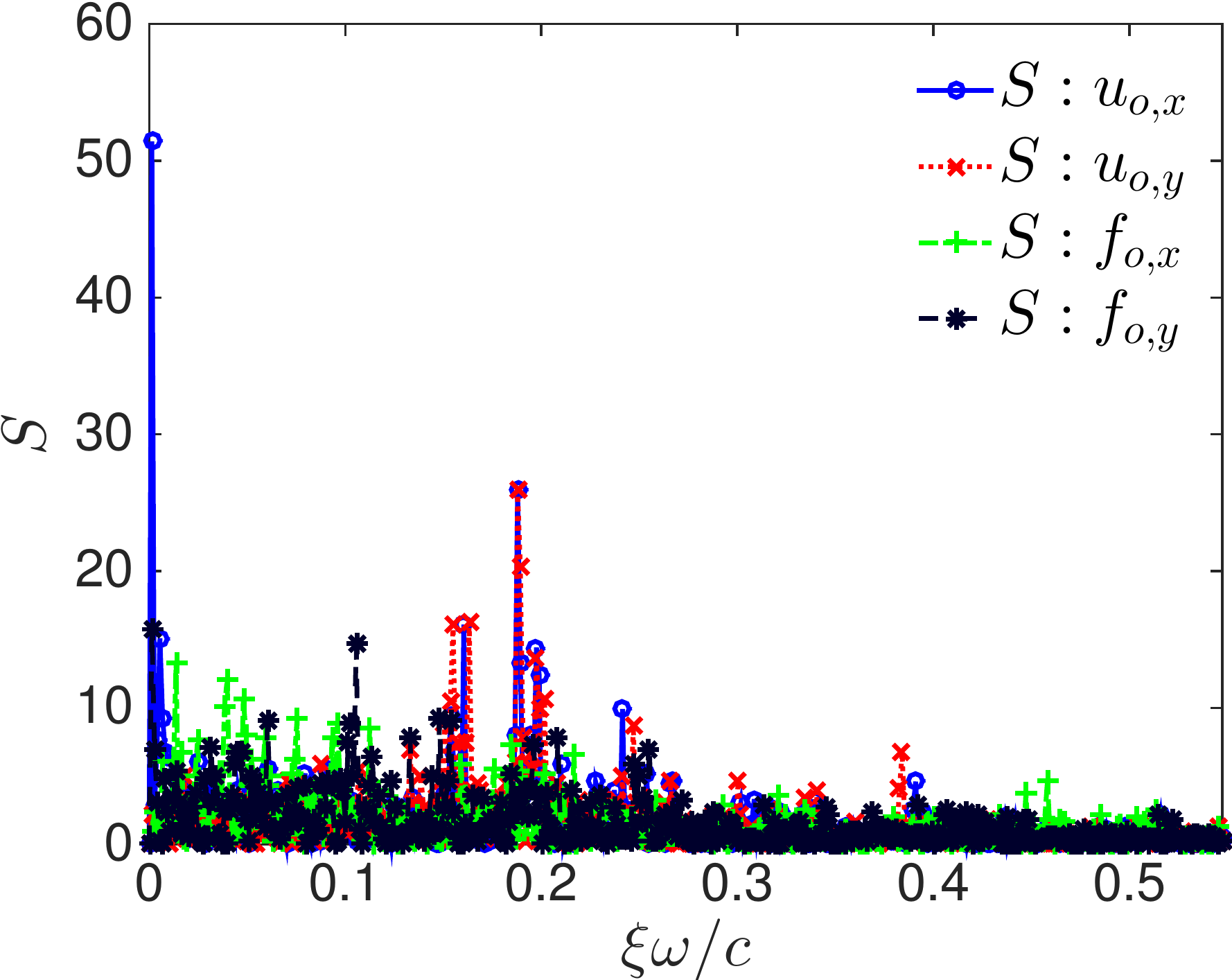}
\put(-75,35){\bf (i)}
}
\caption{(Color online) 
	Plots of the Cartesian components of: 
	(a) velocity $u_{\rm o,x}$ and $u_{\rm o,y}$,
	(b) force $f_{\rm o,x}$ and  $f_{\rm o,y}$;
	(c) power spectra of the quantities time series in (a) and (b),
	for the \textbf{heavy} particle ($\mathcal{M}=374$, 
	$\mathbf{F}_{\rm ext}= 0.28\,c^2\xi\rho_0\,\hat{\mathbf{x}}$).
	Plots in (d), (e), and (f)  and (g), (h), and (i) are the analogs of 
	plots in (a), (b), and (c), for the \textbf{neutral}\
	($\mathcal{M}=1$, $\mathbf{F}_{\rm ext}= 0.14\,c^2\xi\rho_0\,\hat{\mathbf{x}}$)
	and \textbf{light} ($\mathcal{M}=0.0374$, 
	$\mathbf{F}_{\rm ext}= 0.14\,c^2\xi\rho_0\,\hat{\mathbf{x}}$) particles, respectively.
	 Power spectra, denoted generically by
         $S(\omega)$, of the time series of $u_{\rm o,x}$ , $u_{\rm o,y}$, 
	 $f_{\rm o,x}$, and  $f_{\rm o,y}$ are plotted versus the 
	 angular frequency $\omega$ for the above three cases.
 }	
\label{fig:velfor.pdfpectraIC1f1}
\end{figure*}

{\bf Heavy particle:} We apply an external force $\mathbf{F}_{\rm ext}=
0.28\,c^2\xi\rho_0\,\hat{\mathbf{x}}$ on the heavy particle. In Fig.~\ref{fig:pdIC1f1H} we show
pseudocolor plots of $\rho(\mathbf{r})$, at different instants of time, to
illustrate its spatiotemporal evolution; the particle and vortices appear as
blue disks in which $\rho=0$, with large and small diameters, respectively.  In
Fig.~\ref{fig:velfor.pdfpectraIC1f1}~(a) we show the temporal evolution of the
$x$- and the $y$-components of the particle velocity $u_{\rm o,x}$ (blue solid
curve) and $u_{\rm o,y}$ (red dashed curve), respectively. 
The particle starts to
move from rest through the superfluid, without disturbing the latter, until the
critical velocity $u_{c}/c\simeq 0.47$ is reached at $ct/\xi= 3.313\times10^4$. When the particle
velocity $\simeq u_c$, a vortex-antivortex pair emerges, with a positive vortex
at the top and a negative vortex at the bottom of the particle, but both still
attached to, and co-moving with, the particle; the particle slows down
slightly Fig.~\ref{fig:pdIC1f1H}(b). Subsequently, the vortex-antivortex pair gets detached from the
particle; it is oriented perpendicular to, and moves along, the $x$-direction
at a much reduced velocity compared to that of the particle Fig.~\ref{fig:pdIC1f1H}(c). 
Given its large
velocity, the particle moves ahead of the slowly moving vortex-antivortex pair;
and, because of the periodic boundary conditions we use, it comes back and
approaches the vortex-antivortex pair from behind. The particle passes through
the vortex-antivortex pair, during which passage the positive and the negative
vortices glide, repectively, along the upper- and lower-half of the
circumference of the particle; such an interaction is also associated with the
initial increase (which is followed by a decrease) in the particle velocity
because of the reinforcing nature of the velocity field in the region in
between the vortex and the antivortex that constitute the pair. Moreover, this
interaction of the particle and the vortex-antivortex pair leads to the
generation of sound waves.  The particle subsequently sheds another
vortex-antivortex pair Fig.~\ref{fig:pdIC1f1H}(d) and (e); and 
then it interacts with two vortex-antivortex pairs;
this is accompanied by an even greater emission of sound waves than in the case
of one vortex-antivortex pair.  Afterwards, the presence of sound waves during
the interaction of the particles with the vortex-antivortex pairs, results in
deflections of the latter from their trajectories Fig.~\ref{fig:pdIC1f1H}(h); 
at the same time, small
fluctuations are induced in the particle velocity (see
Fig.~\ref{fig:velfor.pdfpectraIC1f1}~(a)). The subsequent motions of the
particle and the vortex-antivortex pairs become complicated Fig.~\ref{fig:pdIC1f1H}(i)-(l).  
Many more
vortex-antivortex pairs are shed by the particle; and, at several instances of
the shedding of a vortex-antivortex pair, one of the vortices is trapped on the
particle for a short duration of time; now the vortex-antivortex pairs are
emitted at $u_{\rm o,x}>u_{c}$. Moreover, the vortices and antivortices
frequently annihilate and produce sound waves during this annihilation.
Figure~\ref{fig:velfor.pdfpectraIC1f1}~(b) shows that the force exerted by the
superfluid on the particle (see Eq.~(\ref{eq:frcbyfluid}) and~(\ref{eq:frcbyfluidGT})) exhibits large
fluctuations, after the critical velocity is reached.  In
Fig.~\ref{fig:velfor.pdfpectraIC1f1}~(c) we plot the power spectra of the time
series of th Cartesian components $f_{\rm o,x}$ (green solid curve) and $f_{\rm
o,y}$ (black dashed curve); these show that many frequencies appear in these spectra.
The video M1 in the Supplemental Material~\cite{suppmat} illustrates the
spatiotemporal evolution of a forced heavy particle in a superfluid. This
video, the time series of $u_{\rm o,x}$, $u_{\rm o,y}$, $f_{\rm o,x}$, and
$f_{\rm o,y}$ and their power spectra
(Figs.~\ref{fig:velfor.pdfpectraIC1f1}~(c), (f) and (i)) show that, after the first
vortex-antivortex pair has been emitted, the motion of the particle can be
temporally chaotic, for heavy particle.

\begin{figure*}
\centering
\begin{overpic}
[height=4.5cm,unit=1mm]{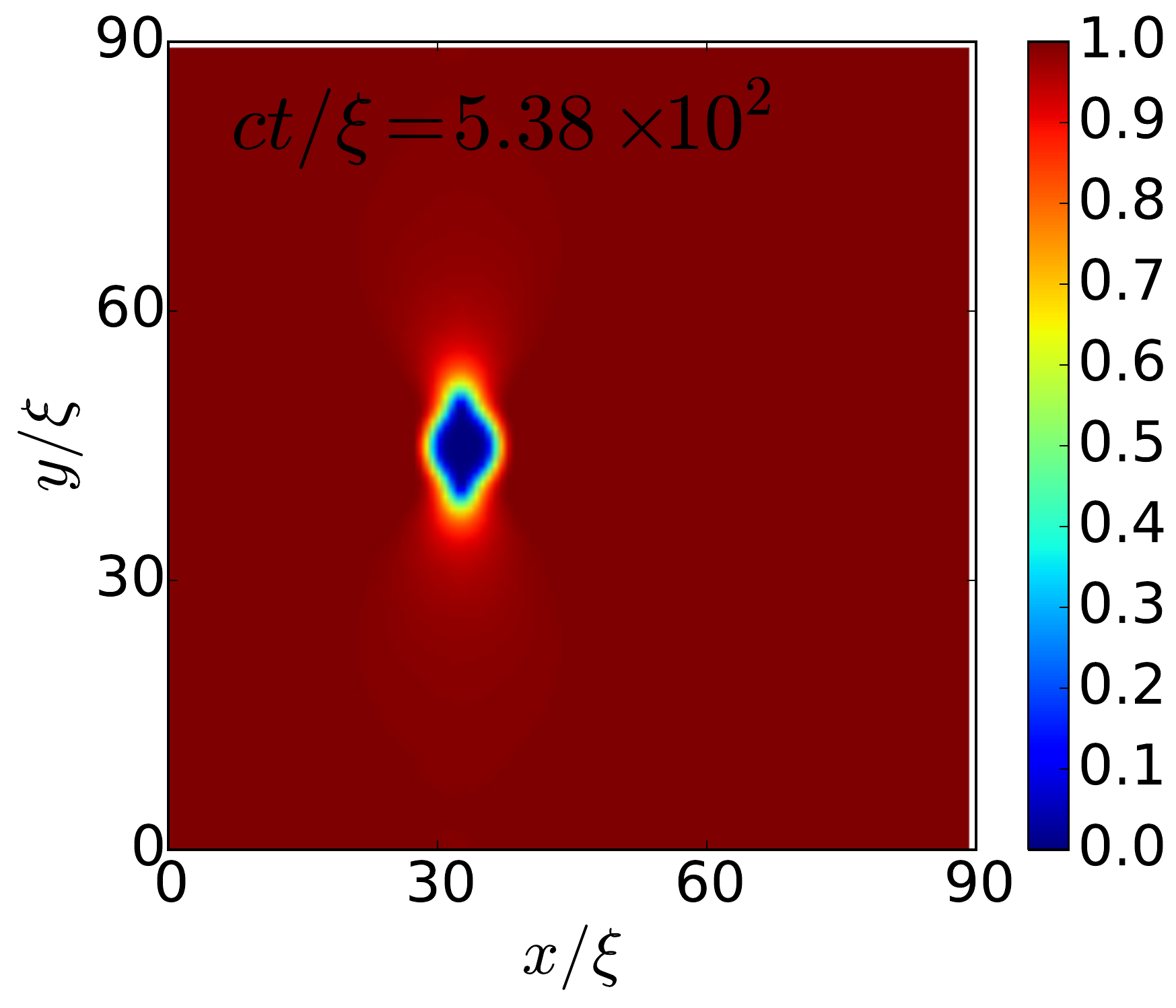}
\put(10.,10){\large{\bf (a)}}
\end{overpic}
\begin{overpic}
[height=4.5cm,unit=1mm]{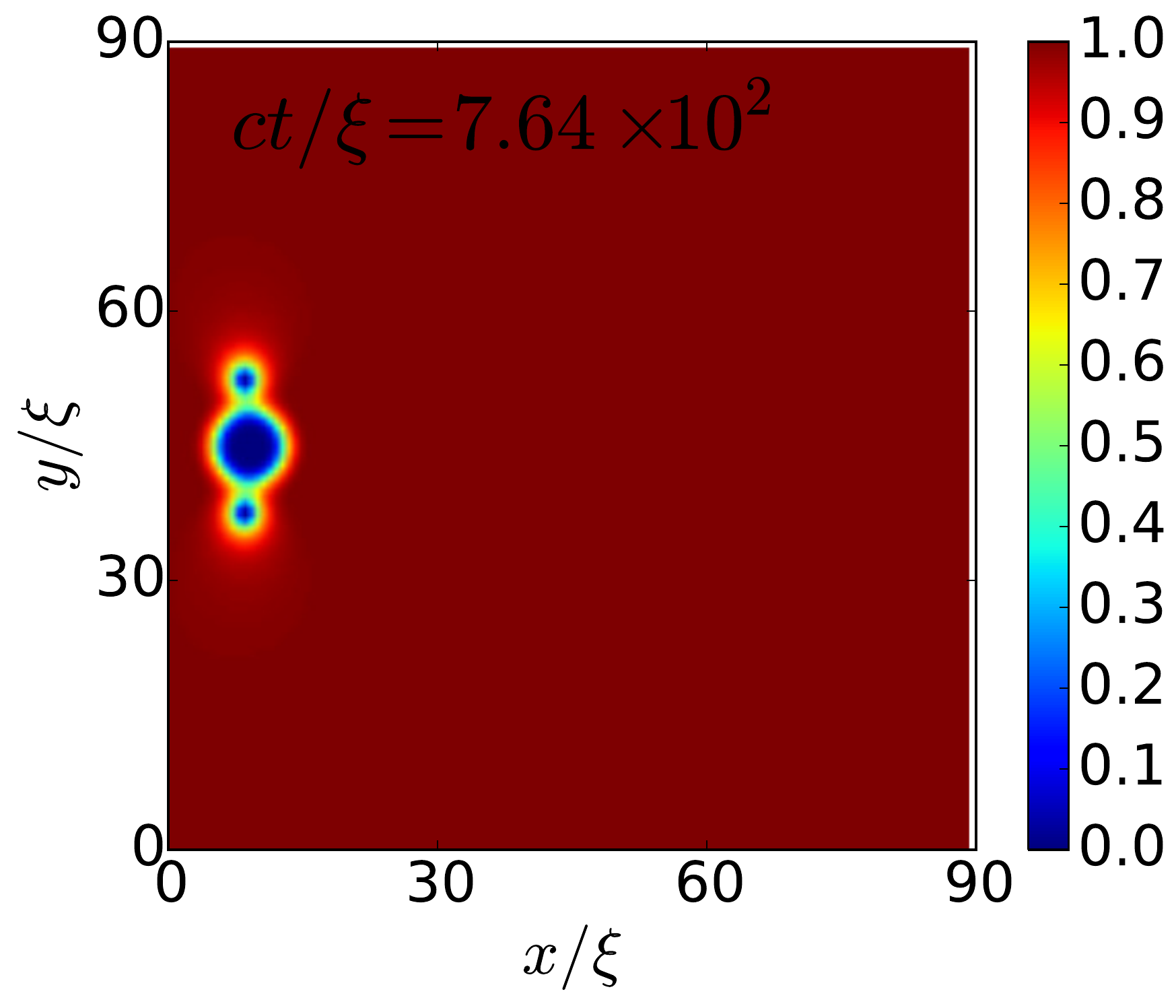}
\put(10,10){\large{\bf (b)}}
\end{overpic}
\begin{overpic}
[height=4.5cm,unit=1mm]{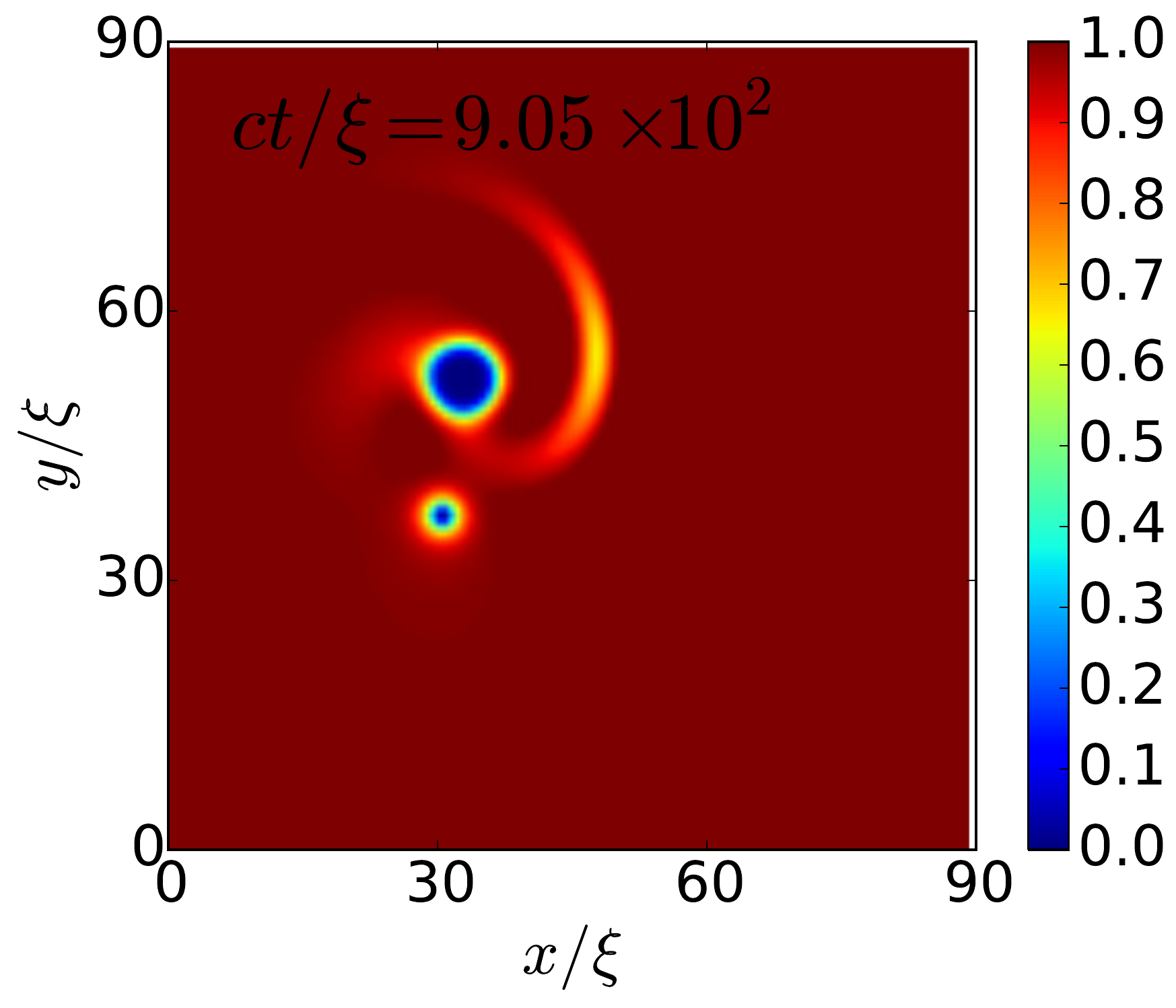}
\put(10,10){\large{\bf (c)}}
\end{overpic}
\\
\begin{overpic}
[height=4.5cm,unit=1mm]{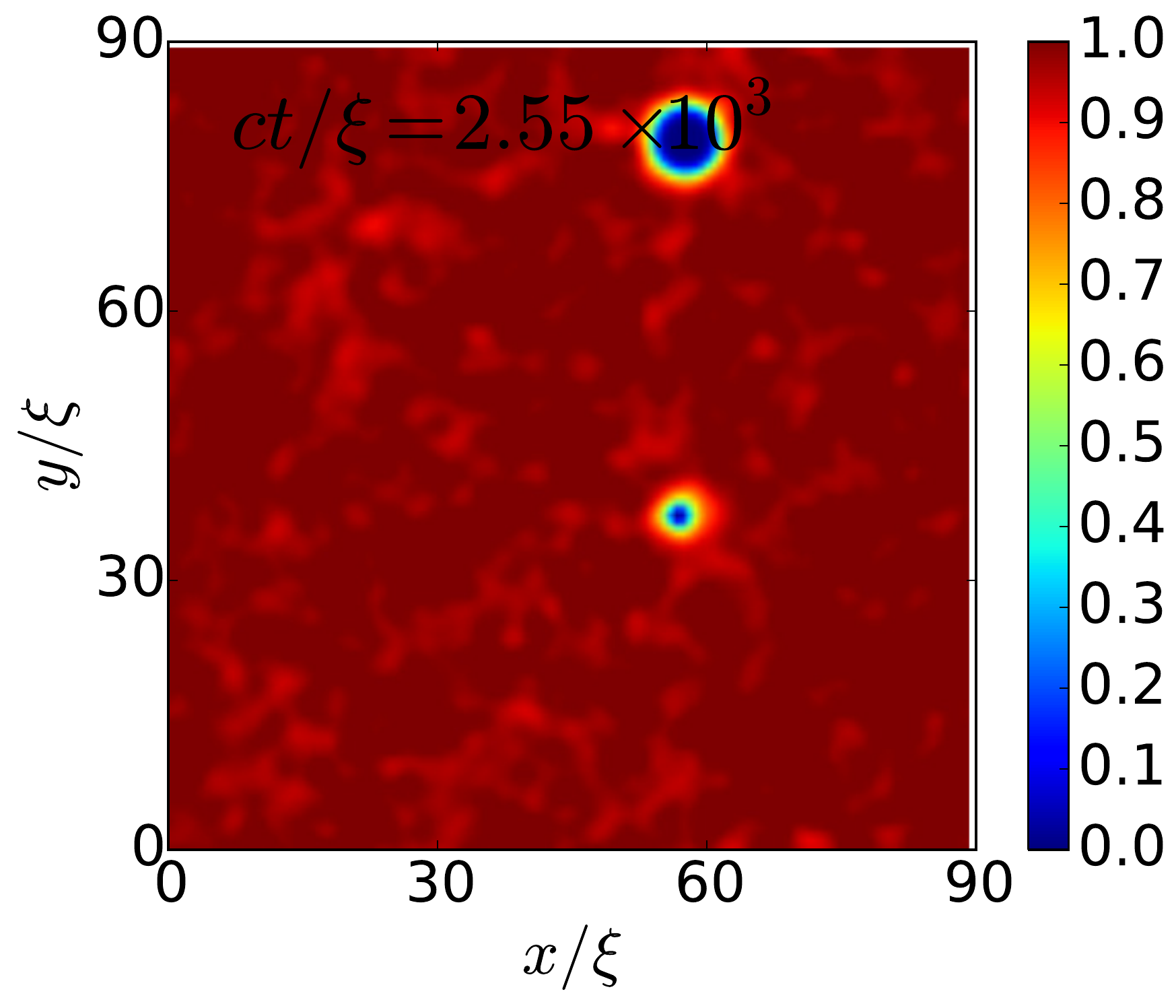}
\put(10.,10){\large{\bf (d)}}
\end{overpic}
\begin{overpic}
[height=4.5cm,unit=1mm]{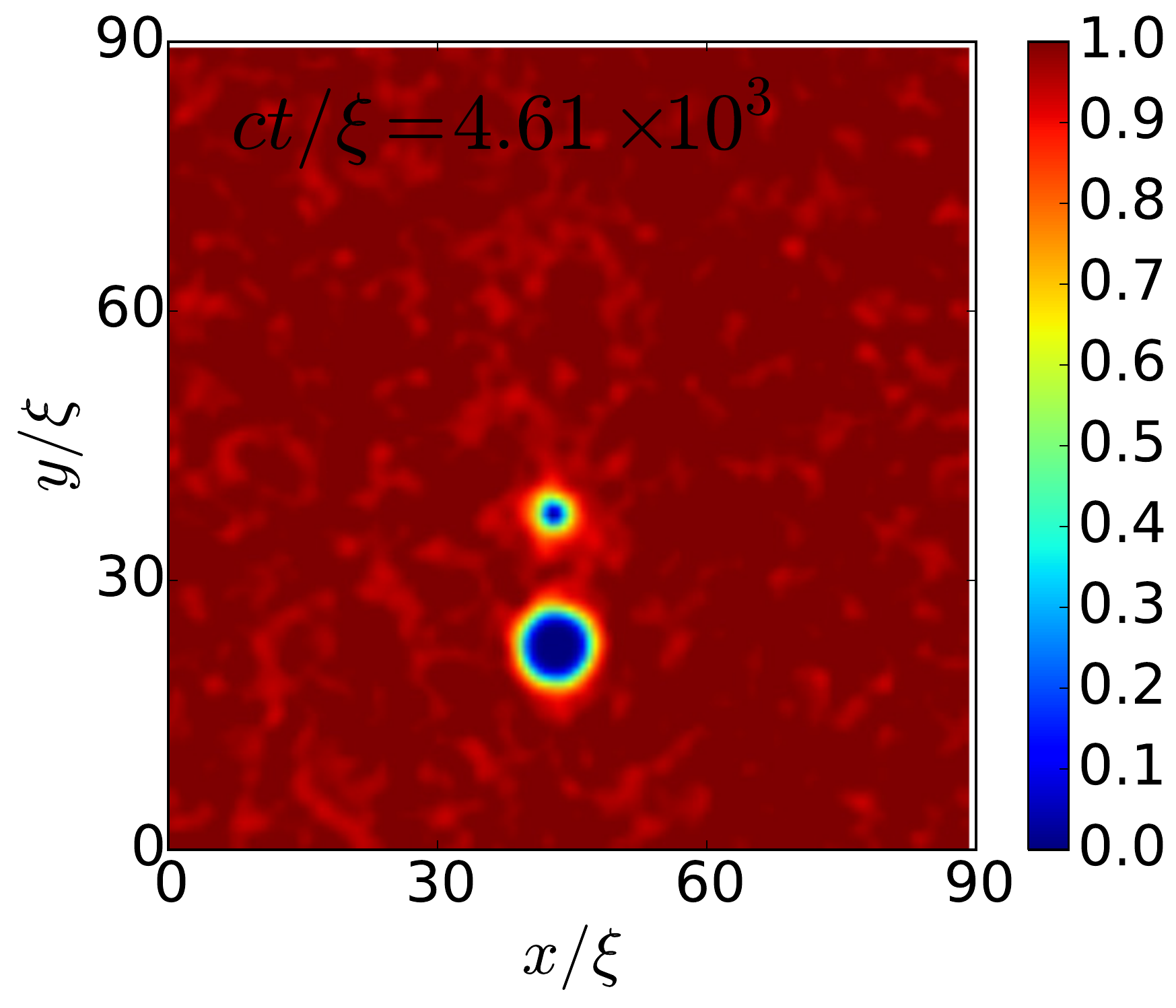}
\put(10,10){\large{\bf (e)}}
\end{overpic}
\begin{overpic}
[height=4.5cm,unit=1mm]{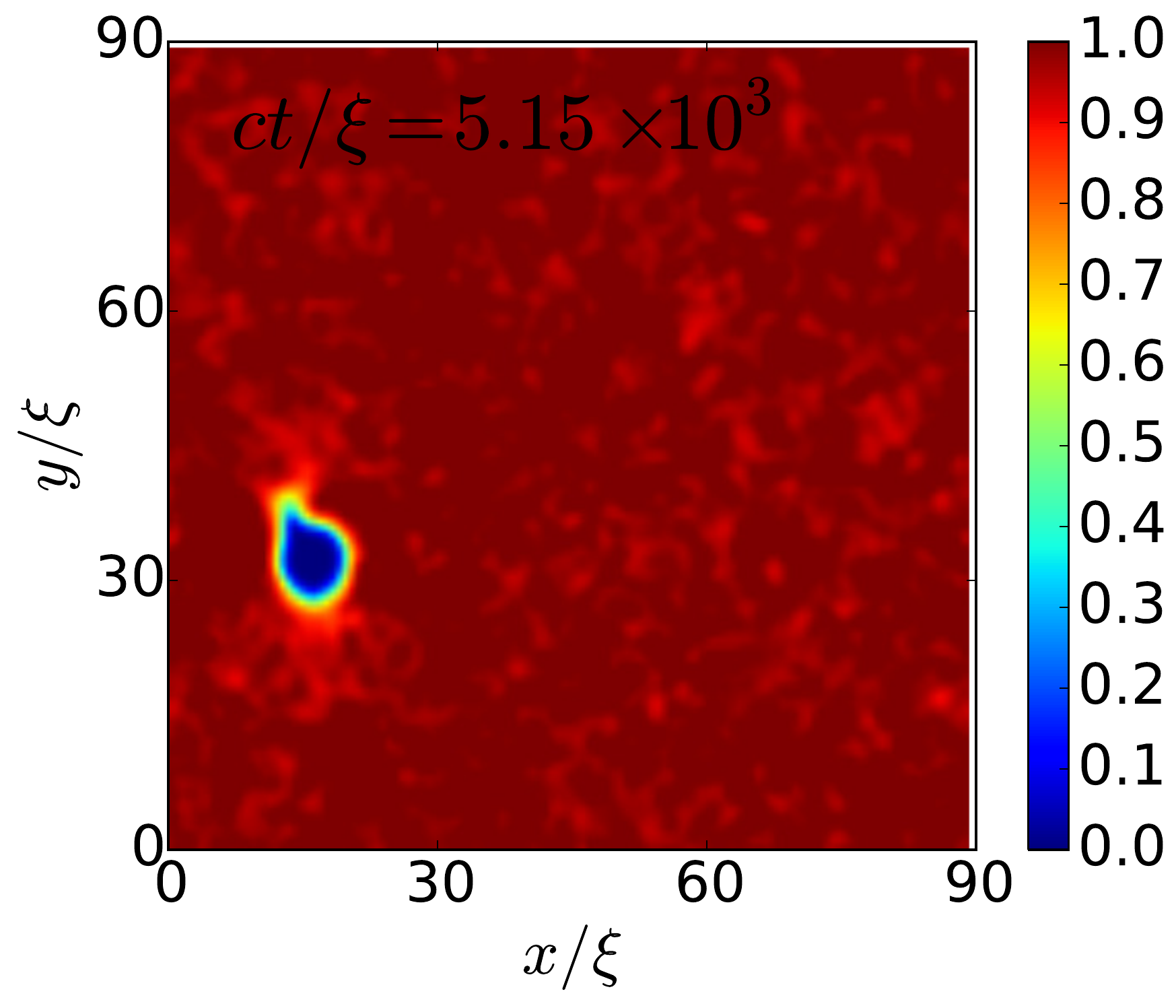}
\put(10,10){\large{\bf (f)}}
\end{overpic}
\\
\begin{overpic}
[height=4.5cm,unit=1mm]{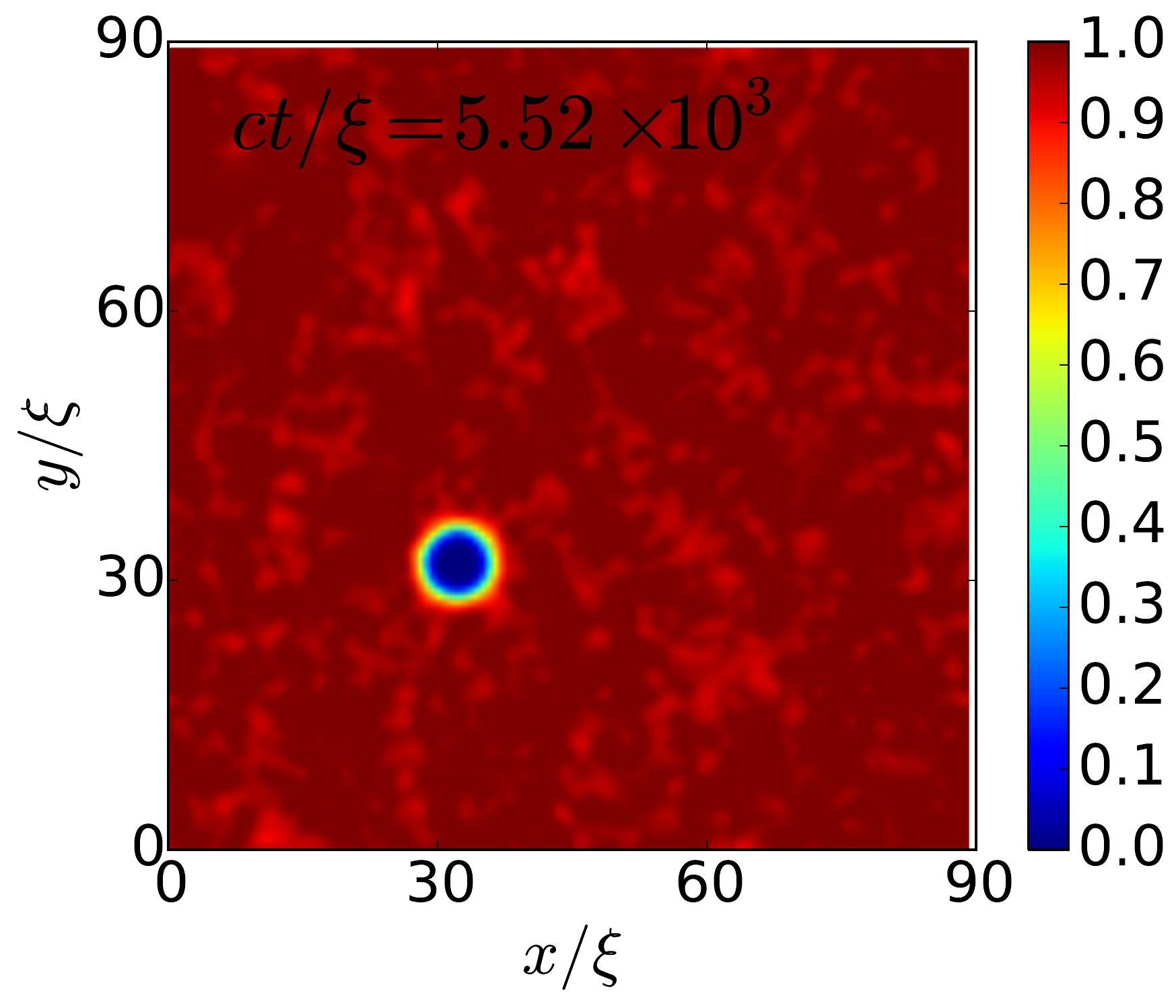}
\put(10.,10){\large{\bf (g)}}
\end{overpic}
\begin{overpic}
[height=4.5cm,unit=1mm]{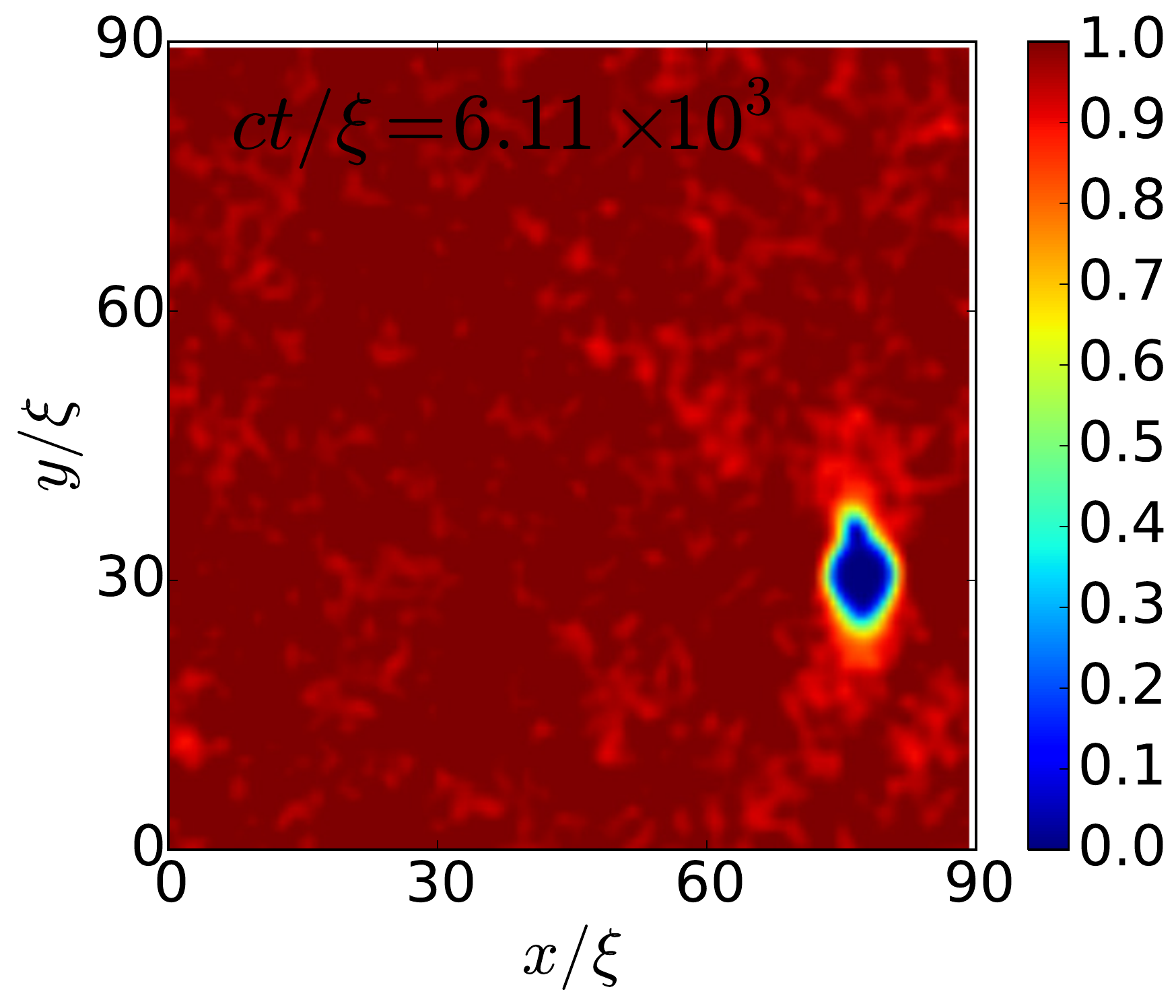}
\put(10,10){\large{\bf (h)}}
\end{overpic}
\begin{overpic}
[height=4.5cm,unit=1mm]{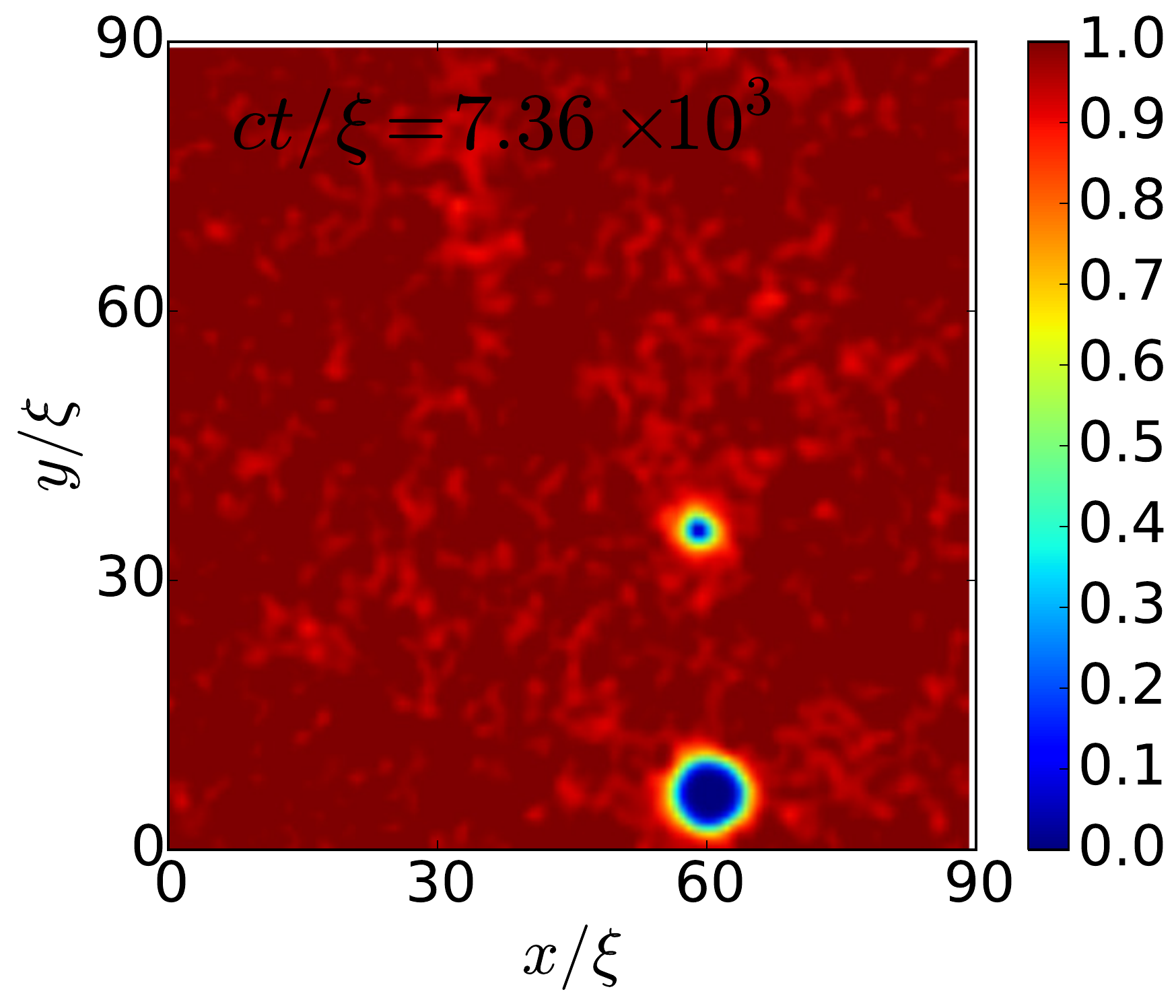}
\put(10,10){\large{\bf (i)}}
\end{overpic}
\caption{\small (Color online) Spatiotemporal evolution of the density field
$\rho(\mathbf{r},t)$ shown via pseudocolor plots, illustrating the dynamics
of a neutral particle, when a constant external force $\mathbf{F}_{\rm
ext}=0.14\,c^2\xi\rho_0\,\hat{\mathbf{x}}$ acts on it (initial configuration $\tt ICP1$).  The
particle appears as a large blue patch and the vortices as blue dots (for
details see text, subsection~\ref{subsec:1partextforce}).}
\label{fig:pdIC1f1N}
\end{figure*}

{\bf Neutral particle:} 
To study the dynamics of a neutral particle in the
superfluid, we use the initial configuration $\tt ICP1$.  We apply an external
force $\mathbf{F}_{\rm ext}= 0.14\,c^2\xi\rho_0\,\hat{\mathbf{x}}$ on the particle. In
Figs.~\ref{fig:pdIC1f1N}~(a)-(i) we show, via pseudocolor plots, the
spatiotemporal evolution of $\rho(\mathbf{r},t)$; the particle appears as a
large blue patch on these plots.  The particle accelerates under the influence
of the external force and its velocity reaches a maximum at $ct/\xi\simeq 467$,
before starting to decrease (see Fig.~\ref{fig:velfor.pdfpectraIC1f1}~(d));
this maximum of velocity is also the critical velocity $u_c\simeq0.47\,c$, where a
vortex-antivortex pair is formed.  Figure~\ref{fig:pdIC1f1N}~(a) shows the
vortex-antivortex pair still attached to the particle at $ct/\xi=5.38\times10^2$, as an
extension along the $y$-direction of the particle (blue patch). Subsequently,
the vortex-antivortex pair is detached from the particle
(Fig.~\ref{fig:pdIC1f1N}~(b)); but the particle and the vortex-antivortex-pair
assembly (henceforth PVA complex) becomes unstable, with respect to motion
transverse to the direction of $\mathbf{F}_{\rm ext}$ at $ct/\xi\simeq 7.64\times10^2$, and
oscillatory modes are excited, as we show in
Fig.~\ref{fig:velfor.pdfpectraIC1f1}~(d), where the Cartesian components
$u_{\rm o,x}$ and $u_{\rm o,y}$ of the particle velocity exhibit modulated
oscillations.  In Fig.~\ref{fig:velfor.pdfpectraIC1f1}~(f), the power spectra
of both $u_{\rm o,x}$ and $u_{o,y}$ show one large peak and two or three small
peaks; the former is associated with the main temporal oscillation and the
latter with the modulation.  Figure~\ref{fig:velfor.pdfpectraIC1f1}~(e) shows
that similar oscillations are present in the Cartesian components of the force
exerted by the superfluid on the particle.  As a result of this instability, at
$ct/\xi\simeq 9.05\times 10^2$ the particle is trapped on the positive (upper) vortex
(Fig.~\ref{fig:pdIC1f1N}~(c)); this is accompanied by an intense emission of
sound waves.  The particle, trapped on the positive vortex, and the negative
(lower) vortex move together, with both aligned roughly perpendicular to the
direction of motion (Fig.~\ref{fig:pdIC1f1N}~(d)); the separation between the
two increases, principally because the positive vortex moves away and takes the
trapped particle along with it.  Because of our periodic boundary conditions,
the positive vortex (and the particle trapped by it) comes back from below (the
$y$-direction).  The direction of the velocity field, in the small region in
between the particle and the vortex, generated by the vortex-antivortex pair,
is reversed (from $+\hat{\mathbf{x}}$ to $-\hat{\mathbf{x}}$).  When the
negative vortex and the positive vortex (and the trapped particle) are close
enough to each other, then the field generated by the pair is so strong that
the PVA complex reverses its direction of motion at $ct/\xi\simeq4.61\times10^3$
(Fig.~\ref{fig:pdIC1f1N}~(e)).  When the two
vortices are very close together (see Fig.~\ref{fig:pdIC1f1N}~(f)), they
annihilate at $ct/\xi=5.15\times10^3$, while the particle moves predominantly 
in the $-\hat{\mathbf{x}}$
direction.  Soon thereafter, the acceleration, because of the external force,
reverses the direction of motion once again and the particle begins to move
predominantly in the $\hat{\mathbf{x}}$ direction as we show in
Fig.~\ref{fig:pdIC1f1N}~(g) at $ct/\xi=5.52\times10^3$.  The particle velocity increases,
reaches a maximum value $u_{\rm o,x}=0.57\,c>u_{c}$, and again a vortex-antivortex
pair is formed at $t=6.11\times10^3$, initially this pair is attached to the particle. At
$ct/\xi\simeq6.14\times10^3$ the particle gets trapped on the negative (lower) vortex and the
cycle of particle and vortex motions, described above, is repeated again (see
Fig.~\ref{fig:velfor.pdfpectraIC1f1}~(d) for $t\gtrsim 6.14\times10^3$ and
Fig.~\ref{fig:pdIC1f1N}(i)).  Video M2~\cite{suppmat} gives the complete
spatiotemporal evolution of the particle and $\rho(\mathbf{r},t)$.

{\bf Light particle:} 
We now describe the dynamics of a light particle in the
superfluid by using the initial configuration $\tt ICP1$; the particle is
accelerated by applying an external force $\mathbf{F}_{\rm ext}=
0.14\,c^2\xi\rho_0\,\hat{\mathbf{x}}$ on it.  
Figures~\ref{fig:pdIC1f1L}~(a)-(i) in the Appendix and Video
M3~\cite{suppmat} summarize the spatiotemporal evolution of the field
$\rho(\mathbf{r},t)$ for the light-particle case.  A comparison of the
$x$-component of the particle velocity $u_{\rm o,x}$ (purple curve) in
Figs.~\ref{fig:velfor.pdfpectraIC1f1}~(d) and (g), and a comparison of
the field $\rho(\mathbf{r},t)$ in Figs.~\ref{fig:pdIC1f1N} and
\ref{fig:pdIC1f1L} in the Appendix, shows that
the dynamics of the light particle is similar to, but not exactly the same as,
that of the neutral particle. A major feature which distinguishes their
dynamics is the presence of a broad range of frequencies in the power spectra
of $\mathbf{u}_{\rm o}$ and $\mathbf{f}_{\rm o}$ of the former case; this
indicates that the motion of the light particle is chaotic, in contrast to that
of the neutral particle, whose dynamical evolution is periodic in time.

The chaotic nature of the particle dynamics is enhanced when we increase the
amplitude of the external force acting on the particle, as we show in
Fig.~\ref{fig:velfor.pdfpectraIC1f2} in the Appendix; 
in particular, now the power spectra of the Cartesian
components of $\mathbf{u}_{\rm o}$ and $\mathbf{f}_{\rm o}$ have a broad range
of frequencies for all the three types of particles. We can understand the
motion of the particles by using the concept of the hydrodynamical mass (or
effective mass) $m_*=\partial P_{\rm ext}/\partial u_{\rm o}$ ($P_{\rm
ext}=F_{\rm ext}t$).  To begin with, there is no drag-force on the particle;
but the particle still transfers momentum to the fluid by virtue of the
increase in its effective mass; this becomes very large (ideally infinity) at
$u_{\rm o}=u_{c}$, where a vortex-antivortex pair is formed. Note that, when a
vortex-antivortex pair is formed, the plot of $u_{\rm o,x}$ versus time has a
maximum (see Fig.~\ref{fig:velfor.pdfpectraIC1f1}(d)), i.e., the acceleration
in the $x$ direction vanishes, even though the force $F_{\rm ext}$ is nonzero.
Therefore, the effective mass $m_*$ diverges when the vortex-antivortex pair is
created. After this, $m_*$ becomes negative and the particle slows down; this
is more apparent in the cases of the neutral and the light particles than for a
heavy particle.

\subsection{Interaction of particles with a translating vortex-antivortex pair}
\label{subsec:1partvortexpair}

\begin{figure}
\centering
\resizebox{\linewidth}{!}{
\includegraphics[height=4.5cm,unit=1mm]{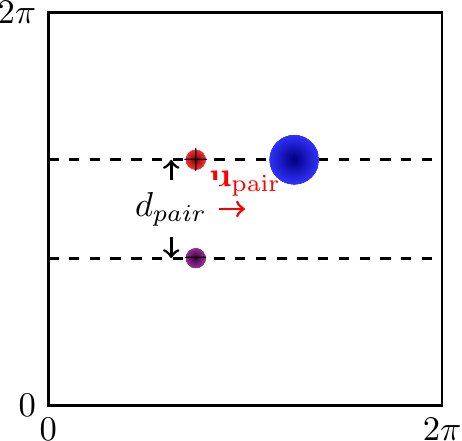}
\put(-75,20){\bf (a)}
\includegraphics[height=4.5cm,unit=1mm]{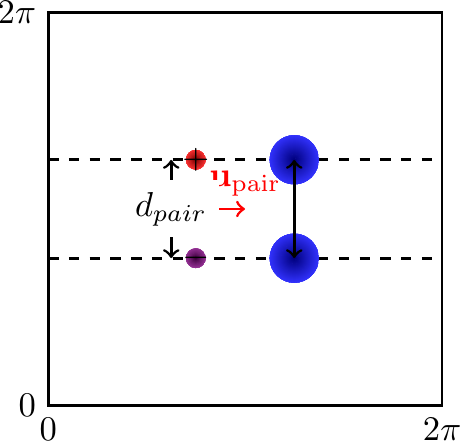}
\put(-75,20){\bf (b)}
}
\caption{\small (Color online) 
The schematic diagrams (a) and (b) illustrate the initial configurations
ICP2A (single particle case) and ICP2B (two particles case), respectively, 
which we use to study the interaction of the particles (blue disks)
with the superfluid field, by placing them in the path of a
translating vortex-antivortex pair (represented by small red and purple disks).  
}
\label{fig:schem1partpairtransl}
\end{figure}

\textbf{Single particle} Next we study the interaction of a single particle with 
a vortex-antivortex pair for heavy, neutral, and light particles by using the 
initial configuration $\tt ICP2A$, in which we place a particle in the path of the positive
vortex of the translating vortex-antivortex pair. In order to implement this, 
we prepare a state with a
stationary particle at $(1.5\pi/\xi,1.257\pi/\xi)$ and then combine it with a state
corresponding to a vortex-antivortex pair, of size $d_{\rm pair}\simeq 23\,\xi$
and which translates with a velocity $\mathbf{u}_{\rm
pair}=0.074\,c\,\hat{\mathbf{x}}$; see the schematic diagram in
Fig.~\ref{fig:schem1partpairtransl} (a) and Appendix~\ref{app:vortpair} for preparation
details.

\begin{figure*}
\centering
\begin{overpic}
[height=4.5cm,unit=1mm]{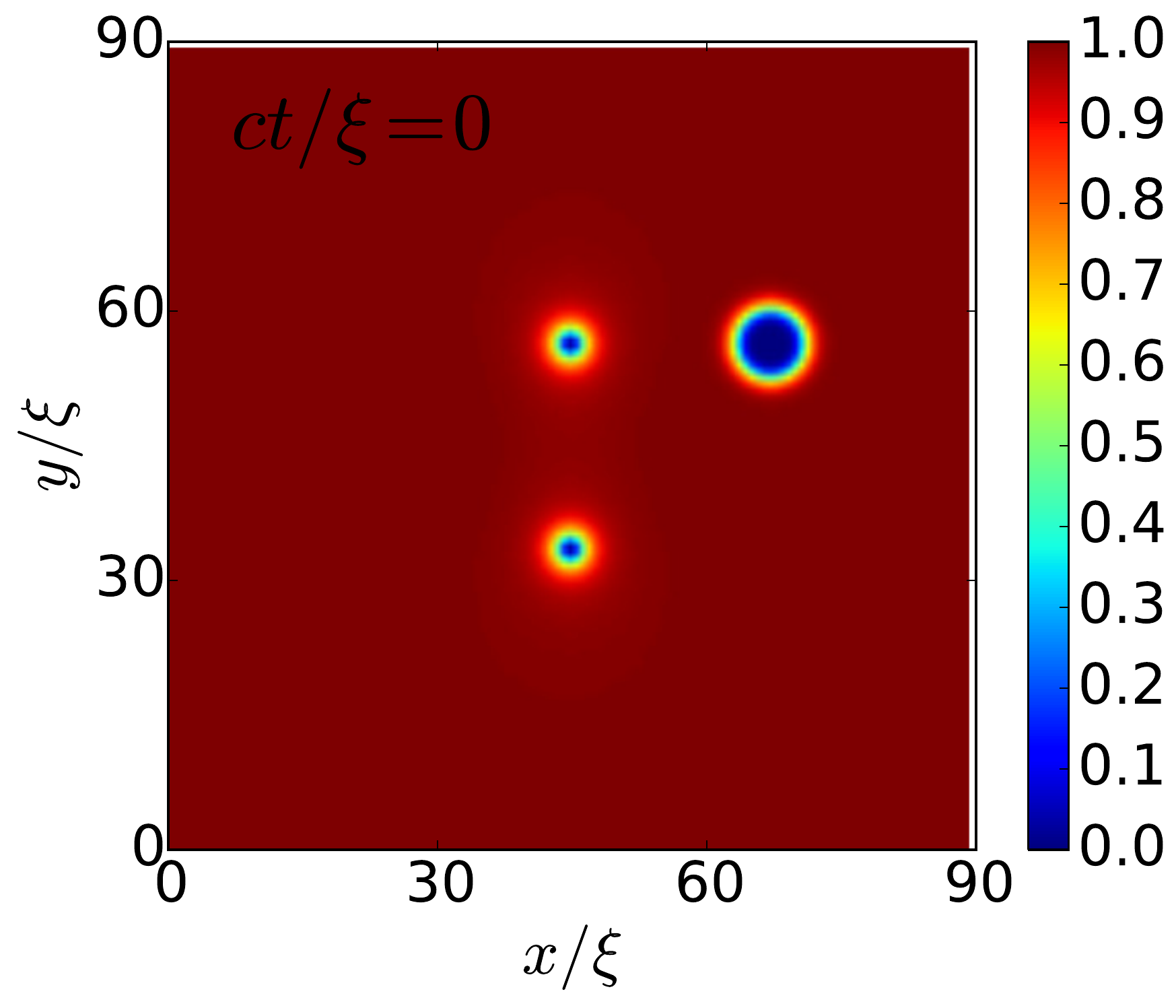}
\put(10.,10){\large{\bf (a)}}
\end{overpic}
\begin{overpic}
[height=4.5cm,unit=1mm]{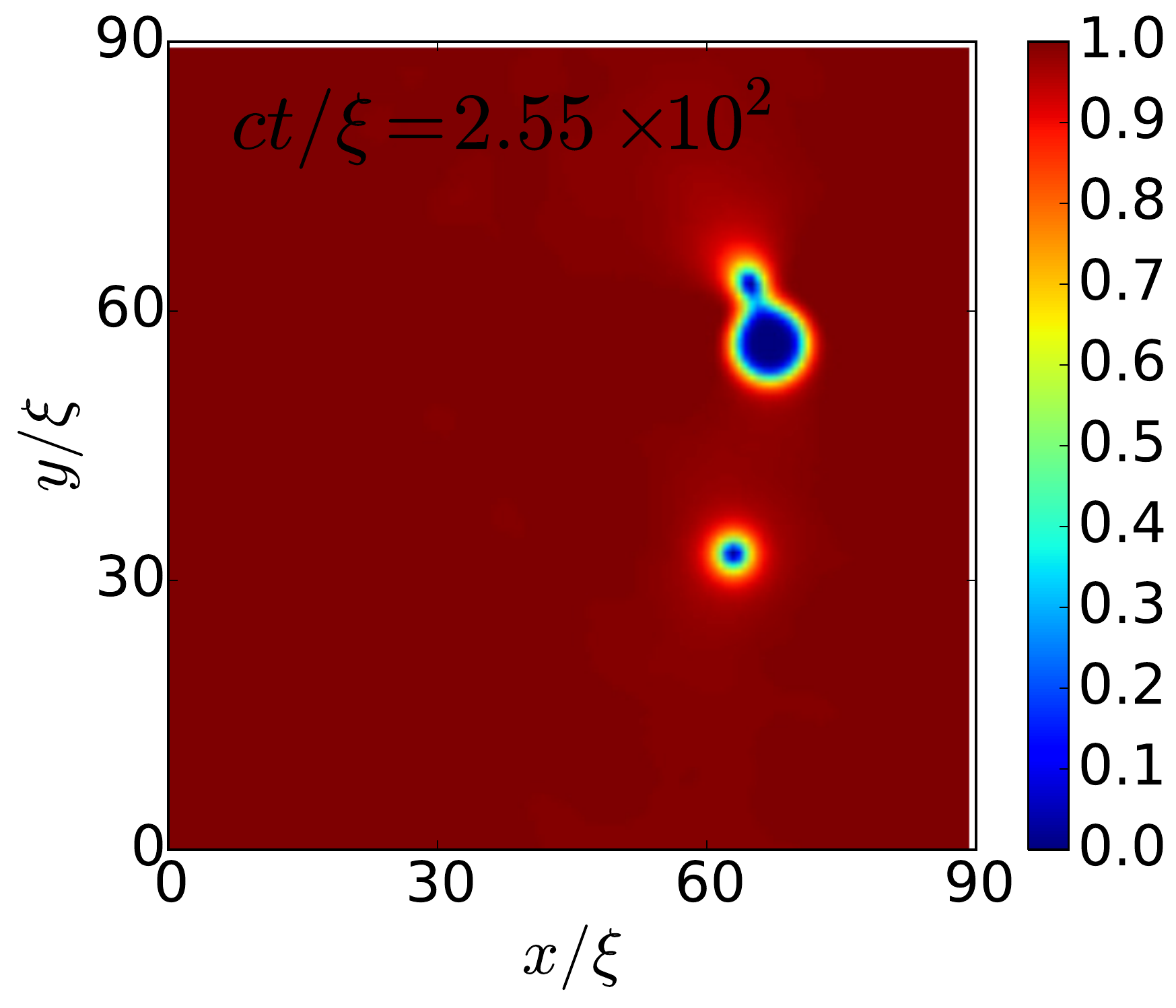}
\put(10,10){\large{\bf (b)}}
\end{overpic}
\begin{overpic}
[height=4.5cm,unit=1mm]{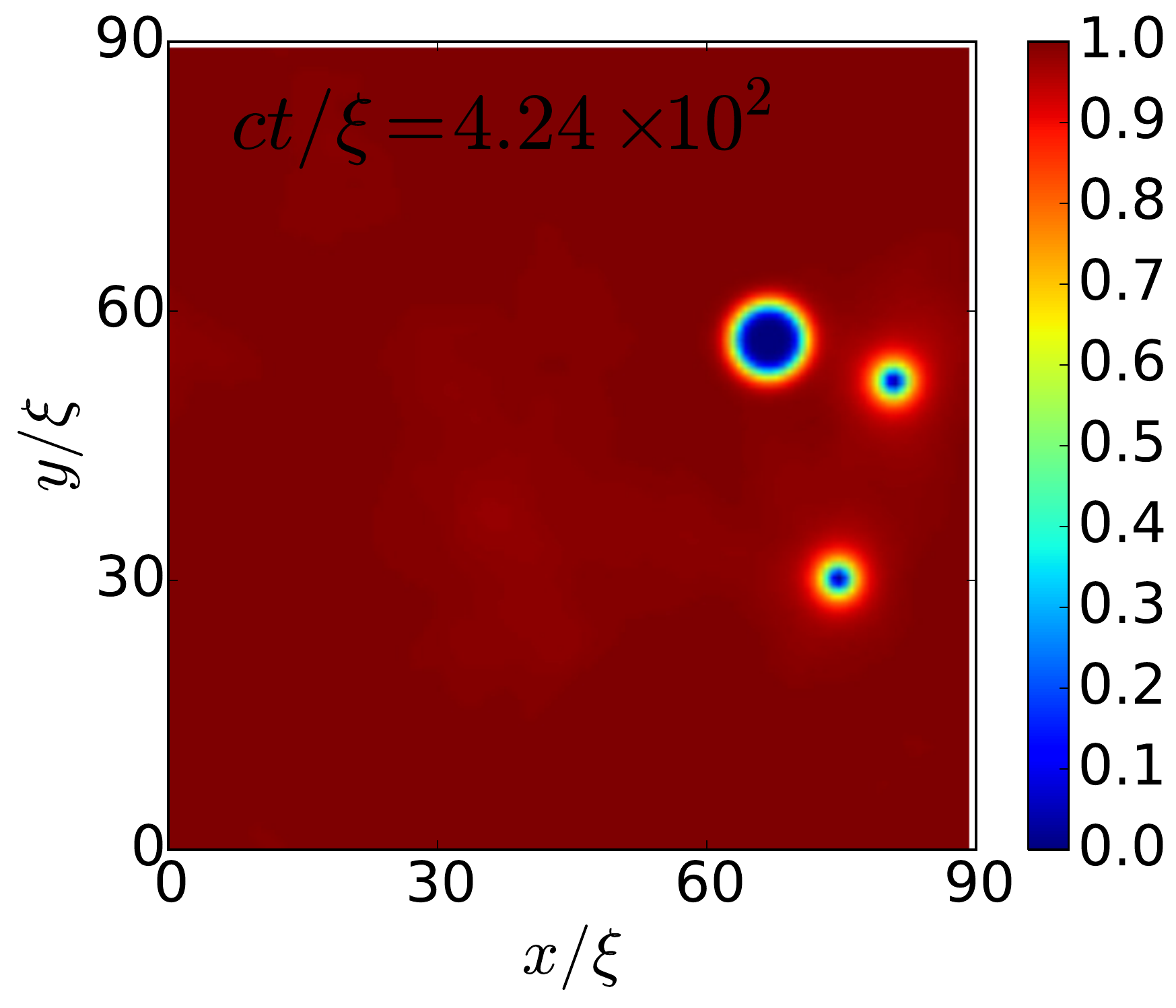}
\put(10,10){\large{\bf (c)}}
\end{overpic}
\\
\begin{overpic}
[height=4.5cm,unit=1mm]{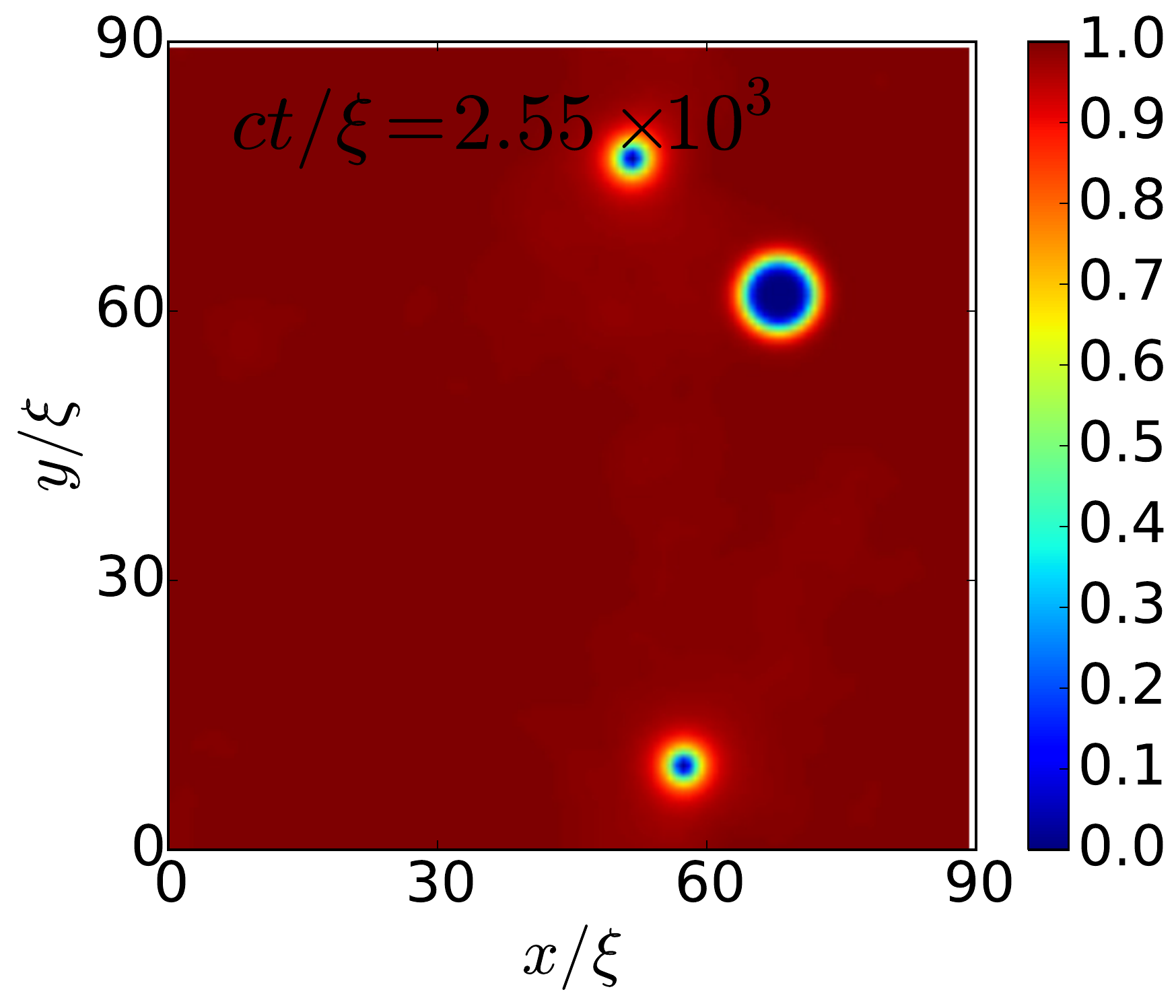}
\put(10.,10){\large{\bf (d)}}
\end{overpic}
\begin{overpic}
[height=4.5cm,unit=1mm]{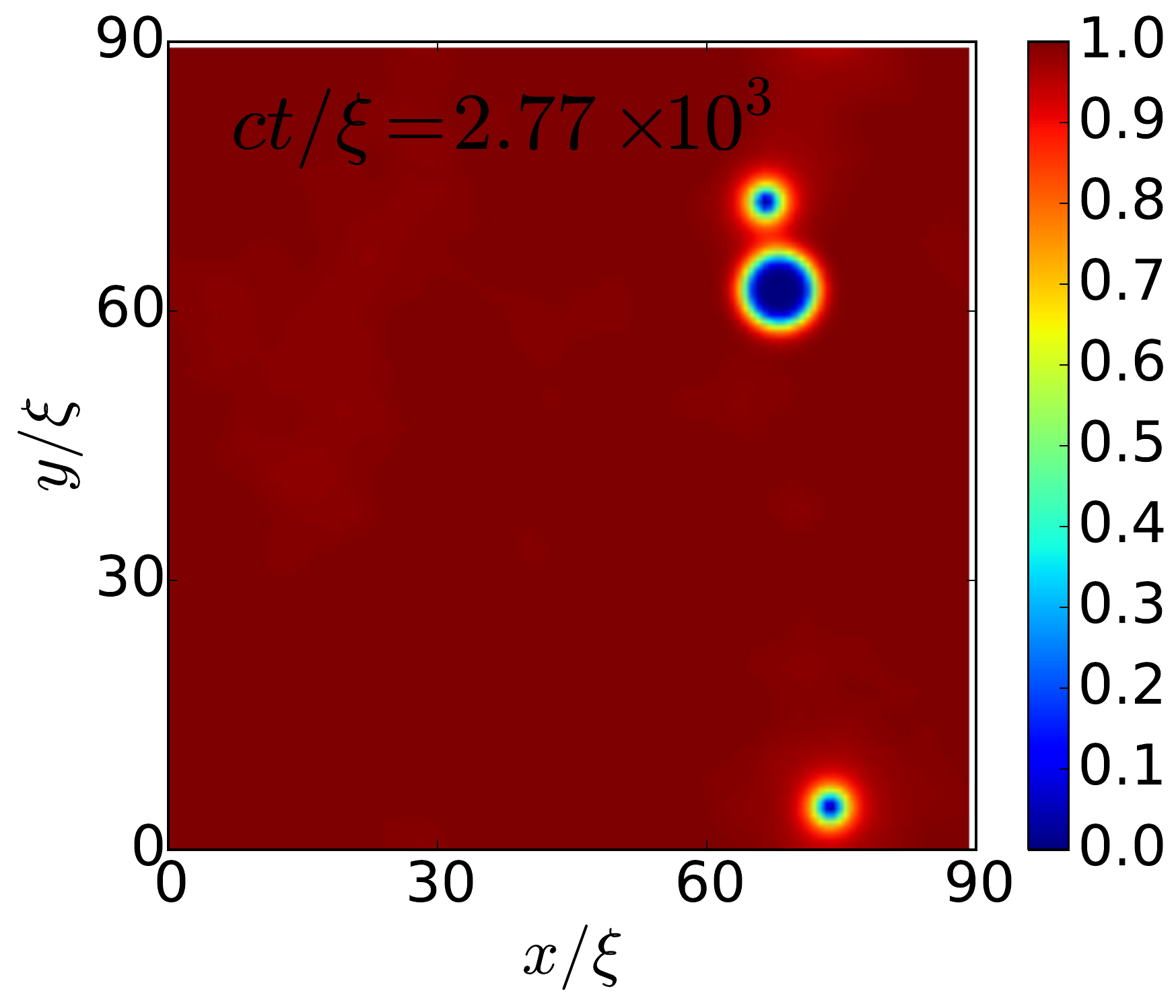}
\put(10,10){\large{\bf (e)}}
\end{overpic}
\begin{overpic}
[height=4.5cm,unit=1mm]{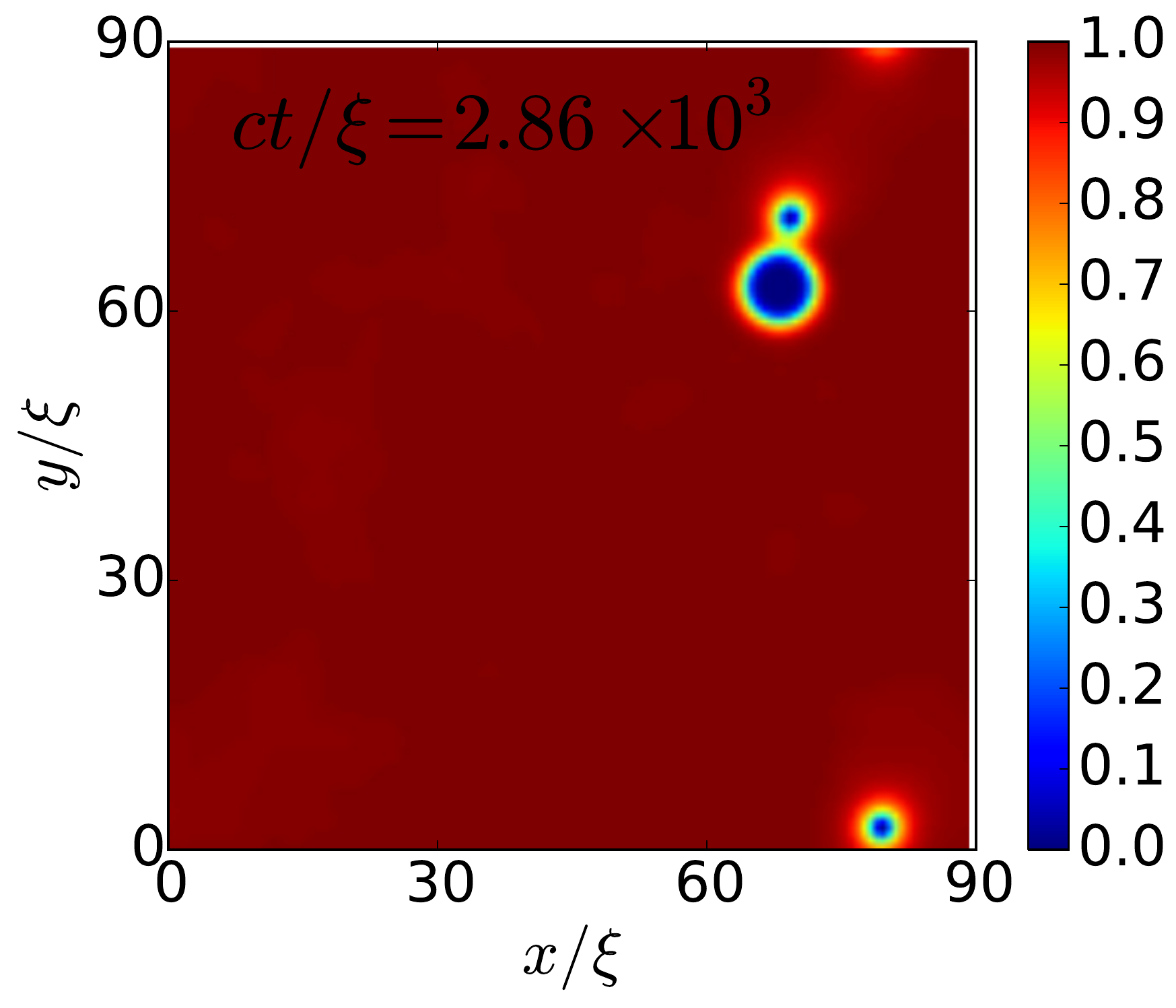}
\put(10,10){\large{\bf (f)}}
\end{overpic}
\\
\begin{overpic}
[height=4.5cm,unit=1mm]{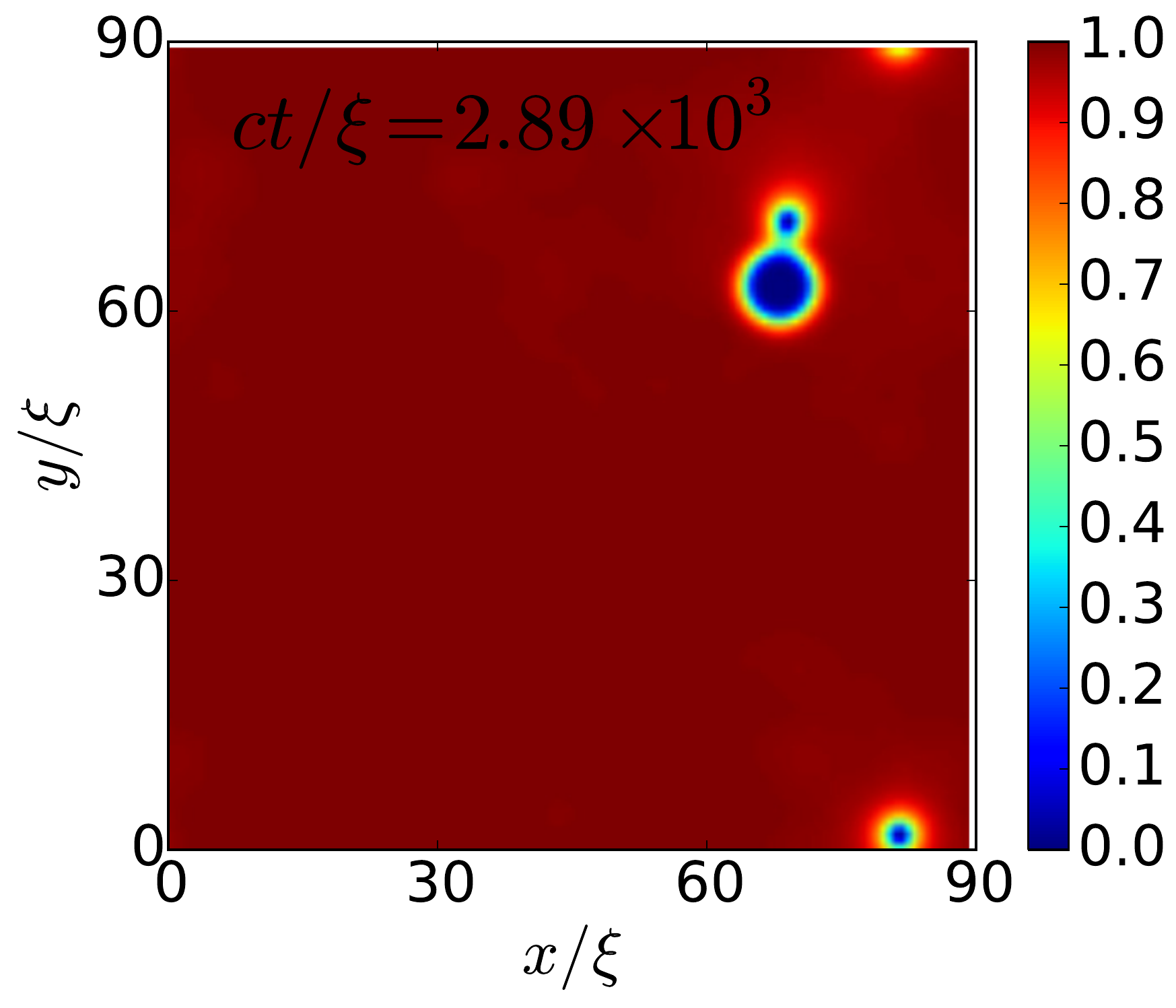}
\put(10.,10){\large{\bf (g)}}
\end{overpic}
\begin{overpic}
[height=4.5cm,unit=1mm]{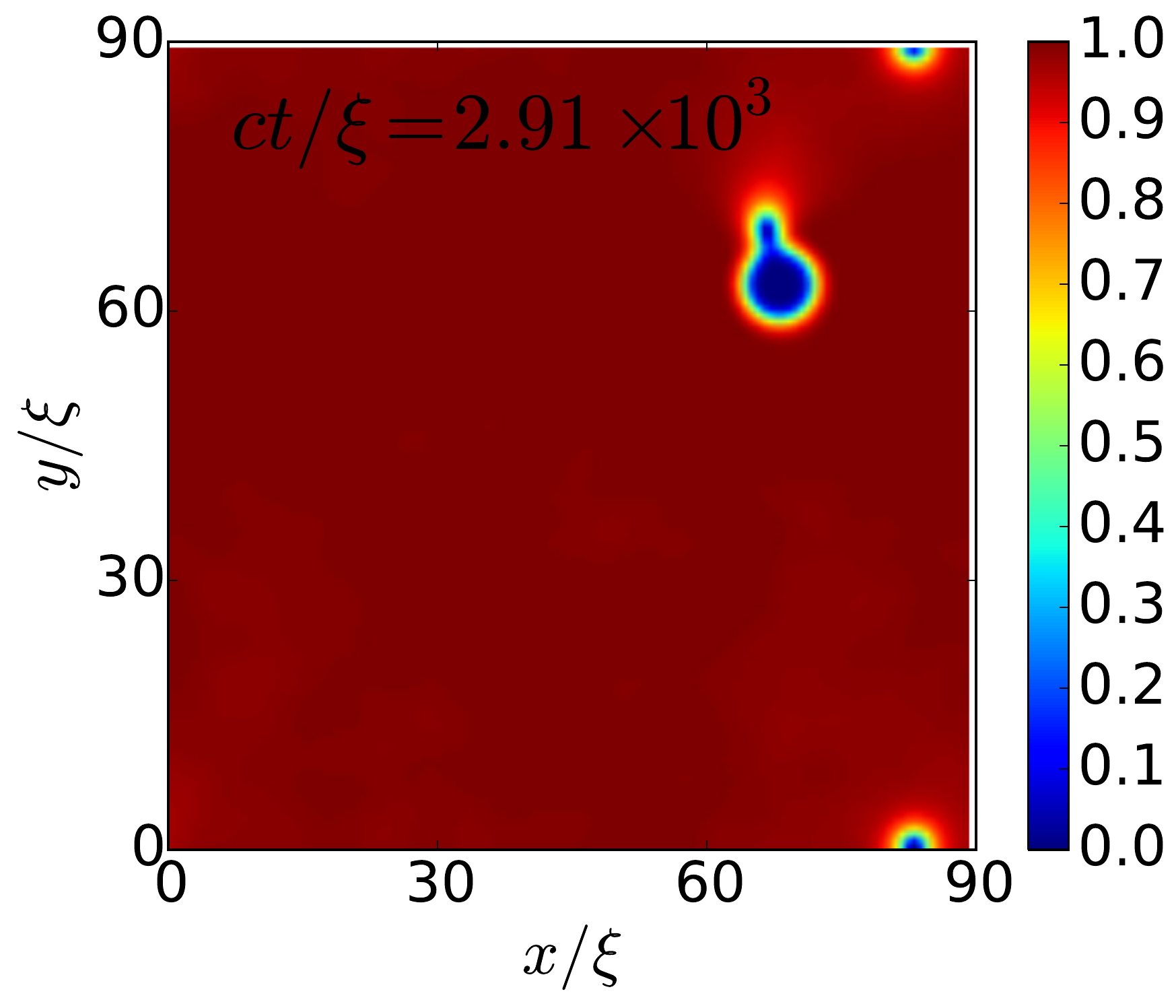}
\put(10,10){\large{\bf (h)}}
\end{overpic}
\begin{overpic}
[height=4.5cm,unit=1mm]{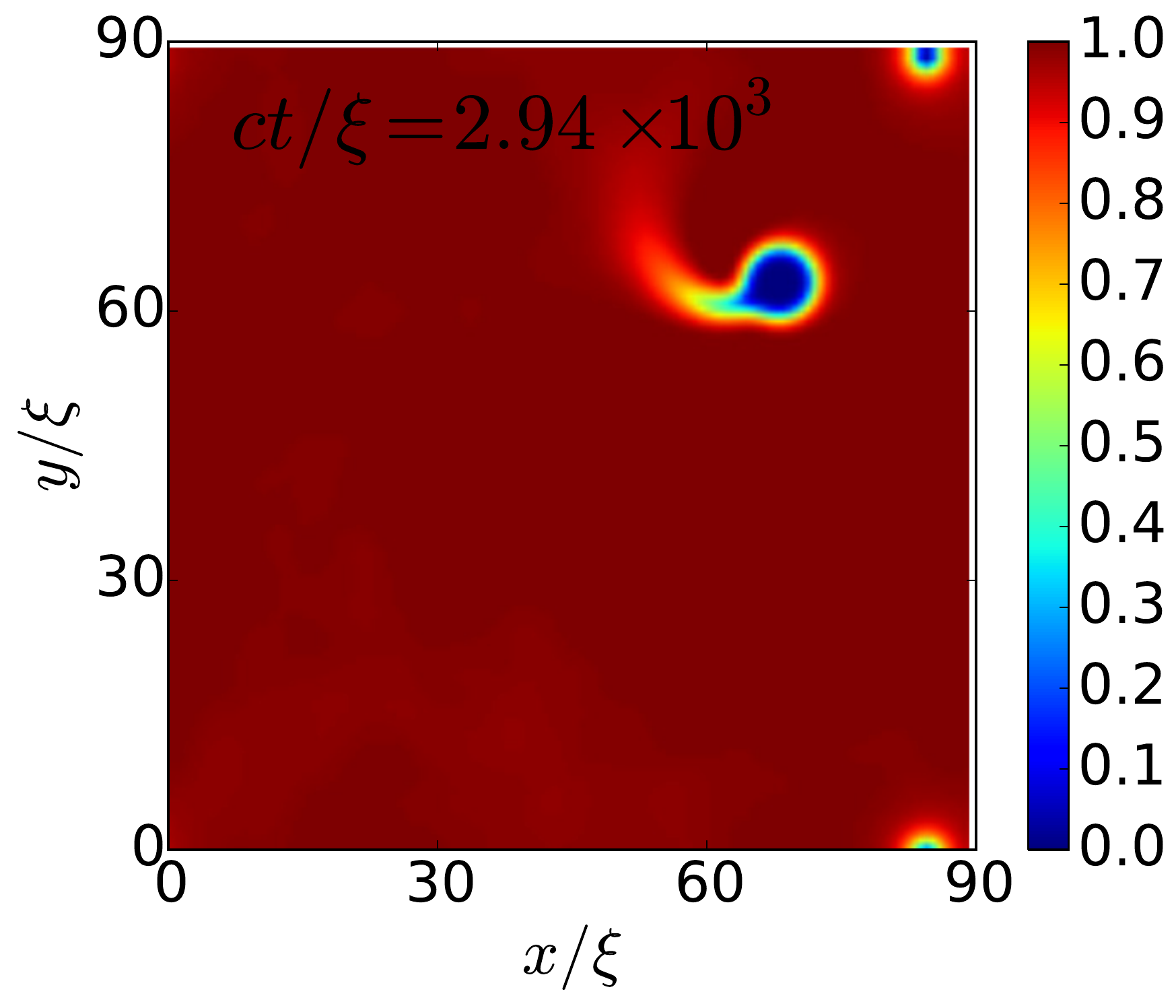}
\put(10,10){\large{\bf (i)}}
\end{overpic}
\\
\begin{overpic}
[height=4.5cm,unit=1mm]{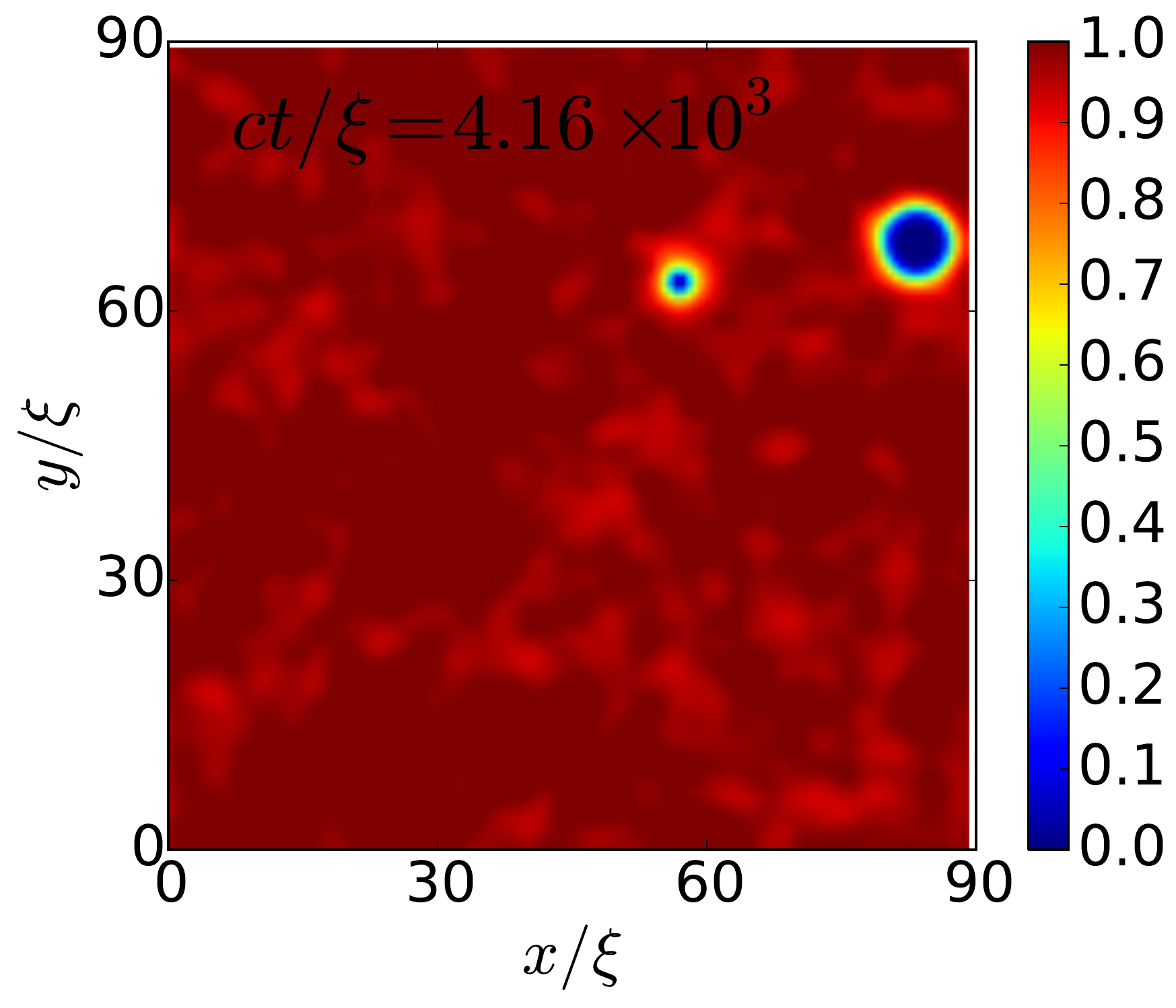}
\put(10.,10){\large{\bf (j)}}
\end{overpic}
\begin{overpic}
[height=4.5cm,unit=1mm]{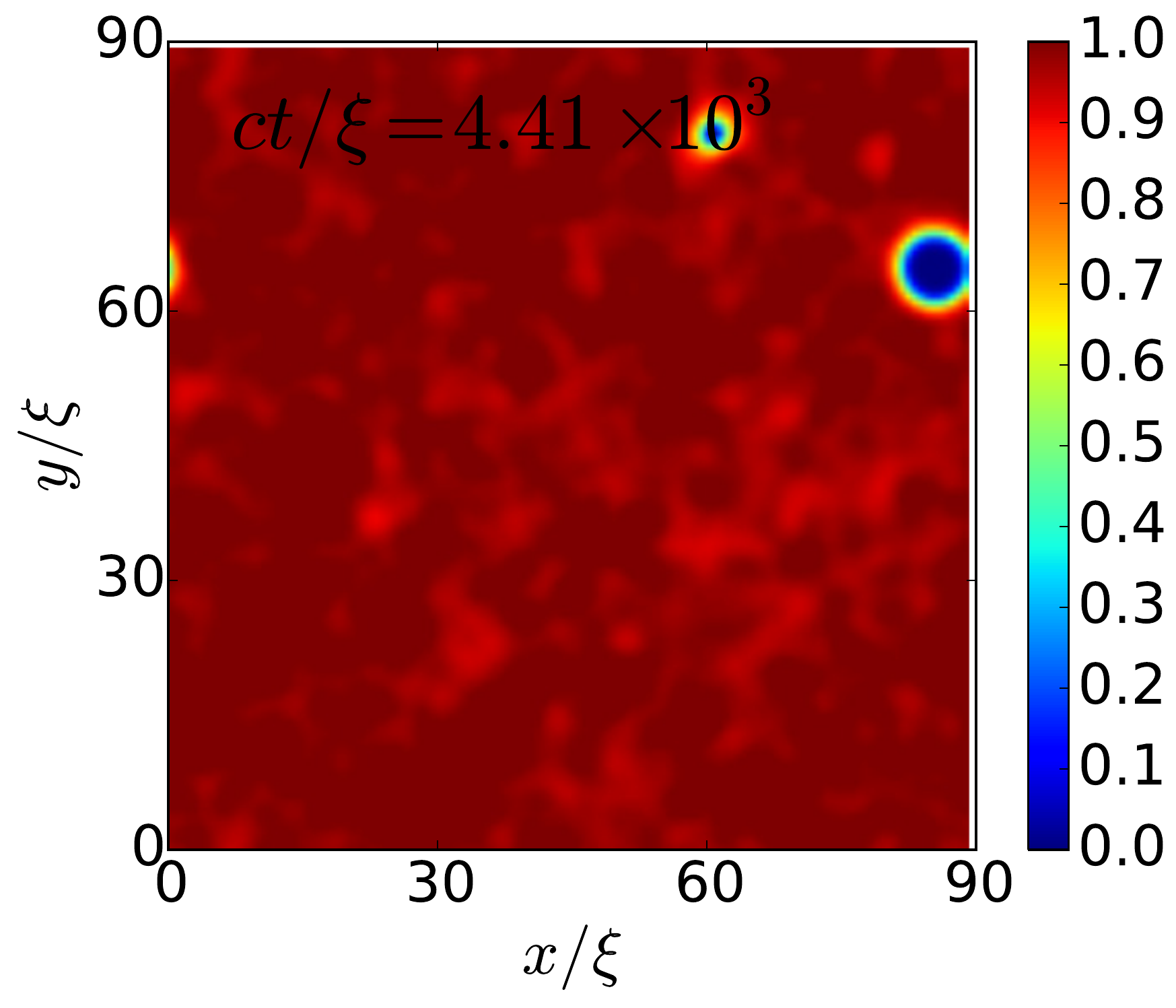}
\put(10,10){\large{\bf (k)}}
\end{overpic}
\begin{overpic}
[height=4.5cm,unit=1mm]{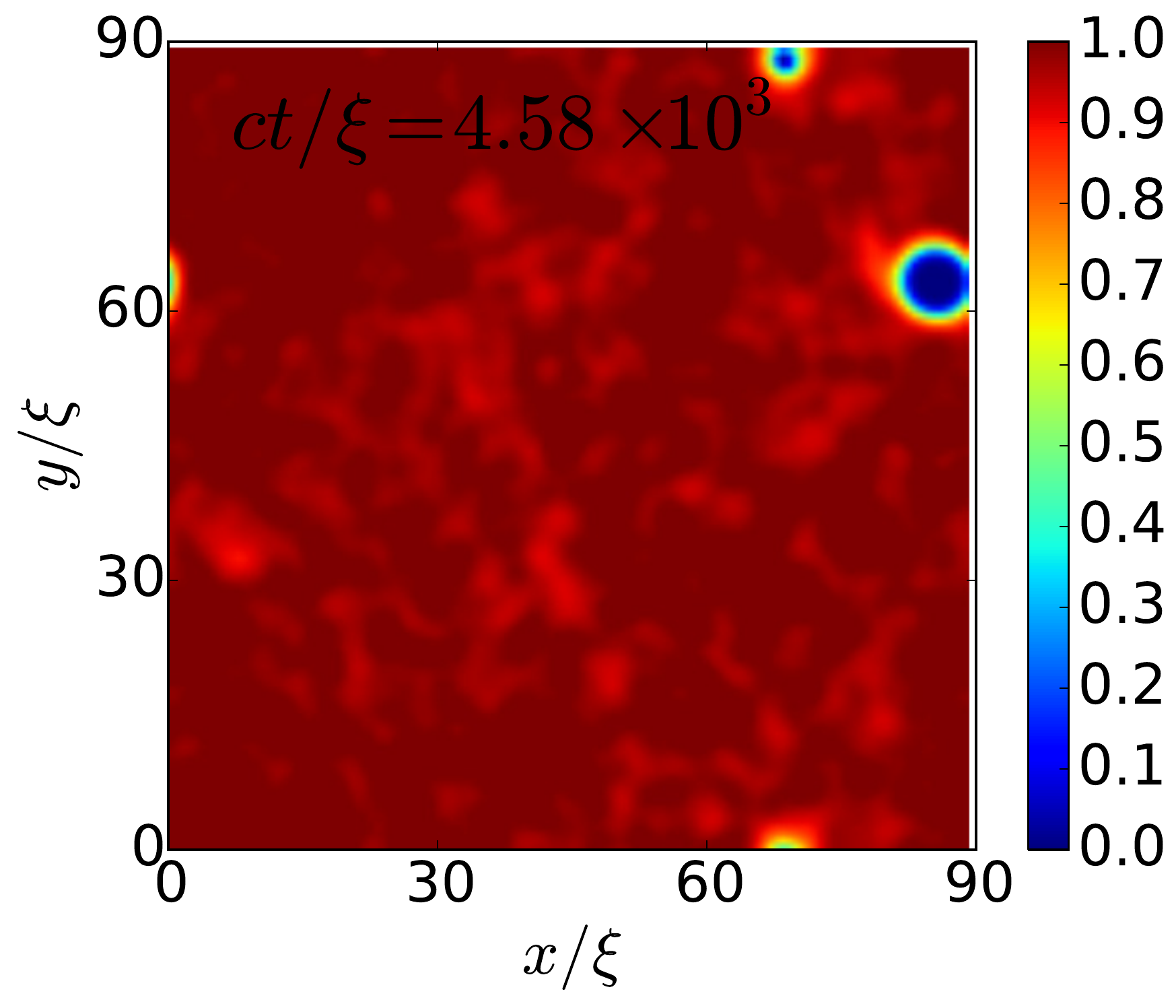}
\put(10,10){\large{\bf (l)}}
\end{overpic}
\caption{\small (Color online) Spatiotemporal evolution of the density field
$\rho(\mathbf{r},t)$ shown via pseudocolor plots, for a heavy particle
placed in the path of the positive (upper) vortex of a translating
vortex-antivortex pair (initial configuration $\tt ICP2A$).}
\label{fig:1partpairtranslpdH}
\end{figure*}

\begin{figure*}[h]
\centering
\resizebox{\linewidth}{!}{
\includegraphics[height=4.5cm,unit=1mm]{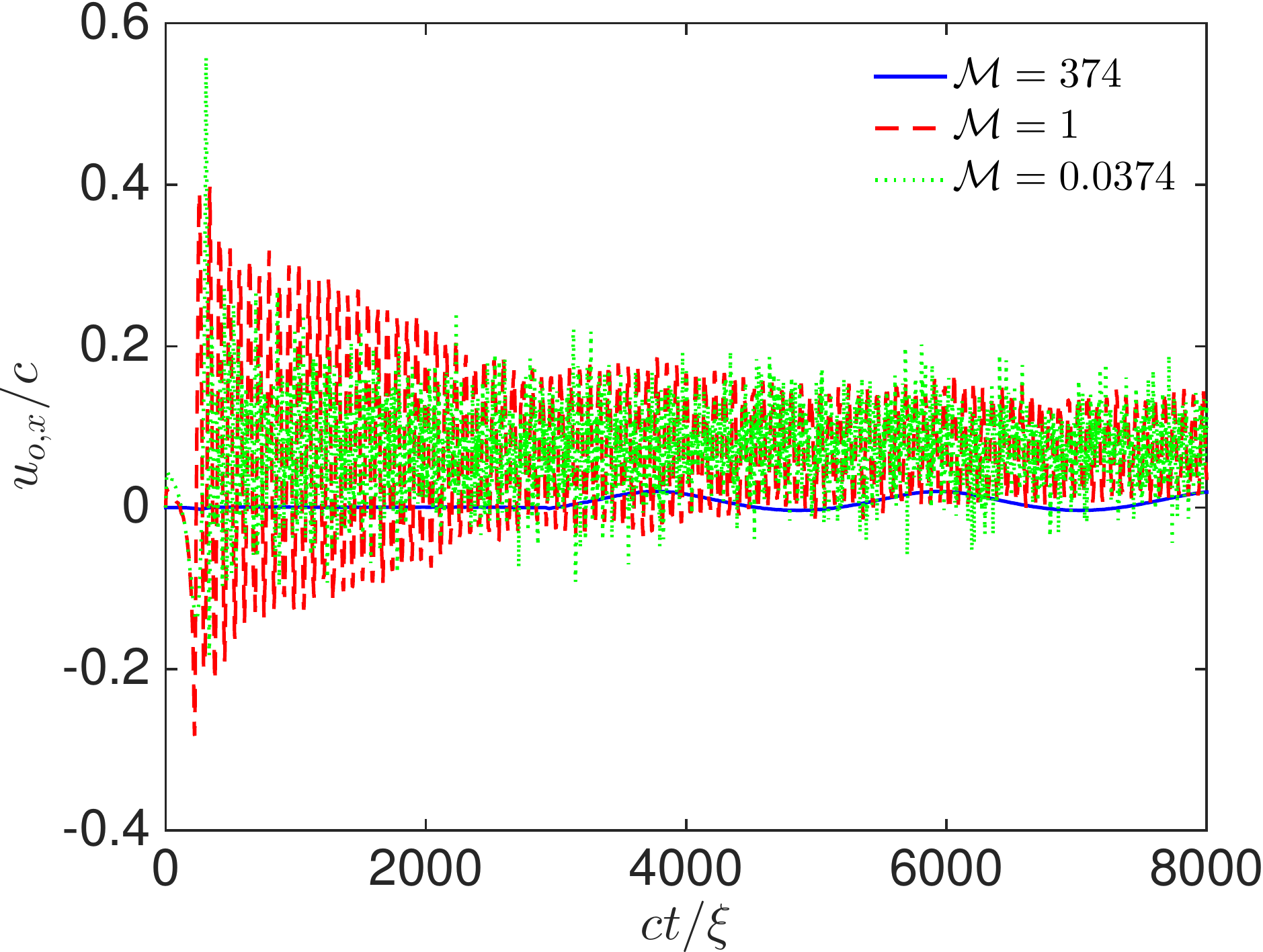}
\put(-85,20){\bf (a)}
\hspace{0.10 cm}
\includegraphics[height=4.5cm,unit=1mm]{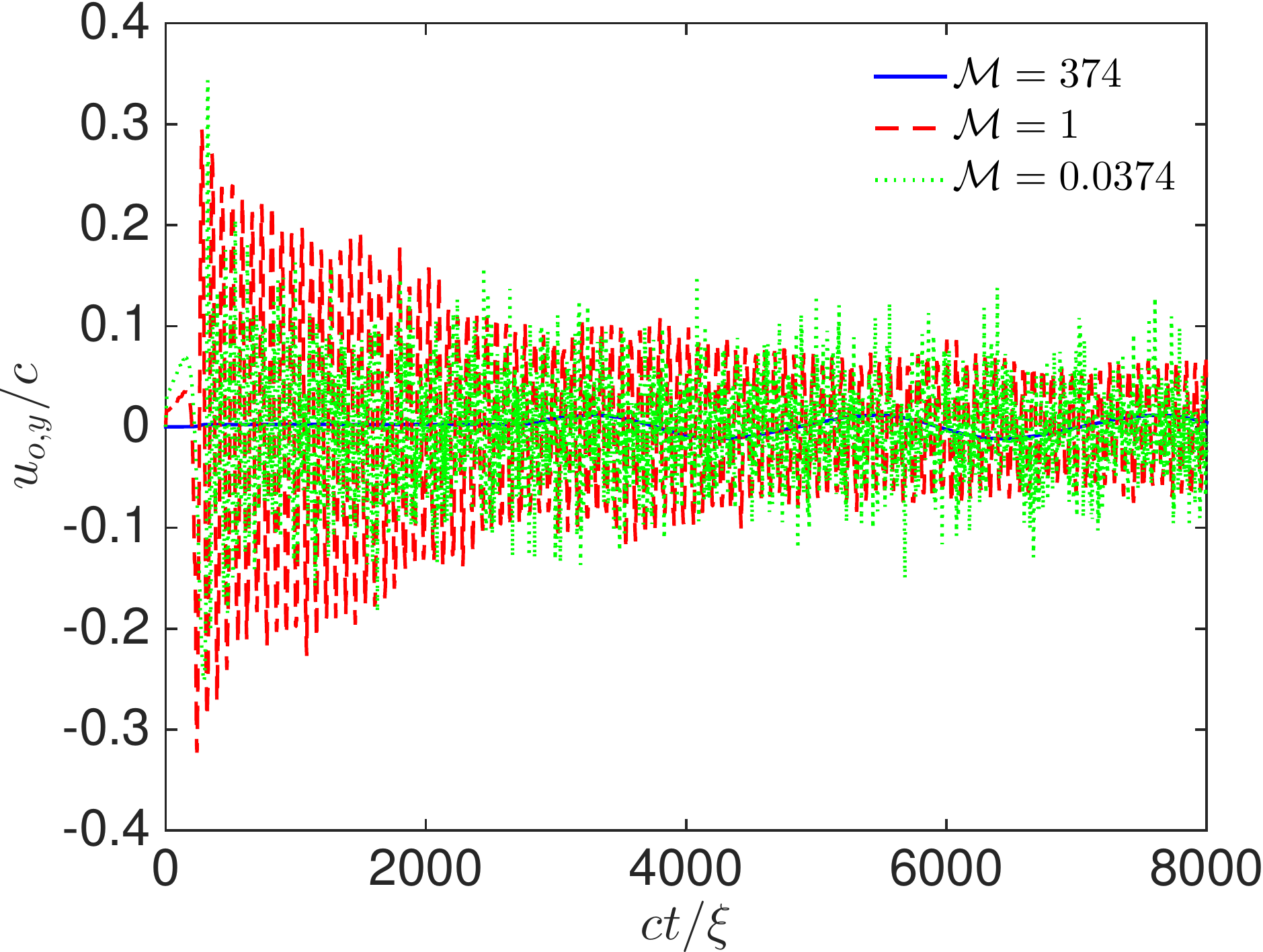}
\put(-85,20){\bf (b)}
}
\\
\vspace{0.3cm}
\resizebox{\linewidth}{!}{
\includegraphics[height=4.5cm,unit=1mm]{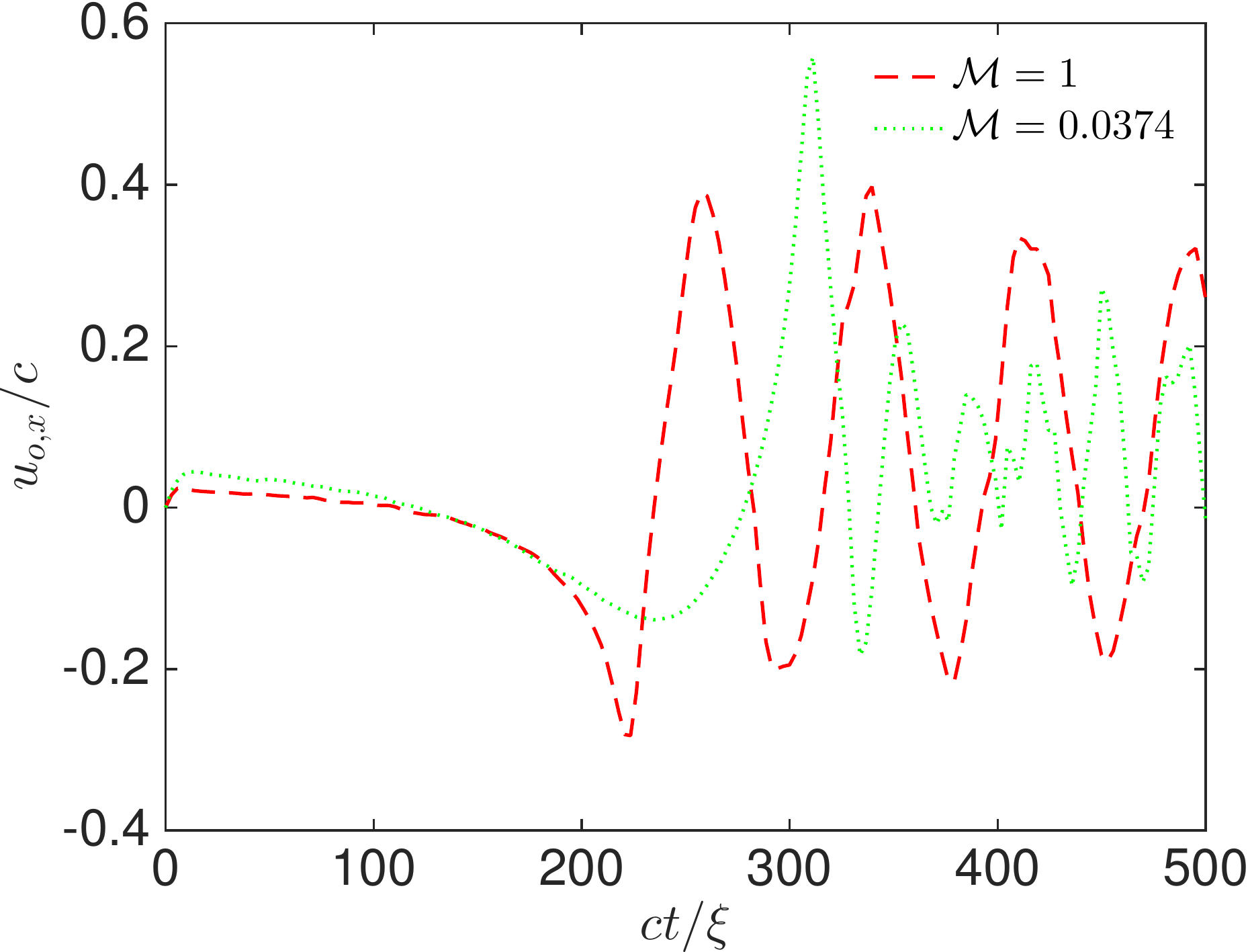}
\put(-130,20){\bf (c)}
\put(-143,72){\includegraphics[scale=0.11]{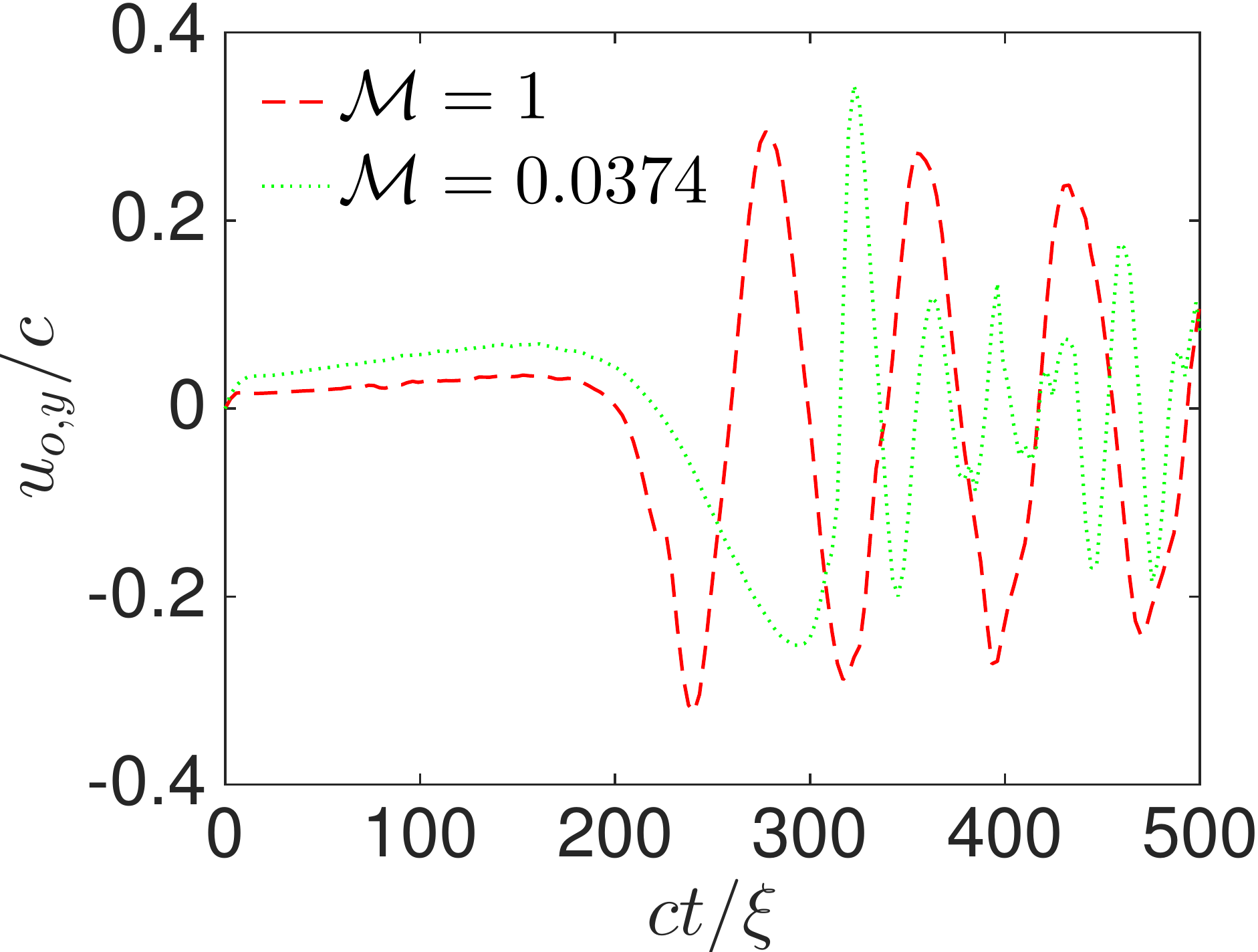}}
\hspace{0.25 cm}
\includegraphics[height=4.5cm,unit=1mm]{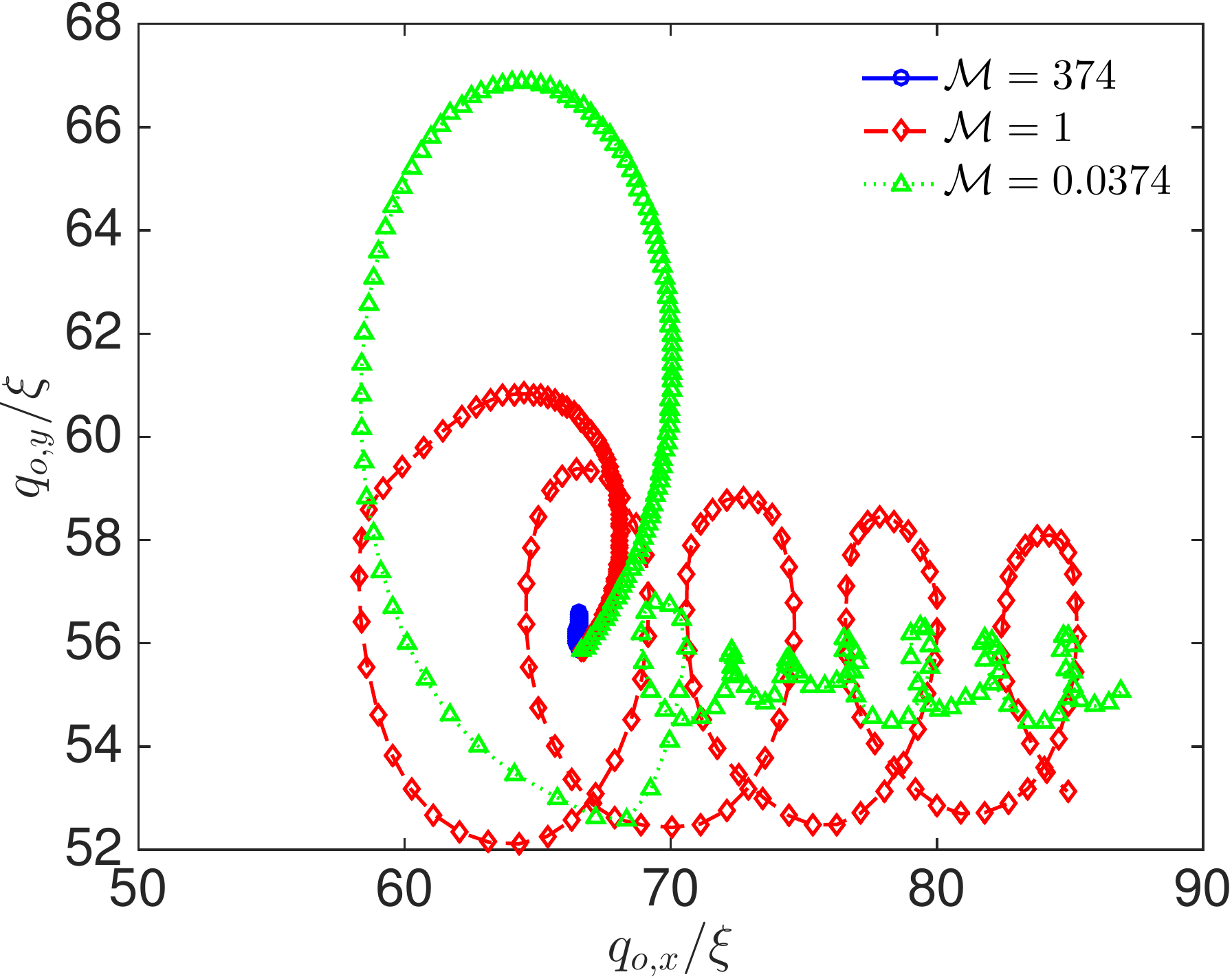}
\put(-130,20){\bf (d)}
}
\caption{\small (Color online) Plots versus time $t$ of 
(a) $u_{\rm o,x}$ and (b) $u_{\rm o,y}$	for heavy
($\mathcal{M}=374$, blue curve), neutral ($\mathcal{M}=1$, red curve), and
light ($\mathcal{M}=0.0374$, green curve) particles, placed in the path of
the positive (upper) vortex of a translating vortex-antivortex pair (initial
configuration $\tt ICP2A$). The initial stages of motion is emphasized by
separately plotting $u_{\rm o,x}$ in (c) and $u_{\rm o,y}$ in the inset for
the neutral and light particles; (d) contains the trajectories 
$(q_{\rm o,x}/\xi,q_{\rm o,y}/\xi)$ for the three particle types
during this time.}
\label{fig:1partpairtranslqu}
\end{figure*}

When the
vortex-antivortex pair approaches the heavy particle, the positive (upper)
vortex glides over the particle, which leads to an exchange of momentum, so the
vortex-antivortex pair is deflected from its path, and it acquires a small
velocity in the $-\hat{\mathbf{y}}$ direction. This interaction leads to the
production of sound waves.  Subsequently, when the negative vortex (of the
vortex-antivortex pair) comes near the heavy particle and, in the presence of
sound waves, it is finally trapped on the negative vortex. During
the trapping of the particle on the negative vortex, a large amount of acoustic
energy is released into the system. This sequence of events is illustrated by
the pseudocolor plots of Figs.~\ref{fig:1partpairtranslpdH}~(a)-(l) and Video
M4~\cite{suppmat}.  The trapped particle executes oscillatory motion while
drifting (see Fig.~\ref{fig:1partpairtranslqu} purple curve for $t\gtrsim 200$)
and the positive vortex now revolves around the particle that is trapped on the
negative vortex.

Figures~\ref{fig:1partpairtranslqu}~(a) and (b) show plots of $u_{\rm o,x}$
and $u_{\rm o,y}$ versus time, respectively, for heavy, neutral, and light particles.

\begin{figure*}
\centering
\begin{overpic}
[height=4.5cm,unit=1mm]{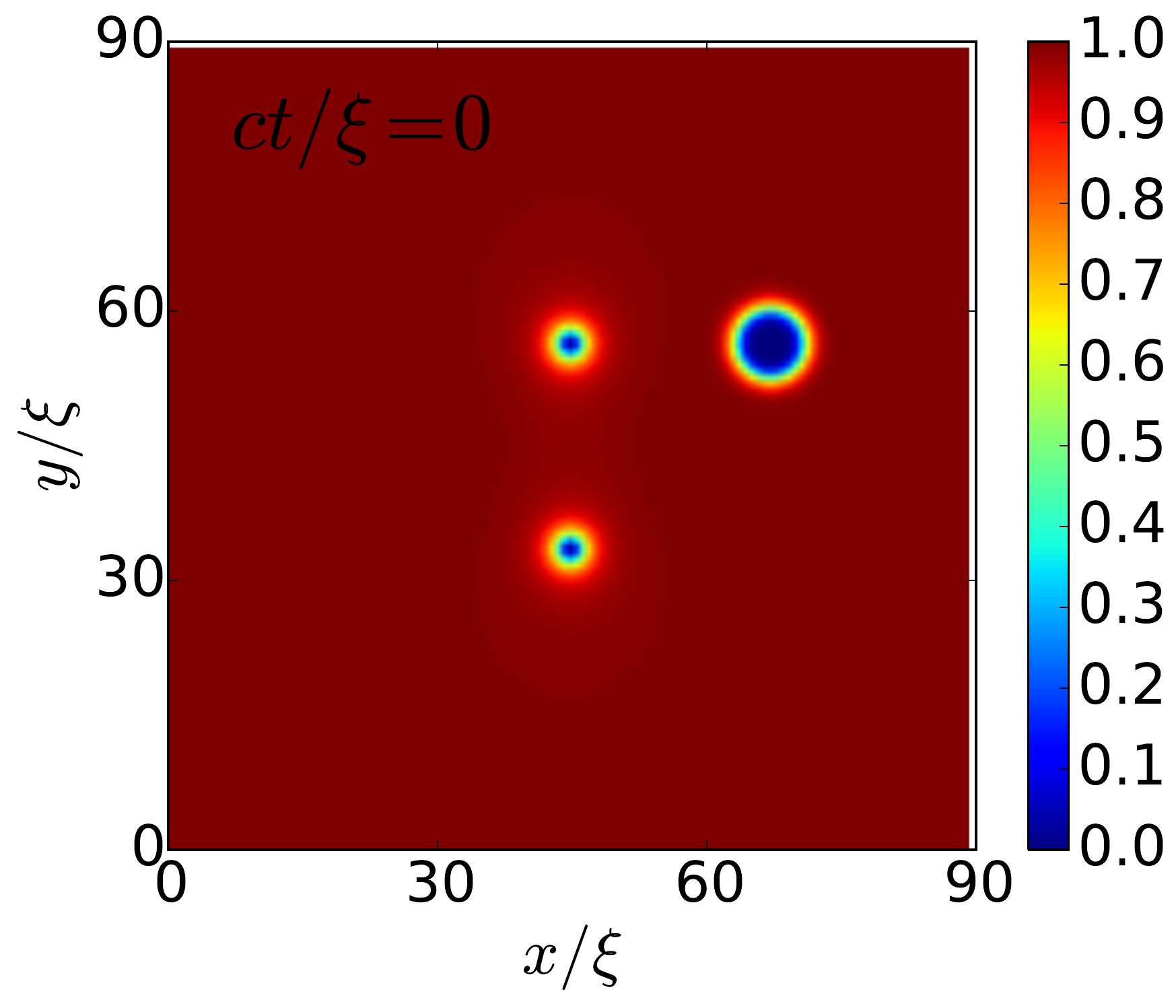}
\put(10.,10){\large{\bf (a)}}
\end{overpic}
\begin{overpic}
[height=4.5cm,unit=1mm]{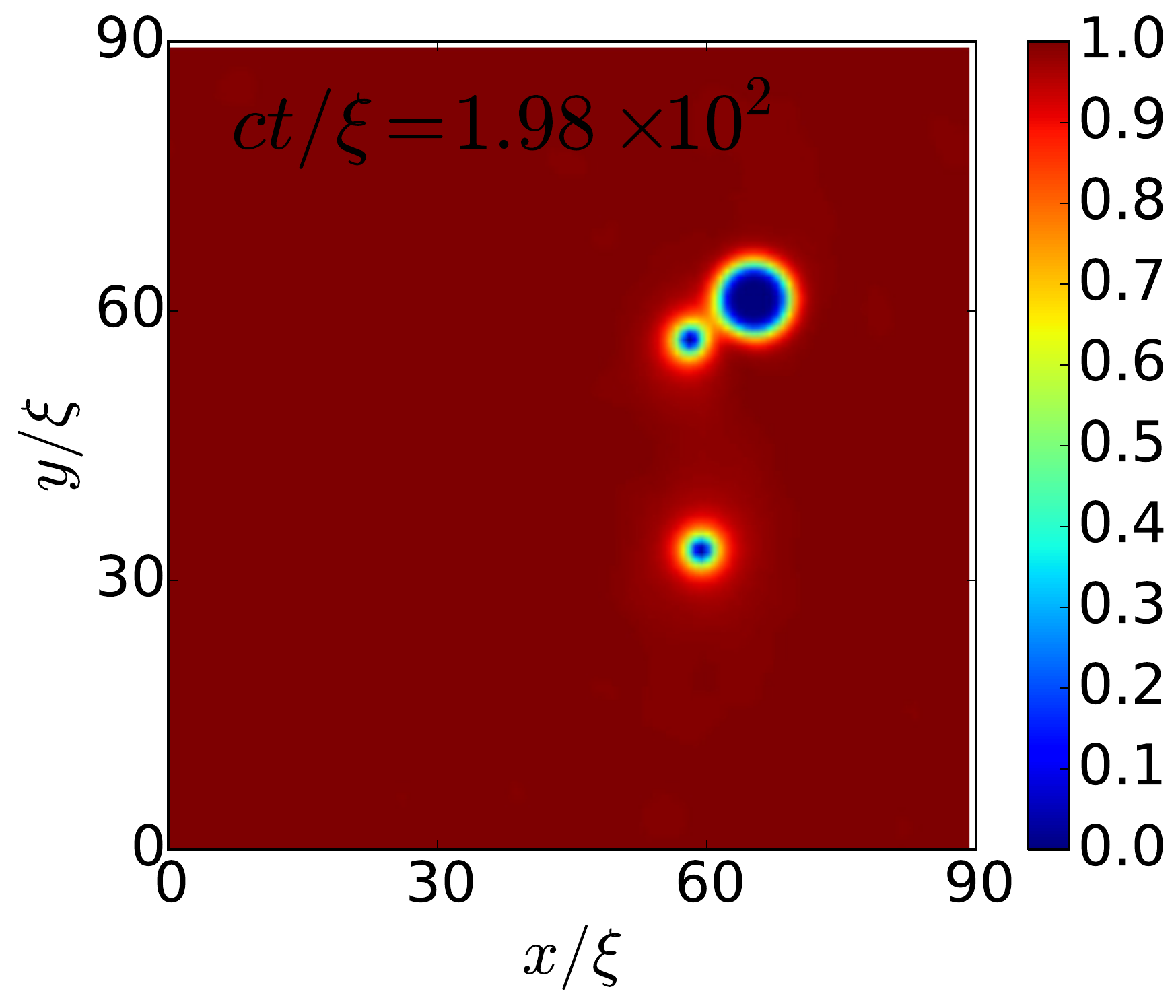}
\put(10,10){\large{\bf (b)}}
\end{overpic}
\begin{overpic}
[height=4.5cm,unit=1mm]{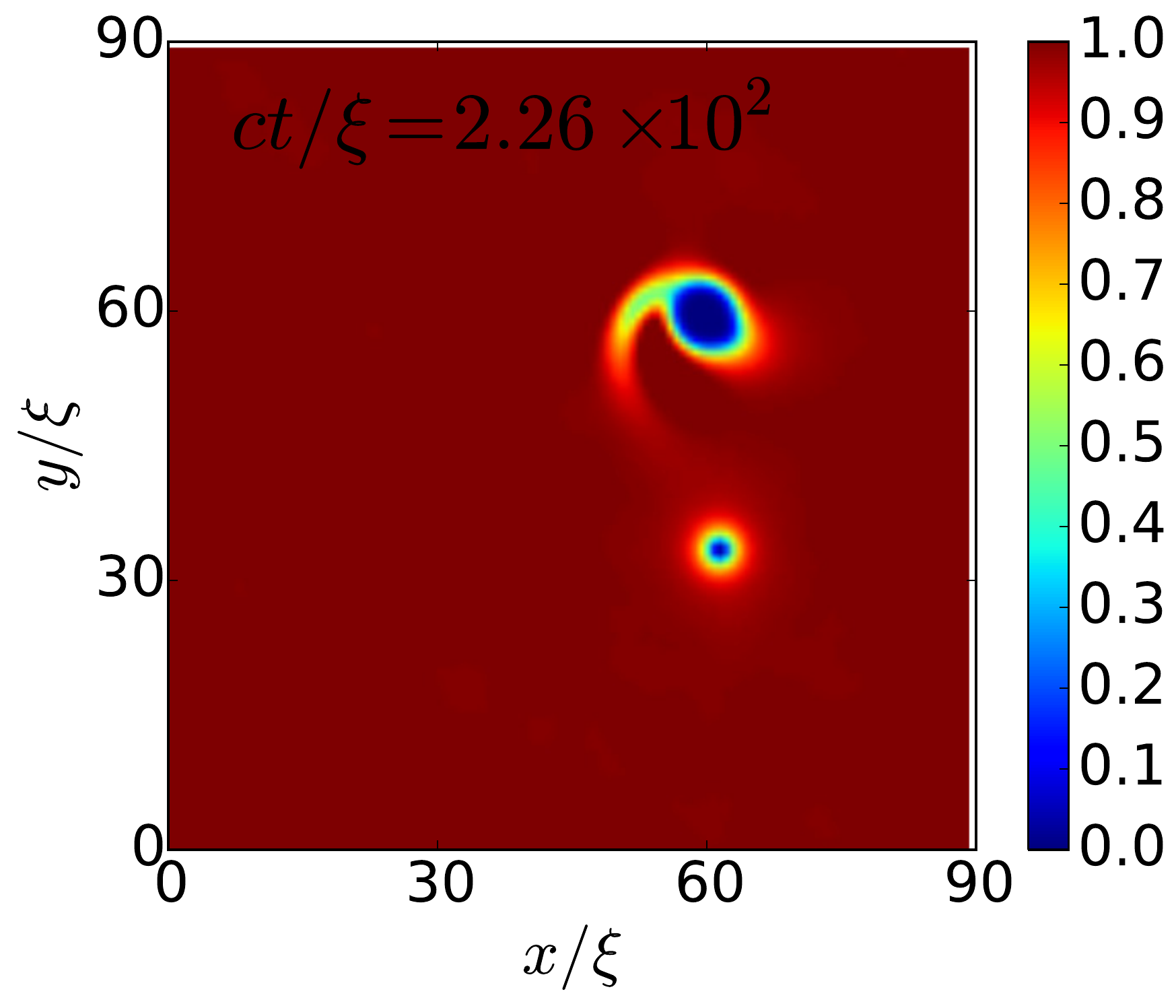}
\put(10,10){\large{\bf (c)}}
\end{overpic}
\\
\begin{overpic}
[height=4.5cm,unit=1mm]{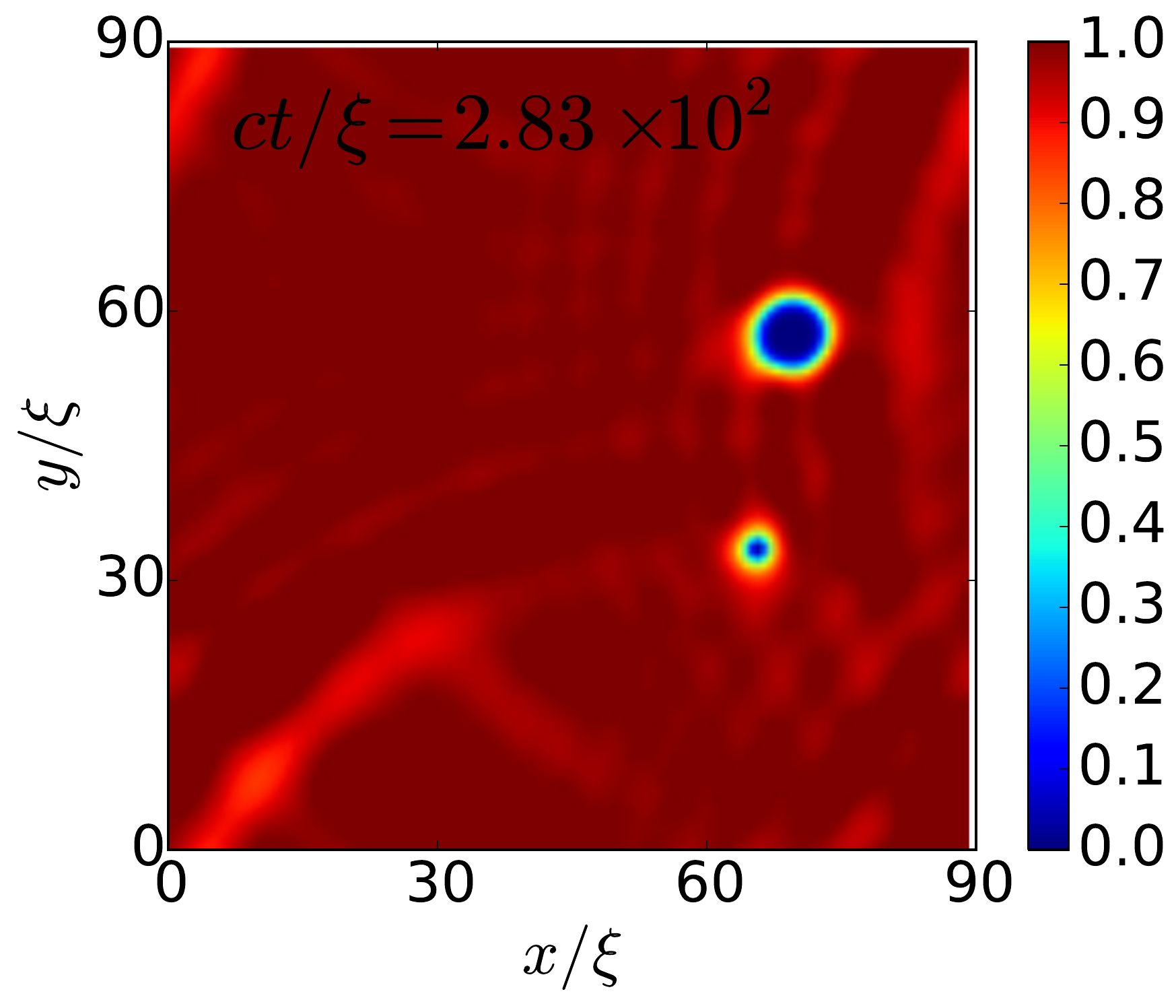}
\put(10.,10){\large{\bf (d)}}
\end{overpic}
\begin{overpic}
[height=4.5cm,unit=1mm]{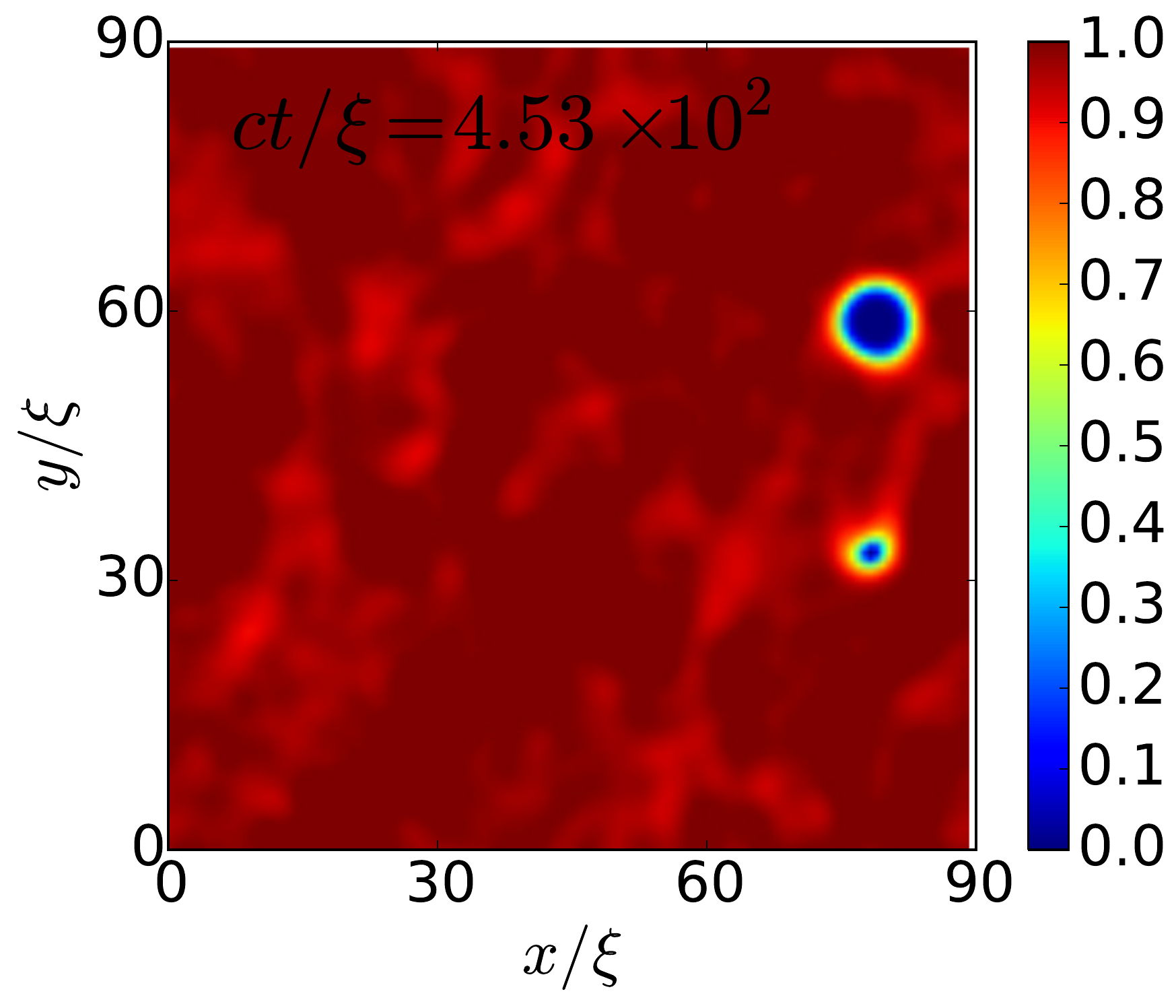}
\put(10,10){\large{\bf (e)}}
\end{overpic}
\begin{overpic}
[height=4.5cm,unit=1mm]{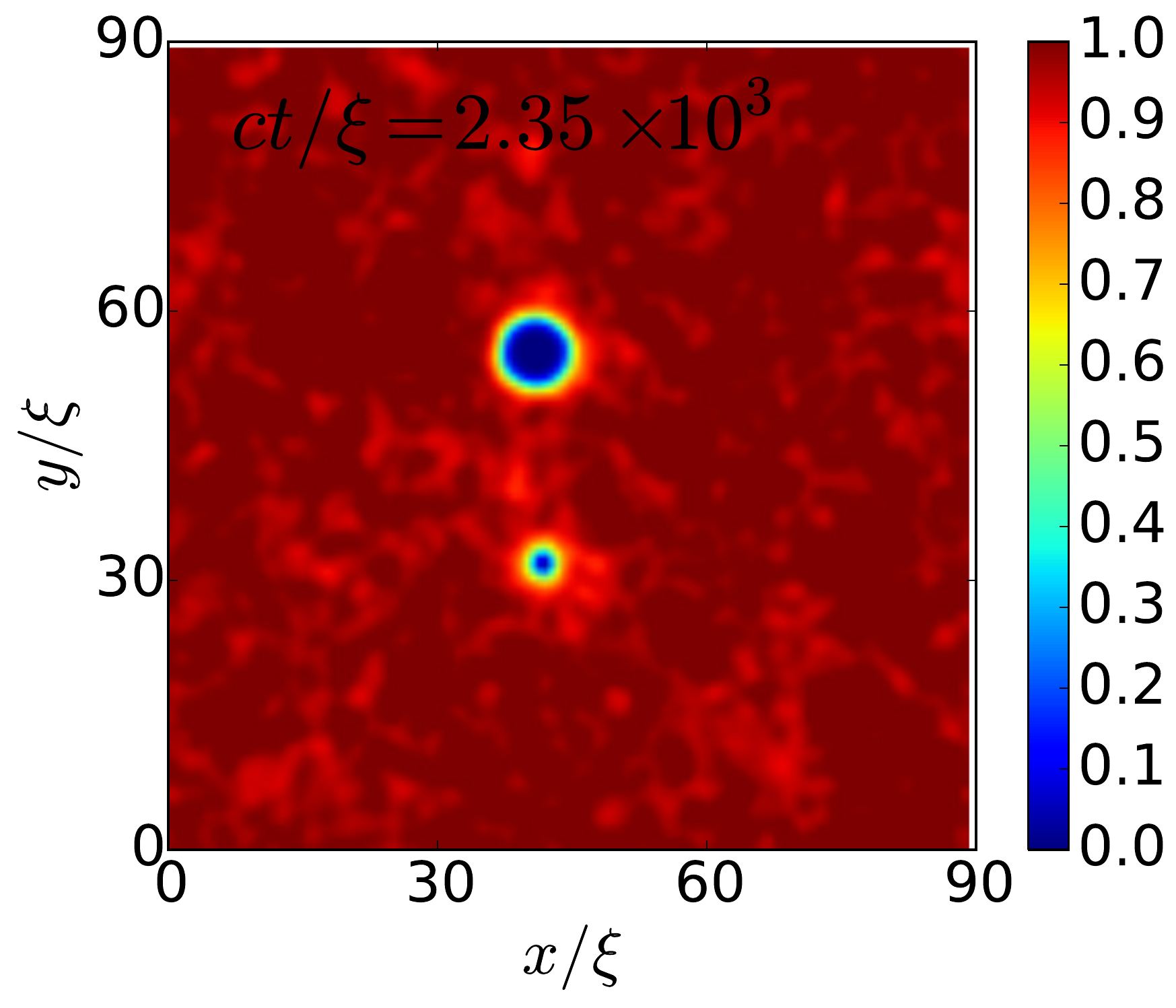}
\put(10,10){\large{\bf (f)}}
\end{overpic}
\caption{\small (Color online) Spatiotemporal evolution of the density field
$\rho(\mathbf{r},t)$ shown via pseudocolor plots, for a neutral particle
placed in the path of the positive (upper) vortex of a translating
vortex-antivortex pair (initial configuration $\tt ICP2A$).}
\label{fig:1partpairtranslpdN}
\end{figure*}

\begin{figure*}
\centering
\resizebox{\linewidth}{!}{
\includegraphics[height=4.5cm,unit=1mm]{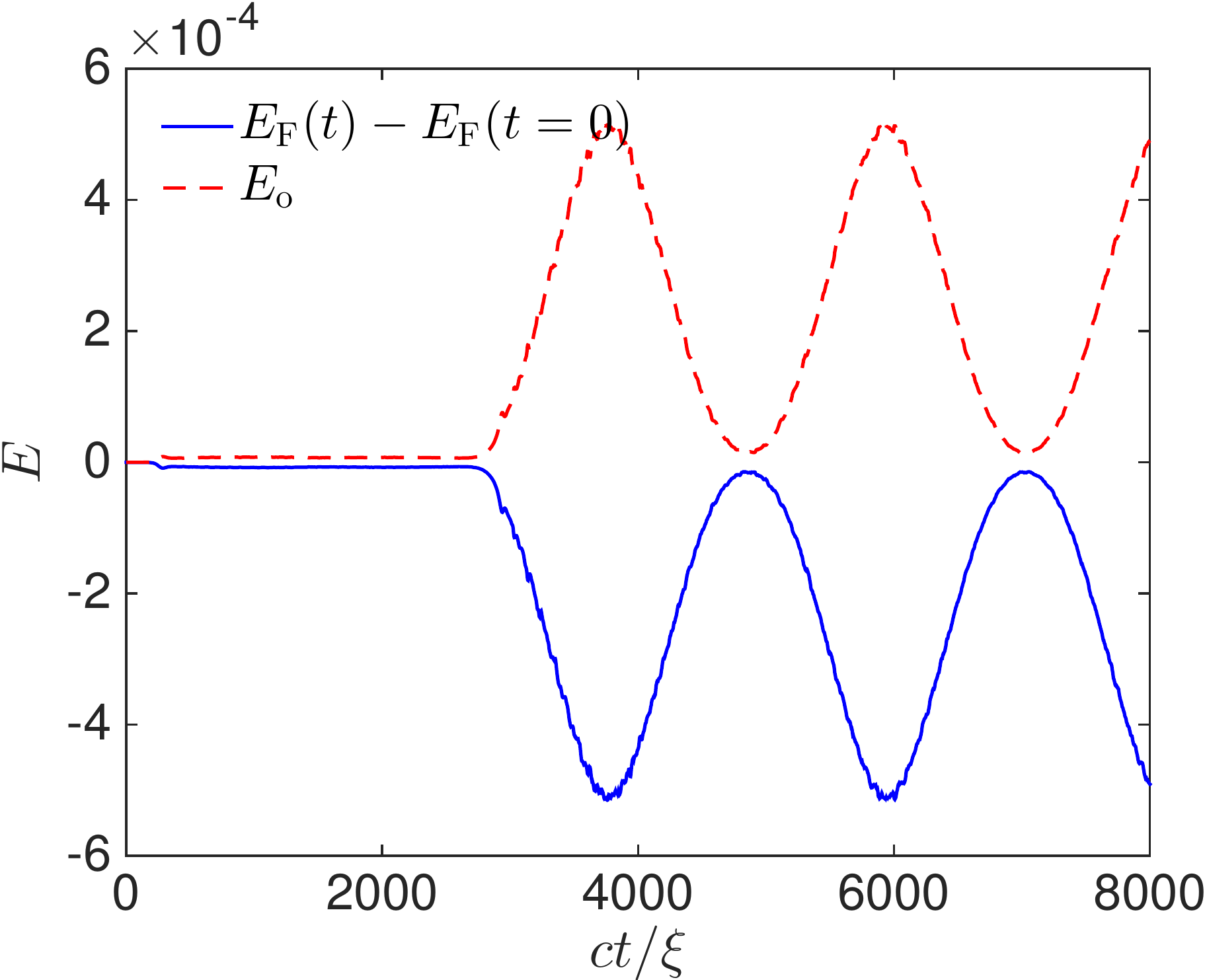}
\put(-120,30){\bf (a)}
\hspace{0.15 cm}
\includegraphics[height=4.5cm,unit=1mm]{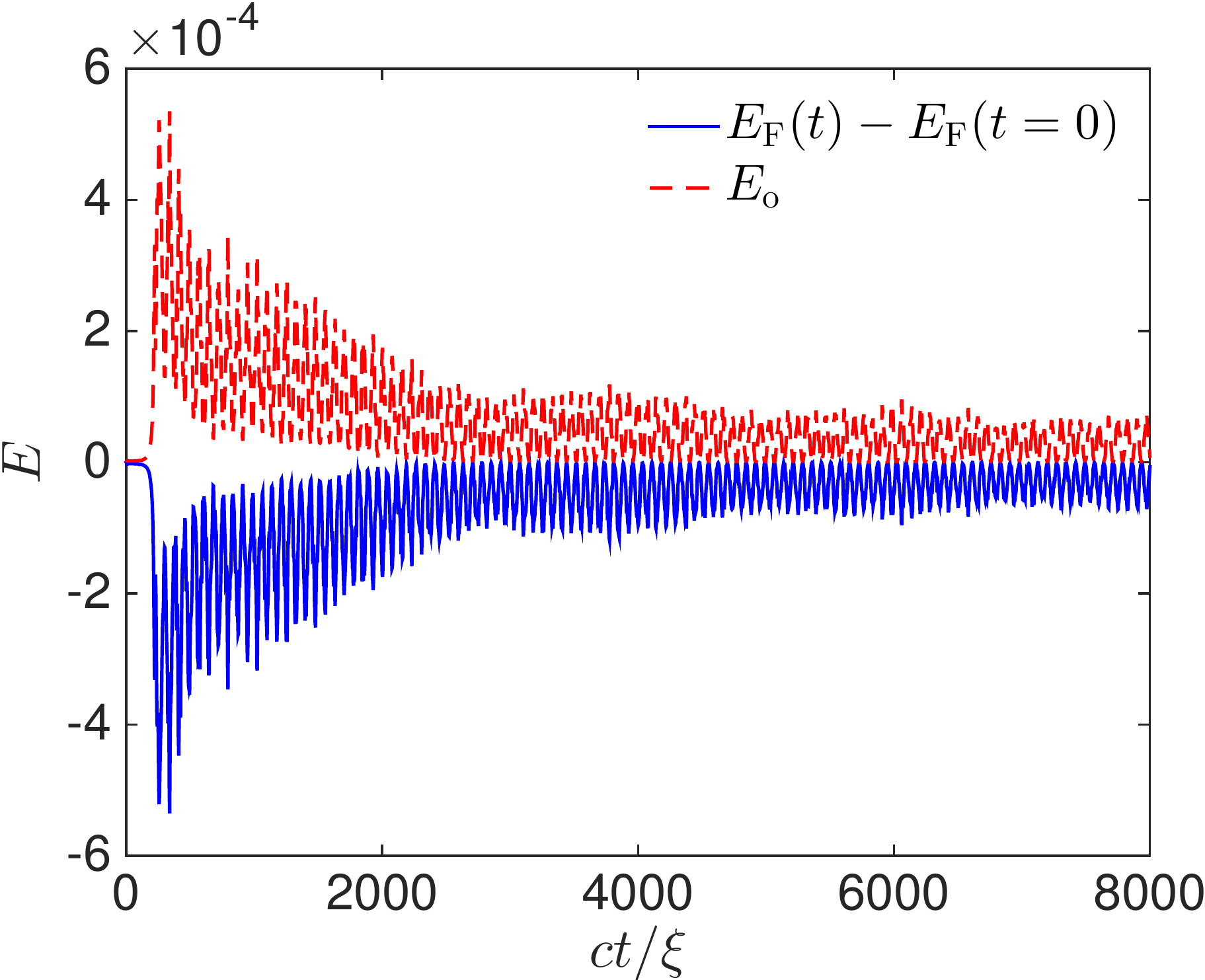}
\put(-120,30){\bf (b)}
\hspace{0.15 cm}
\includegraphics[height=4.5cm,unit=1mm]{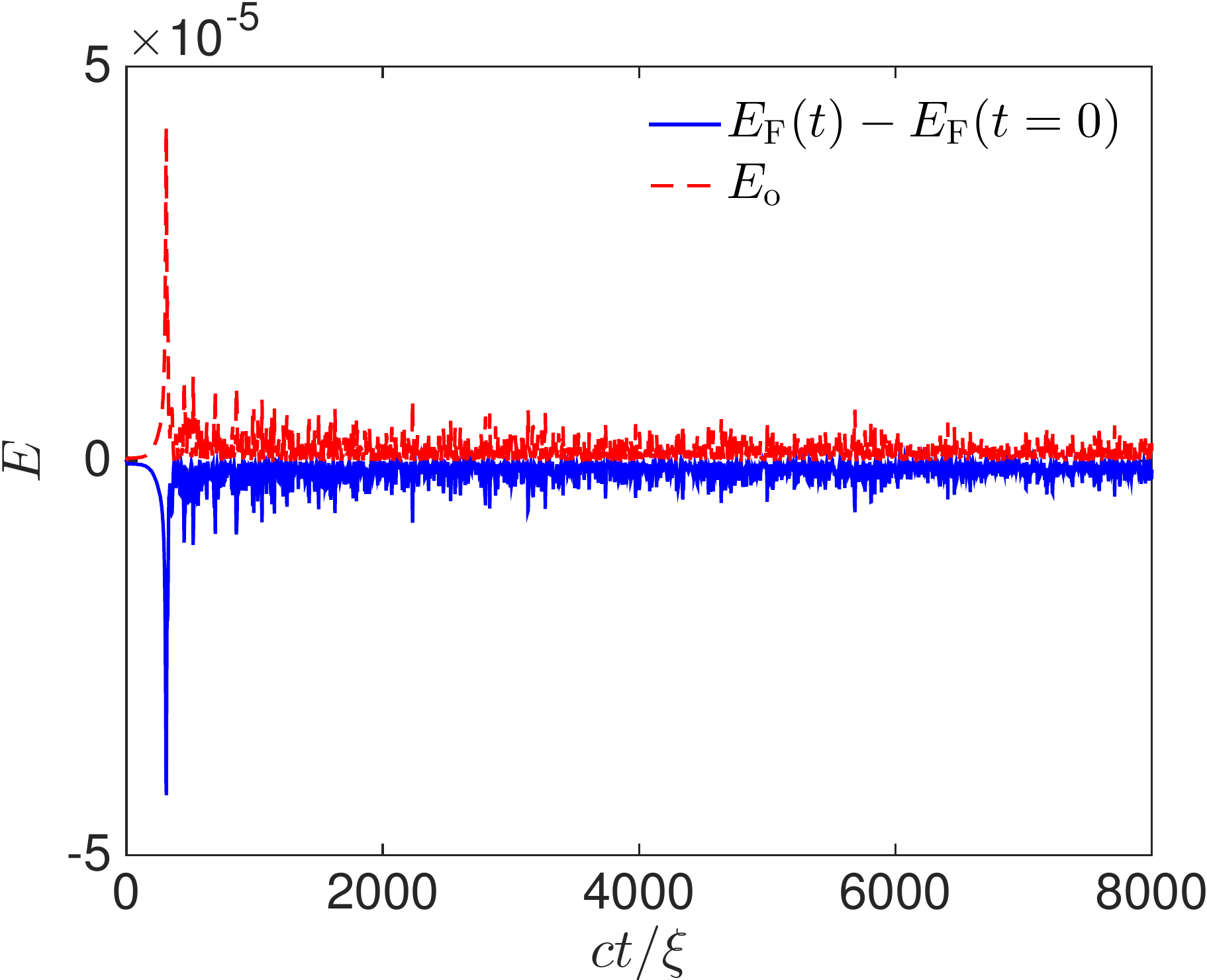}
\put(-120,30){\bf (c)}
}
\caption{\small (Color online) Plots versus time $t$ of the energy components
$\delta E_{\rm field}=E_{\rm F}(t)-E_{\rm F}(t=0)$ (blue solid curve) and 
$E_{\rm o}$ (red dashed curve) for (a) heavy, (b) neutral, and (c) light
particles, placed in the path of the positive (upper) vortex
of a translating vortex-antivortex pair (initial configuration $\tt ICP2A$).
}
\label{fig:1partpairtranslEcomps}
\end{figure*}

When a translating vortex-antivortex pair approaches neutral or light
particles, they feel the flow around the positive (upper) vortex more strongly
than did the heavy particle. The neutral and light particles are pushed out and
they move around the positive vortex before getting trapped on the positive
vortex; in Fig.~\ref{fig:1partpairtranslqu} (c), inset of (c) and (d) we show
this initial motion by plotting the
particle velocity components $u_{\rm o,x}$, $u_{\rm o,y}$ and the trajectories,
respectively.
The response of the light particle is most dramatic:
while moving around the positive vortex, 
it is pushed almost to the back of the vortex ($ct/\xi\simeq 2.83\times 10^2$), 
before getting trapped on it.
The pseudocolor plots of Figs.~\ref{fig:1partpairtranslpdN}~(a)-(f) (and
\ref{fig:1partpairtranslpdL}~(a)-(i) in the Appendix)  and
the Videos M5 and M6 summarize the spatiotemporal evolutions of the neutral
and light particles, respectively. When the neutral and the light particles get
trapped on the positive vortex, there is a sudden change in their velocities,
as we show in Figs.~\ref{fig:1partpairtranslqu}~(a)-(c) at $ct/\xi\simeq 2.00\times10^2$;
this results in large fluctuations.  Neutral particles exhibit modulated
oscillations, whereas light particles display chaotic temporal evolution.  In
Figs.~\ref{fig:1partpairtranslEcomps}~(a), (b), and (c) we plot the energy time
series for the heavy, neutral and light particles, respectively. These plots 
illustrate the continual exchange of energy between the particle and the
superfluid, which is very much representative of the motion executed by the particles.

\begin{figure*}
\centering
\begin{overpic}
[height=4.5cm,unit=1mm]{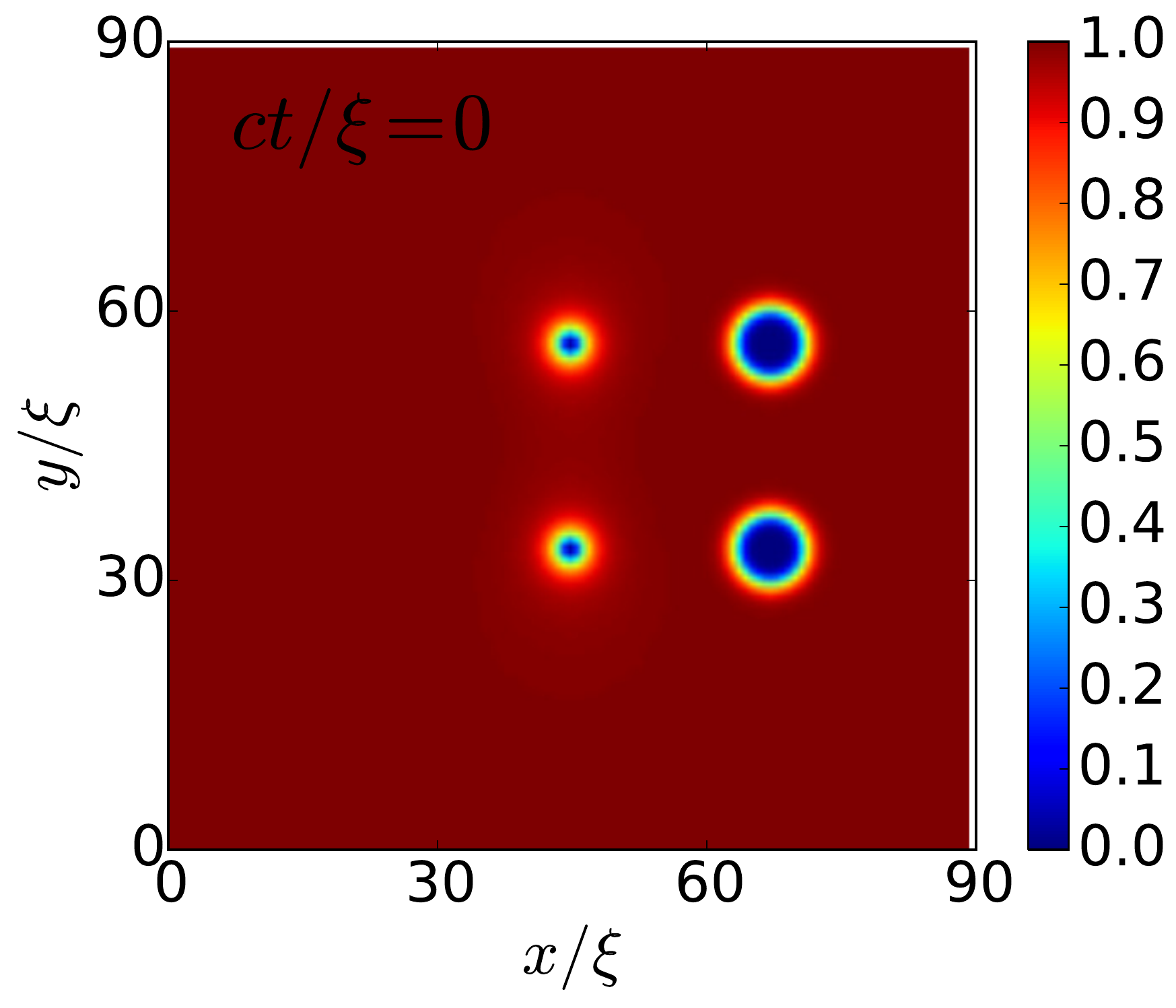}
\put(10.,10){\large{\bf (a)}}
\end{overpic}
\begin{overpic}
[height=4.5cm,unit=1mm]{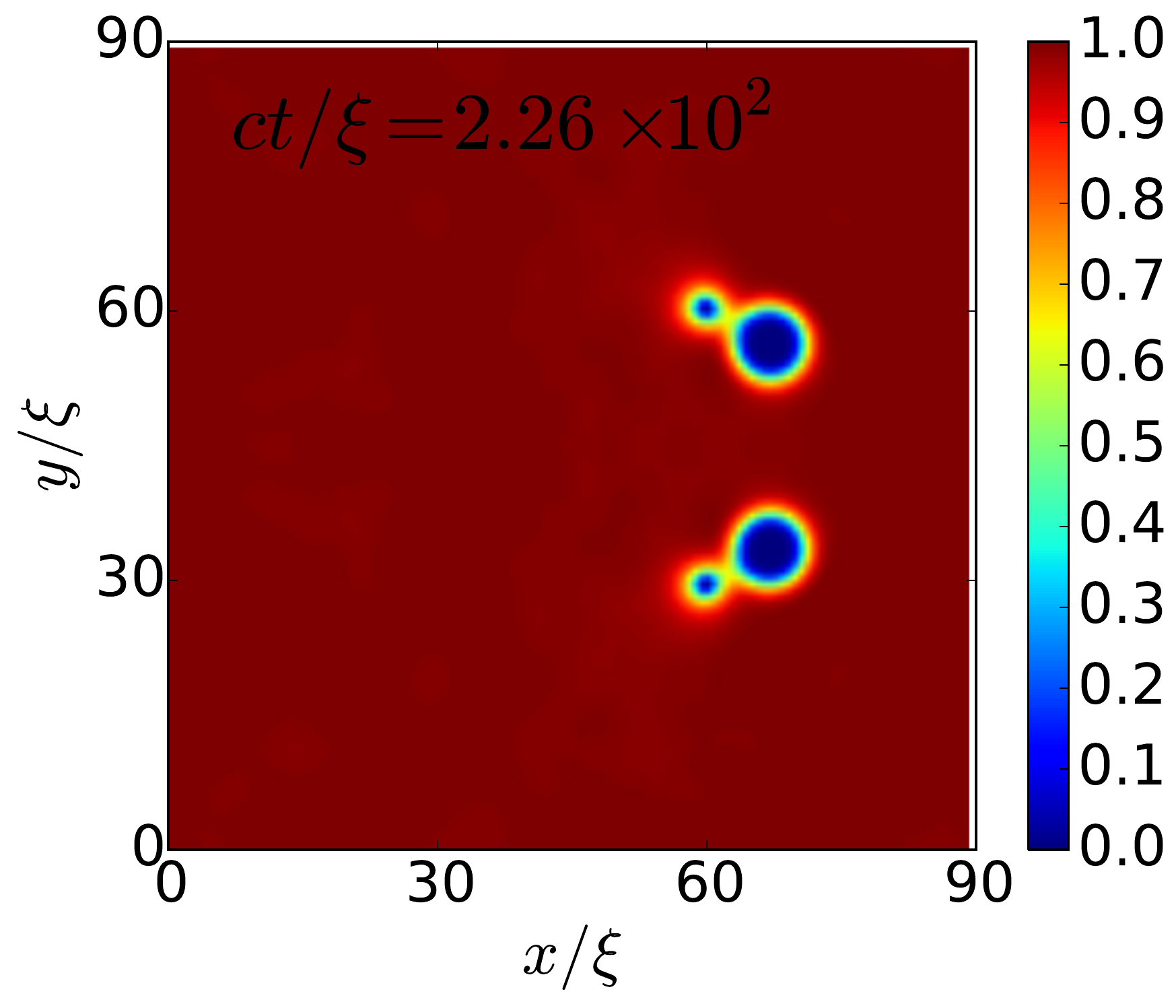}
\put(10,10){\large{\bf (b)}}
\end{overpic}
\begin{overpic}
[height=4.5cm,unit=1mm]{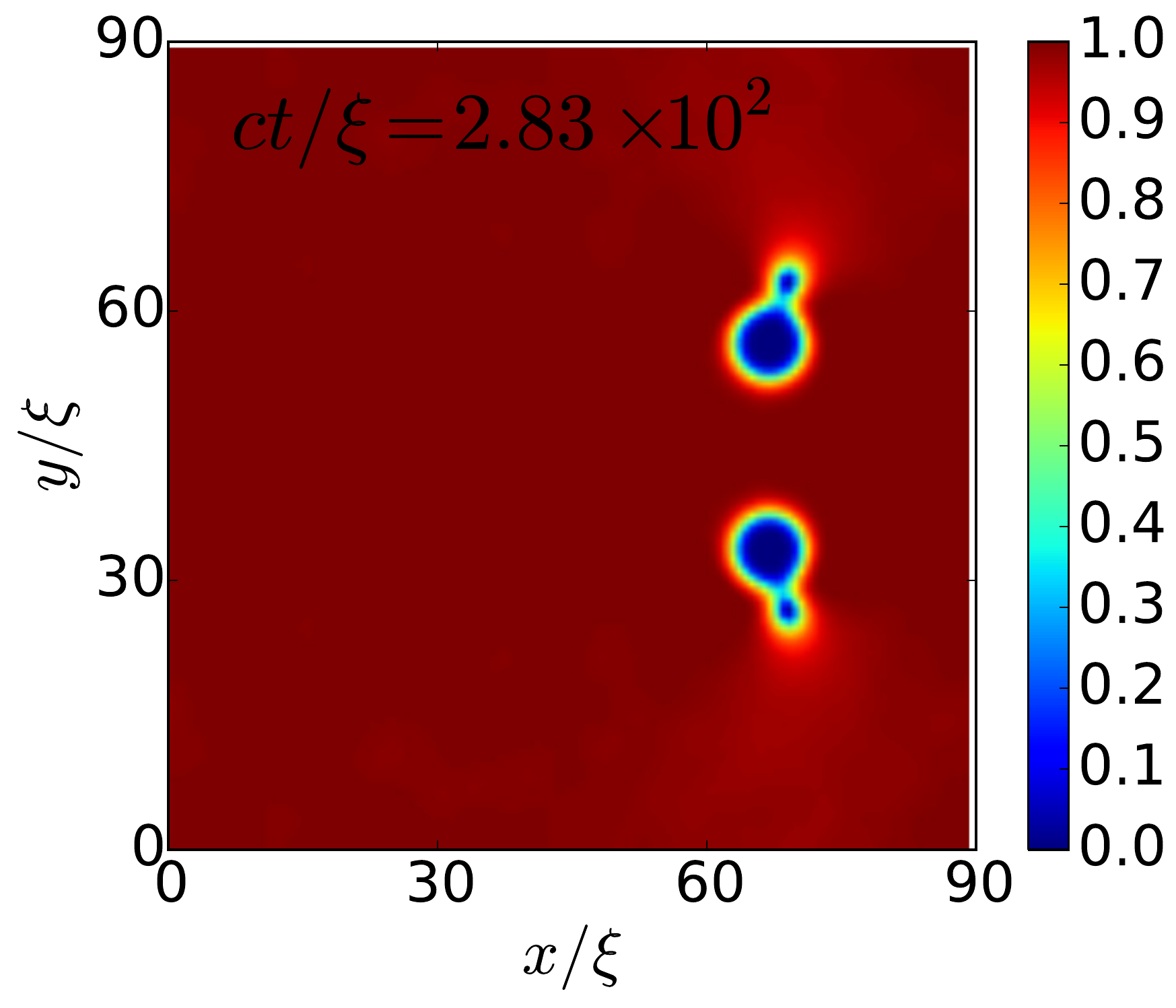}
\put(10,10){\large{\bf (c)}}
\end{overpic}
\\
\begin{overpic}
[height=4.5cm,unit=1mm]{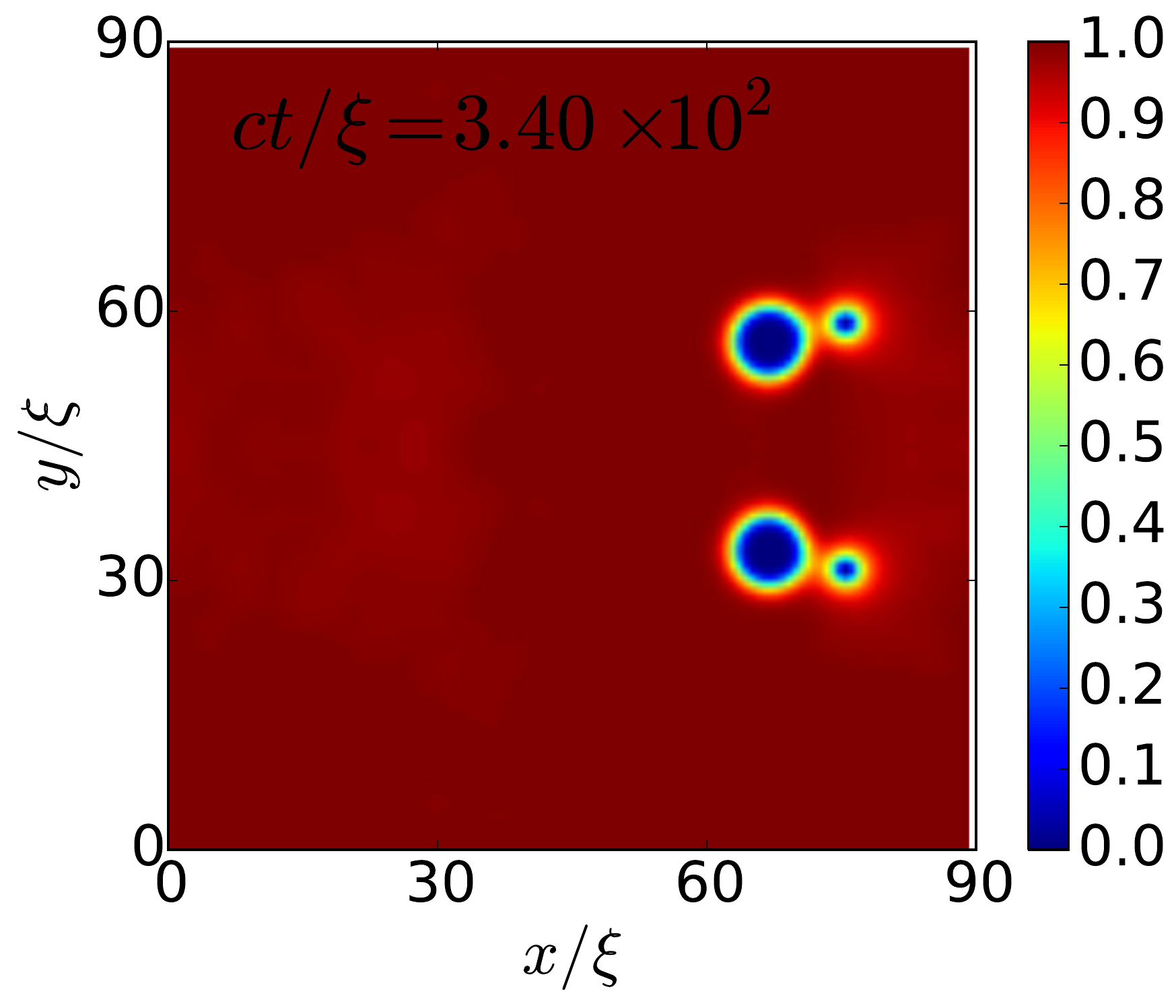}
\put(10.,10){\large{\bf (d)}}
\end{overpic}
\begin{overpic}
[height=4.5cm,unit=1mm]{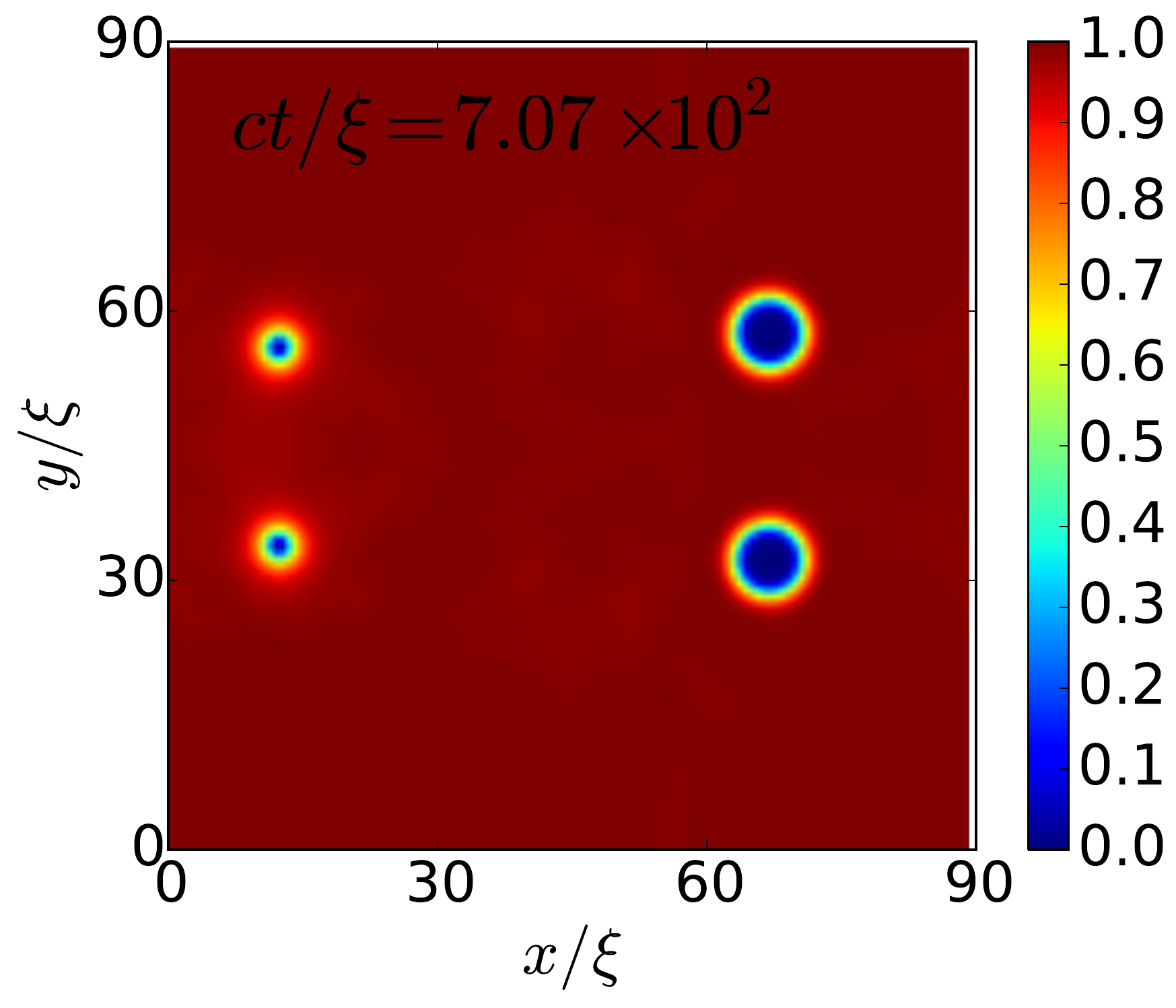}
\put(10,10){\large{\bf (e)}}
\end{overpic}
\begin{overpic}
[height=4.5cm,unit=1mm]{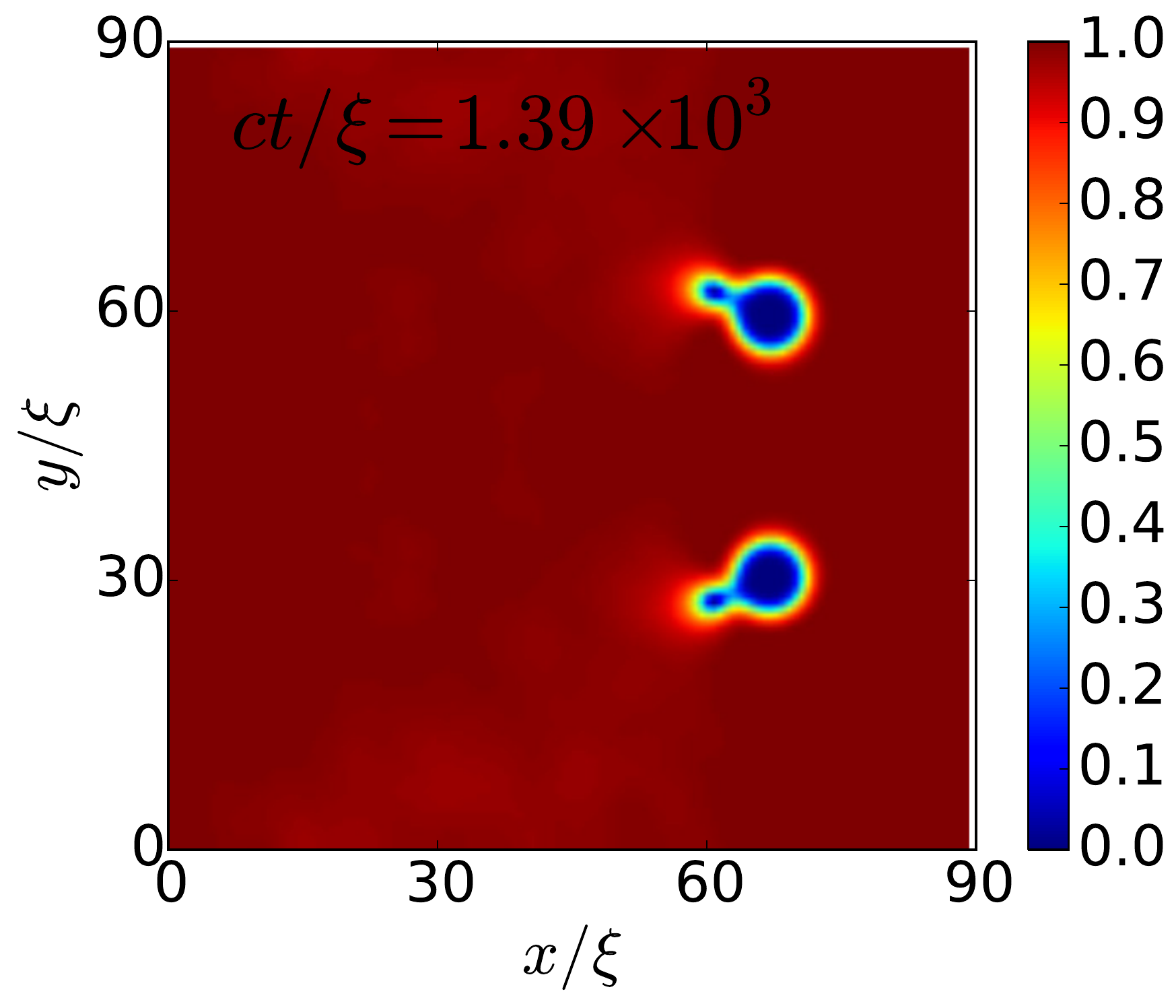}
\put(10,10){\large{\bf (f)}}
\end{overpic}
\\
\begin{overpic}
[height=4.5cm,unit=1mm]{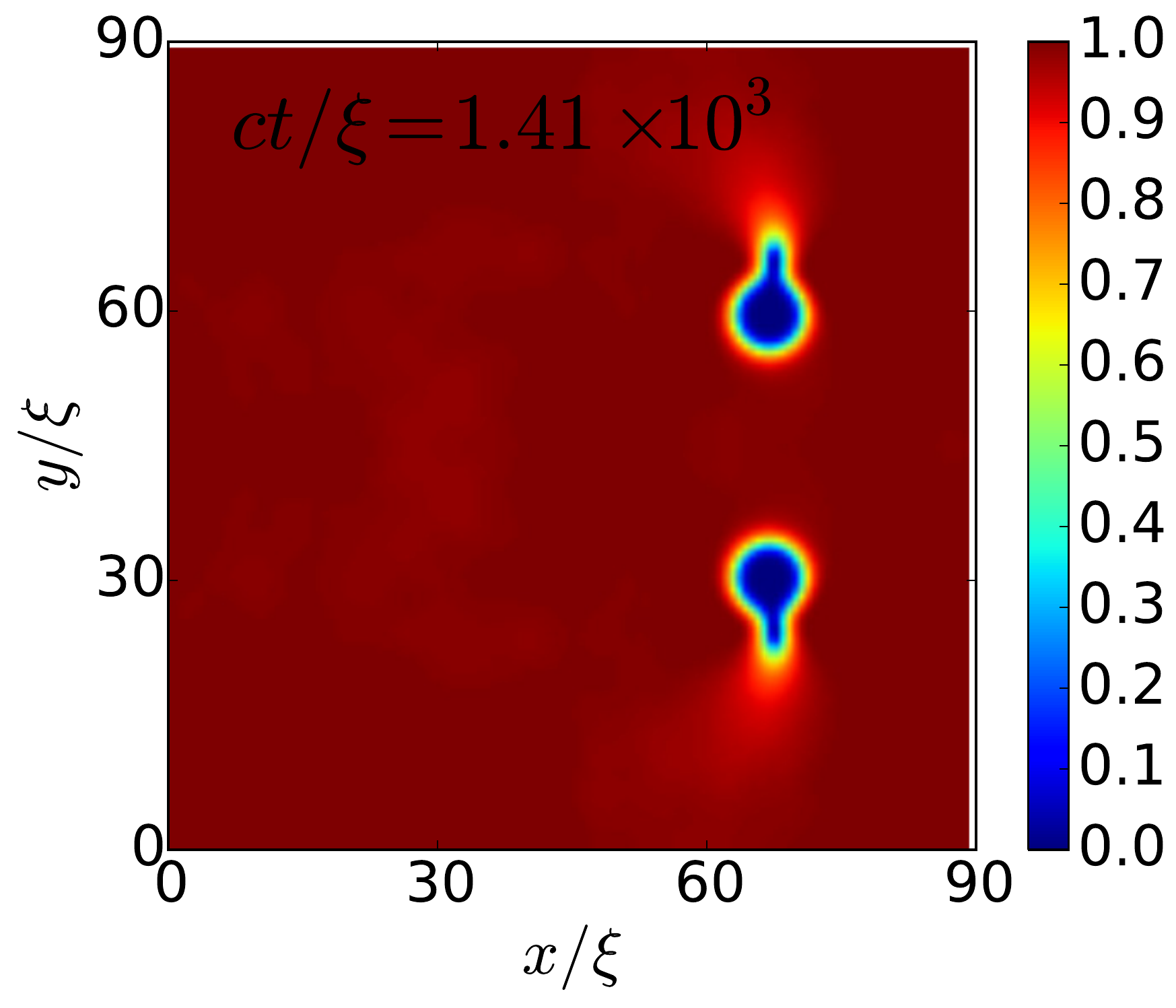}
\put(10.,10){\large{\bf (g)}}
\end{overpic}
\begin{overpic}
[height=4.5cm,unit=1mm]{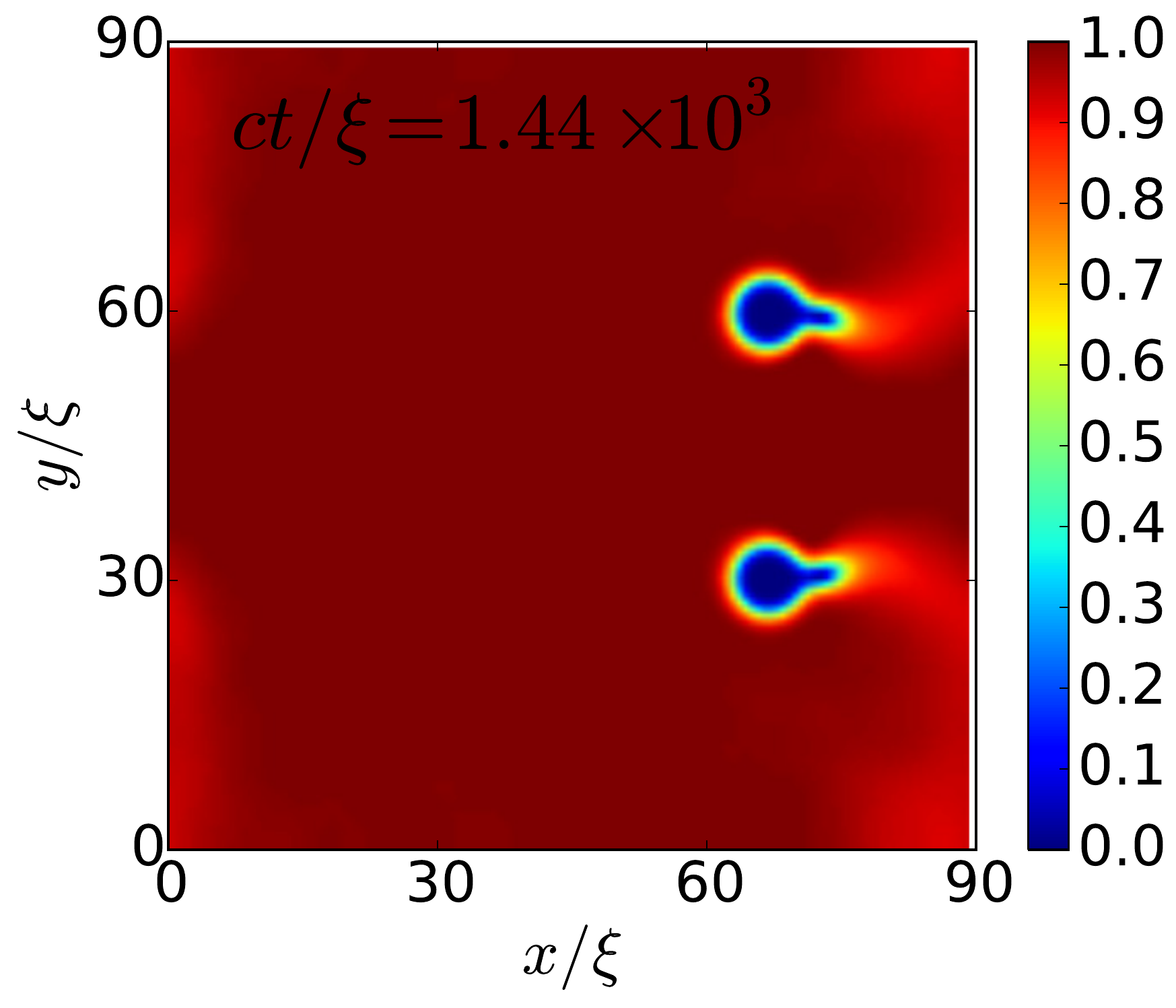}
\put(10,10){\large{\bf (h)}}
\end{overpic}
\begin{overpic}
[height=4.5cm,unit=1mm]{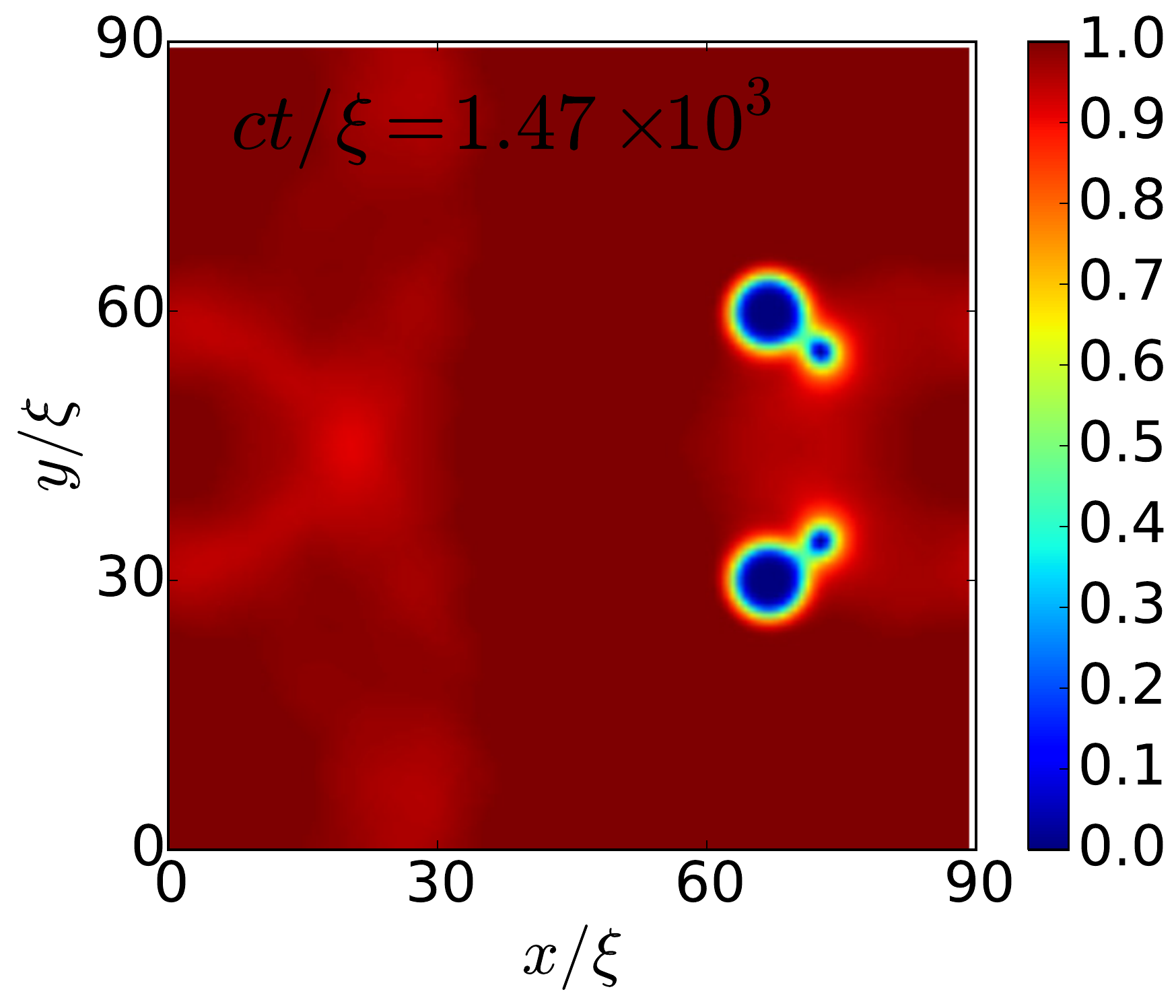}
\put(10,10){\large{\bf (i)}}
\end{overpic}
\\
\begin{overpic}
[height=4.5cm,unit=1mm]{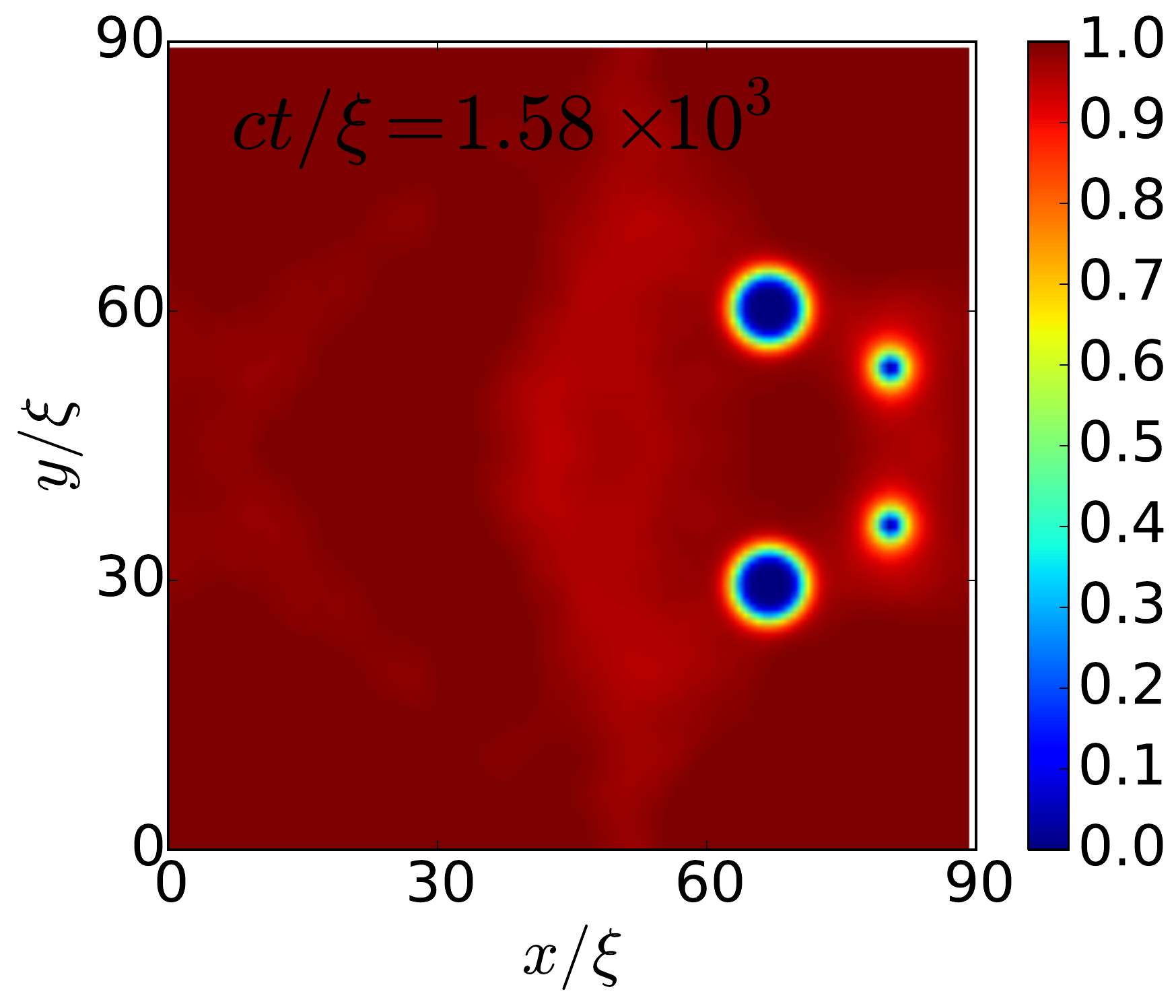}
\put(10.,10){\large{\bf (j)}}
\end{overpic}
\begin{overpic}
[height=4.5cm,unit=1mm]{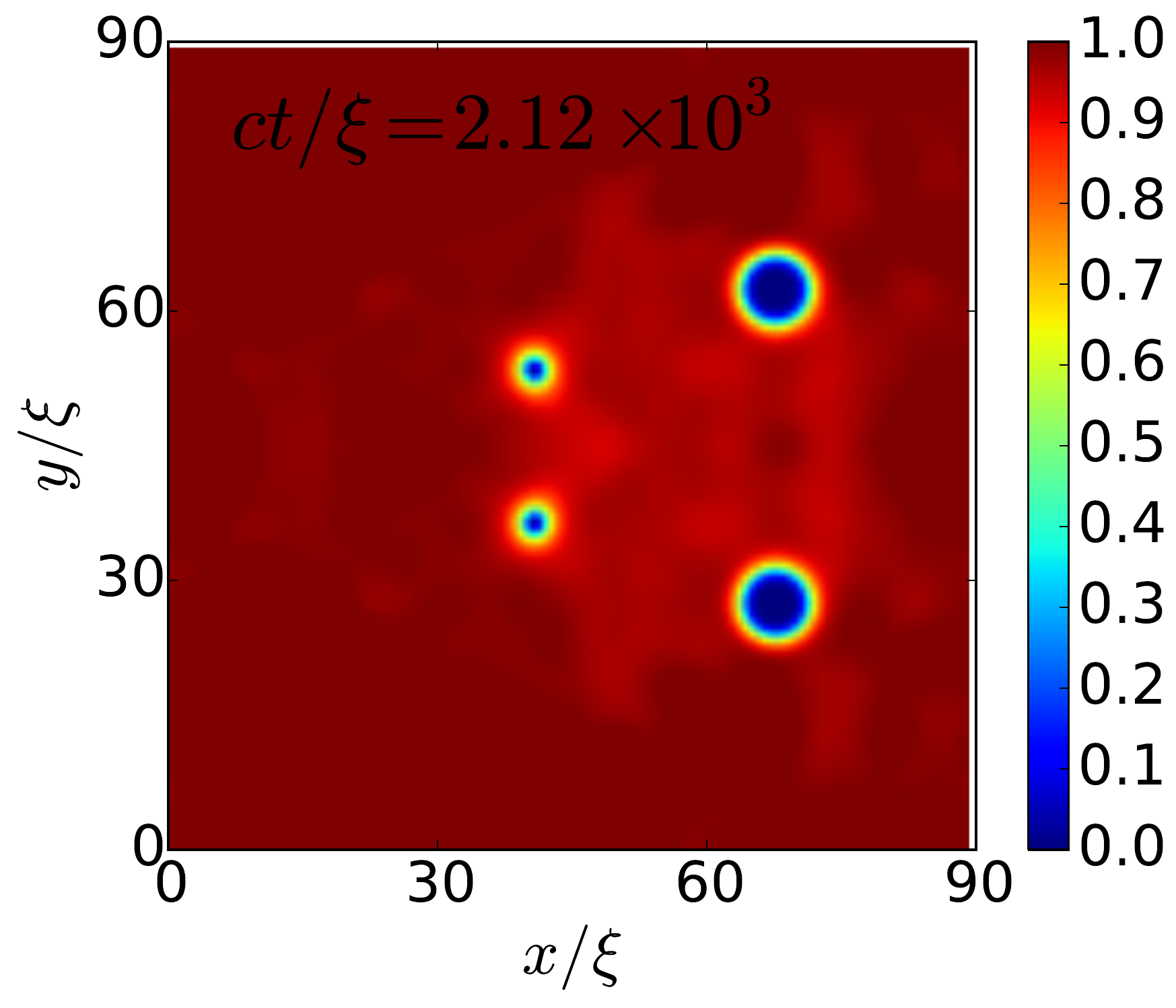}
\put(10,10){\large{\bf (k)}}
\end{overpic}
\begin{overpic}
[height=4.5cm,unit=1mm]{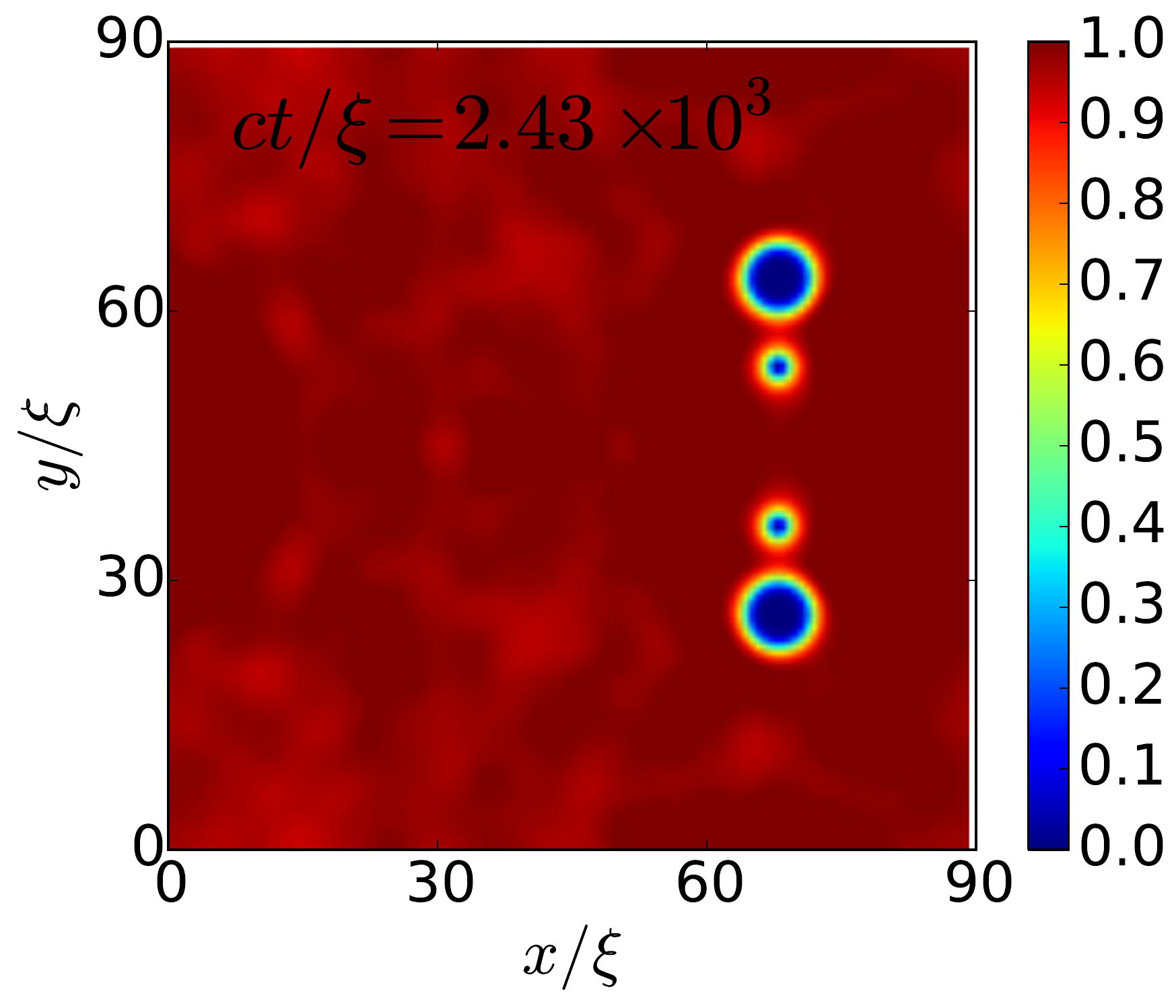}
\put(10,10){\large{\bf (l)}}
\end{overpic}
\caption{\small (Color online) Spatiotemporal evolution of the density field
$\rho(\mathbf{r},t)$ shown via pseudocolor plots, for two heavy particle
placed in the path of the positive (upper) and negative (lower) vortices,
respectively, of a translating vortex-antivortex pair (initial configuration
$\tt ICP2B$).}
\label{fig:2partpairtranslpdH}
\end{figure*}

\begin{figure*}
\centering
\resizebox{\linewidth}{!}{
\includegraphics[height=4.5cm,unit=1mm]{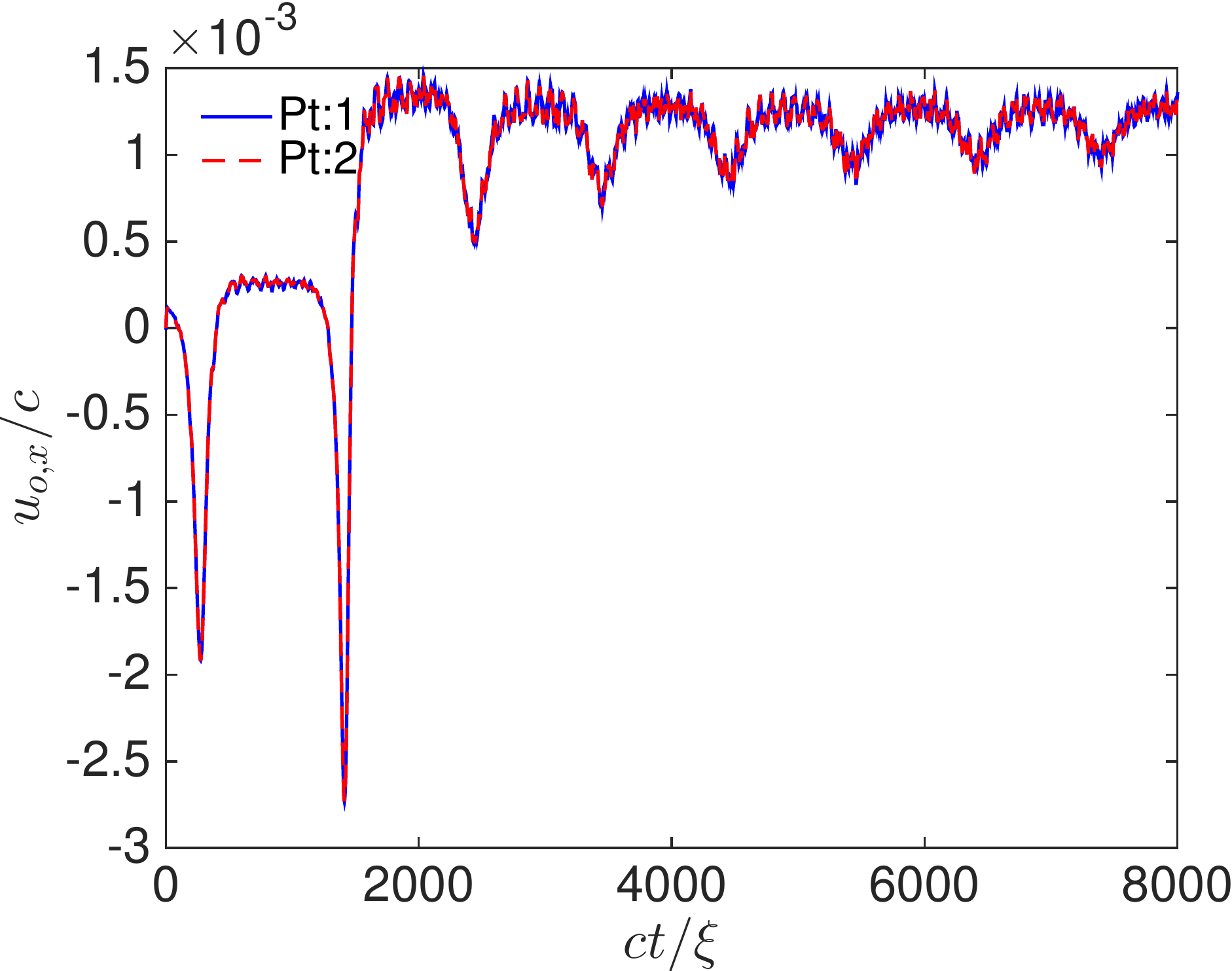}
\put(-90,20){\bf(a)}
\put(-70,60){\includegraphics[scale=0.1]{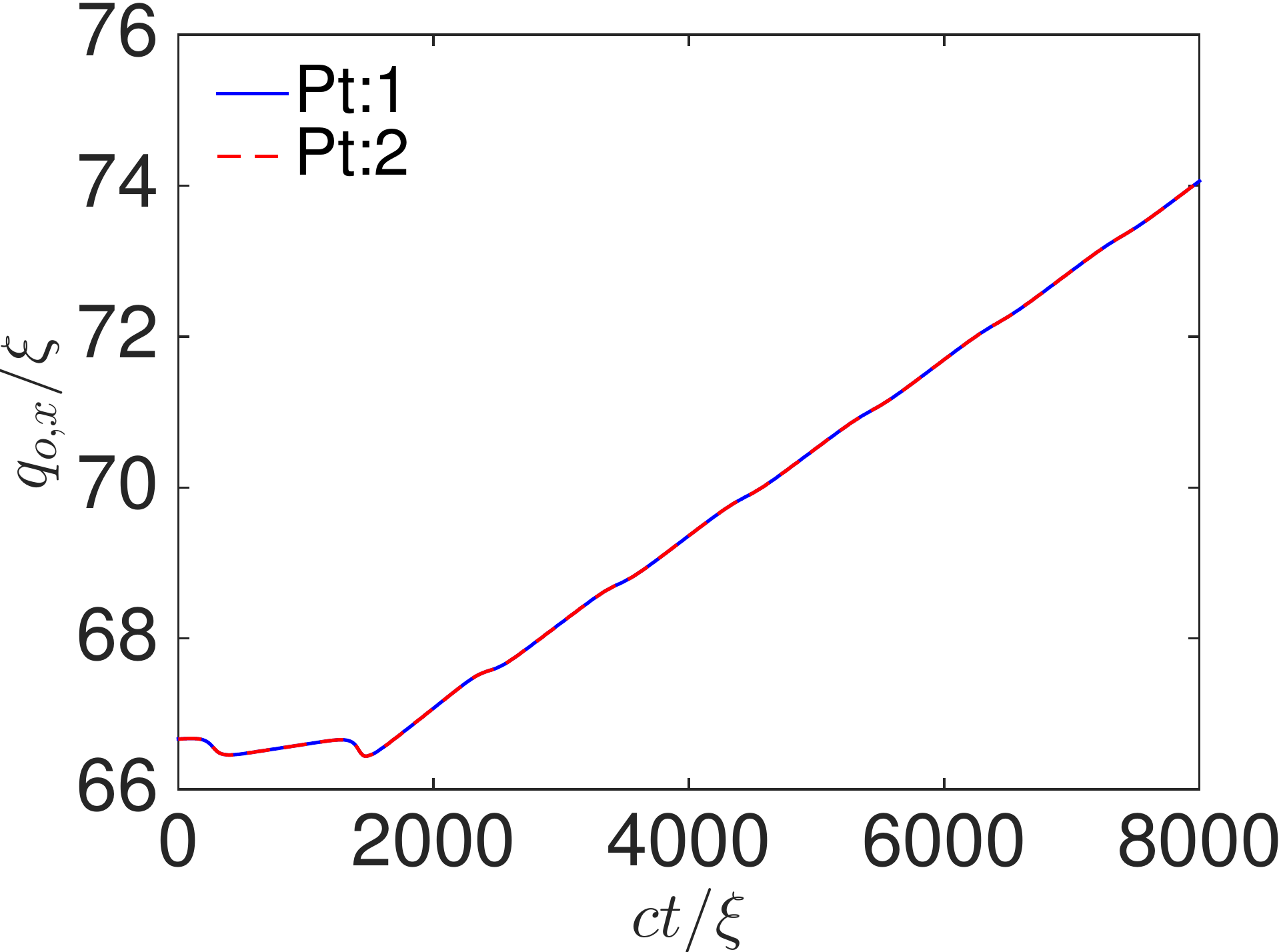}\put(-30,10){\tiny{(a.1)}}}
\put(-70,16){\includegraphics[scale=0.1]{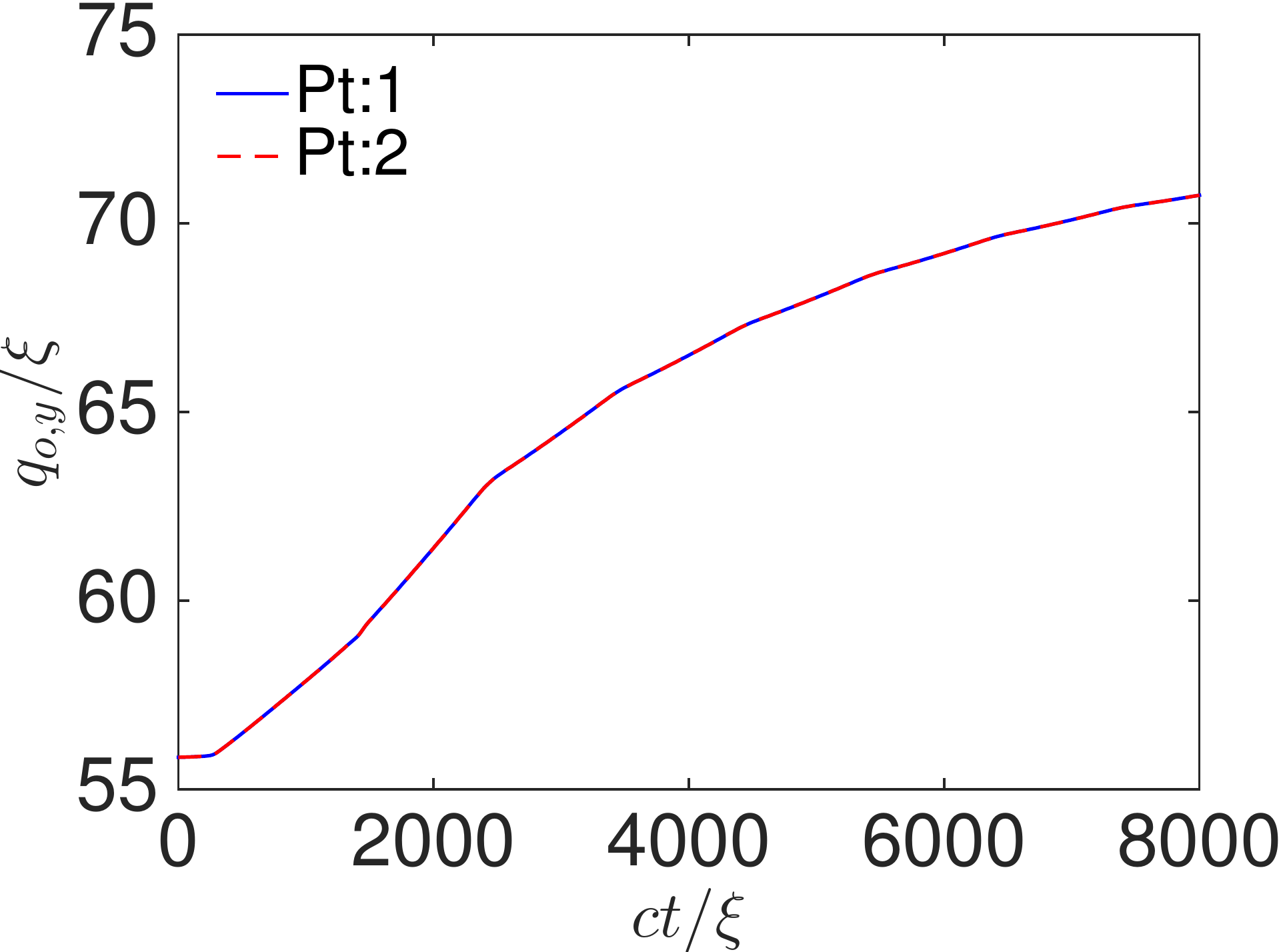}\put(-30,10){\tiny{(a.2)}}}
\hspace{0.25cm}
\includegraphics[height=4.5cm,unit=1mm]{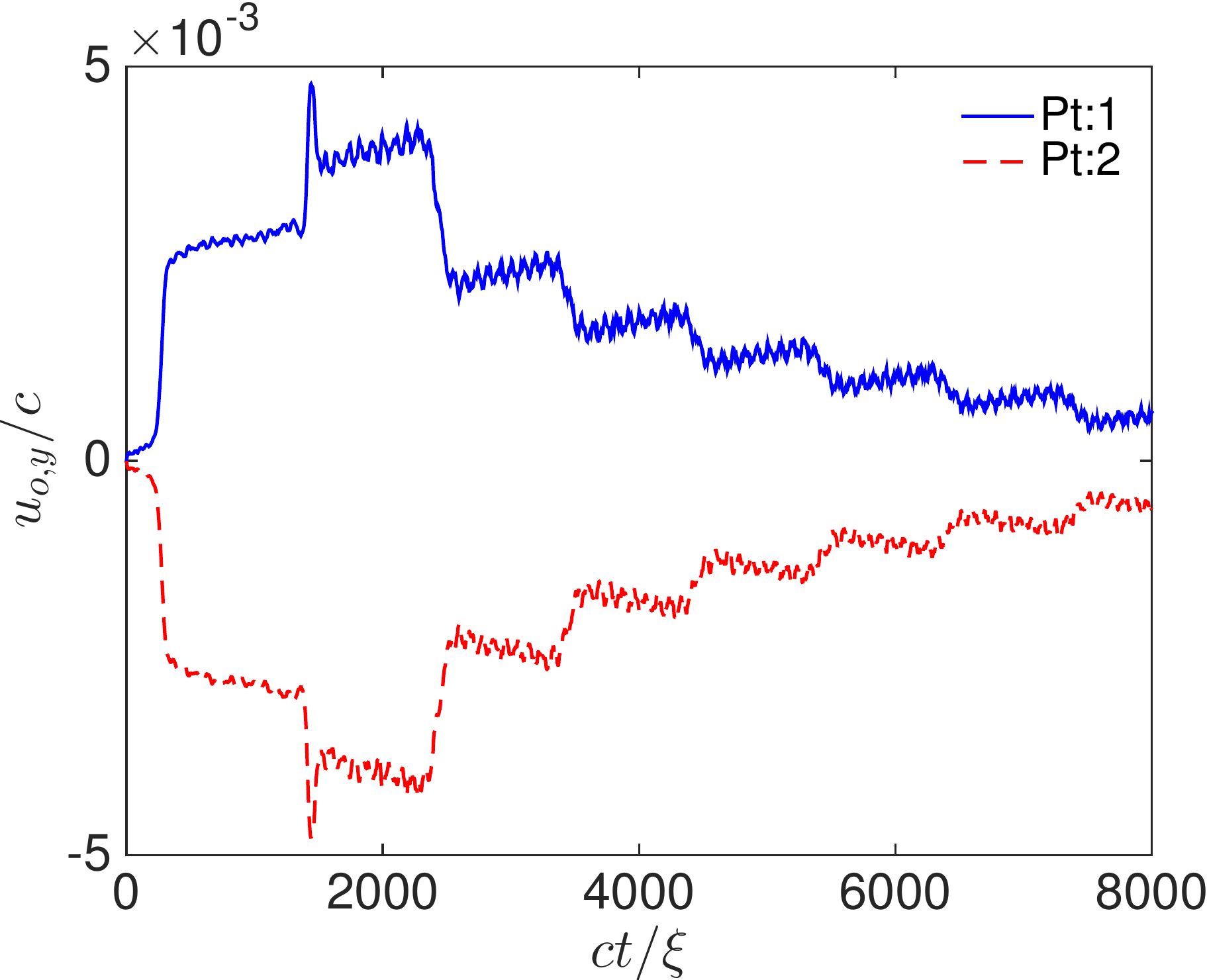}
\put(-90,20){\bf(b)}
}
\caption{\small (Color online) 
Plots of (a) $u_{\rm o,x}$ and (b) $u_{\rm o,y}$ versus time for two heavy particles
$Pt:1$ (blue solid curve) and $Pt:2$ (red dashed curve), placed in the path of the
positive (upper) and negative (lower) vortices, respectively, of a translating
vortex-antivortex pair (initial configuration $\tt ICP2B$). 
Insets: Plots of (a.1) $q_{\rm o,x}$ and (a.2) $q_{\rm o,y}$ versus time. 
The values of $q_{\rm o,x}$ and $q_{\rm o,y}$ are not mod$2\pi$; i.e., if particle
goes around our periodic simulation domain once, say in the $\hat{\mathbf{x}}$
direction, then the values of $q_{\rm o,x}$ is its value in the box plus
$2\pi$.}
\label{fig:2partpairtranslquH}
\end{figure*}

\textbf{Two-particles}: 
We now investigate the interaction of two particles with a vortex-antivortex
pair. For this we use the initial configuration $\tt ICP2B$, in which we place
the two particles $Pt:1$ and $Pt:2$ at a distance, in front of the positive and
the negative vortices of a translating vortex-antivortex pair (see the
schematic diagram in Fig.~\ref{fig:schem1partpairtransl} (b)).  As in the case of 
a single particle, we use the ARGLE to prepare a state with two stationary particles
$Pt:1$ and $Pt:2$ at $(1.5\pi/\xi,1.257\pi/\xi)$ and $(1.5\pi/\xi,0.743\pi/\xi)$, respectively;
this state is then combined with a state corresponding to a vortex-antivortex
pair of size $d_{\rm pair}\simeq 23\,\xi$ which translates with a velocity
$\mathbf{u}_{\rm pair}=0.074\,c\,\hat{\mathbf{x}}$
(see Appendix~\ref{app:vortpair} for preparation details).  We use the above initial
configuration to study the interaction of the vortex-antivortex pair with
heavy, neutral, and light particles.

In Figs.~\ref{fig:2partpairtranslpdH}~(a)-(e) we show that, when the
vortex-antivortex pair approaches the two symmetrically placed heavy particles,
the positive (upper) vortex and the negative (lower) vortex glide along the
circumferences of $Pt:1$ and $Pt:2$, respectively; thereafter, the
vortex-antivortex pair continues to translate in the $\hat{\mathbf{x}}$
direction. The interaction of the vortex-antivortex pair with these particles
leads to the transfer of momentum to the latter; and these particles start
moving slowly (see Figs.~\ref{fig:2partpairtranslquH}~(a)-(d) for 
$ct/\xi\simeq 2.26\times10^2$.
Because of our periodic boundary conditions, the translating vortex-antivortex
pair comes back and again glides along the particle boundaries, which are still
in the path of translation of this pair (see
Figs.~\ref{fig:2partpairtranslpdH}~(f)-(i)).  The particles move away from the
vortices, as the vortex-antivortex pair moves beyond them (see
Figs.~\ref{fig:2partpairtranslquH}~(a)-(d) for $ct/\xi\gtrsim 1.41\times10^3$).  At later
times, the separation between the particles is wide enough for the
vortex-antivortex pair to pass through the region in between the particles
without any significant obstruction (see
Figs.~\ref{fig:2partpairtranslpdH}~(j)-(l)).  However, the plots of the
particle-velocity components versus times show jumps when the vortex-antivortex
pair passes through the region in between the particles.  The Video
M7~\cite{suppmat} illustrates the complete spatiotemporal evolution of the
particles and the density field $\rho(\mathbf{r},t)$.

When the translating vortex-antivortex pair approaches symmetrically placed
neutral or light particles, the particles $Pt:1$ and $Pt:2$ get trapped on the
positive and the negative vortices, respectively. The trapping of the two
neutral (light) particles here is similar to the trapping of a single neutral
(light) particle placed in front of a translating vortex-antivortex pair.  
After the particles are trapped on
the vortices, the two-particle-vortex-antivortex-pair complex continues to
translate in the $\hat{\mathbf{x}}$ direction, but the particles now exhibit
fluctuations. The pseudocolor plots of
Figs.~\ref{fig:2partpairtranslpdN}~(a)-(f) and
\ref{fig:2partpairtranslpdL}~(a)-(f) in the Appendix and the Videos M8 and M9 in the 
Supplemental Material~\cite{suppmat}
summarize the spatiotemporal evolution of the density field $\rho(\mathbf{r},t)$
for the neutral and the light particles, respectively.  The fluctuations in the
case of the neutral particle are temporally periodic, with some modulation,
whereas those in the case of the light particle are chaotic(see
Figs.~\ref{fig:2partpairtranslquN}~(a)-(d) (neutral particle) and
\ref{fig:2partpairtranslquL}~(a)-(d) (light particle) in the Appendix 
~\ref{app:additionalplots} for details).

\subsection{Single particle dynamics in the presence of counter-rotating vortex clusters}
\label{subsec:1partcrotN}
\begin{figure*}
\centering
\begin{overpic}
[height=4.5cm,unit=1mm]{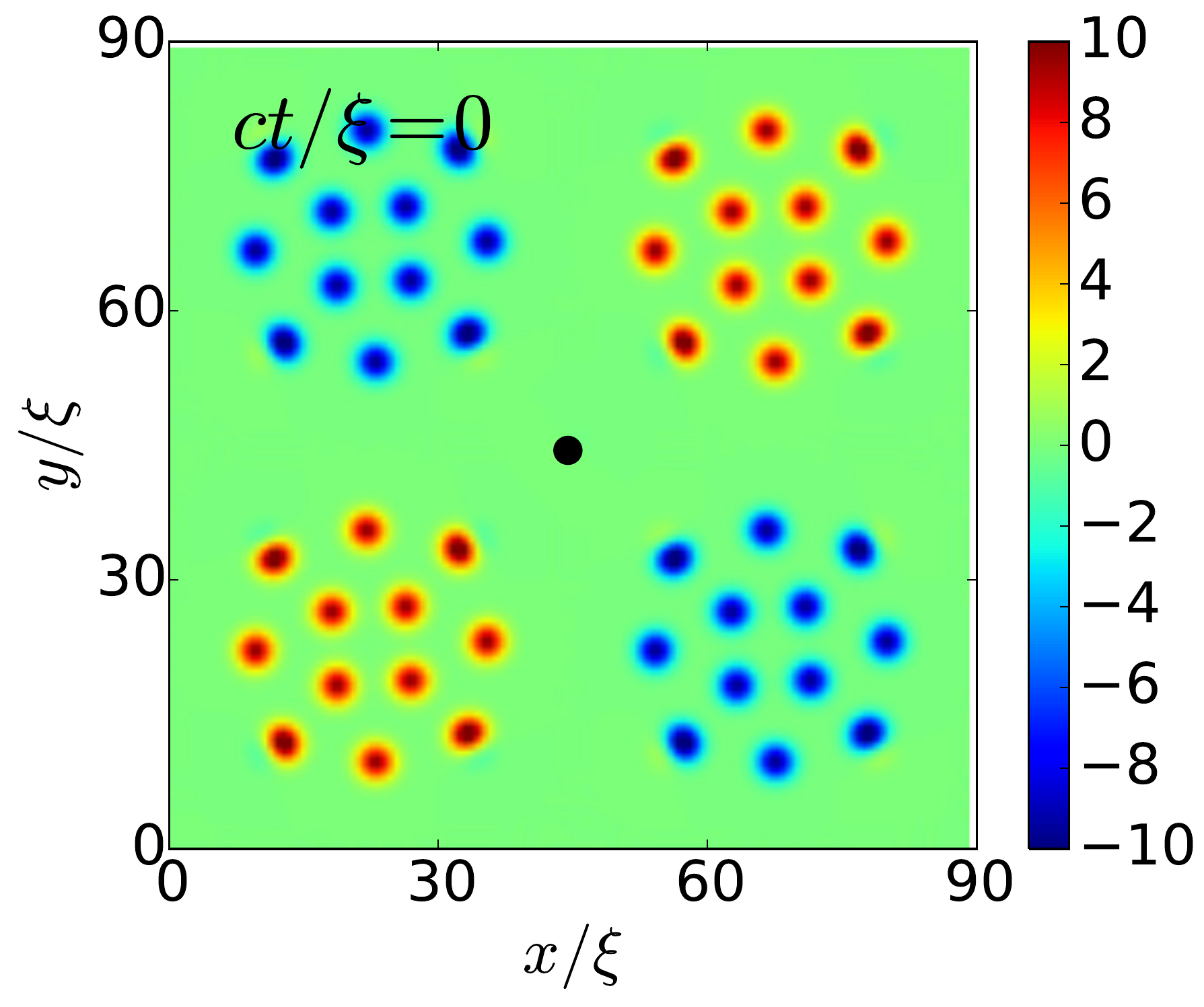}
\put(10.,10){\large{\bf (a)}}
\end{overpic}
\begin{overpic}
[height=4.5cm,unit=1mm]{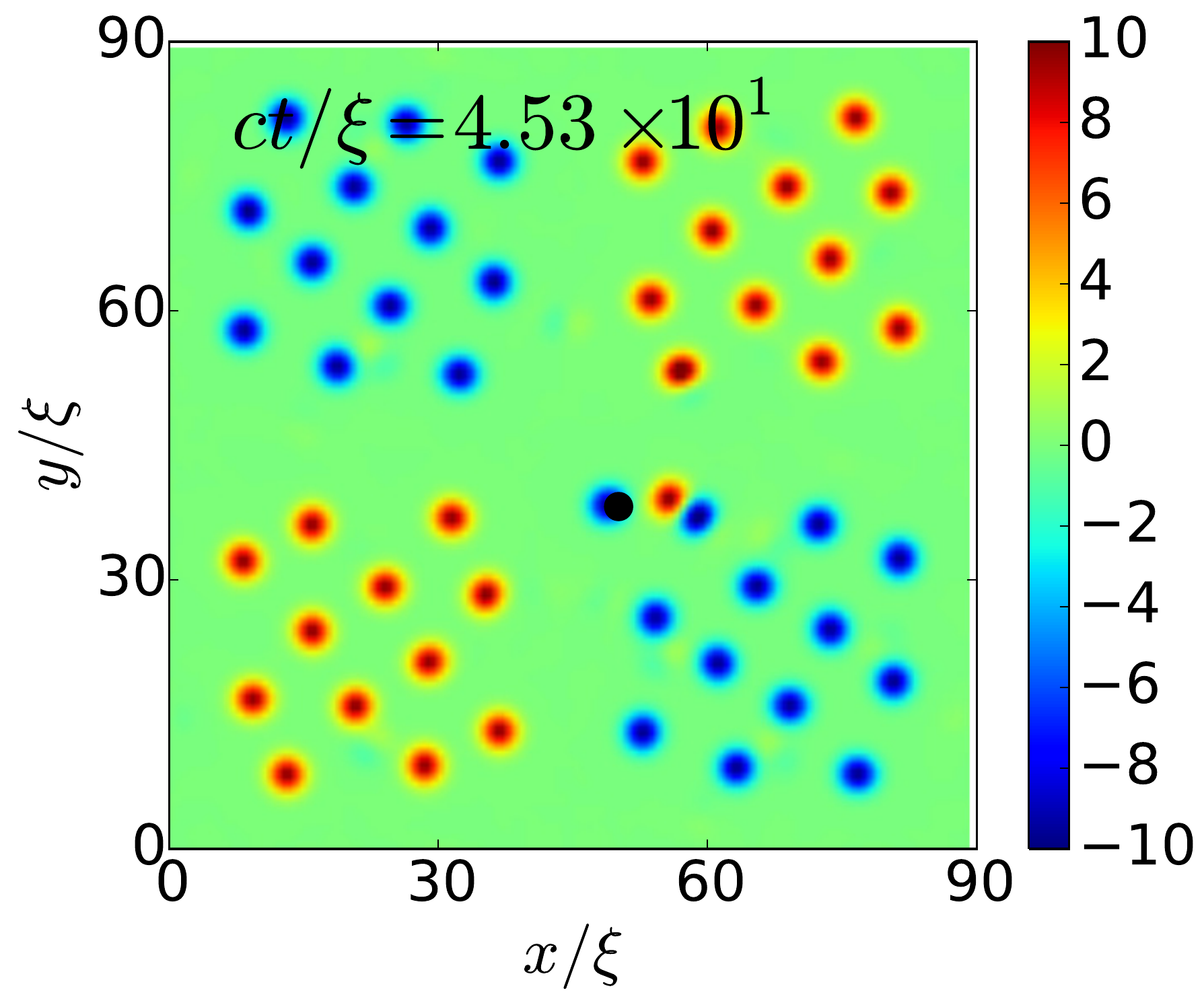}
\put(10.,10){\large{\bf (b)}}
\end{overpic}
\begin{overpic}
[height=4.5cm,unit=1mm]{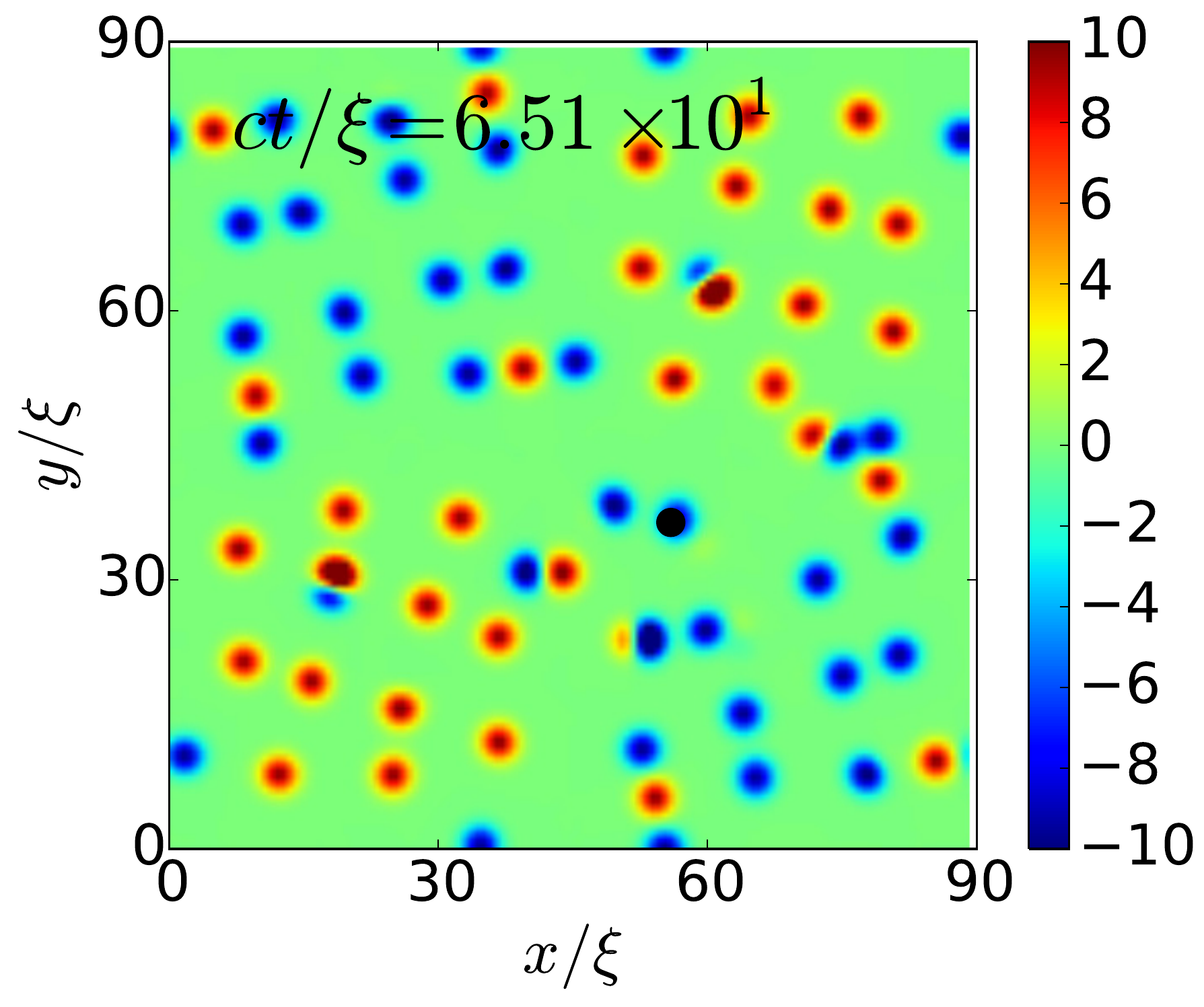}
\put(10.,10){\large{\bf (c)}}
\end{overpic}
\\
\begin{overpic}
[height=4.5cm,unit=1mm]{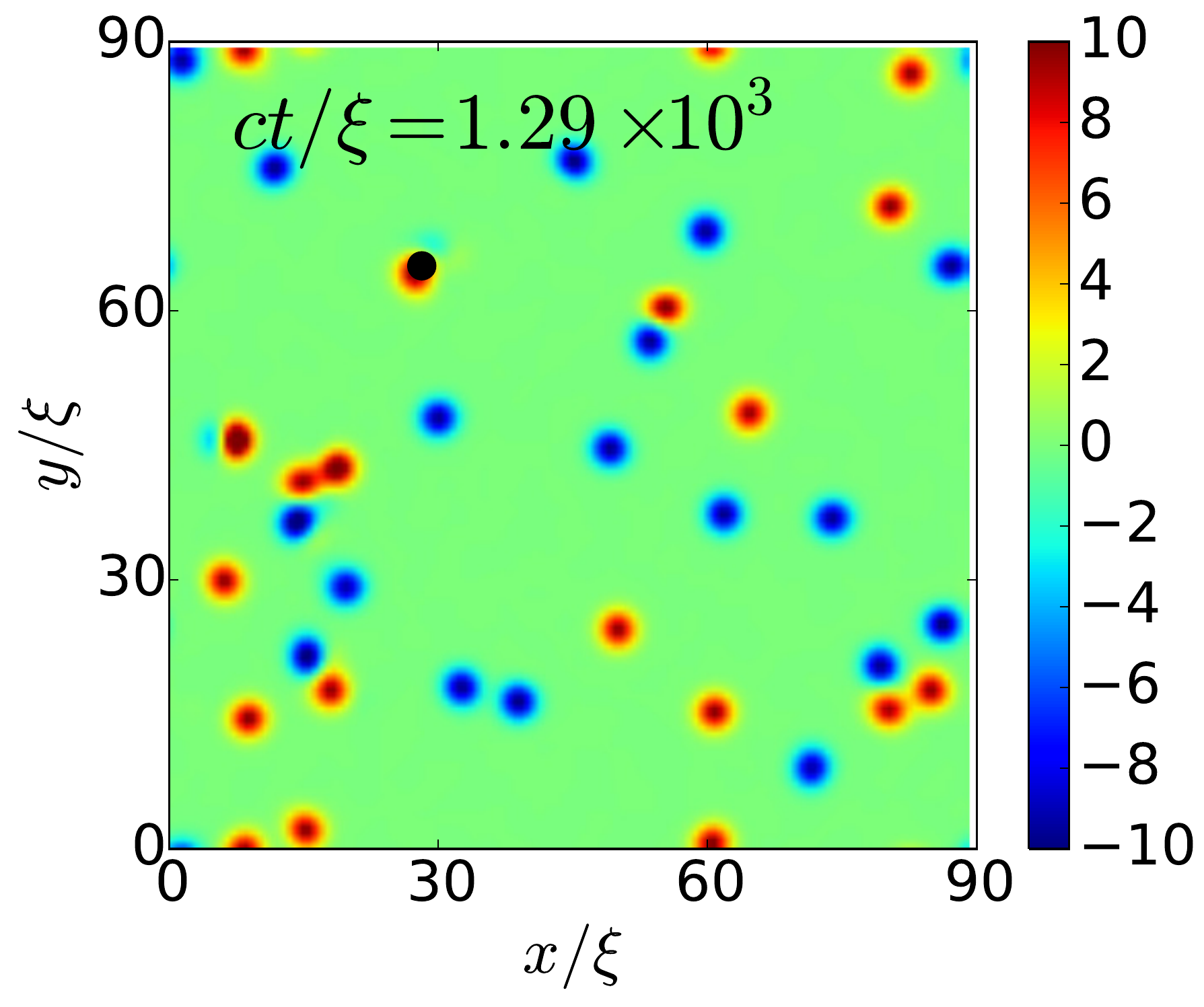}
\put(10.,10){\large{\bf (d)}}
\end{overpic}
\begin{overpic}
[height=4.5cm,unit=1mm]{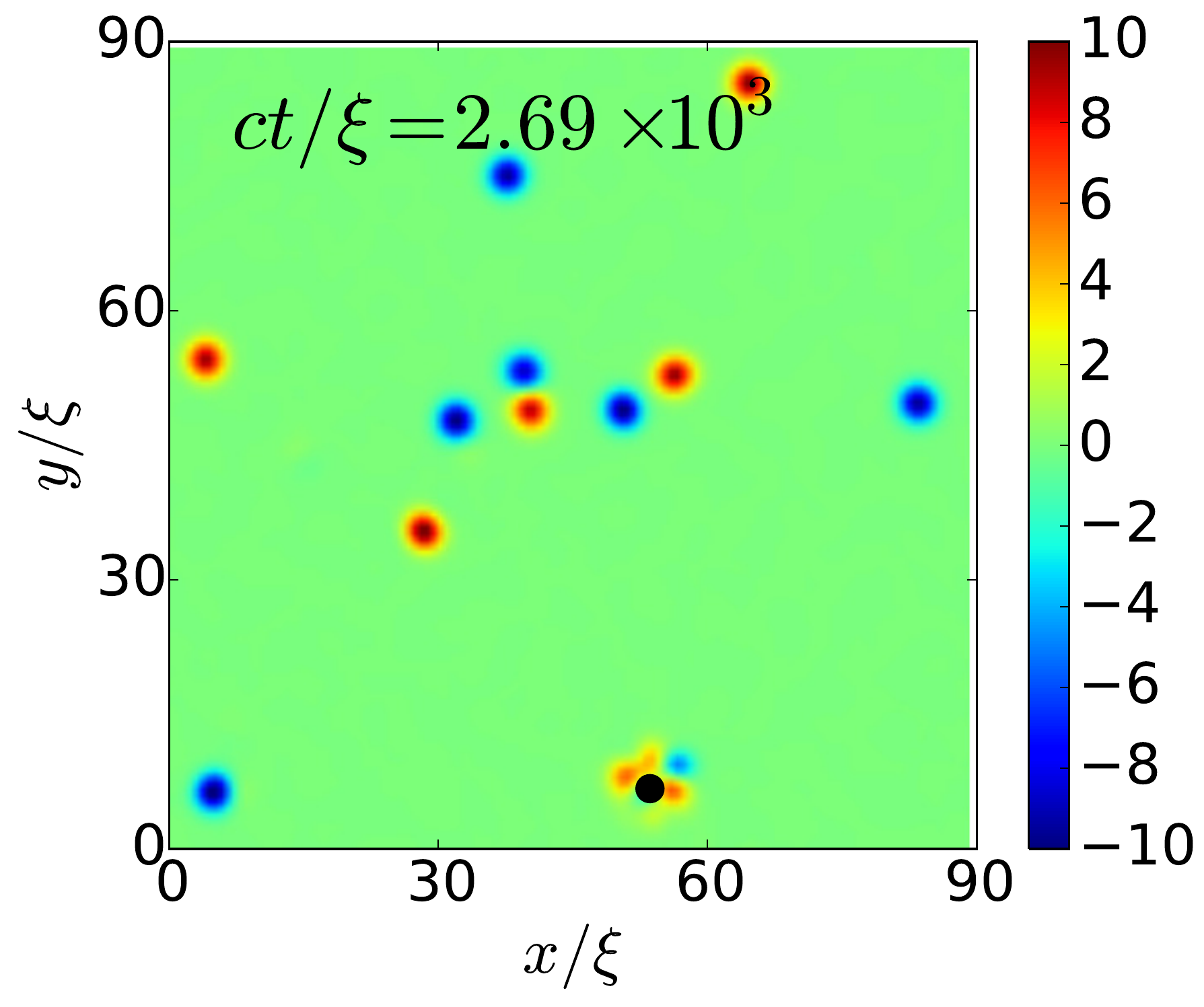}
\put(10.,10){\large{\bf (e)}}
\end{overpic}
\begin{overpic}
[height=4.5cm,unit=1mm]{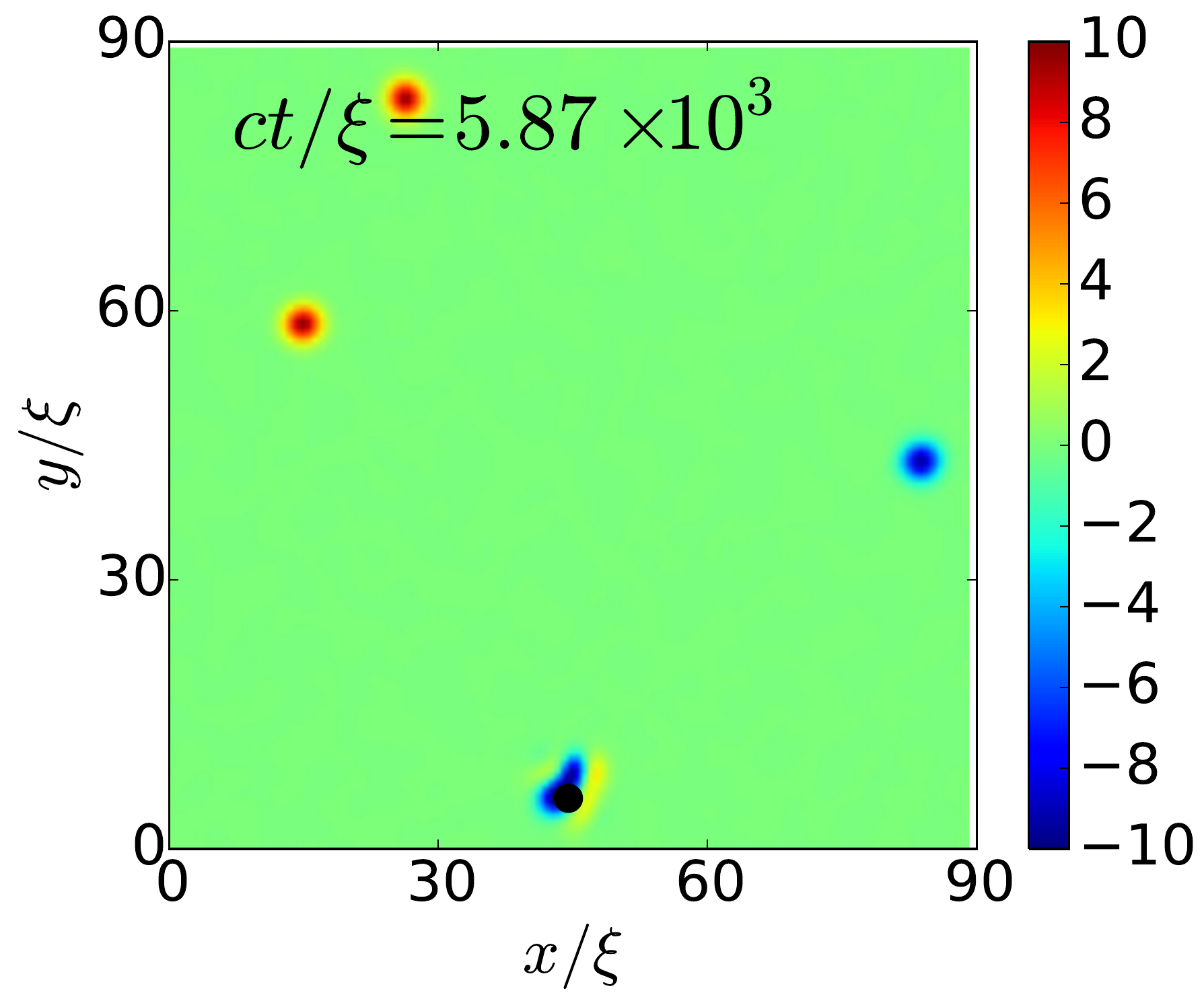}
\put(10.,10){\large{\bf (f)}}
\end{overpic}
\caption{\small (Color online) Spatiotemporal evolution of the filtered
vorticity field (derived from the incompressible velocity field), for the
neutral particle initially in the presence of counter-rotating vortex clusters
(initial configuration $\tt ICP3A$). The instantaneous position of the particle
is shown by a black disk.}
\label{fig:1partcrotclustersvort}
\end{figure*}

To study the dynamics of a single neutral particle in the presence of
counter-rotating vortex clusters we generate the initial configuration $\tt
ICP3A$ in three s.pdf: (1) we use the ARGLE to prepare a state with a particle at
$(\pi/\xi,\pi/\xi)$ moving with velocity $\mathbf{u}_{o}=0.1(1/\sqrt{2},-1/\sqrt{2})\,c$ 
(see Appendix~\ref{app:counterrotvort} for details); (2) we then use the ARGLE to
prepare two positive and two negative vortex clusters, where each cluster has
$12$ vortices of the same sign, and the positive and the negative clusters
rotate in opposite directions (for preparation details see the Appendix); (3)
the states obtained in the s.pdf (1) and (2) are combined together, by
multiplying their wave functions. The initial configuration so prepared is then
used in the GPE. By design, we prepare the state with counter-rotating vortex clusters in
a state that is not the ground state; therefore, under the TGPE dynamics, the
clusters expand and interact with neighboring clusters; this results in a flow
with a complex distribution of vortices.  Thus, this initial state allows us to
study a neutral particle in a state that displays superfluid turbulence.

\begin{figure}
\centering
\resizebox{\linewidth}{!}{
\includegraphics[height=4.cm,unit=1mm]{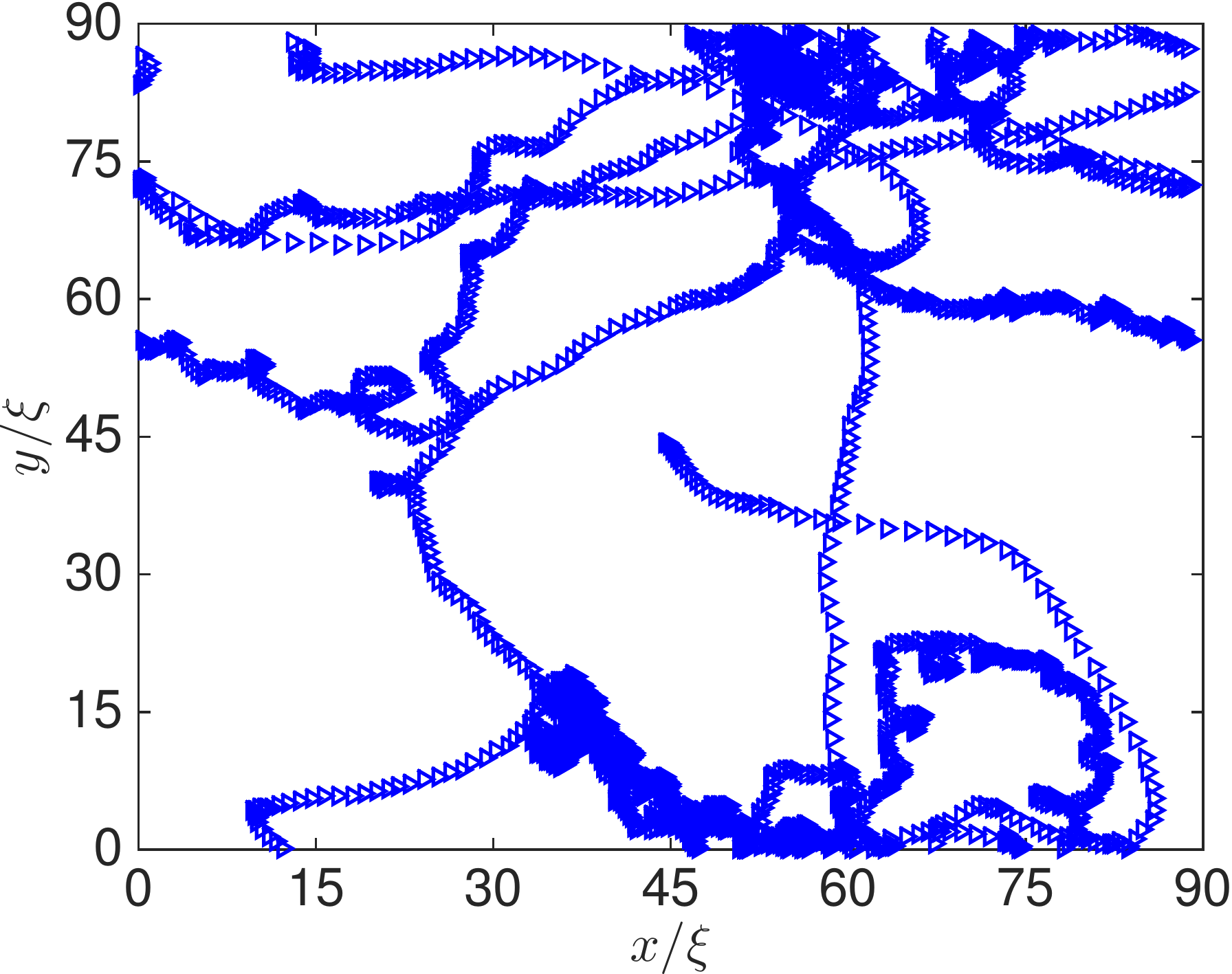}
}
\caption{\small (Color online) Trajectory of a neutral particle (denoted by
blue triangles), initially in the presence of counter-rotating vortex clusters
(initial configuration $\tt ICP3A$).}
\label{fig:1partcrotclusterstraject}
\end{figure}

\begin{figure*}
\centering
\resizebox{\linewidth}{!}{
\includegraphics[height=4.cm,unit=1mm]{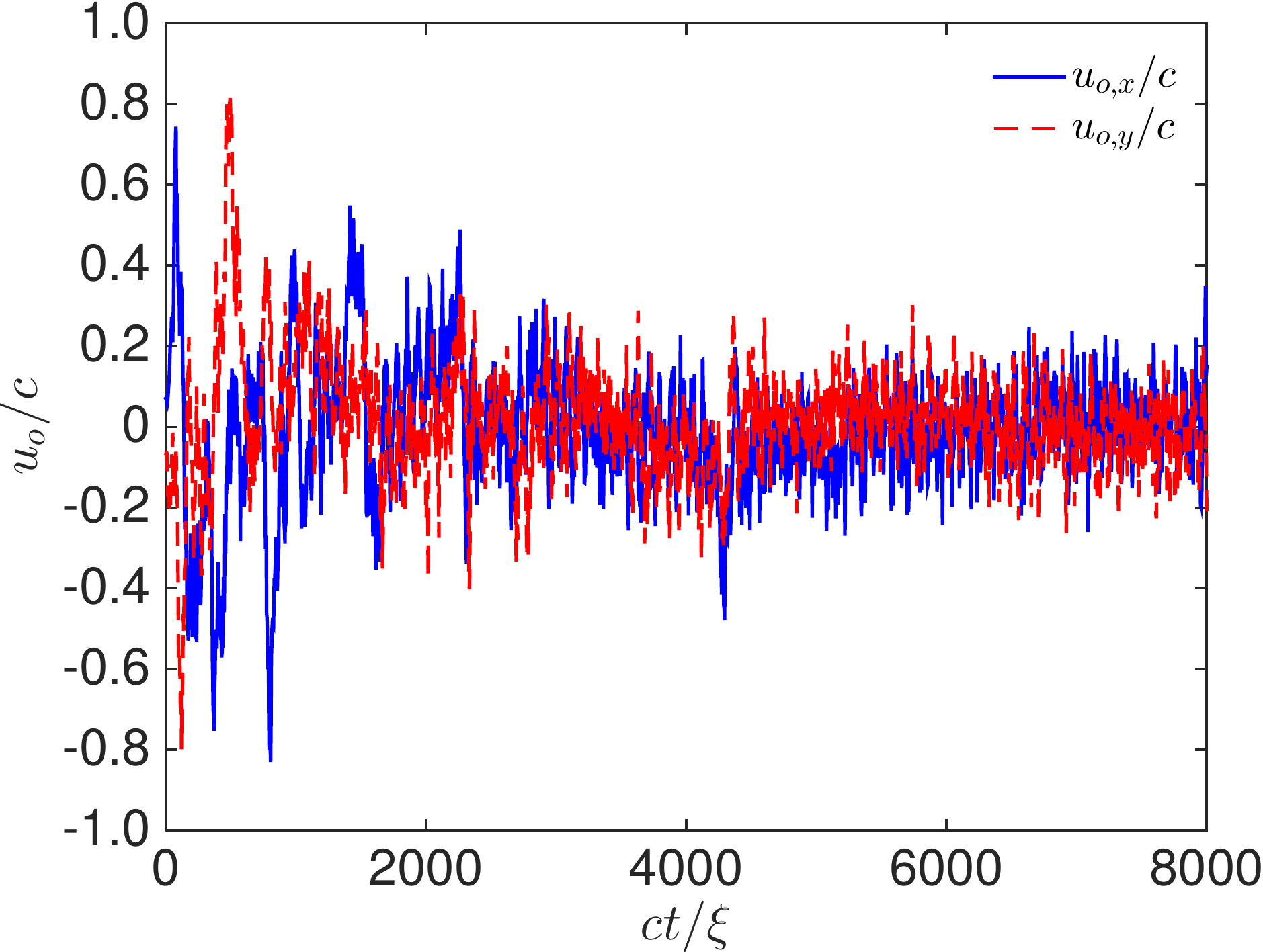}
\put(-75,20){\bf (a)}
\hspace{0.25cm}
\includegraphics[height=4.cm,unit=1mm]{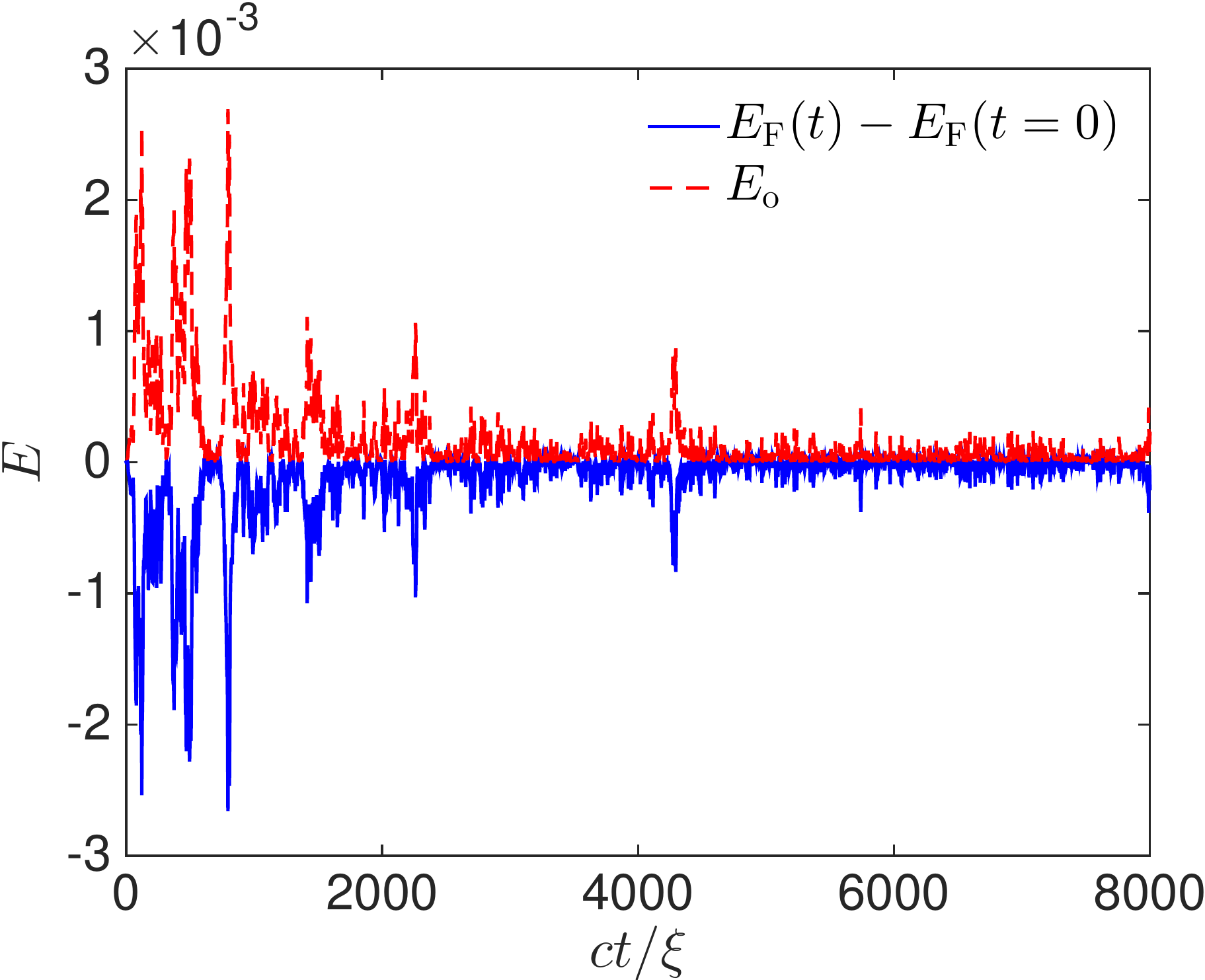}
\put(-75,20){\bf (b)}
}
\caption{\small (Color online) Plots versus time $t$ of (a) $u_{\rm o,x}$
(blue solid curve) and $u_{\rm o,y}$ (red dashed curve); 
(b) $\delta E_{\rm field}=E_{\rm F}(t)-E_{\rm F}(t=0)$ (blue solid curve) and
$E_{\rm o}$ (red dashed curve); obtained from the dynamical evolution of the neutral
particle in the presence of counter-rotating vortex clusters (initial
configuration $\tt ICP3A$).  }
\label{fig:1partcrotclustersuE}
\end{figure*}

In Figs.~\ref{fig:1partcrotclustersvort}~(a)-(f) we show the spatiotemporal
evolution of the filtered vorticity field; the particle is represented by a
black disk here. Figures~\ref{fig:1partcrotclustersvort}~(a)-(c) show that the
vortex clusters expand quickly and interact with their neighboring clusters. At
$ct/\xi\simeq 4.53\times10^1$ the particle sheds a vortex-antivortex pair, while moving towards
the negative vortex cluster in the right bottom corner of the simulation
domain; the particle gets trapped on a nearby negative vortex and it is dragged
inside the cluster; at this time, its velocity shoots up to $u_{\rm o}\sim
0.8\,c$.  The vortex density decreases as the system evolves because of the
annihilation of the vortices and the anitvortices (see
Figs.~\ref{fig:1partcrotclustersvort}~(d)-(e)). The Video M10~\cite{suppmat}
illustrates the dynamics of a neutral particle in the presence of
counter-rotating vortex clusters. Figure~\ref{fig:1partcrotclusterstraject}
shows that the trajectory of the particle (denoted by a series of blue
triangles), in the presence of vortices, is complex.  The spacing between
successive triangles is large (small) when the particle velocity is large
(small). During the motion the particle switches from one vortex to another and
its direction of motion k.pdf changing because of its interactions with
neighboring vortices. The regions with a high density of circles on the
trajectory plot in Fig.~\ref{fig:1partcrotclusterstraject} occur when the area
around the particle is free of vortices or at late times when the overall
vortex density has decreased considerably.
Figure~\ref{fig:1partcrotclustersuE}~(a) shows that $u_{\rm o,x}$ and $u_{\rm
o,y}$ exhibit chaotic fluctuations; in Fig.~\ref{fig:1partcrotclustersuE}~(b)
we show plots of the energies that illustrate the exchange of energy between
the particle and the superfluid. Energy spectra of the incompressible kinetic
energy, plot not included here, shows the spread of kinetic energy over the full
range of scales, similar to the case of a turbulent flow~\cite{vmrnjp13}.


\subsection{Many-particle dynamics in the presence of counter-rotating vortex
clusters} \label{subsec:manypartcrotN}

We study the dynamics of four particles in the presence of small,
counter-rotating vortex clusters as an illustrative example of many-particle
dynamics in the presence of vortices. We generate the initial configuration
$\tt IPC3B$ for this purpose in three s.pdf: (1) we use the ARGLE
to prepare a minimum-energy state with two clusters of
positive and negative vortices; each cluster has $4$ vortices of the same sign,
and the positive and the negative clusters rotate in opposite directions (for
preparation details see Appendix~\ref{app:counterrotvort}).  (2) We prepare a state
with four stationary particles $Pt:1$, $Pt:2$, $Pt:3$, and $Pt:4$ at the
coordinates $(3\pi/2\xi,\pi/2\xi)$, $(\pi/2\xi,\pi/2\xi)$, $(\pi/2\xi,3\pi/2\xi)$, and
$(3\pi/2\xi,3\pi/2\xi)$, respectively, which correspond to the centers of vortex
clusters; (3) the states obtained in the s.pdf (1) and (2) are combined
together, by multiplying their wave functions, for the initial configuration
$\tt ICP3A$ that is then used in the GPE to study the dynamics of this system.
Note that to study the particle-field dynamics for this initial configuration
we have set the speed of sound $c=2$.

\begin{figure*}
\centering
\begin{overpic}
[height=4.5cm,unit=1mm]{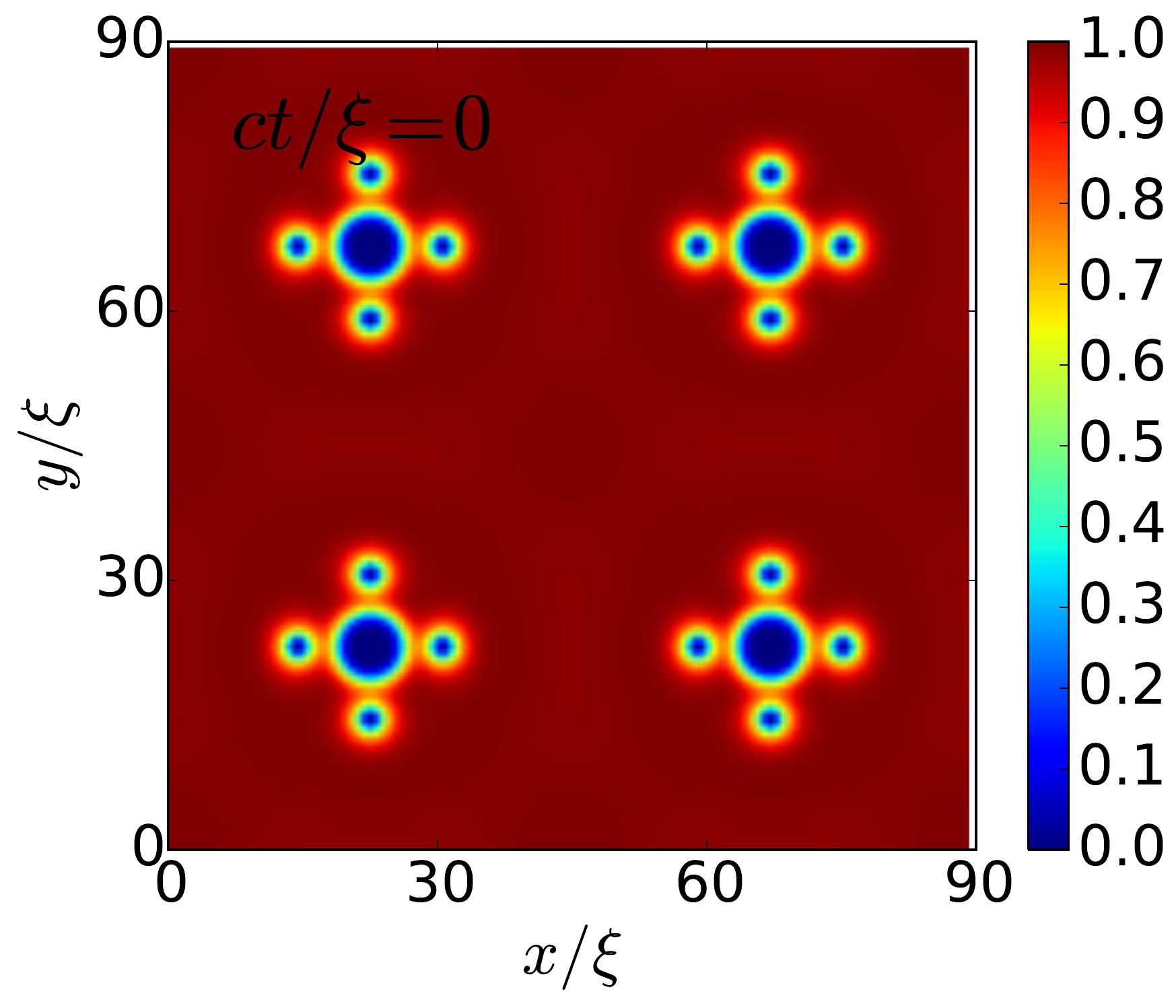}
\put(10.,10){\large{\bf (a)}}
\end{overpic}
\begin{overpic}
[height=4.5cm,unit=1mm]{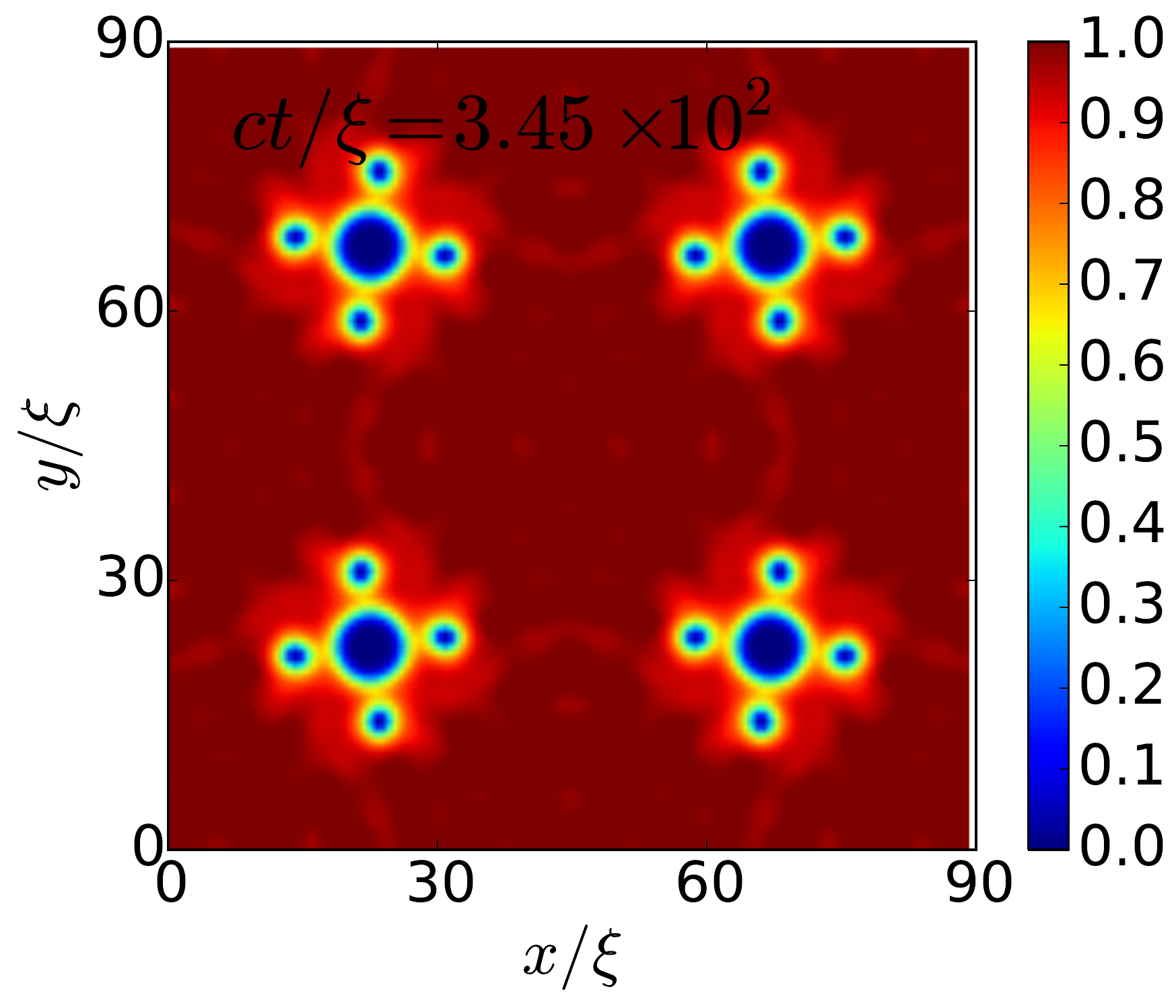}
\put(10.,10){\large{\bf (b)}}
\end{overpic}
\begin{overpic}
[height=4.5cm,unit=1mm]{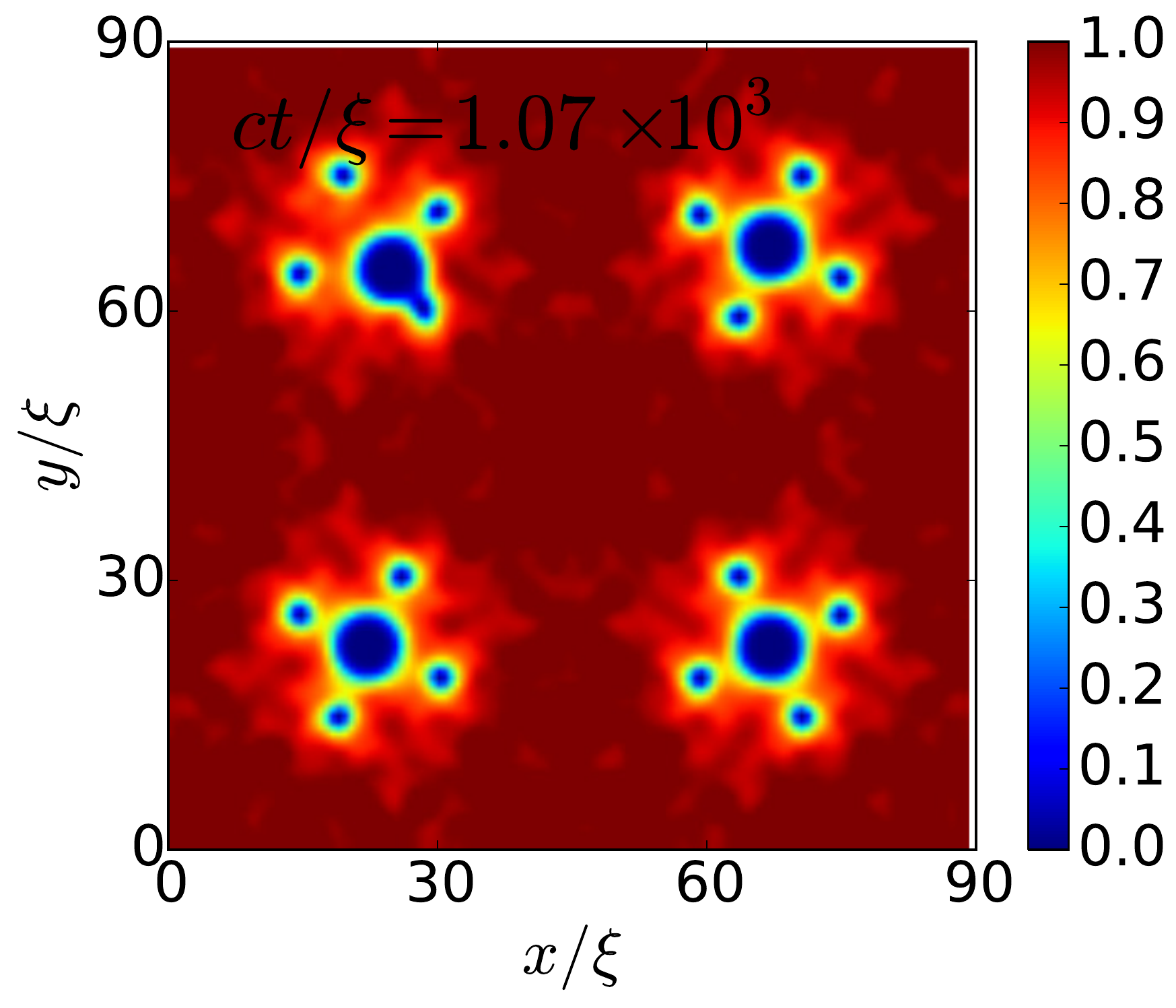}
\put(10.,10){\large{\bf (c)}}
\end{overpic}
\\
\begin{overpic}
[height=4.5cm,unit=1mm]{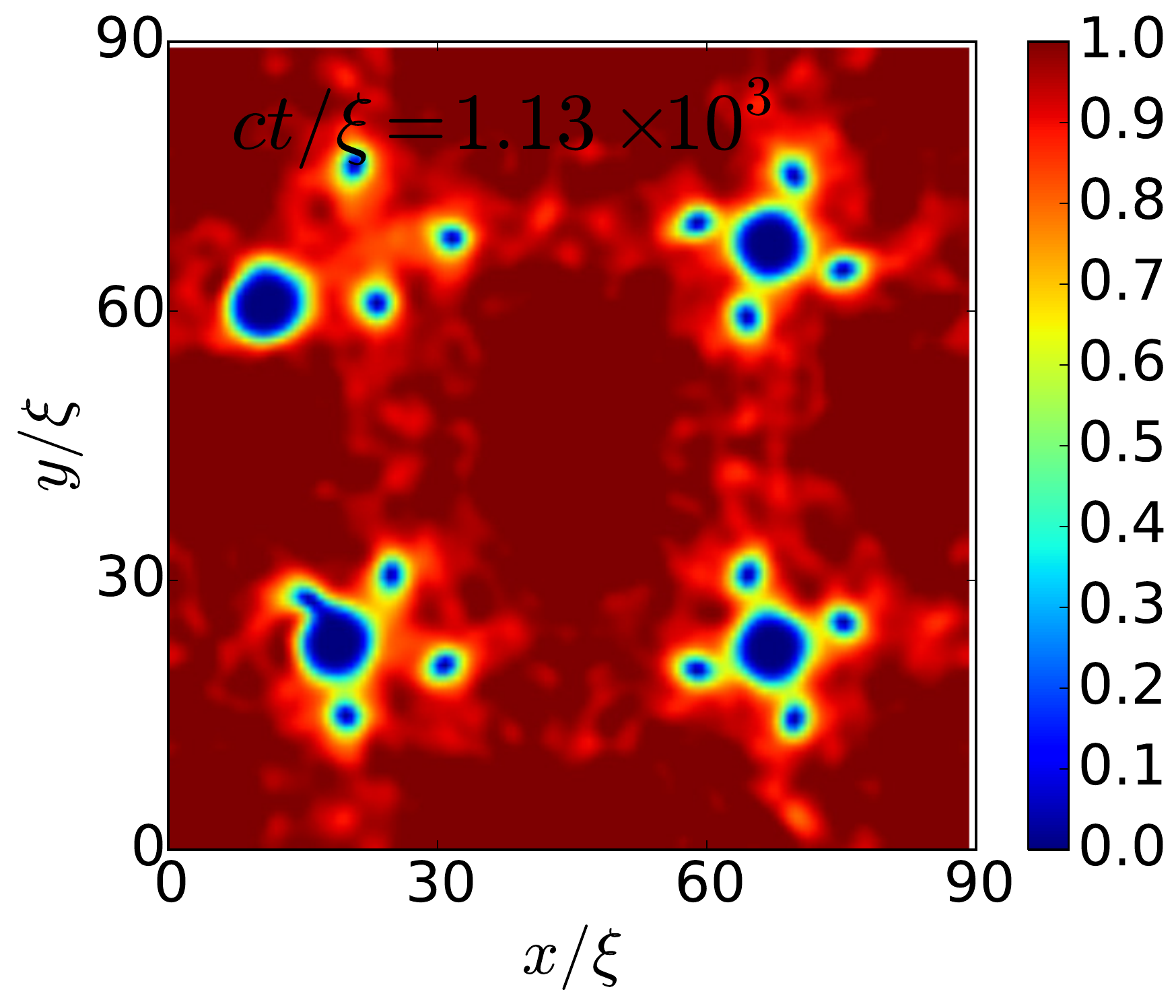}
\put(10.,10){\large{\bf (d)}}
\end{overpic}
\begin{overpic}
[height=4.5cm,unit=1mm]{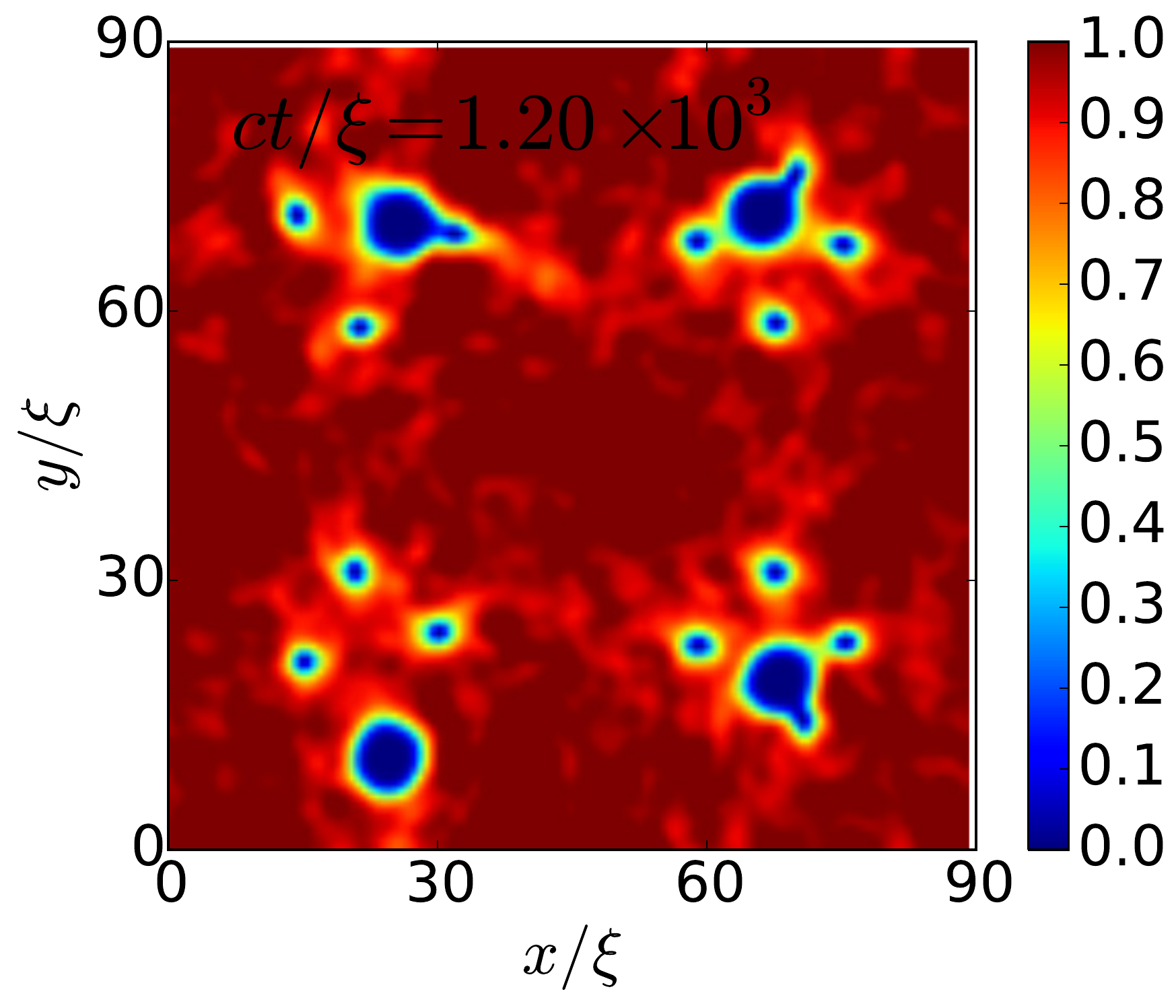}
\put(10.,10){\large{\bf (e)}}
\end{overpic}
\begin{overpic}
[height=4.5cm,unit=1mm]{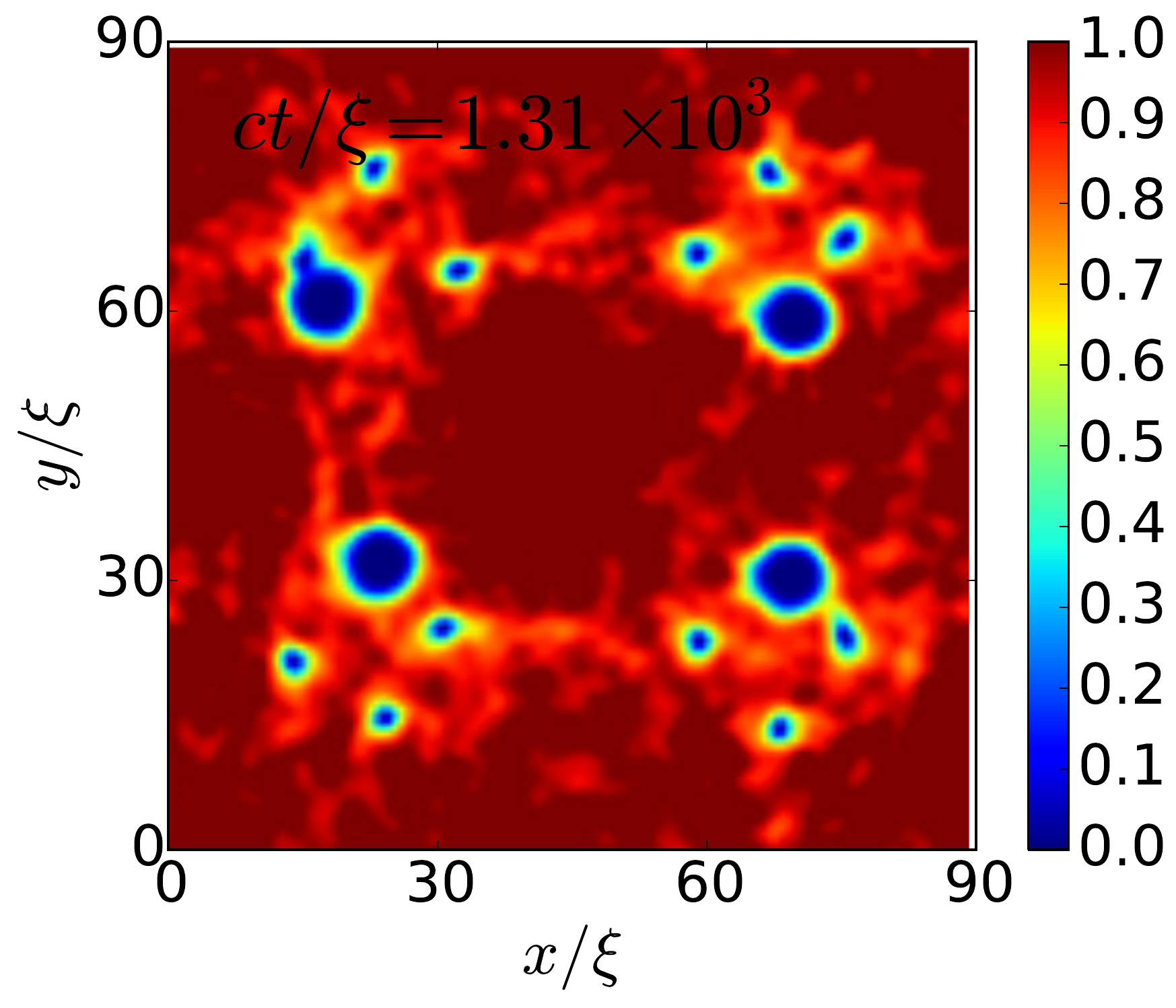}
\put(10.,10){\large{\bf (f)}}
\end{overpic}
\\
\begin{overpic}
[height=4.5cm,unit=1mm]{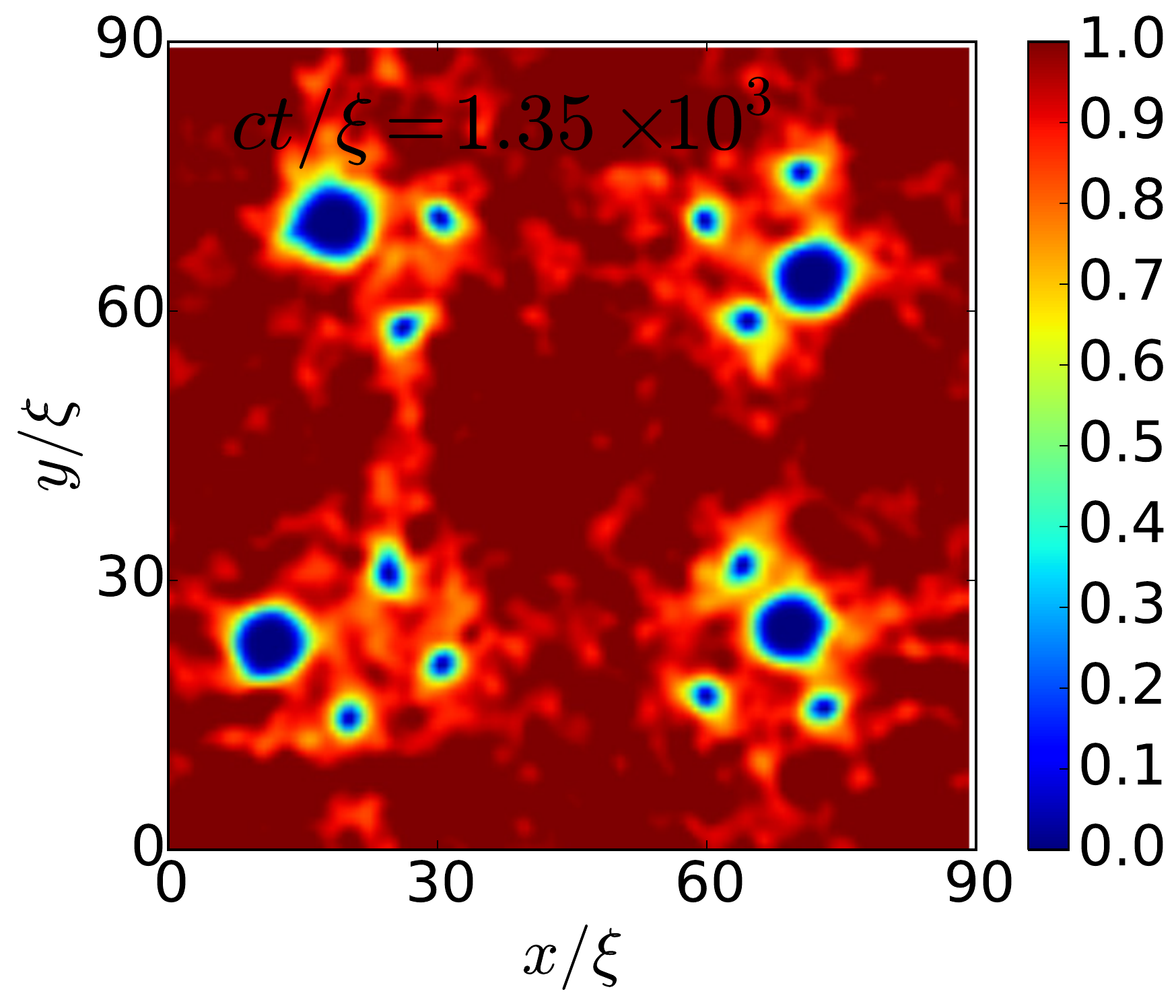}
\put(10.,10){\large{\bf (g)}}
\end{overpic}
\begin{overpic}
[height=4.5cm,unit=1mm]{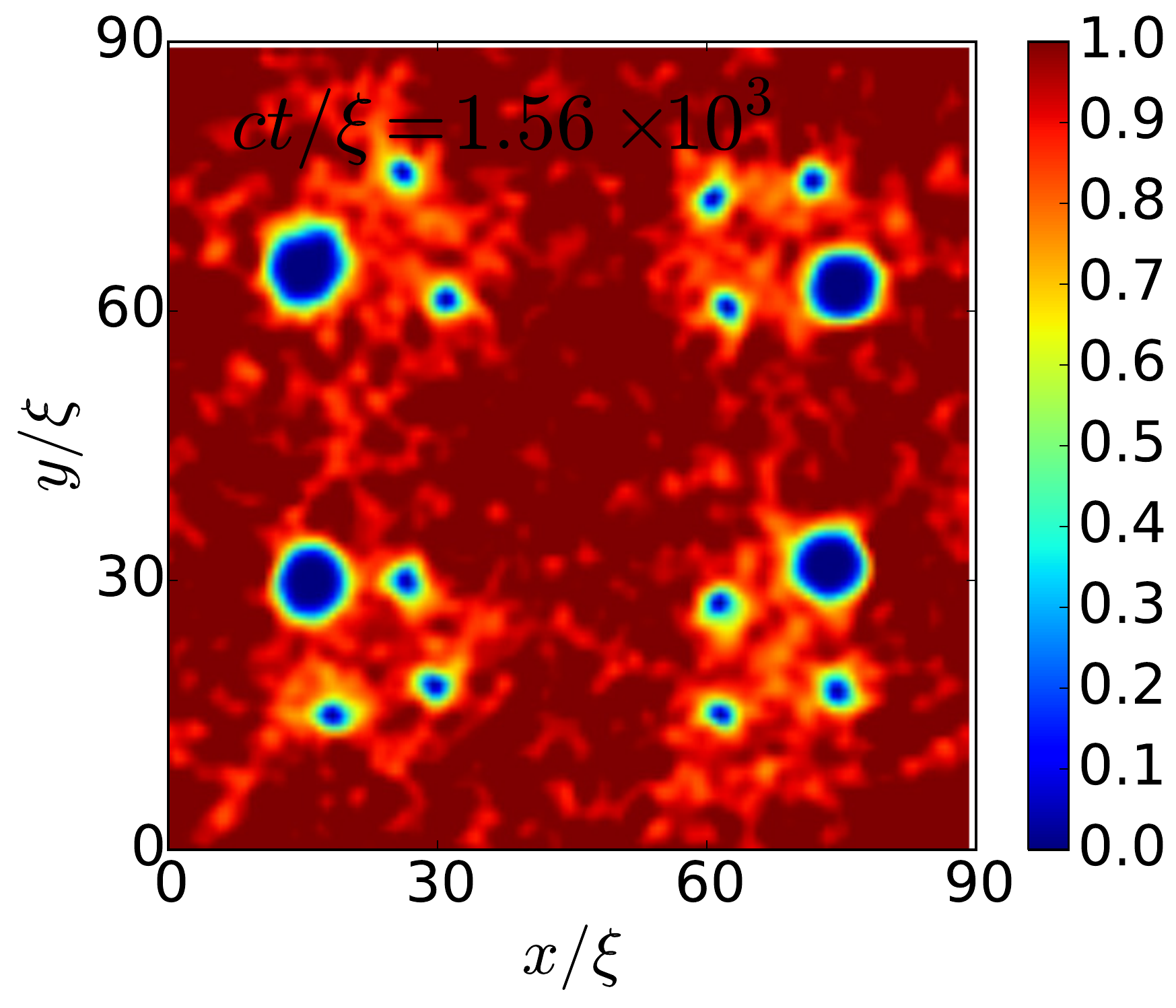}
\put(10.,10){\large{\bf (h)}}
\end{overpic}
\begin{overpic}
[height=4.5cm,unit=1mm]{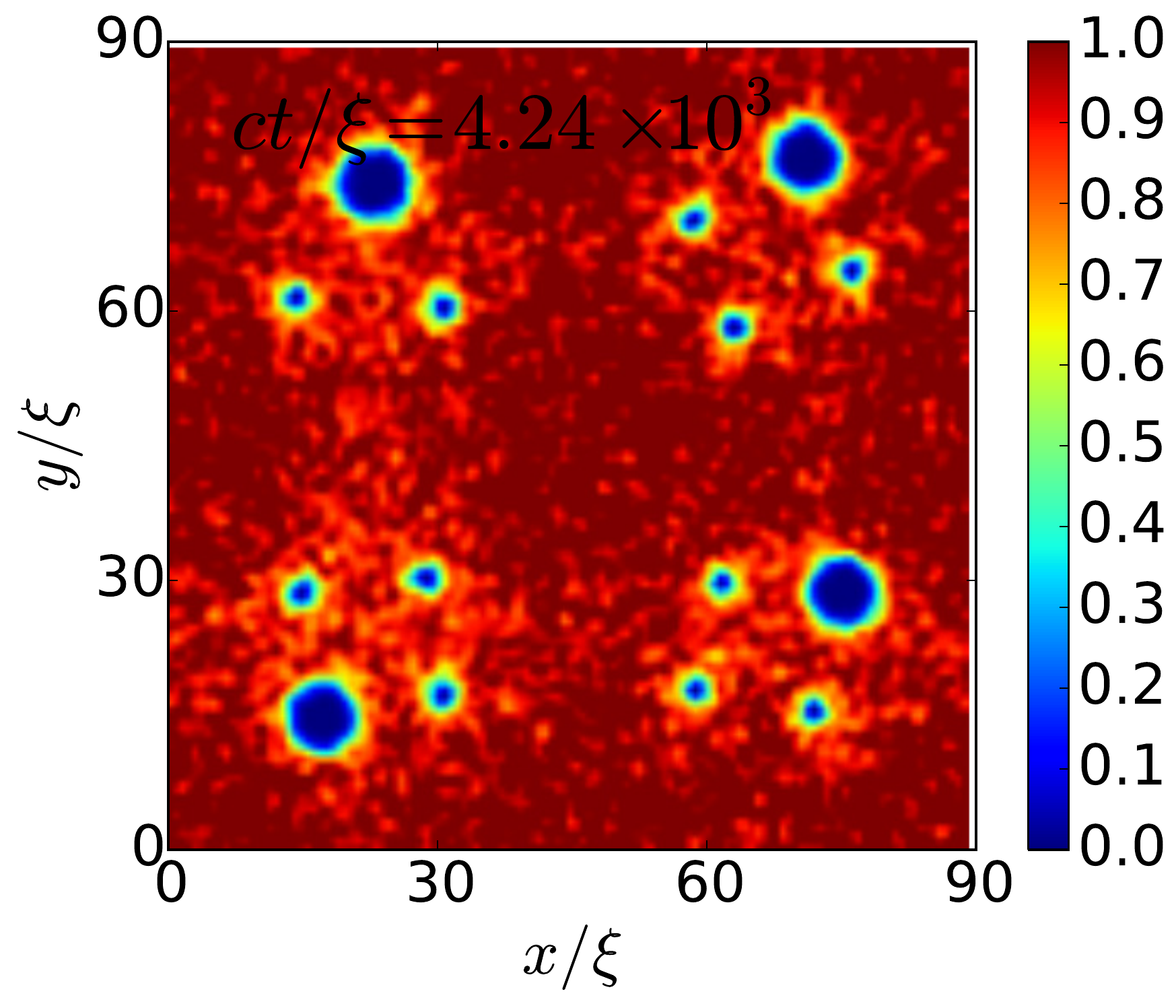}
\put(10.,10){\large{\bf (i)}}
\end{overpic}
\caption{\small (Color online) Spatiotemporal evolution of the density field
$\rho(\mathbf{r},t)$ shown via pseudocolor plots, for the four neutral
particles (large blue patches), initially placed at the centers of the
counter-rotating vortex clusters (initial configuration $\tt ICP3B$).}
\label{fig:4partcrotpd}
\end{figure*}

\begin{figure*}
\centering
\resizebox{0.9\linewidth}{!}{
\includegraphics[height=4.5cm,unit=1mm]{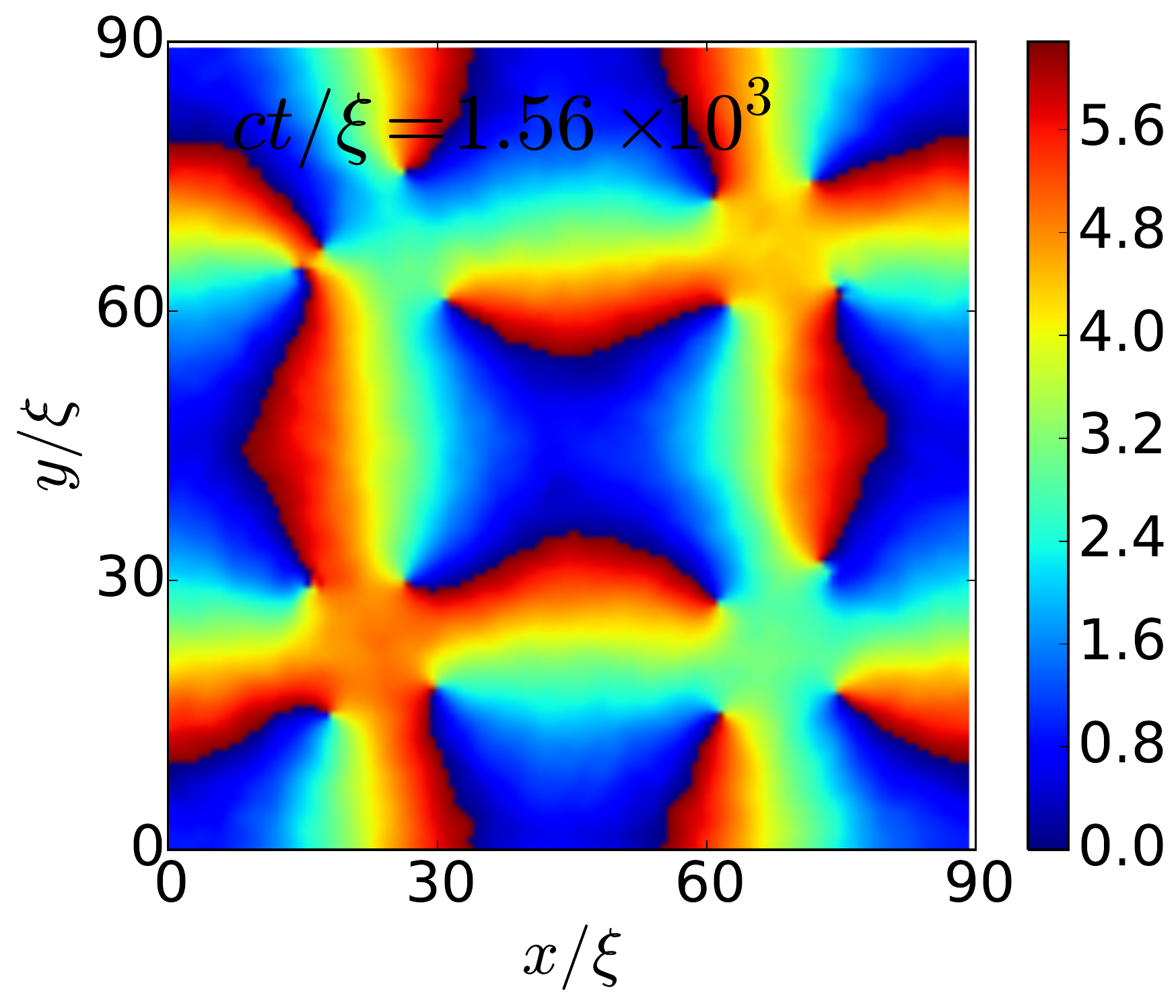}
\put(-75,20){\bf (a)}
\hspace{0.25 cm}
\includegraphics[height=4.5cm,unit=1mm]{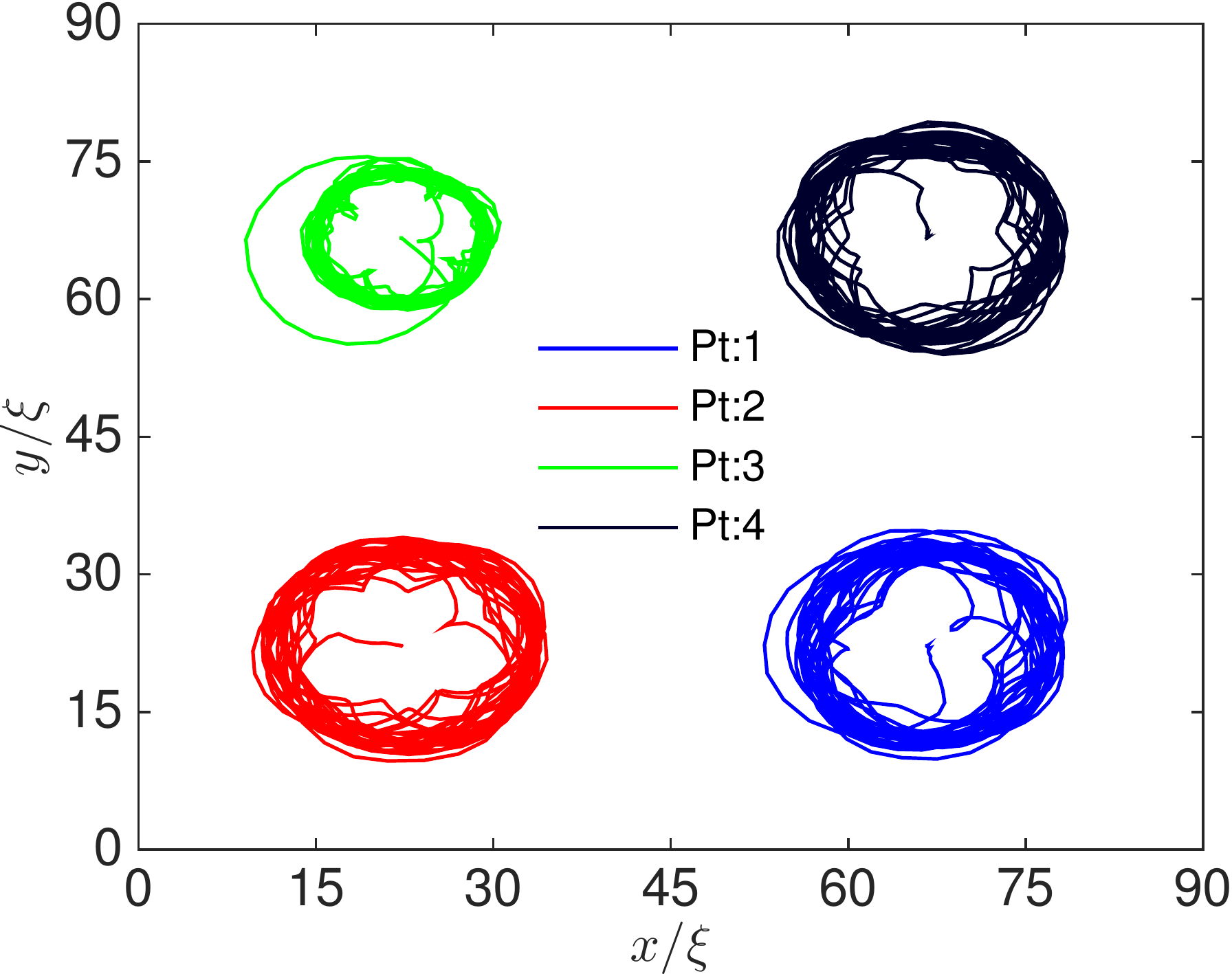}
\put(-75,20){\bf (b)}
}
\caption{\small (Color online) (a) Pseudocolor plot of the phase of
$\psi(\mathbf{x},t)$; (b) trajectories of four neutral particles $Pt:1$ (blue
curve, right bottom),  $Pt:2$ (red curve, left bottom) $Pt:3$ (green curve, left top), 
and $Pt:4$ (black curve, right top), in the presence of counter-rotating vortex clusters (initial
configuration ICP3B).}
\label{fig:4partcrotphasetraject}
\end{figure*}

\begin{figure*}
\centering
\resizebox{\linewidth}{!}{
\includegraphics[height=4.5cm,unit=1mm]{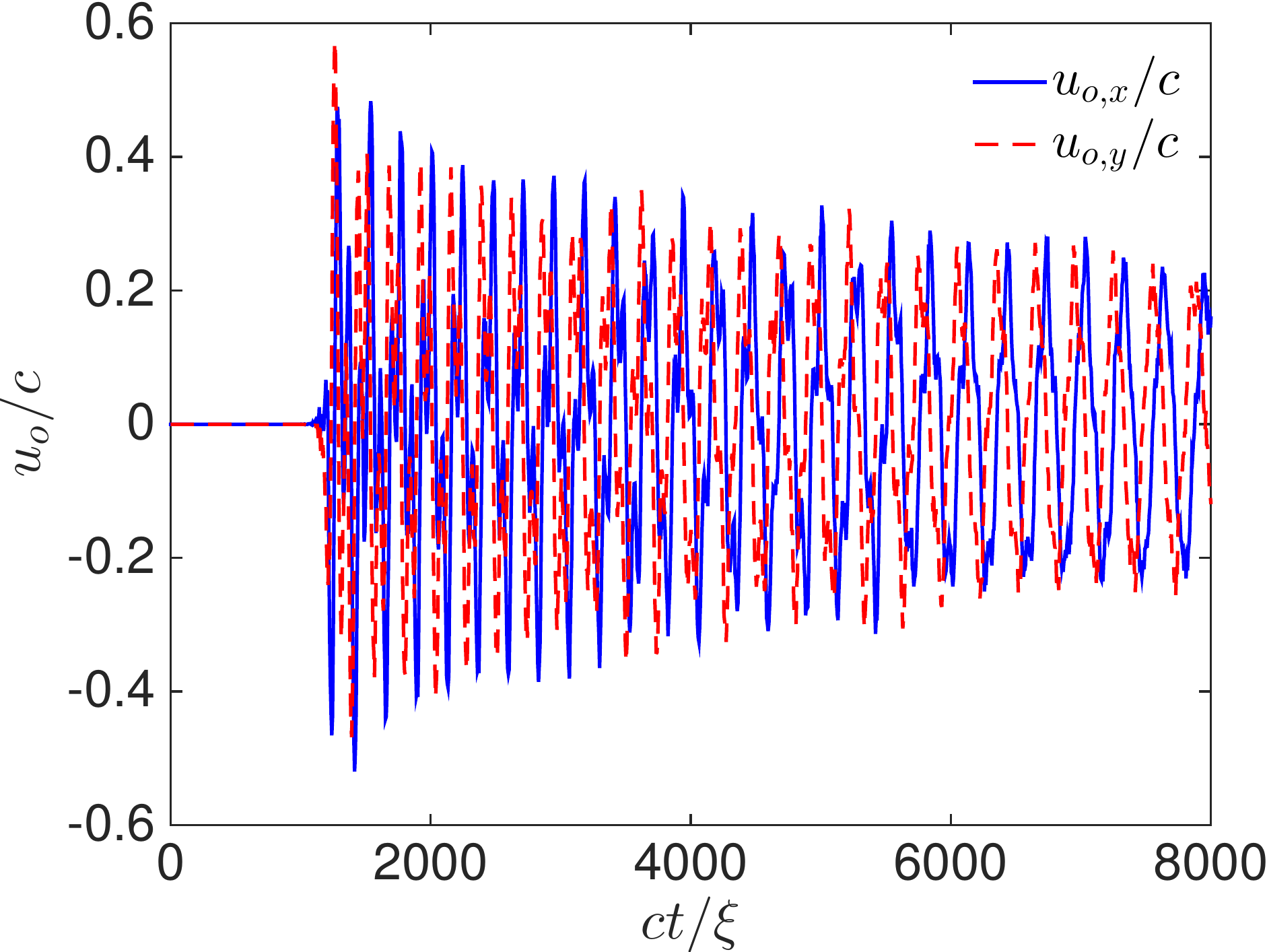}
\put(-75,20){\bf (a)}
\hspace{0.25cm}
\includegraphics[height=4.5cm,unit=1mm]{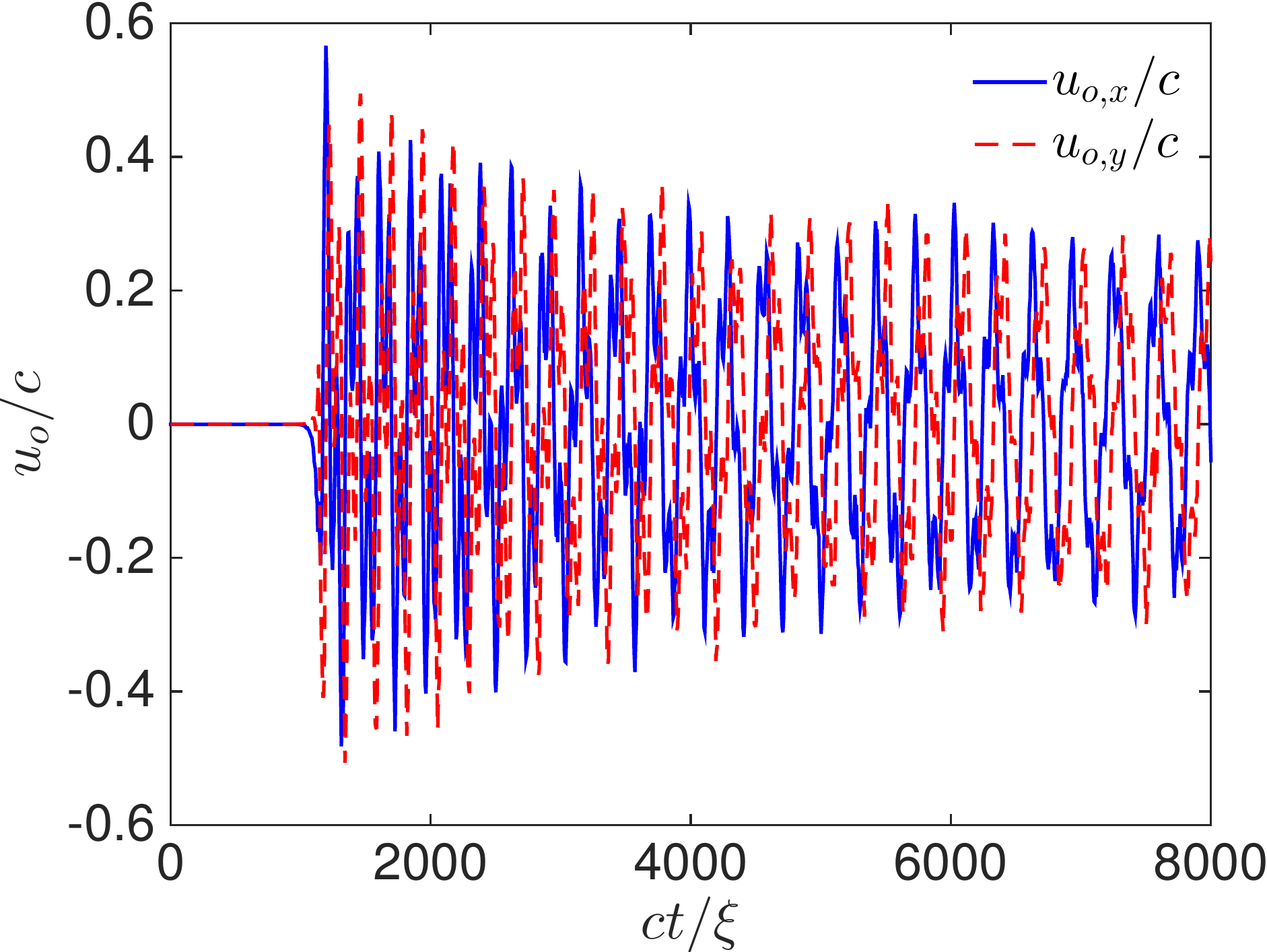}
\put(-75,20){\bf (b)}
}
\\
\vspace{0.30cm}
\resizebox{\linewidth}{!}{
\includegraphics[height=4.5cm,unit=1mm]{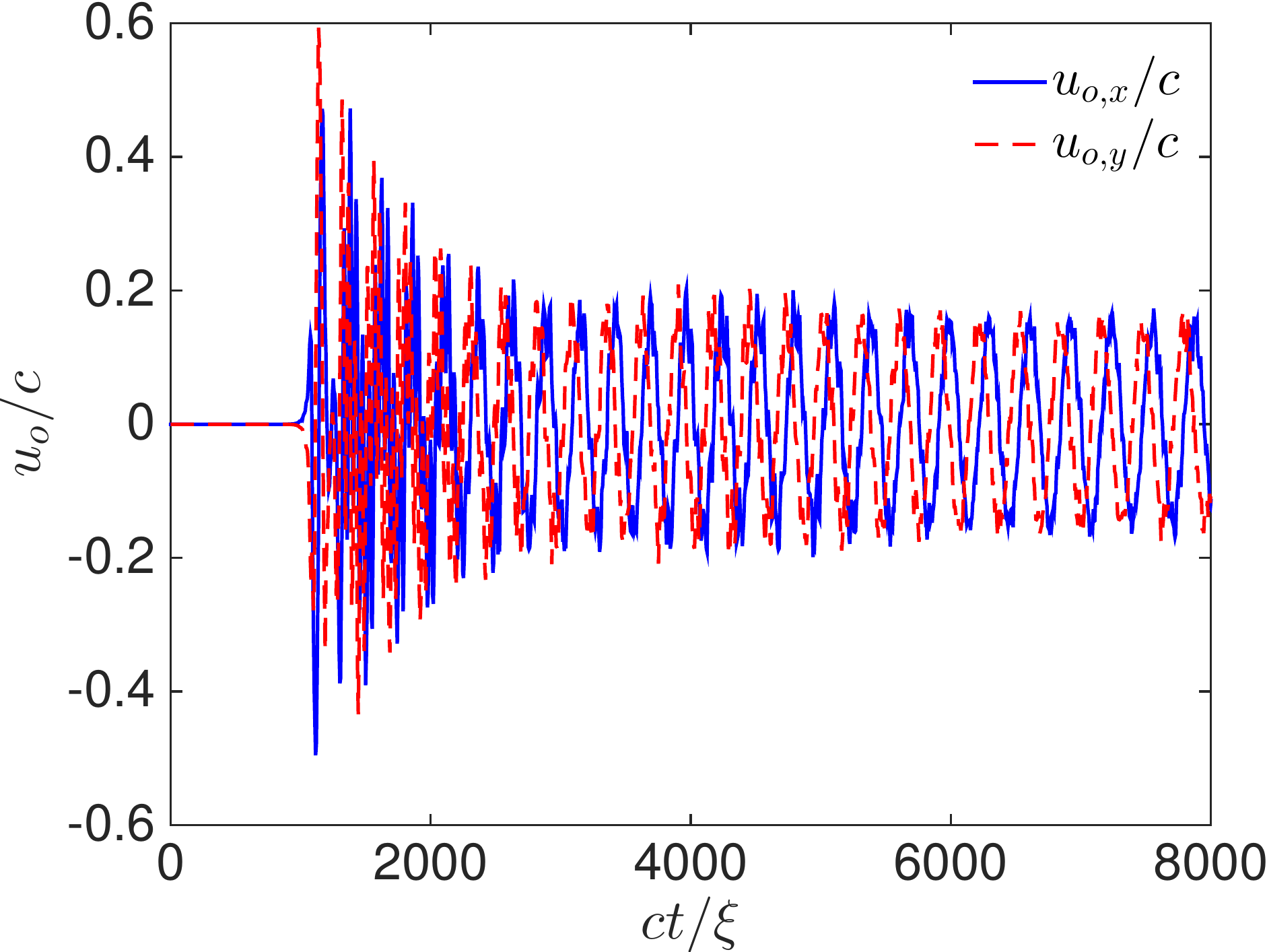}
\put(-75,20){\bf (c)}
\hspace{0.25cm}
\includegraphics[height=4.5cm,unit=1mm]{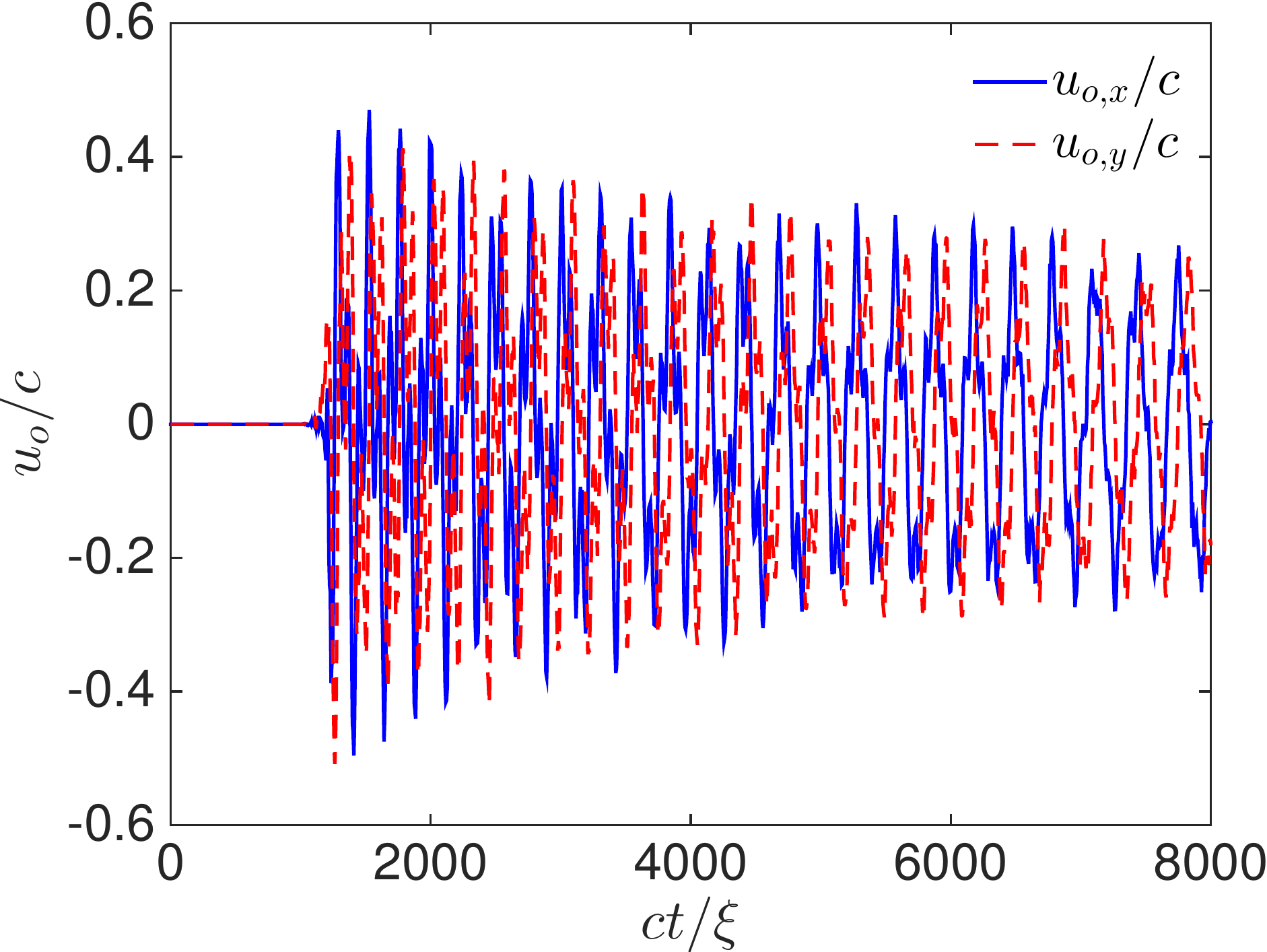}
\put(-75,20){\bf (d)}
}
\caption{\small (Color online) Plots versus time $t$ of $u_{\rm o,x}$ (blue solid
curve) and $u_{\rm o,y}$ (red dashed curve) for the four neutral particles (a)
$Pt:1$, (b) $Pt:2$, (c) $Pt:3$, and (d) $Pt:4$, in the presence of
counter-rotating vortex clusters (initial configuration ICP3B).  }
\label{fig:4partcrotu}
\end{figure*}

\begin{figure*}
\centering
\resizebox{\linewidth}{!}{
\includegraphics[height=4.5cm,unit=1mm]{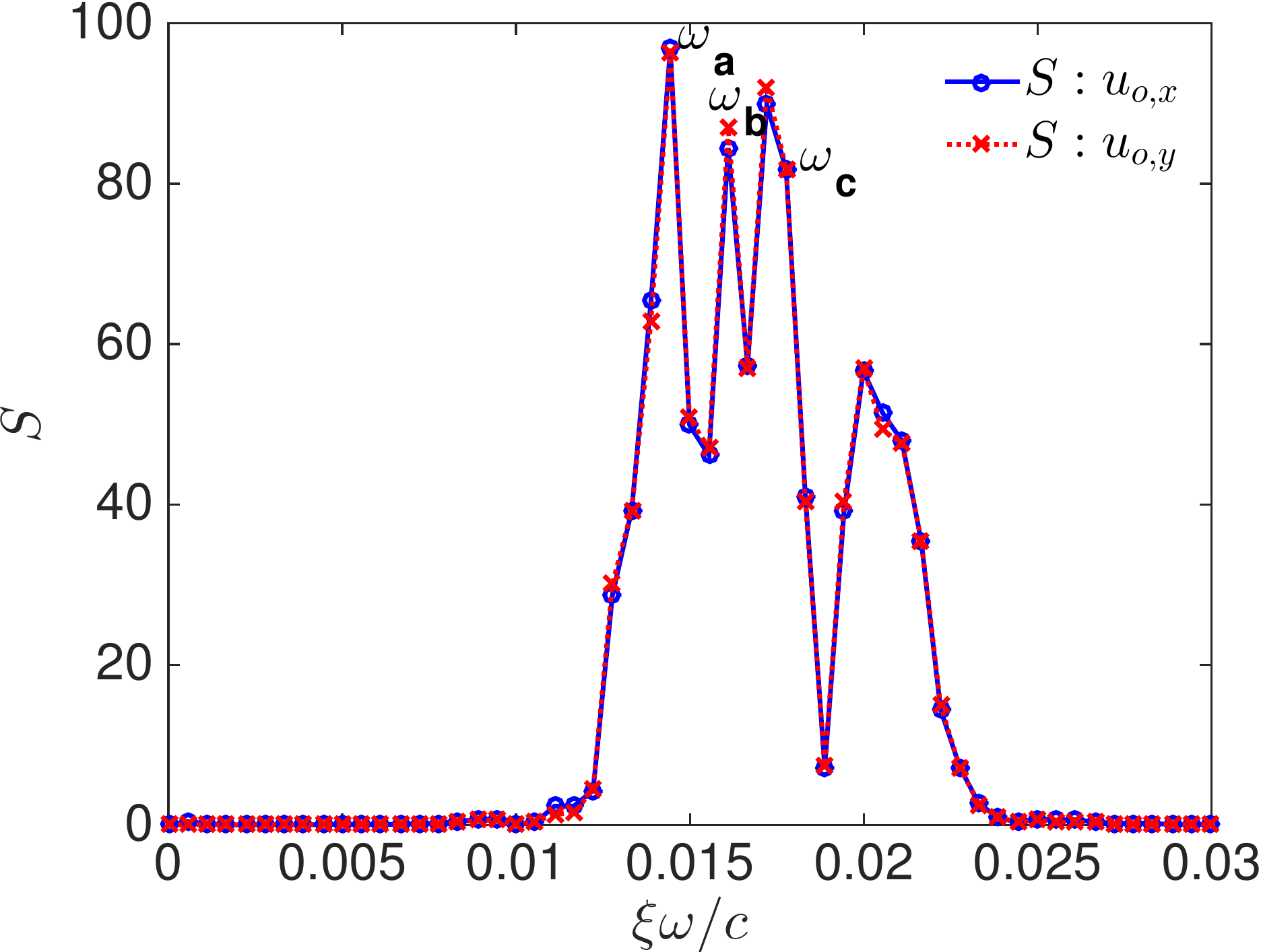}
\put(-120,30){\bf (a)}
\hspace{0.25cm}
\includegraphics[height=4.5cm,unit=1mm]{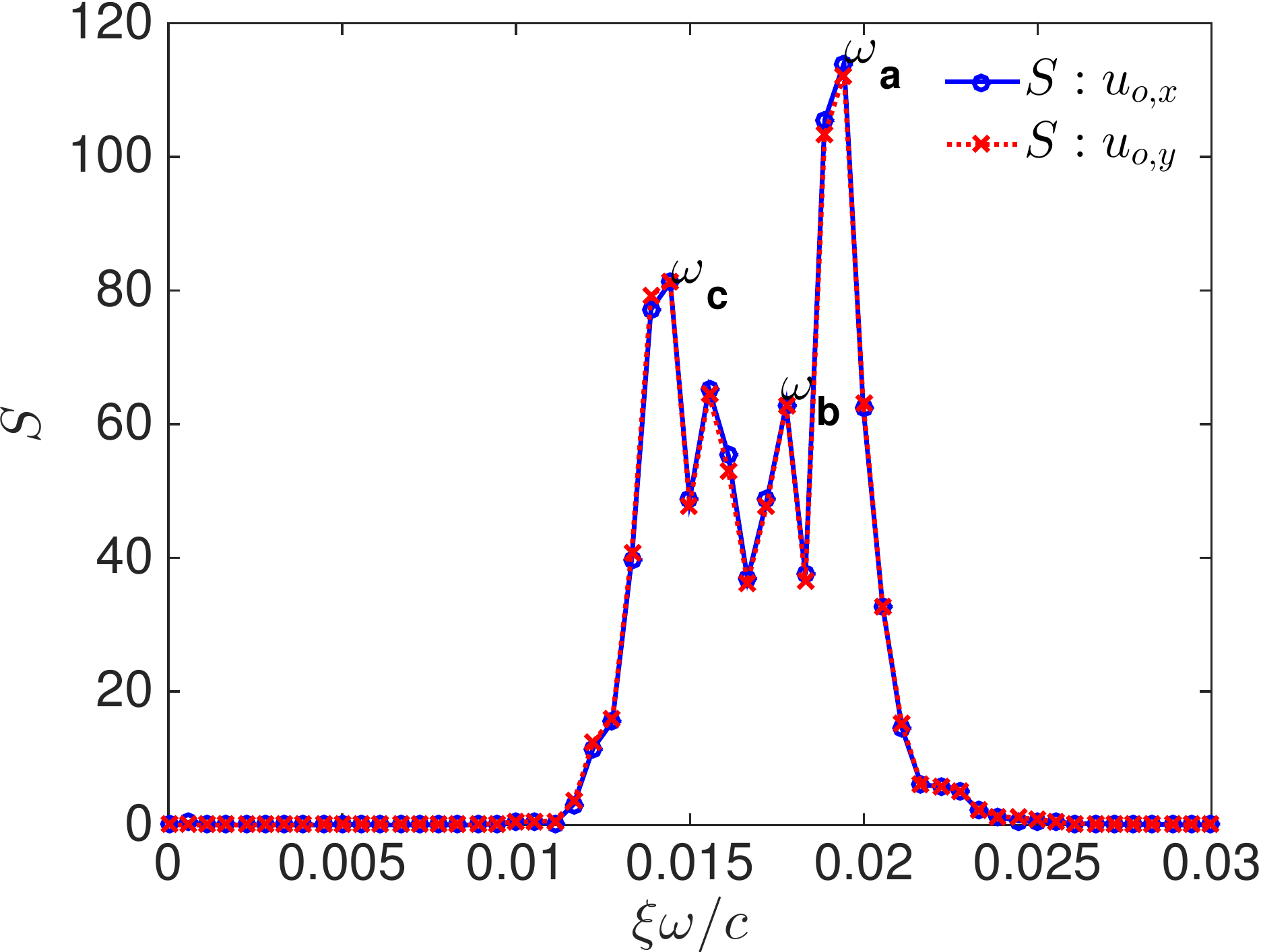}
\put(-120,30){\bf (b)}
}
\\
\vspace{0.30cm}
\resizebox{\linewidth}{!}{
\includegraphics[height=4.5cm,unit=1mm]{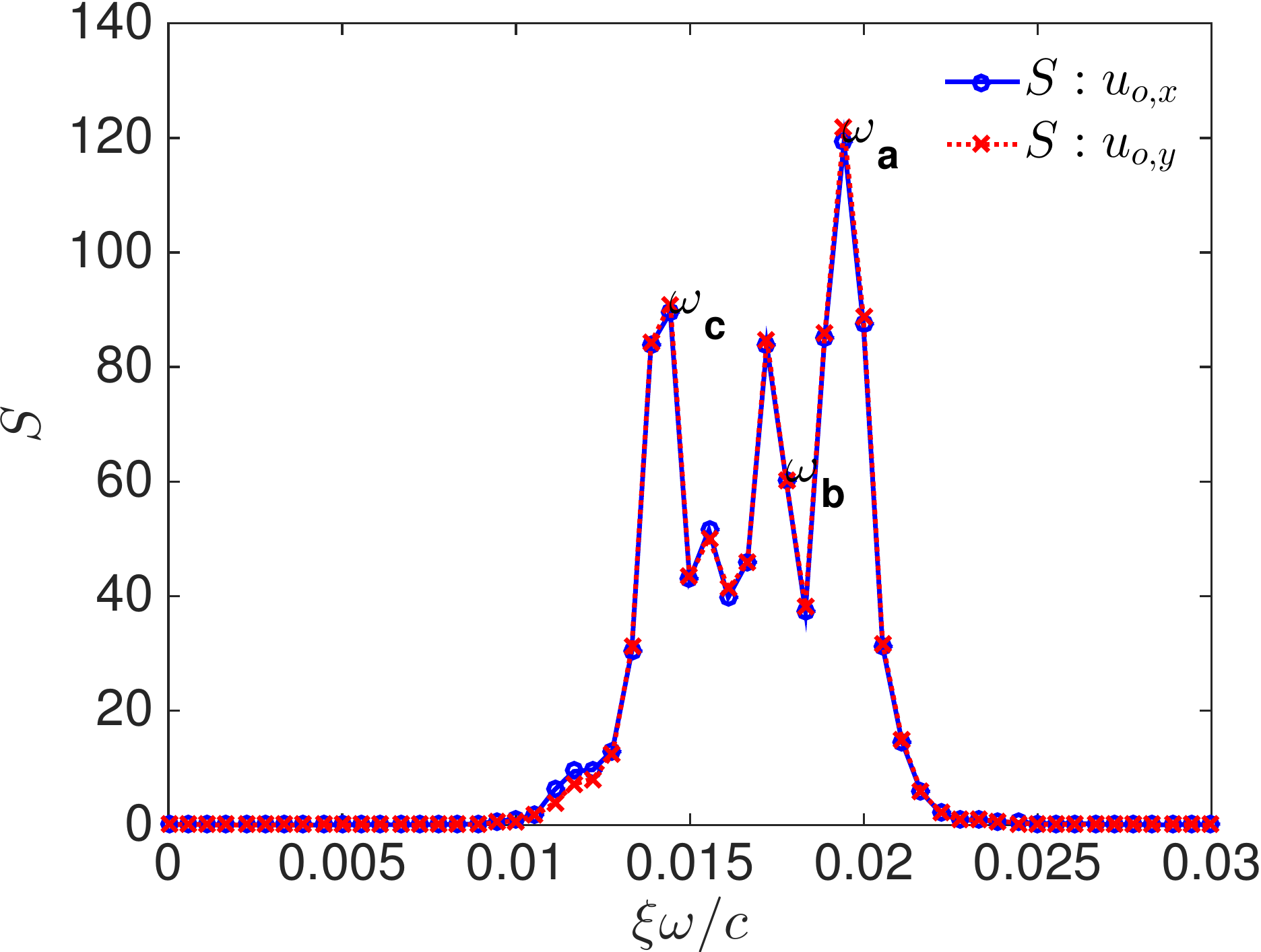}
\put(-120,30){\bf (c)}
\hspace{0.25cm}
\includegraphics[height=4.5cm,unit=1mm]{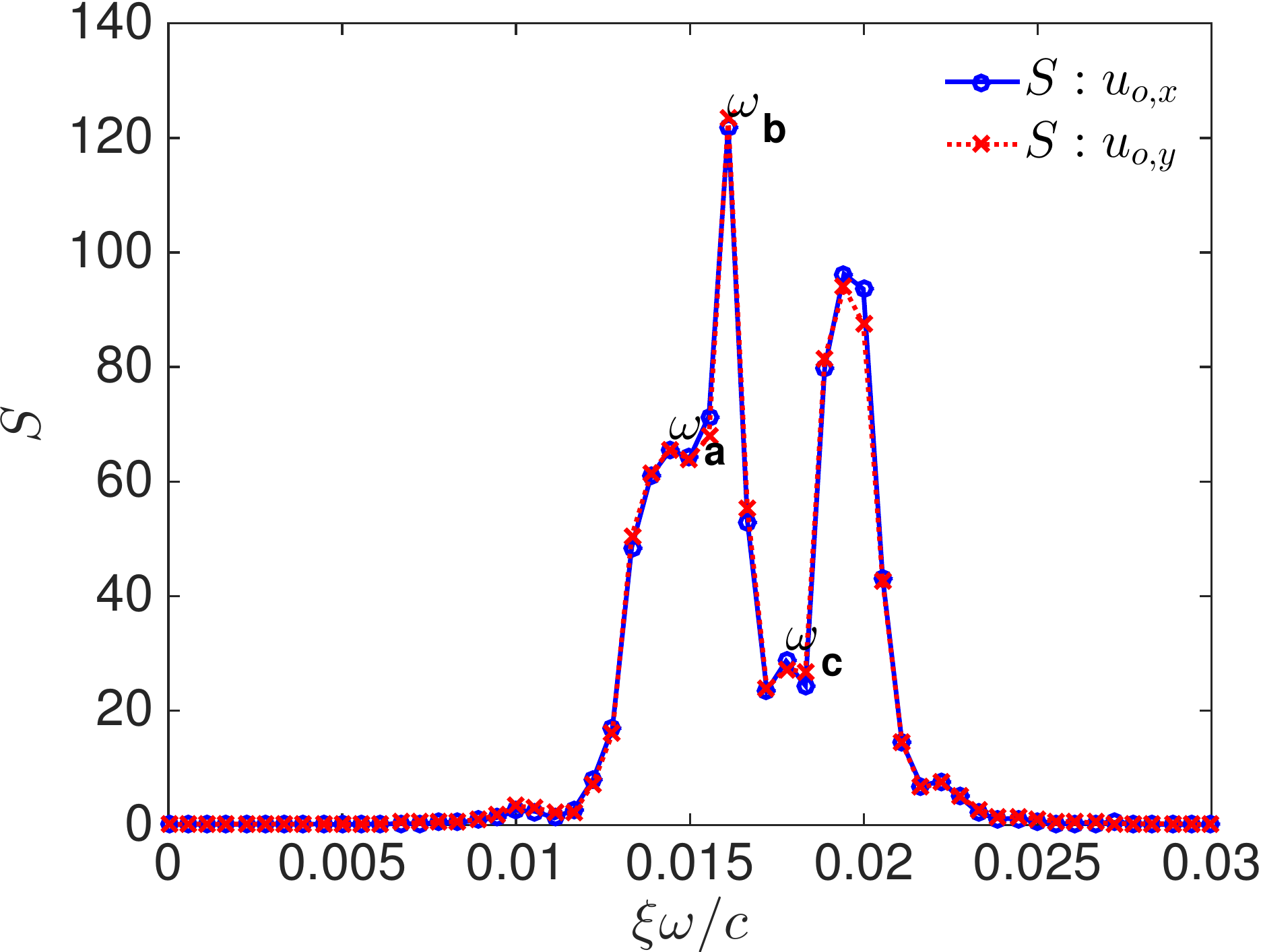}
\put(-120,30){\bf (d)}
}
\caption{\small (Color online) Plots of power spectra, denoted generically by
$S(\omega)$, of the time series of $u_{\rm o,x}$ (blue solid curve with circles), $u_{\rm o,y}$
(red dotted curve with cross) (see Figs.~\ref{fig:4partcrotu}) for the four neutral particles
(initial configuration $\tt ICP3B$) 
(a) $Pt:1$, $\omega_c=2\omega_b-\omega_a$
($\omega_a=0.01443$, $\omega_b=0.0161$, and $\omega_c=0.01777$); 
(b) $Pt:2$, $\omega_c=3\omega_b-2\omega_a$ 
($\omega_a=0.01943$, $\omega_b=0.01777$, and $\omega_c=0.01443$); 
(c) $Pt:3$, $\omega_c=3\omega_b-2\omega_a$
($\omega_a=0.01943$, $\omega_b=0.01777$, and $\omega_c=0.01443$); 
(d) $Pt:4$, $\omega_c=2\omega_b-\omega_a$ 
($\omega_a=0.01443$, $\omega_b=0.0161$, and $\omega_c=0.01777$).  
The frequencies $\omega_a$, $\omega_b$, and $\omega_c$
that we give are associated with $S(\omega)$ for $u_{\rm o,x}$ (blue curves with circles).
}
\label{fig:4partcrotpsd}
\end{figure*}

In Figs.~\ref{fig:4partcrotpd}~(a)-(i) we show the spatiotemporal evolution of
the density field $\rho(\mathbf{r},t)$. In the initial stages of the dynamical
evolution of the system, the particles $Pt:1$-$Pt:4$ remain stationary at the
respective centers of the rotating-vortex clusters, because our initial
configuration has four-fold symmetry ($C_4$) and we prepare it by using the
ground state of the vortex clusters.  (By contrast, in
subsection~\ref{subsec:1partcrotN}, our initial configuration is not the ground
of the vortex clusters.) However, at around $ct/\xi=1.05\times10^3$, an instability sets in as a
result of which the particle $Pt:3$ starts to move out; it is then trapped by
the negative vortex in front of it (see Fig.~\ref{fig:4partcrotpd}~(c),
top-left vortex cluster); the trapped particle now rotates along with the other
three vortices.  Similarly, the motion of the other particles also becomes
unstable and they are trapped in vortices, in their respective clusters;
Fig.~\ref{fig:4partcrotpd}~(d) shows that the particle $Pt:2$ is trapped by a
vortex at $ct/\xi\simeq 1.13\times10^3$ and the particles $Pt:1$ and $Pt:4$ are trapped by
vortices at $ct/\xi\simeq1.20\times10^3$ (see Fig.~\ref{fig:4partcrotpd}~(e)). Moreover, at
$ct/\xi\simeq1.31\times10^3$ the trapped particle $Pt:3$ and its vortex form a complex by
including another negative vortex of the cluster (see
Fig.~\ref{fig:4partcrotpd}~(f), top-left vortex cluster).  In
Fig.~\ref{fig:4partcrotphasetraject}~(a) we show the phase of the wave function
$\psi$; in such a plot, the vortices are the points around which the phase
changes from $0$ to $2\pi$; the top-left quadrant still has four vortices,
although two on the left are very close by and are held together in a
particle-two-vortex complex. The particle-two-vortex complex and the trapped
particles continue rotating along with the other vortices, in their respective
clusters, over the rest of the simulation time, as we show in
Figs.~\ref{fig:4partcrotpd}~(g)-(i). The Video M11~\cite{suppmat} illustrates
the spatiotemporal evolution of the four neutral particles and the density field
$\rho(\mathbf{r},t)$. If the simulation time is further extend, the 
vortex-particle clusters will expand and result in a complex motion of
the particles and vortices, which is akin to 2D superfluid turbulence.

We plot the trajectories of the four particles $Pt:1$
(blue curve), $Pt:2$ (red curve), $Pt:3$ (green curve), and $Pt:4$
(black curve) in Fig.~\ref{fig:4partcrotphasetraject}~(b).  These particles move
along roughly circular trajectories; however, their motions are not completely
periodic in time, so these trajectories meander away from perfectly closed
curves and, therefore, fill out two-dimensional areas. This is a signature of
either (a) ergodic behavior, e.g., with quasiperiodic temporal evolution or (b)
chaotic time evolution. To distinguish between (a) and (b), we examine the time
series of $u_{\rm o,x}$ and $u_{\rm o,y}$ for all the four particles in
Figs.~\ref{fig:4partcrotu}~(a)-(d).  The principal peaks in these power spectra
can be indexed as $n_a\omega_a+n_b\omega_b$, where $n_a$ and $n_b$ are integers
and there are two main incommensurate frequencies $\omega_a$ and $\omega_b$
(i.e., $\omega_a/\omega_b$ is an irrational number).  For example, in
Fig.~\ref{fig:4partcrotpsd}~(a) the frequency $\omega_c$, of one of the peaks
in the power spectrum $S(\omega)$, can be written as
$\omega_c=2\omega_b-\omega_a$ (here $\omega_c=0.0.01777$, $\omega_a=0.01443$, and
$\omega_b=0.0161$).  This labelling of peaks indicates clearly that the
temporal evolution of $Pt:1$ is quasi-periodic.

\section{Conclusions}

We have carried out an extensive DNS to investigate the interaction of
particles and fields in the 2D GPE, in both simple and turbulent flows. At the
one-particle level, we explore, for light, neutral, and heavy particles, the
nature of their dynamics in the superfluid, when a constant external force acts
on them; in particular, we show, by a careful consideration of the effects of
the particle mass, how the motion of such particles can become chaotic.  We
demonstrate that the interaction of a particle with vortices leads to dynamics
that depends sensitively on the particle characteristics. The motion of a
single particle through a three-dimensional (3D) GP superfluid has been
explored in Refs.~\cite{winiecki1,winiecki2}. Our work complements and extends
significantly these earlier studies by (a) considering the 2D GPE, (b)
examining the particle-mass-dependence of the dynamics, and (c) showing that,
after the emission of a vortex-antivortex pair, the temporal evolution of the
particle position can be periodic or chaotic.

We extend our studies to assemblies of particles and vortices and demonstrate
that their dynamics show rich, turbulent spatiotemporal evolution.  In
particular, we systematize the spatiotemporal evolution of an initial
configuration in which one particle is placed in front of a translating
vortex-antivortex pair. Here our study goes well beyond the recent work
presented in Ref.~\cite{griffin2017vortex}, which considers vortex scattering by
impurities in a Bose-Einstein condensate; in this study the impurities are
static; by contrast, in our work the particle has nontrivial dynamics.  Note
that, in the limit of large particle mass, our results are akin to those of
Ref.\cite{griffin2017vortex}, with the vortices moving in a superfluid background
with a fixed impurity potential. Moreover, our one-particle studies are of
direct relevance to the experiments of Ref.~\cite{kwon15}, 
in which vortex shedding is examined in a BEC
with a repulsive Gaussian laser beam that is moved through the condensate.
Another system which for which our results may prove to be useful is the
quantum fluid of light in nonlinear optical systems~\cite{RMPCarusotto2013}.

Next we examine the complex motions of two particles placed in front of a
translating vortex-antivortex pair. An important point that emerges from our
study is that, once a particle becomes coincident with a vortex, we can use
this particle as a tracer that can track vortex motion. This is of direct
experimental relevance; indeed, recent experiments have employed frozen
hydrogen particles to track quantized vortices in superfluid
Helium~\cite{bewley2006superfluid,bewley2008characterization,bewleyparticlesdetails,He2excimerT0tracer,mantiaprbaccpdf}.

Finally, our investigations show, for the illustrative examples
of  (a) a single particle moving in the presence of 
complex spatial distribution of vortices
and (b) four particles in the presence of counter-rotating vortex 
clusters, how the interactions of particles and fields in a 2D GP superfluid
can lead to rich, turbulent, and spatiotemporally chaotic evolution. We
hope our work will lead to experimental studies of such spatiotemporal chaos
in superfluid Helium and BECs.

\acknowledgements

We thank the Indo-French Centre for Applied Mathematics (IFCAM) for financial
support.  RP and VS thank the Council of Scientific and Industrial Research
(India), University Grants Commission (India), Department of Science and
Technology (India) for support and the Supercomputing Education and Research
Centre, IISc, India for computational resources. V.S. acknowledges support from
Centre Franco-Indien pour la Promotion de la Recherche Avanc\'ee (CEFIPRA)
Project No.  4904.  VS and RP thank ENS, Paris for hospitality and MB thanks
IISc, Bangalore for hospitality. We are grateful to G. Krstulovic for useful
discussions.

\appendix
\renewcommand\thefigure{\thesection.\arabic{figure}}  

\section{Note on units}
\label{app:units}

The s.pdf given below constitute an easy exercise, however, we give them
to avoid any confusion; moreover, this will form an useful background for
the appendices on initial data preparation.
Consider the field part of the full Lagrangian~\eqref{eq:Lagrangianfull},
\begin{widetext}
\begin{align}
\nonumber
\mathcal{L} &= \int_{\mathcal{A}}\Bigl[\frac{i\hbar}{2}\Bigl(\psi^*(\mathbf{r},t)\frac{\partial \psi(\mathbf{r},t)}{\partial t}
-\psi(\mathbf{r},t)\frac{\partial \psi^*(\mathbf{r},t)}{\partial t}\Bigr)
- \frac{\hbar^2}{2m}\nabla\psi(\mathbf{r},t)\cdot\nabla\psi^*(\mathbf{r},t)
+\mu|\psi(\mathbf{r},t)|^2
- \frac{g}{2}|\psi(\mathbf{r},t)|^4\\
&\quad-\sum^{\mathcal{N}_0}_{i=1}V_{\mathcal{P}}(\mathbf{r}-\mathbf{q}_i)|\psi(\mathbf{r},t)|^2\Bigr]d\mathbf{r},
\end{align}
\end{widetext}
which we rewrite as follows:
\begin{widetext}
\begin{align}
\nonumber
\mathcal{L} &= \int_{\mathcal{A}}\Bigl[\frac{i\hbar}{2\,m}\Bigl(\tilde{\psi}^*(\mathbf{r},t)\frac{\partial\tilde{\psi}(\mathbf{r},t)}{\partial t}
-\tilde{\psi}(\mathbf{r},t)\frac{\partial \tilde{\psi}^*(\mathbf{r},t)}{\partial t}\Bigr)
- \frac{\hbar^2}{2\,m^2}\nabla\tilde{\psi}(\mathbf{r},t)\cdot\nabla\tilde{\psi}^*(\mathbf{r},t)
+\frac{\hbar}{m}\frac{\mu}{\hbar}|\tilde{\psi}(\mathbf{r},t)|^2
- \frac{\hbar}{m}\frac{g}{2\,\hbar\,m}|\tilde{\psi}(\mathbf{r},t)|^4\\
&\quad-\sum^{\mathcal{N}_0}_{i=1}\frac{\hbar}{m}\frac{V_{\mathcal{P}}(\mathbf{r}-\mathbf{q}_i)}{\hbar}|\tilde{\psi}(\mathbf{r},t)|^2\Bigr]d\mathbf{r},\\
\nonumber
	&= \int_{\mathcal{A}} \Bigl[i\alpha_0\Bigl(\tilde{\psi}^*(\mathbf{r},t)\frac{\partial\tilde{\psi}(\mathbf{r},t)}{\partial t}
-\tilde{\psi}(\mathbf{r},t)\frac{\partial \tilde{\psi}^*(\mathbf{r},t)}{\partial t}\Bigr)
- 2\alpha^2_0 \nabla\tilde{\psi}(\mathbf{r},t)\cdot\nabla\tilde{\psi}^*(\mathbf{r},t)
+2\alpha_0\tilde{\mu}|\tilde{\psi}(\mathbf{r},t)|^2 
- 2\alpha_0 \frac{\tilde{g}}{2}|\tilde{\psi}(\mathbf{r},t)|^4\\
&\quad-\sum^{\mathcal{N}_0}_{i=1}2\alpha_0\tilde{V}_{\mathcal{P}}(\mathbf{r}-\mathbf{q}_i)|\tilde{\psi}(\mathbf{r},t)|^2\Bigr]d\mathbf{r}.\label{eq:lagfieldfluidvariable}
\end{align}
\end{widetext}
From the above s.pdf it follows that $\tilde{g}=g/\hbar\,m$, $\tilde{\mu}=\mu/\hbar$,
$\tilde{V}_{\mathcal{P}}=V_{\mathcal{P}}/\hbar$ and 
$|\tilde{\psi}|^2=m|\psi|^2$; thus, $|\tilde{\psi}|^2$ can be directly interpreted
as the mass density of the superfluid. From the Lagrangian Eq.~\eqref{eq:lagfieldfluidvariable}
we obtain the following equation of motion for the field $\tilde{\psi}$
\begin{equation}
2\alpha_0\,i\frac{\partial\tilde{\psi}}{\partial\,t} = 
2\alpha_0\left[-\alpha_0\,\nabla^2
+ \tilde{g}|\tilde{\psi}|^2 - \tilde{\mu} + \tilde{V}_{\mathcal{P}}\right]\tilde{\psi}.
\end{equation}

Hereafter we drop the $\widetilde{}$ symbol for notational convenience in the discussions
of Secs.~\ref{ARGLE}, \ref{app:ARGLEnumerics} and \ref{app:initialstates}.

\section{Advective real Ginzburg-Landau equation (ARGLE)}
\label{ARGLE}

Gross-Pitaevskii hydrodynamics is compressible, so, when vortices are present,
we get dynamics that is dominated by acoustic waves. To minimize these acoustic
emissions, it is useful to prepare initial conditions by using the
advective-real-Ginzburg-Landau equation (ARGLE) given in
Refs.~\cite{nore1997,mfcoeff}. The initial states we want emerge
as the large-time-asymptotic solutions of the following ARGLE: 
\begin{equation} \label{eq:argle}
\frac{\partial \psi}{\partial t} = \alpha_0\nabla^2\psi
-g|\psi|^2\psi \\ 
+ \mu\psi -i\mathbf{u}_{\rm adv}\cdot\nabla\psi 
-\frac{\mathbf{u}^2_{\rm adv}}{4\alpha_0}\psi;
\end{equation}
these states minimize the free-energy functional
\begin{equation}\label{eq:arglefunctional}
\begin{split}
\mathcal{F}_{\rm ARGLE}(\psi,\psi^*)&=\int d^3x 
\Biggl(\alpha_0\left|\nabla\psi-i\frac{\mathbf{u}_{\rm adv}}{2\alpha_0}\psi\right|^2
+\frac{1}{2}g|\psi|^4 \\
&\quad -\mu|\psi|^2\Biggr);
\end{split}
\end{equation}
here, $\mathbf{u}_{\rm adv}$ is the imposed flow velocity.
In Eqs.~\eqref{eq:argle} and \eqref{eq:arglefunctional} a potential term
can be included to prepare initial states with particles.

\section{Numerical implementation of ARGLE}
\label{app:ARGLEnumerics}

For time stepping in the ARGLE we use the following implicit-Euler method:
\begin{equation}
\psi(t+\Delta t) = \frac{\psi(t) + NL(t)\Delta t}{1-L\Delta t},
\end{equation}
where we suppress the spatial argument of $\psi$,
$L=\alpha_0\nabla^2$, and $NL=(\mu-g|\psi|^2)\psi 
-i\mathbf{u}_{\rm adv}\cdot\nabla\psi -\frac{\mathbf{u}^2_{\rm adv}}{4\alpha_0}\psi$.
The field $\psi$ at the time step $(n+1)$ is given by
\begin{equation}
\hat{\psi}_{n+1} = \frac{\hat{\psi}_n+\Delta t(\mu-g\widehat{|\psi_n|^2\psi_n} 
-i\widehat{\mathbf{u}_{\rm adv}\cdot\nabla\psi_n} 
- \widehat{\frac{\mathbf{u}^2_{\rm adv}}{4\alpha_0}\psi_n})}{1-(-\alpha_0 k^2)\Delta t}.
\end{equation}
We use the Newton method to find the stable and the unstable fixed
points  of the above equation; this is equivalent to finding $\psi_*$, such
that
\begin{equation}
F(\psi_*) \equiv \psi_*(t) - \psi_*(t+\Delta t) = 0.
\end{equation}
Every Newton step requires the solution, for $\delta \psi$, of
\begin{equation}
\frac{\delta F}{\delta \psi}\delta \psi = -F(\psi);
\end{equation}
we obtain this by an iterative, bi-conjugate-gradient-stabilized method 
(BiCGSTAB)~\cite{van1992bcgstab}. This
method uses $[\delta F/\delta\psi]$ over an arbitrary field $\phi$:
\begin{equation}
\begin{split}
\frac{\delta F}{\delta \psi}\phi &= \frac{-\Delta t}{1-L\Delta t}
\Bigl[L\phi + g(2|\psi|^2\phi+\psi^2\phi^*)
-i\mathbf{u}_{\rm adv}\cdot\nabla\phi  \\
&-(\mathbf{u}^2_{\rm adv}/4\alpha_0)\phi\Bigr].
\end{split}
\end{equation}

\section{Preparation of initial states with vortices}
\label{app:initialstates}

\subsection{Preparation of a translating vortex-antivortex pair: $\psi_{\rm pair}$}
\label{app:vortpair}

We give below the s.pdf required to prepare $\psi_{\rm pair}$
(see Refs.~\cite{nore1997,mfcoeff}):
\begin{enumerate}

\item Initialize $\psi(x,y)=\exp(ix)$ for $l_{\rm min}<y< l_{\rm max}$ and 
$\psi(x,y)=1$ otherwise.

\item  Evolve $\psi$ by using ARGLE, with $u_{\rm adv}=0$, and allow the
vortex-antivortex pair that is generated to contract until it reaches the
desired value of the pair length $d$.

\item Evolve $\psi$, from the previous step, by using ARGLE, with
$\mathbf{u}_{\rm adv}=u\hat{x}$, so that the contraction of the
vortex-antivortex pair stops.

\item Use Newton's method, coupled with BiCGSTAB, to find the exact state of
the vortex-antivortex pair for $\mathbf{u}_{\rm adv}$ in step $3$ above.  This
Newton method helps to speed up the convergence to the desired solution
(the solution is a saddle point of Eq.~(\ref{eq:argle}); the
ARGLE procedure, if used alone, first converges, but finally ends up
diverging.

\end{enumerate}

\subsection{Preparation of counter-rotating vortex clusters}
\label{app:counterrotvort}

The s.pdf involved in the preparation of $\psi_{\rm cluster}$ are outlined
below:

\begin{enumerate}
\item Initialize $\psi_e(\lambda_1,\lambda_2)=\frac{(\lambda_1+\iota
\lambda_2)}{A} \tanh\bigl(\frac{A}{\sqrt{2}\xi}\bigr)$, where
$\lambda_1=\sqrt{2}\cos x$, $\lambda_2=\sqrt{2}\cos y$, and
$A=\sqrt{\lambda_1^2+\lambda_2^2}$.
\item Prepare
$\psi_4=\psi_e(\lambda_1-\eta,\lambda_2)\psi_e(\lambda_1,\lambda_2-\eta)
\psi_e(\lambda_1+\eta,\lambda_2)\psi_e(\lambda_1,\lambda_2+\eta)$, where
$\eta=1/\sqrt{2}$.
\item Prepare $\psi_{\rm cluster}=(\psi_4)^{[\gamma_d/4]}$, where
$\gamma_d=8/(4\pi\alpha_0)$ and $[\cdot]$ denotes the integer part of a real
number.  
\item Evolve $\psi_{\rm cluster}$ by using ARGLE with $u_{\rm
adv,x}=\sin(x)\cos(y)$, $u_{\rm adv,y}=-\cos(x)\sin(y)$ to minimize acoustic
emission.
\end{enumerate}
For more details on the preparation of a counter-rotating vortex clusters see
Ref.~\cite{nore1997}.


\section{Additional figures}\label{app:additionalplots}
\setcounter{figure}{0}
In this appendix we give additional plots to compliment the discussion
of Sec.~\ref{subsec:1partextforce} and~\ref{subsec:1partvortexpair}.


\begin{figure*}
\centering
\begin{overpic}
[height=4.5cm,unit=1mm]{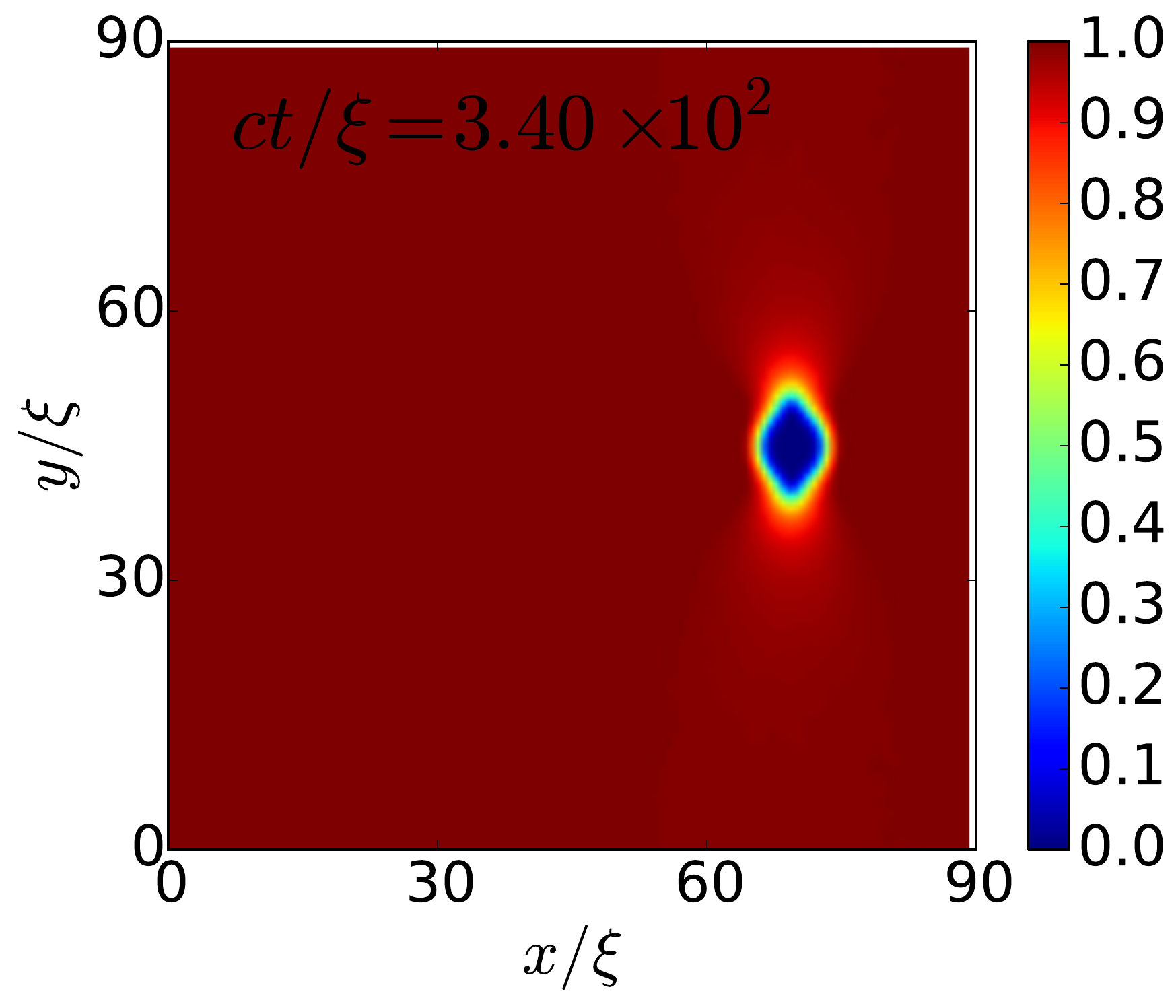}
\put(10.,10){\large{\bf (a)}}
\end{overpic}
\begin{overpic}
[height=4.5cm,unit=1mm]{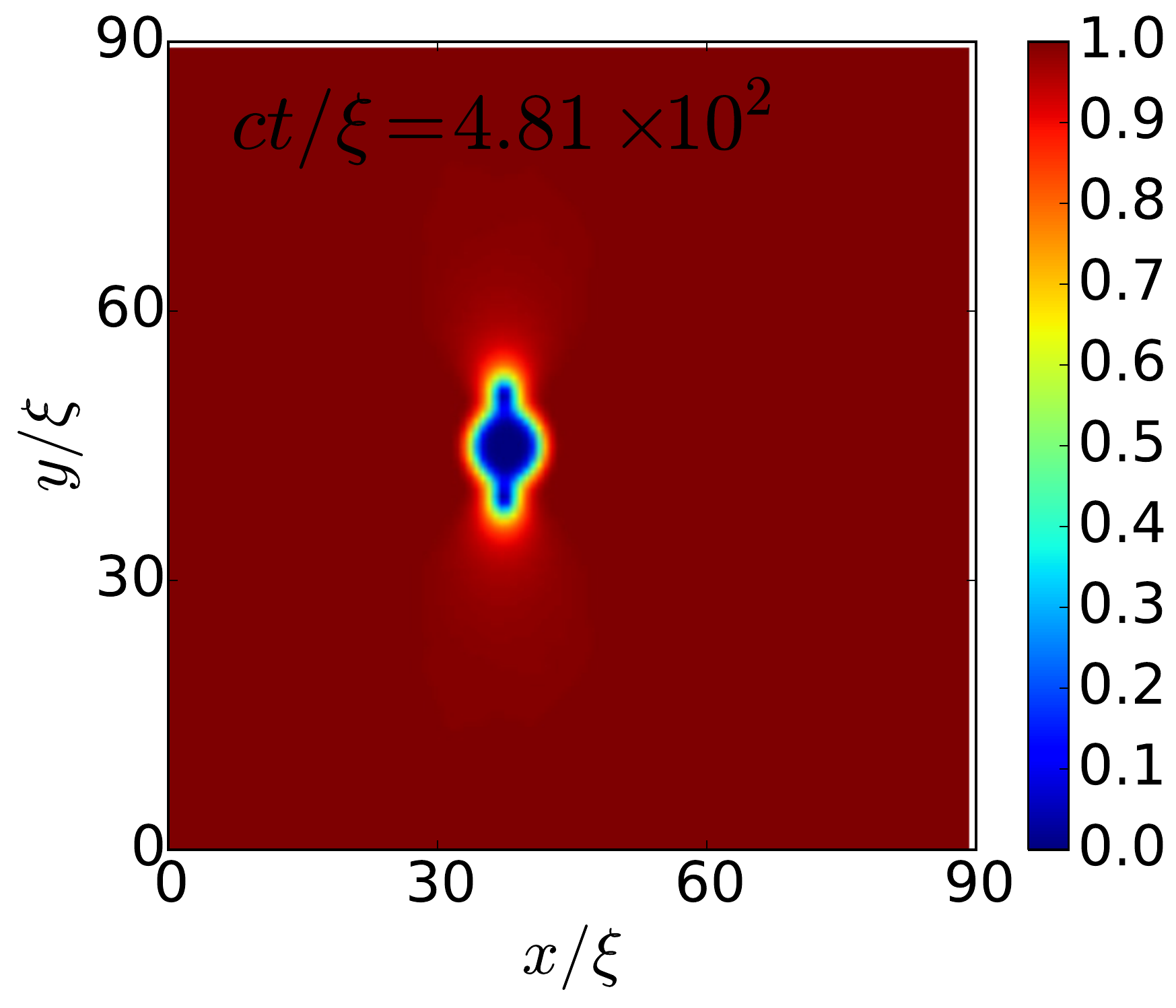}
\put(10,10){\large{\bf (b)}}
\end{overpic}
\begin{overpic}
[height=4.5cm,unit=1mm]{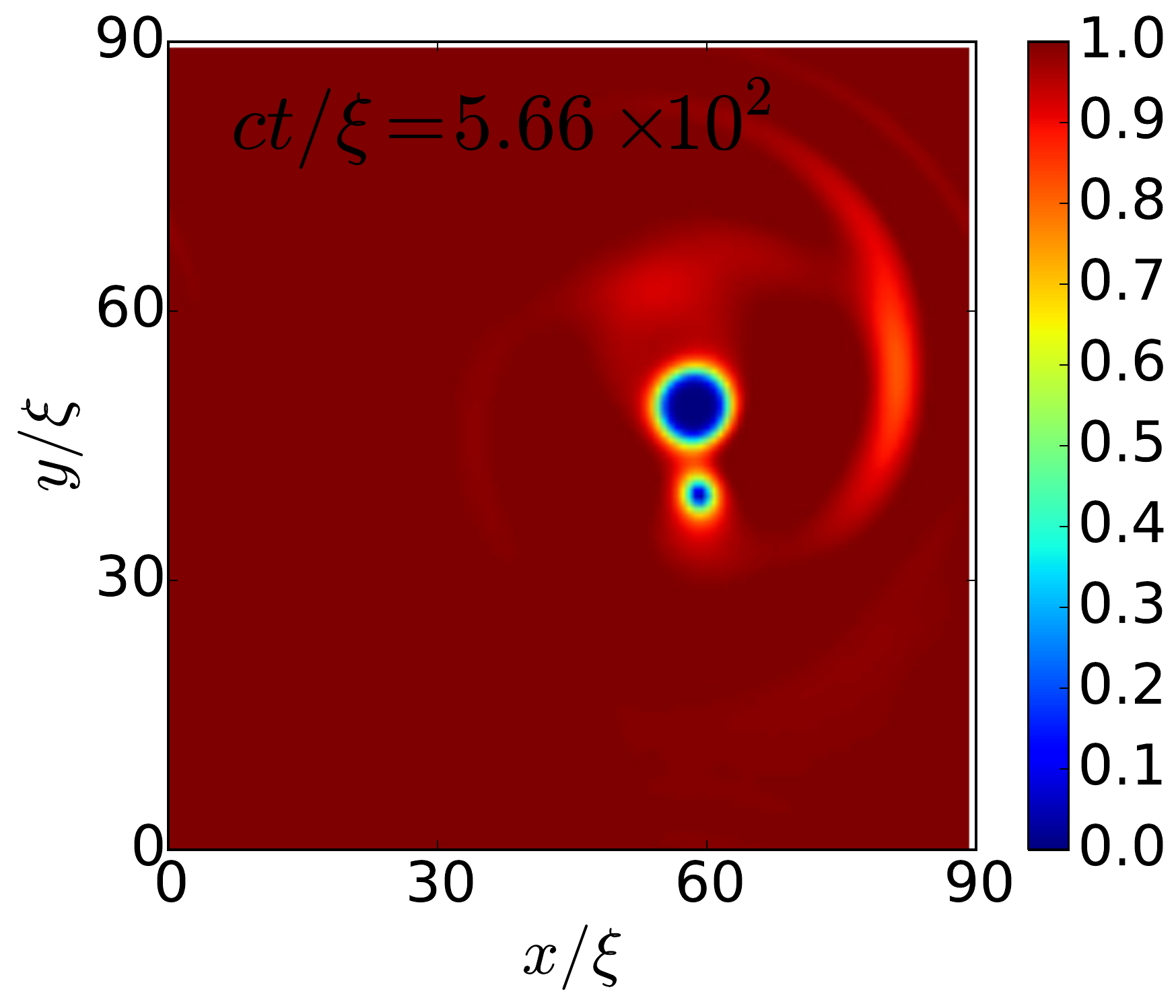}
\put(10,10){\large{\bf (c)}}
\end{overpic}
\\
\begin{overpic}
[height=4.5cm,unit=1mm]{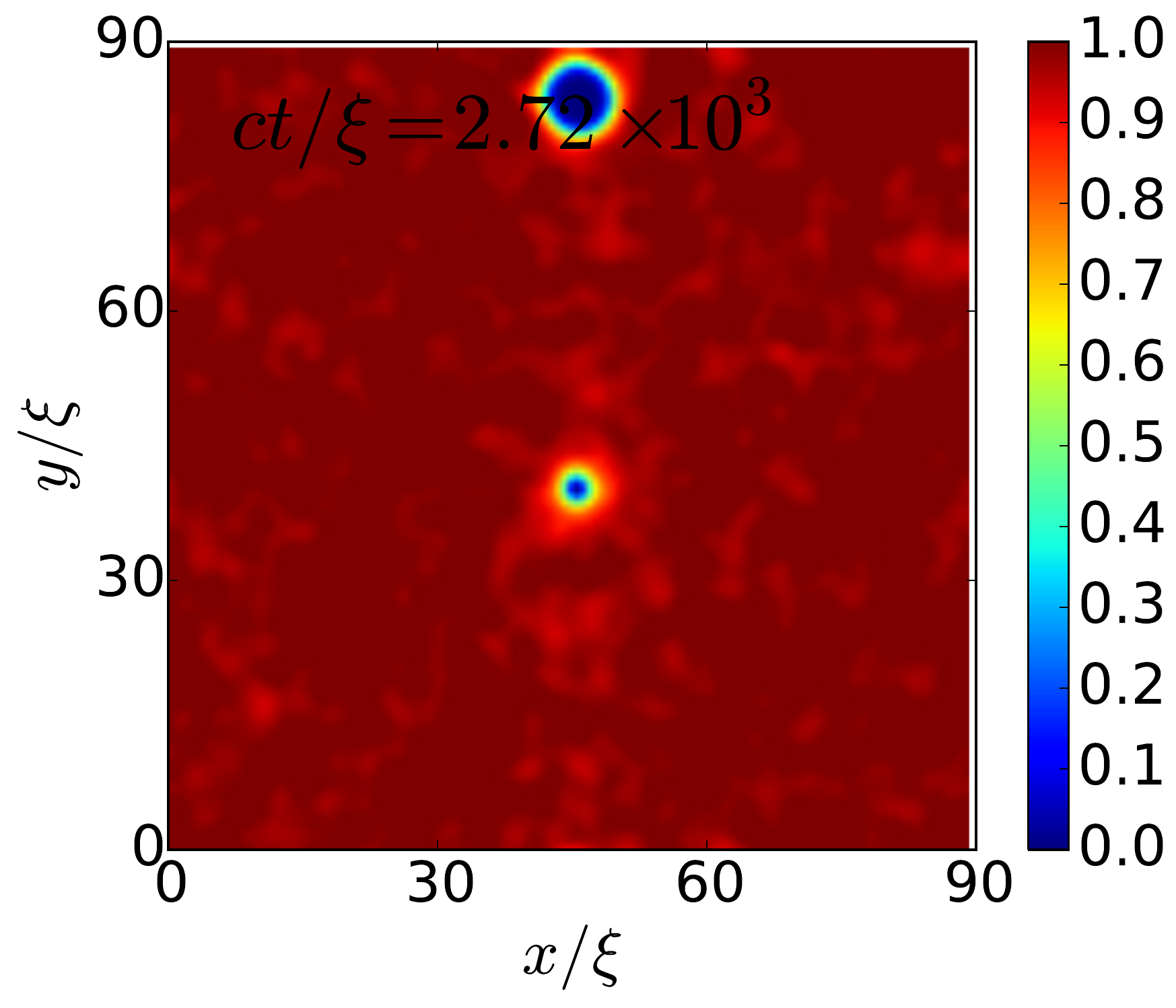}
\put(10.,10){\large{\bf (d)}}
\end{overpic}
\begin{overpic}
[height=4.5cm,unit=1mm]{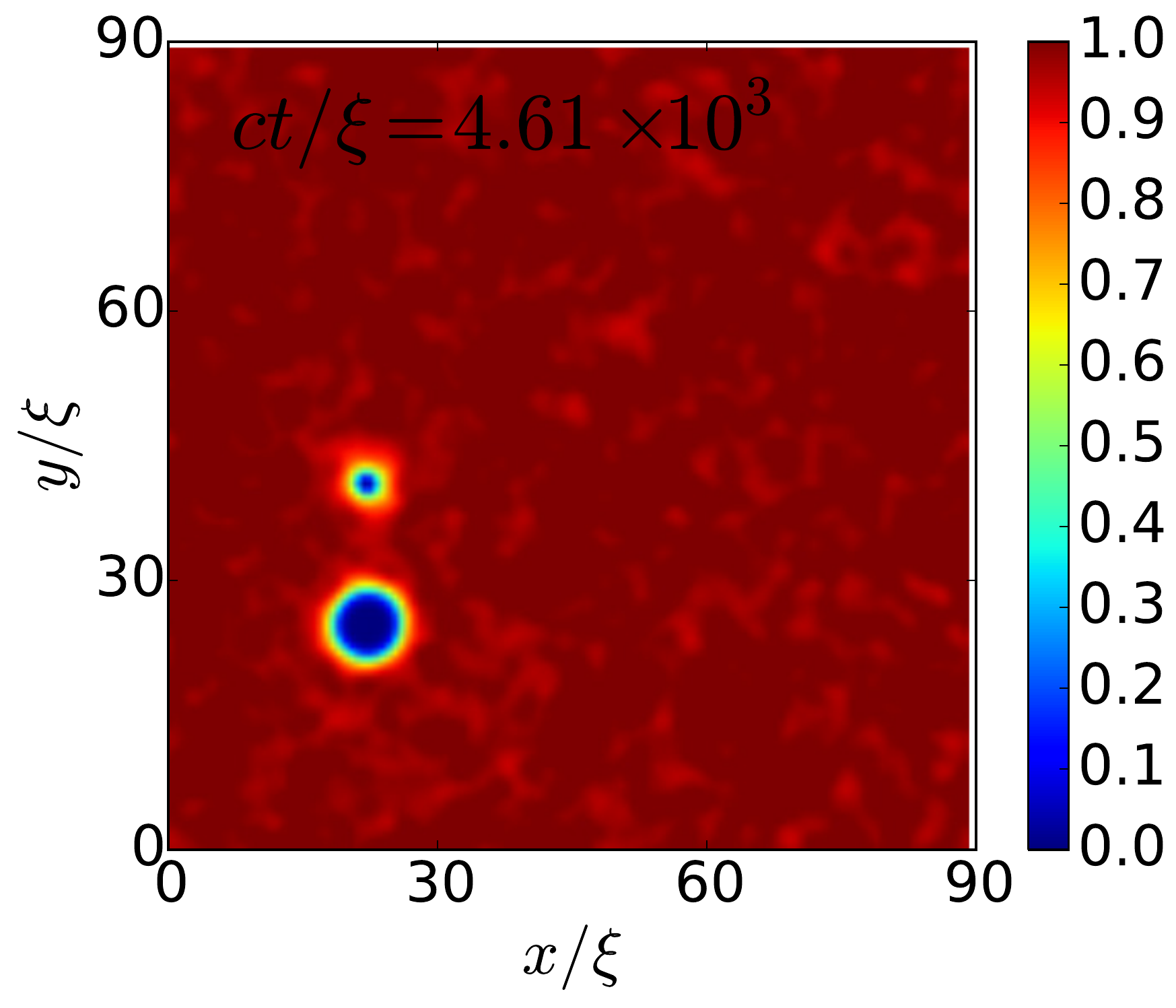}
\put(10,10){\large{\bf (e)}}
\end{overpic}
\begin{overpic}
[height=4.5cm,unit=1mm]{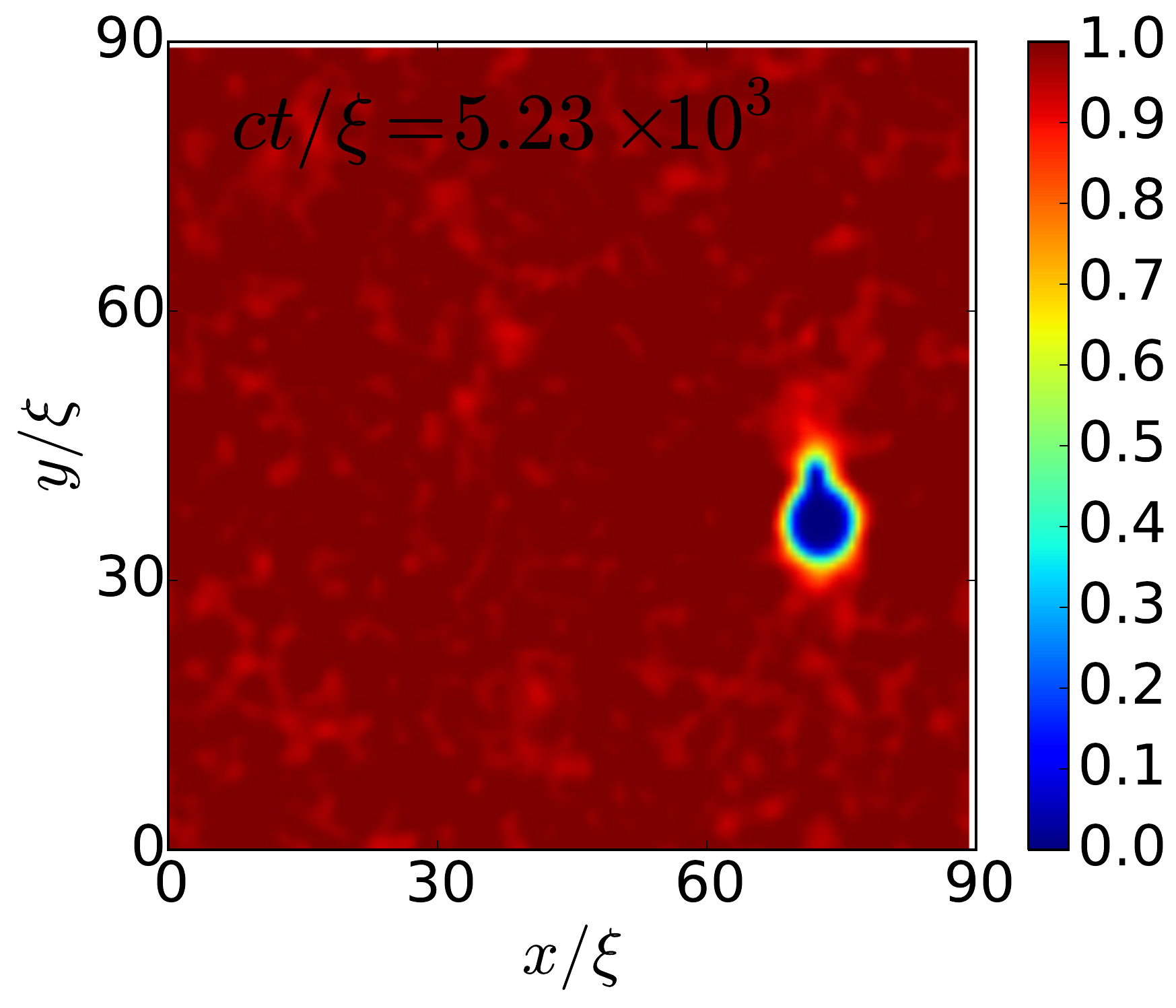}
\put(10,10){\large{\bf (f)}}
\end{overpic}
\\
\begin{overpic}
[height=4.5cm,unit=1mm]{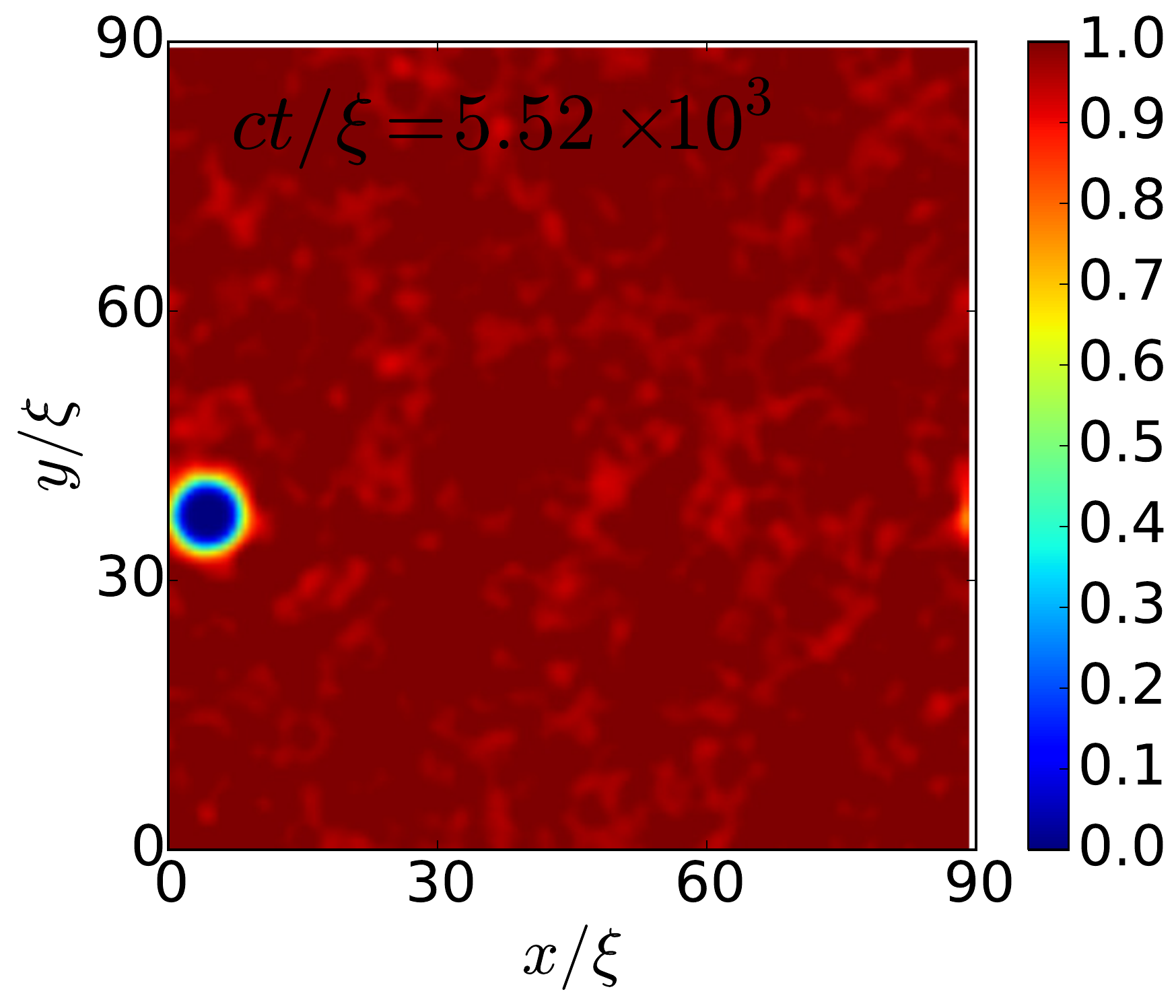}
\put(10.,10){\large{\bf (g)}}
\end{overpic}
\begin{overpic}
[height=4.5cm,unit=1mm]{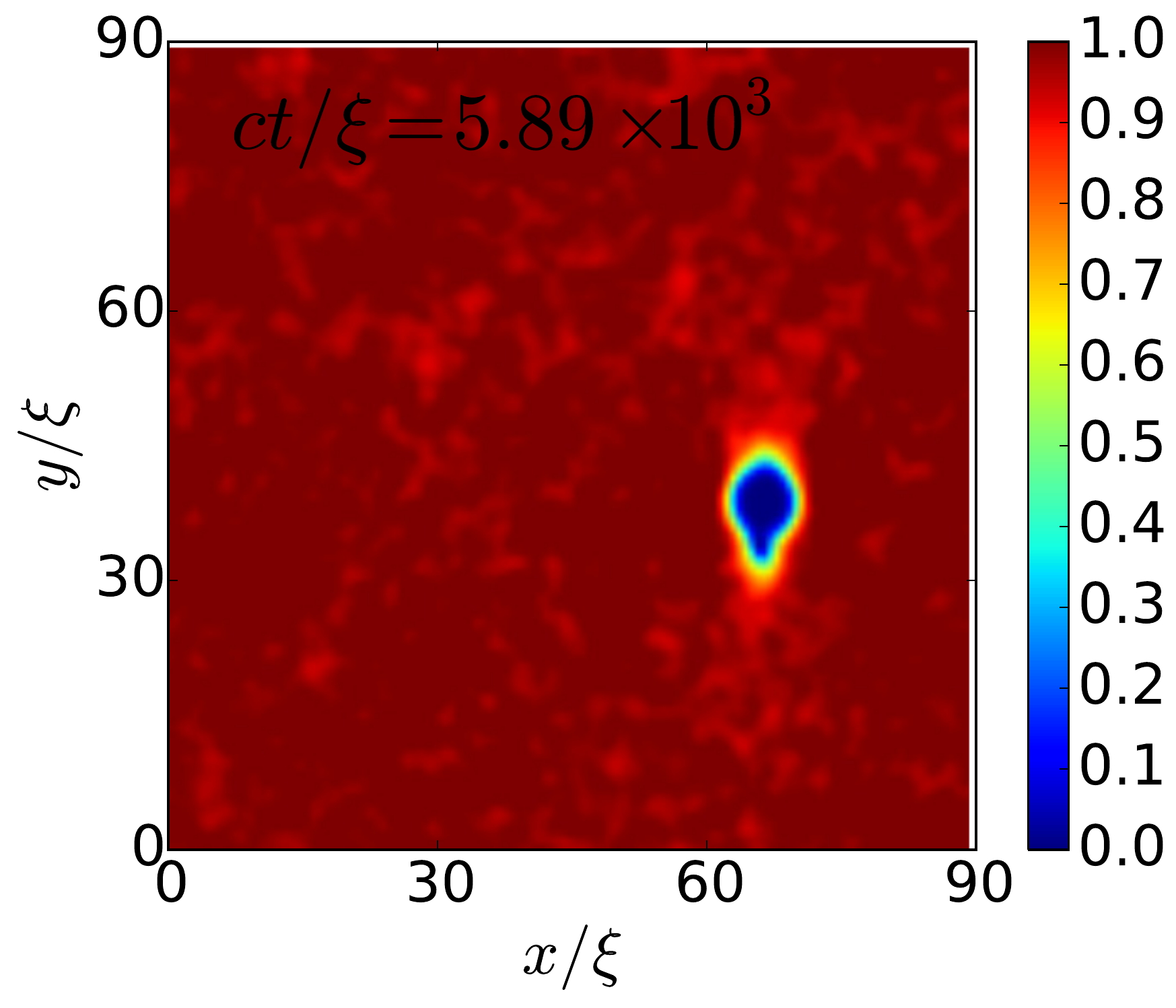}
\put(10,10){\large{\bf (h)}}
\end{overpic}
\begin{overpic}
[height=4.5cm,unit=1mm]{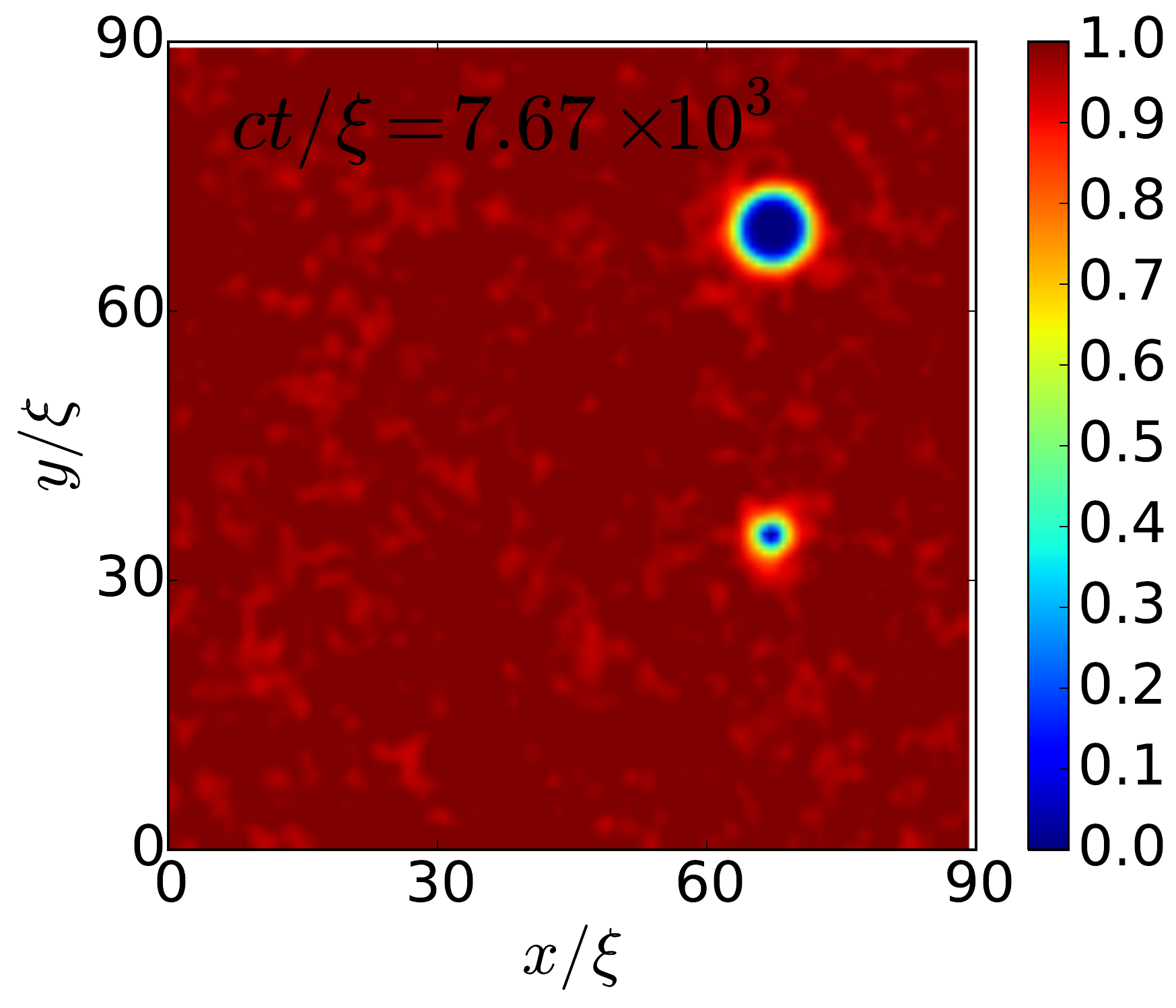}
\put(10,10){\large{\bf (i)}}
\end{overpic}
\caption{\small (Color online) \textbf{Constant external force on a light particle}:
Spatiotemporal evolution of the density field
$\rho(\mathbf{r},t)$ shown via pseudocolor plots, illustrating the dynamics
of a light particle, when a constant external force $\mathbf{F}_{\rm
ext}=0.14\,c^2\xi\rho_0\,\hat{\mathbf{x}}$ acts on it (initial configuration $\tt ICP1$).  The
particle appears as a large blue patch and the vortices as blue dots (for
details see text, subsection~\ref{subsec:1partextforce}).}
\label{fig:pdIC1f1L}
\end{figure*}


\begin{figure*}
\centering
\resizebox{\linewidth}{!}{
\includegraphics[height=4.cm,unit=1mm]{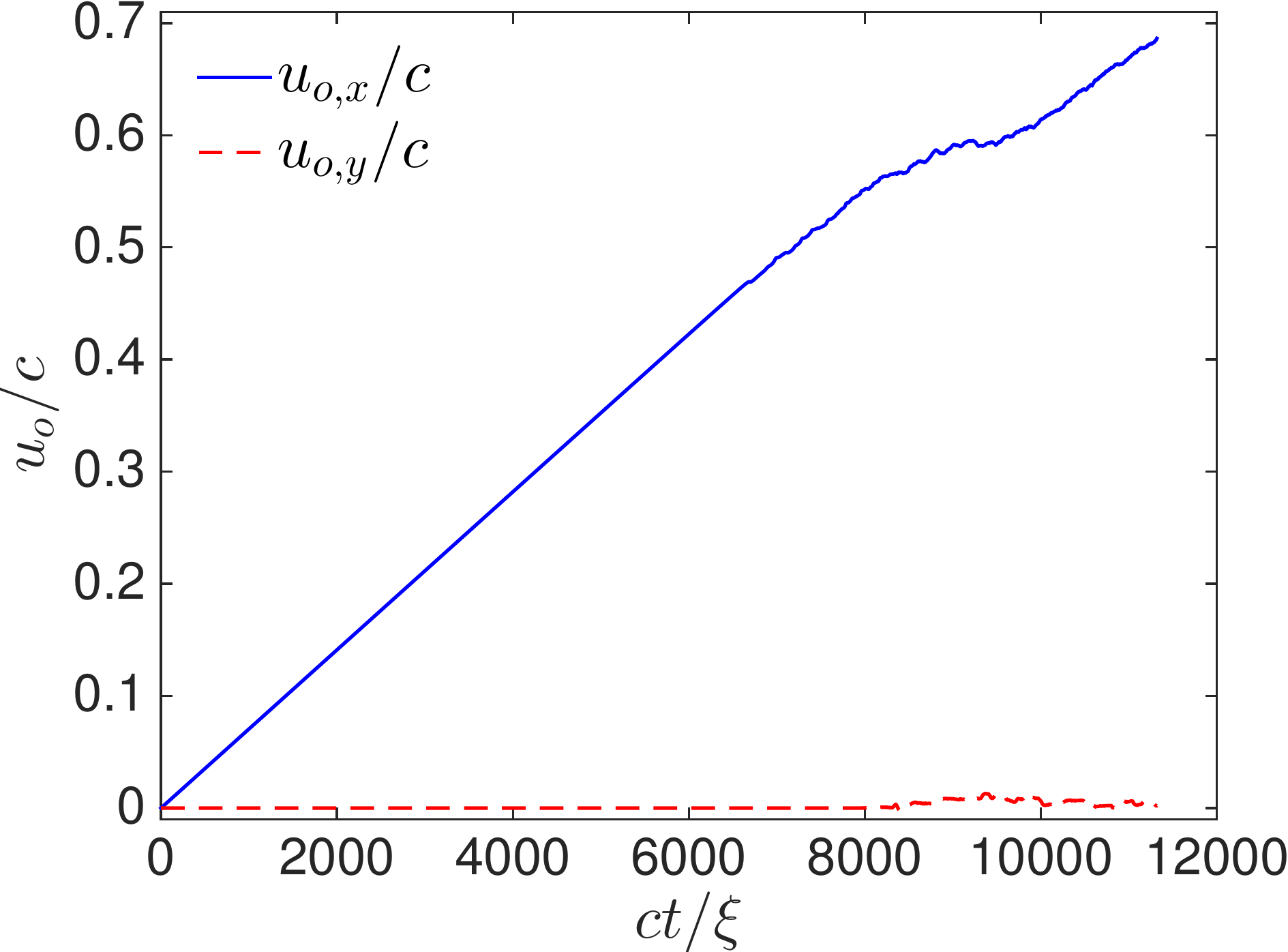}
\put(-75,30){\bf (a)}
\hspace{0.15 cm}
\includegraphics[height=4.cm,unit=1mm]{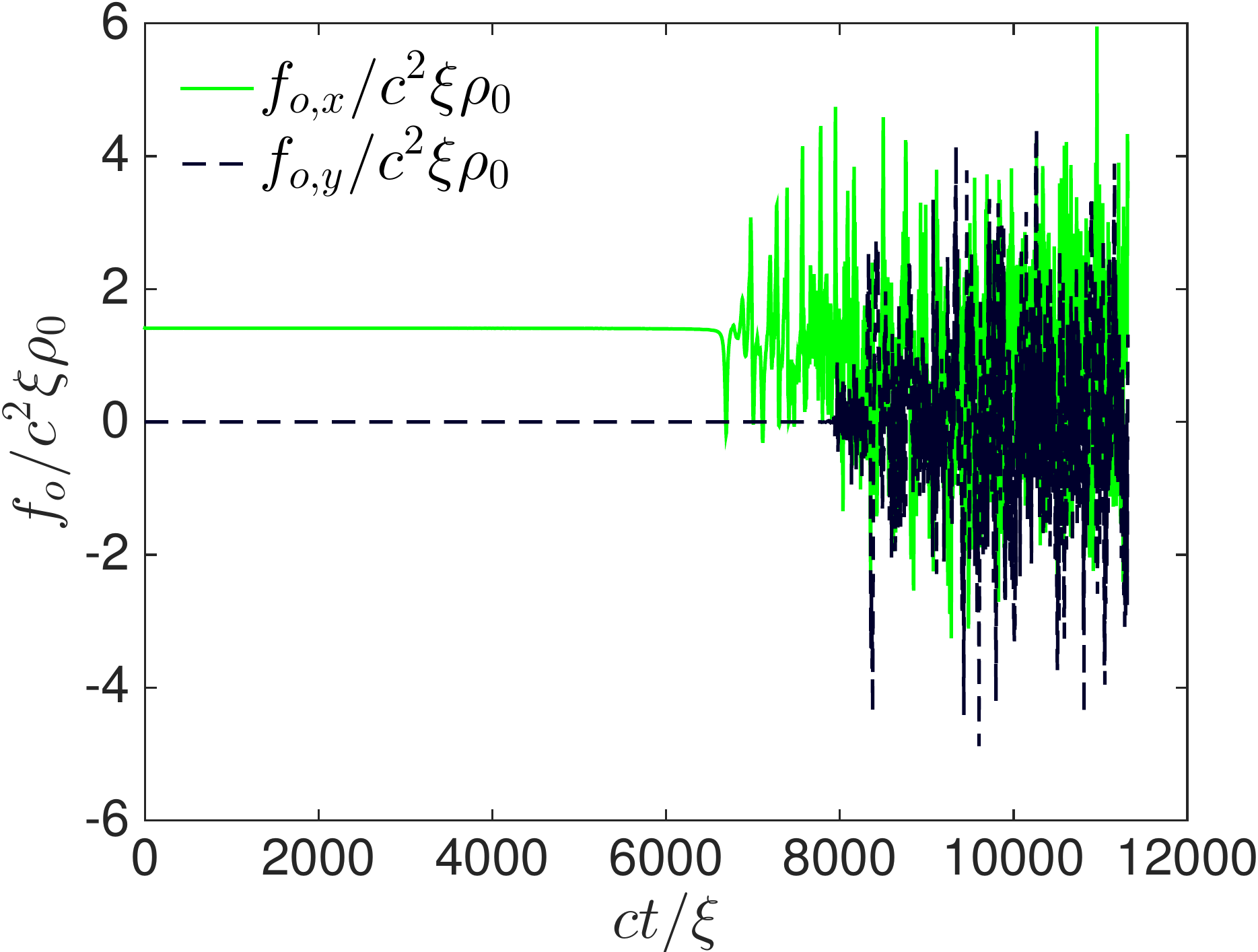}
\put(-75,30){\bf (b)}
\hspace{0.15 cm}
\includegraphics[height=4.cm,unit=1mm]{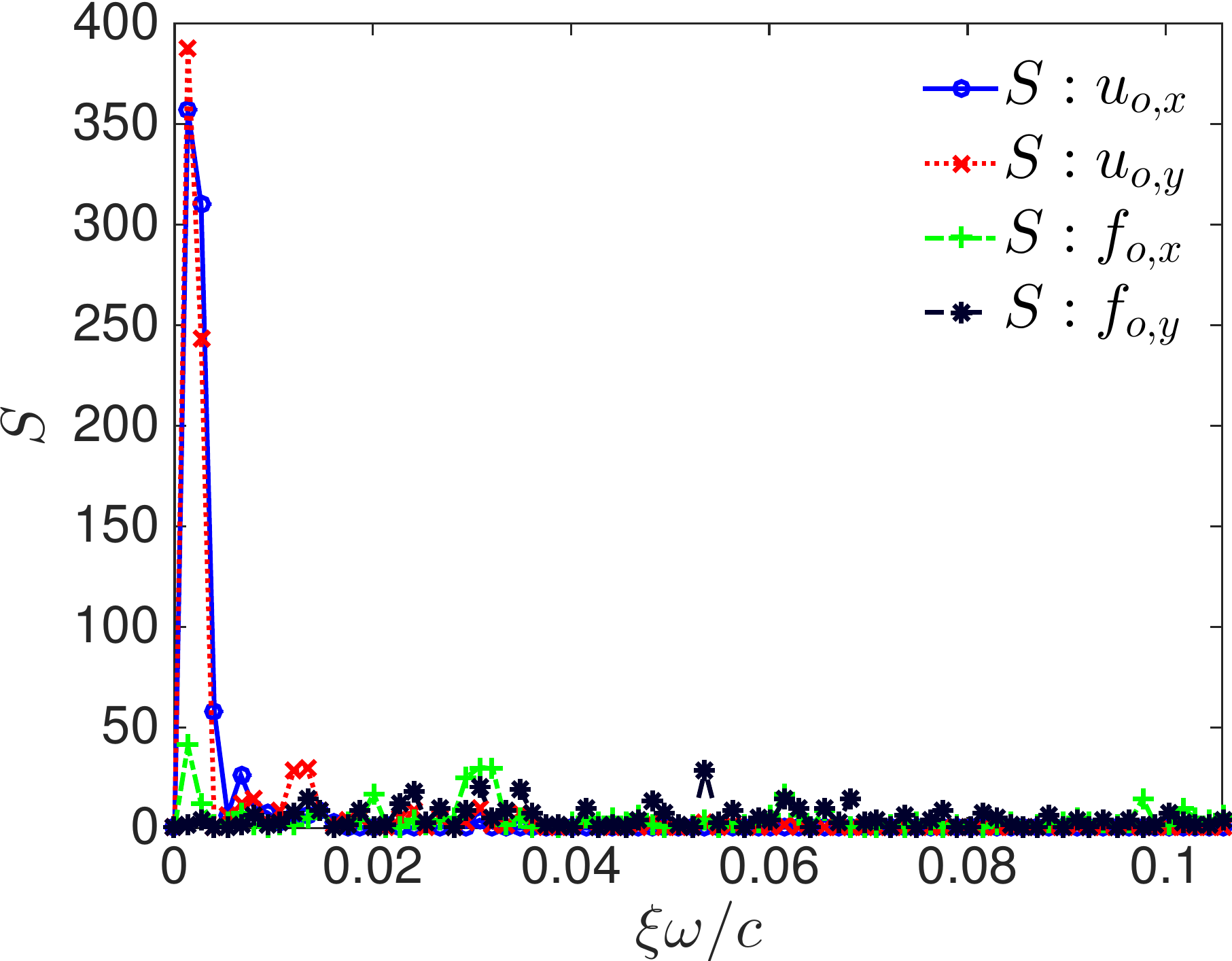}
\put(-75,30){\bf (c)}
}
\\
\vspace{0.25 cm}
\resizebox{\linewidth}{!}{
\includegraphics[height=4.cm,unit=1mm]{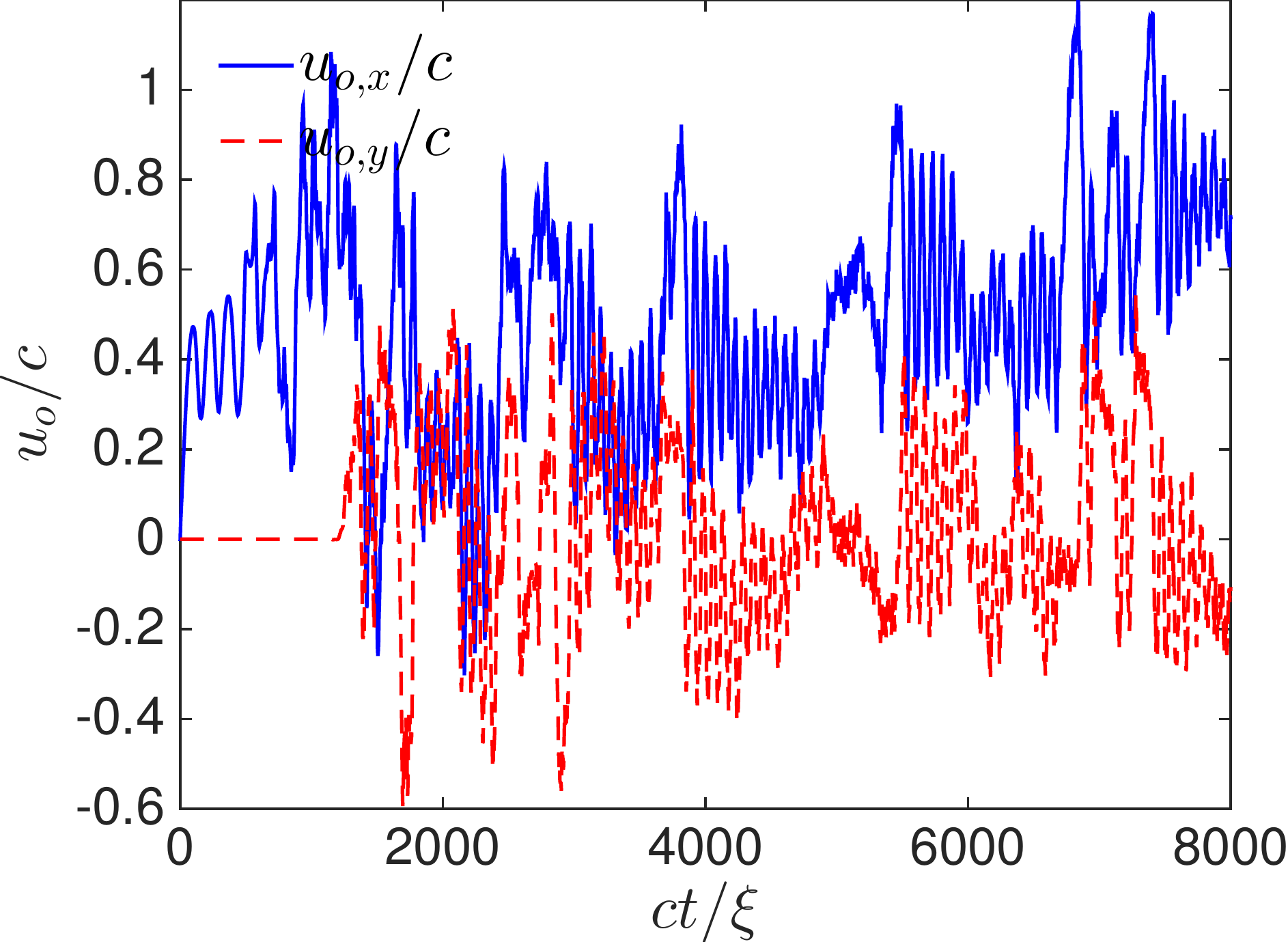}
\put(-75,30){\bf (d)}
\hspace{0.15 cm}
\includegraphics[height=4.cm,unit=1mm]{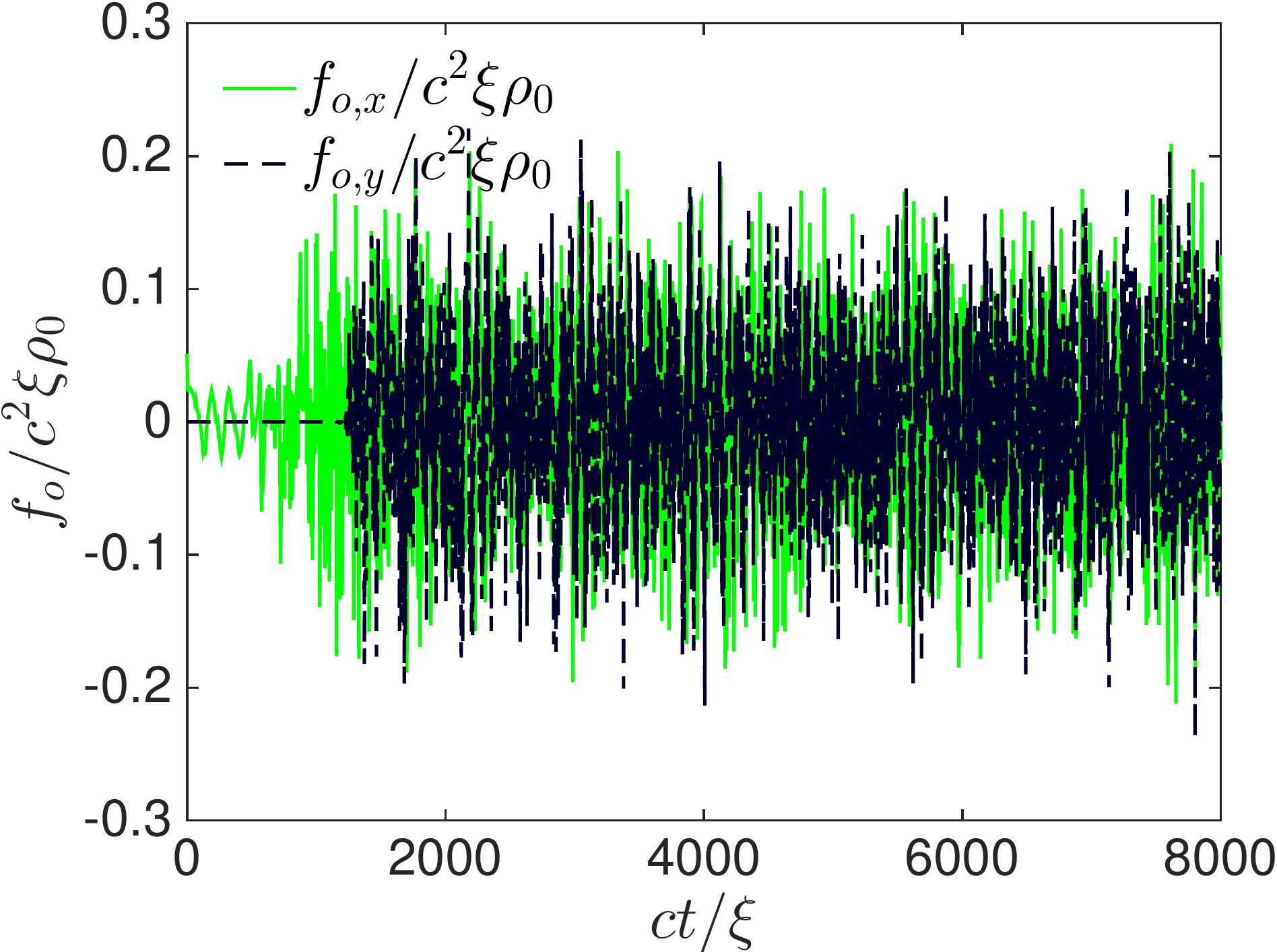}
\put(-75,25){\bf (e)}
\hspace{0.15 cm}
\includegraphics[height=4.cm,unit=1mm]{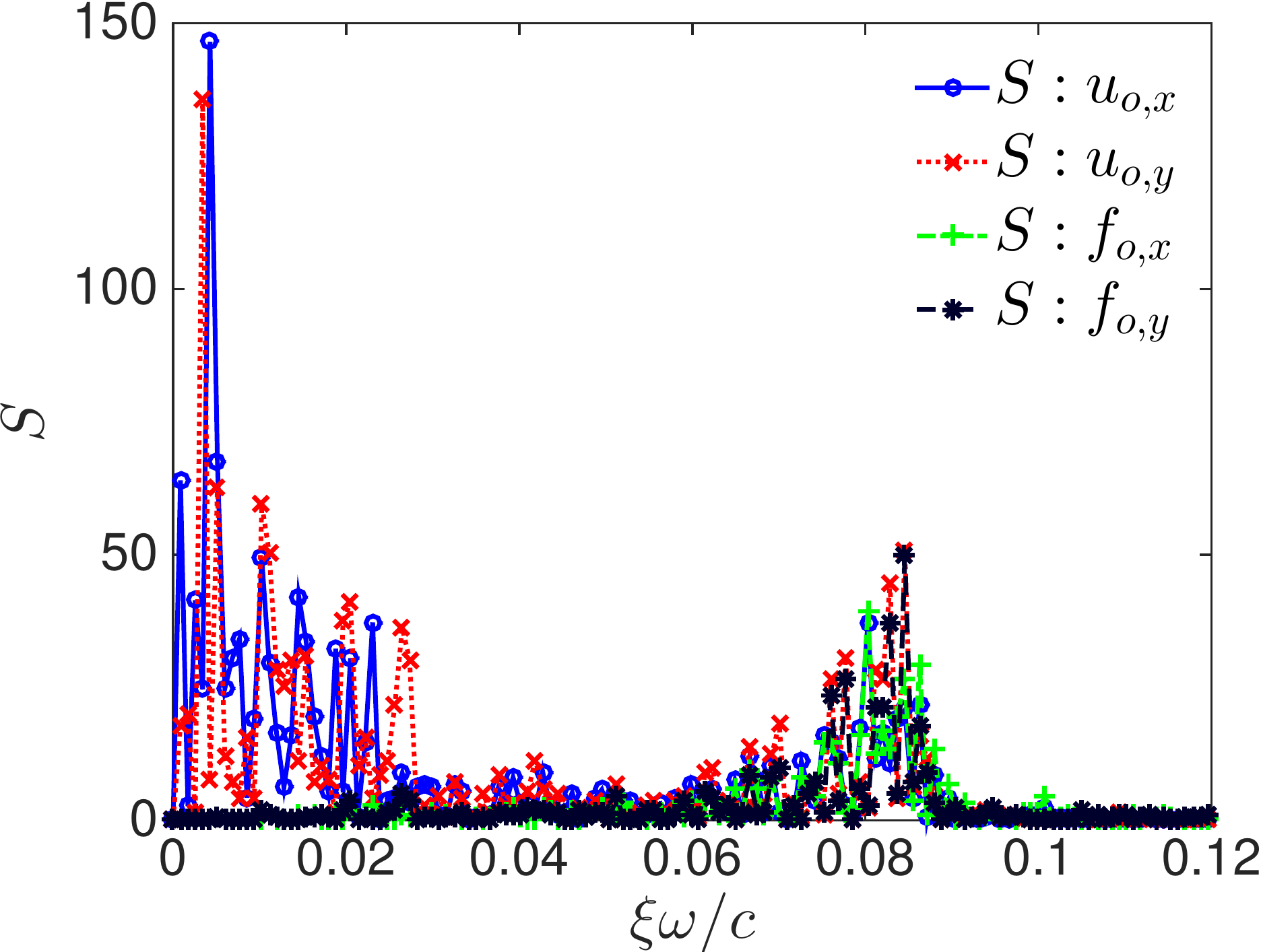}
\put(-75,25){\bf (f)}
}
\\
\vspace{0.25 cm}
\resizebox{\linewidth}{!}{
\includegraphics[height=4.cm,unit=1mm]{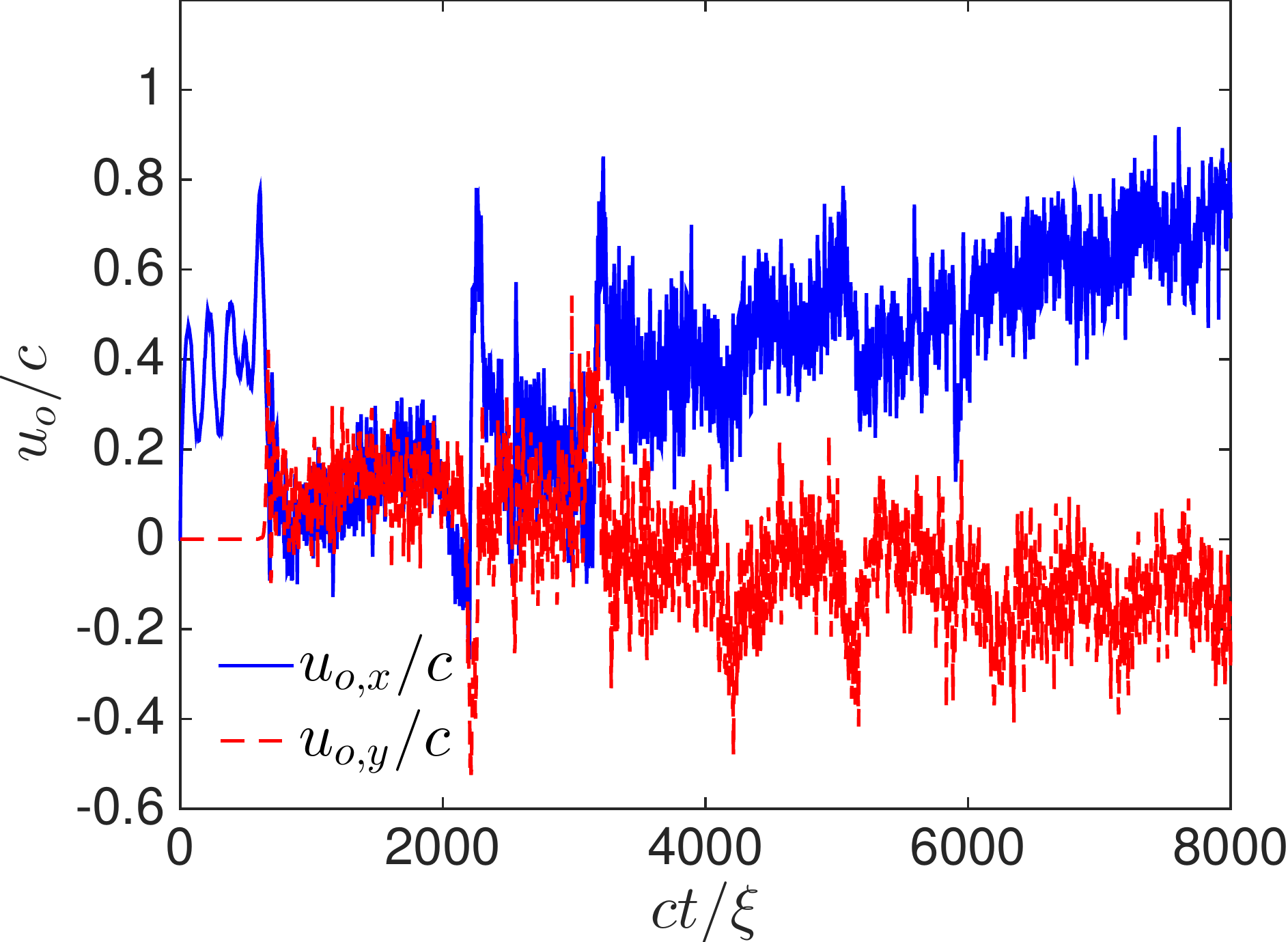}
\put(-75,30){\bf (g)}
\hspace{0.15 cm}
\includegraphics[height=4.cm,unit=1mm]{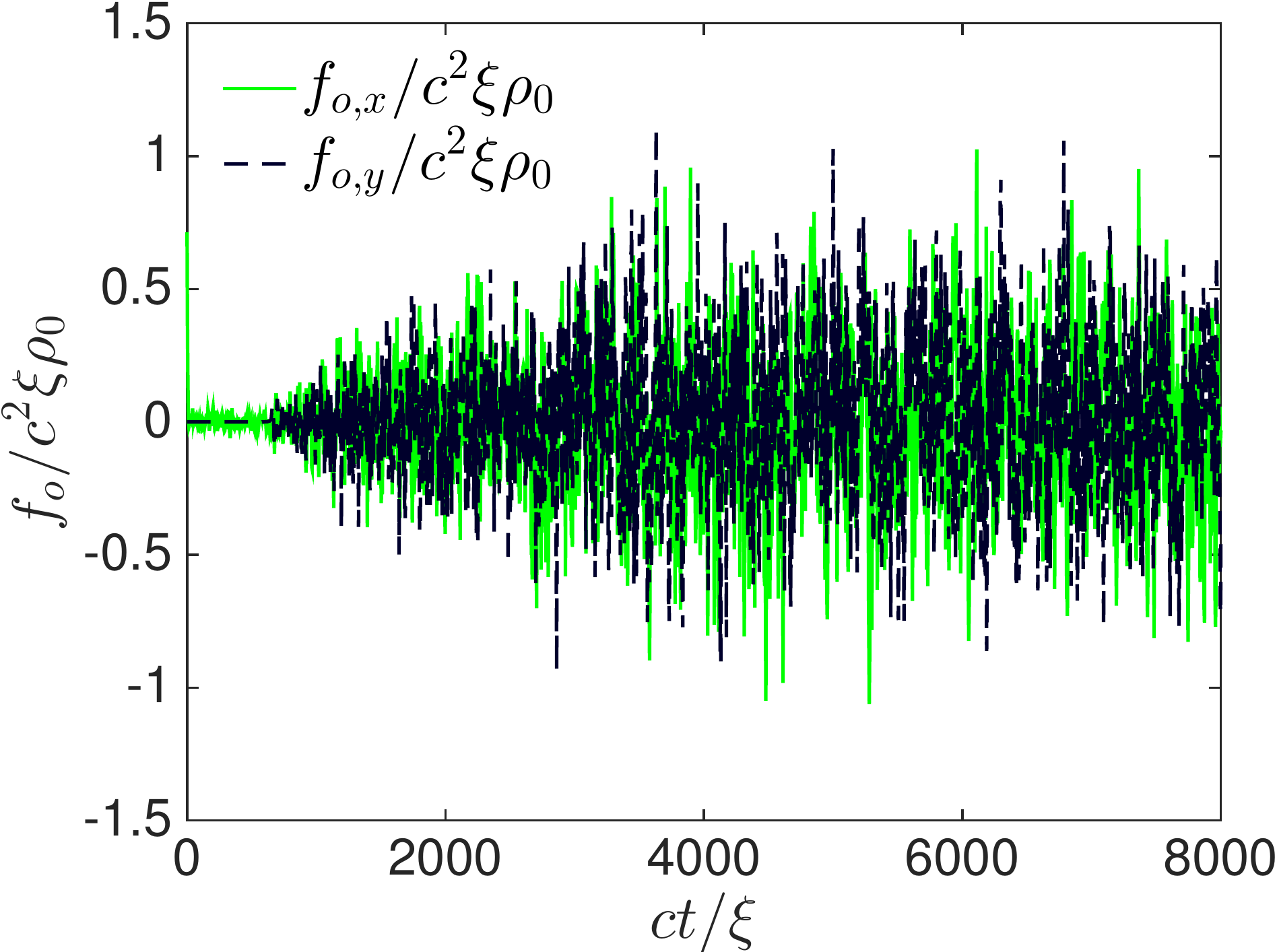}
\put(-75,30){\bf (h)}
\hspace{0.15 cm}
\includegraphics[height=4.cm,unit=1mm]{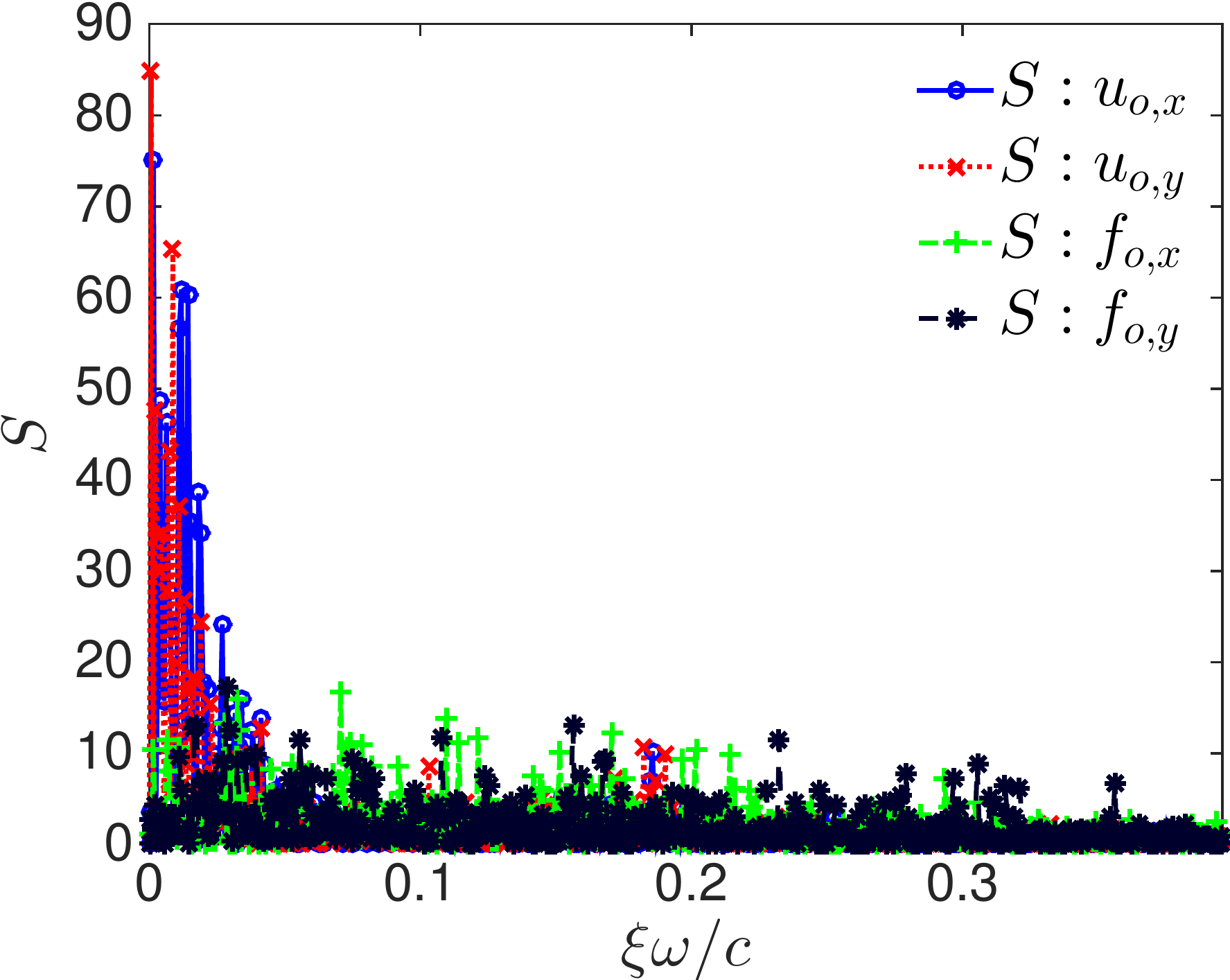}
\put(-75,35){\bf (i)}
}
\caption{(Color online) 
	Plots of the Cartesian components of: 
	(a) velocity $u_{\rm o,x}$ and $u_{\rm o,y}$,
	(b) force $f_{\rm o,x}$ and  $f_{\rm o,y}$;
	(c) power spectra of the quantities time series in (a) and (b),
	for the \textbf{heavy} particle ($\mathcal{M}=374$, 
	$\mathbf{F}_{\rm ext}= 1.42\,c^2\xi\rho_0\,\hat{\mathbf{x}}$).
	Plots in (d), (e), and (f)  and (g), (h), and (i) are the analogs of 
	plots in (a), (b), and (c), for the \textbf{neutral}\
	($\mathcal{M}=1$, $\mathbf{F}_{\rm ext}= 0.71\,c^2\xi\rho_0\,\hat{\mathbf{x}}$)
	and \textbf{light} ($\mathcal{M}=0.0374$, 
	$\mathbf{F}_{\rm ext}= 0.71\,c^2\xi\rho_0\,\hat{\mathbf{x}}$) particles, respectively.
	 Power spectra, denoted generically by
         $S(\omega)$, of the time series of $u_{\rm o,x}$ , $u_{\rm o,y}$, 
	 $f_{\rm o,x}$, and  $f_{\rm o,y}$ are plotted versus the 
	 angular frequency $\omega$ for the above three cases.
 }	
\label{fig:velfor.pdfpectraIC1f2}
\end{figure*}


\begin{figure*}
\centering
\begin{overpic}
[height=4.5cm,unit=1mm]{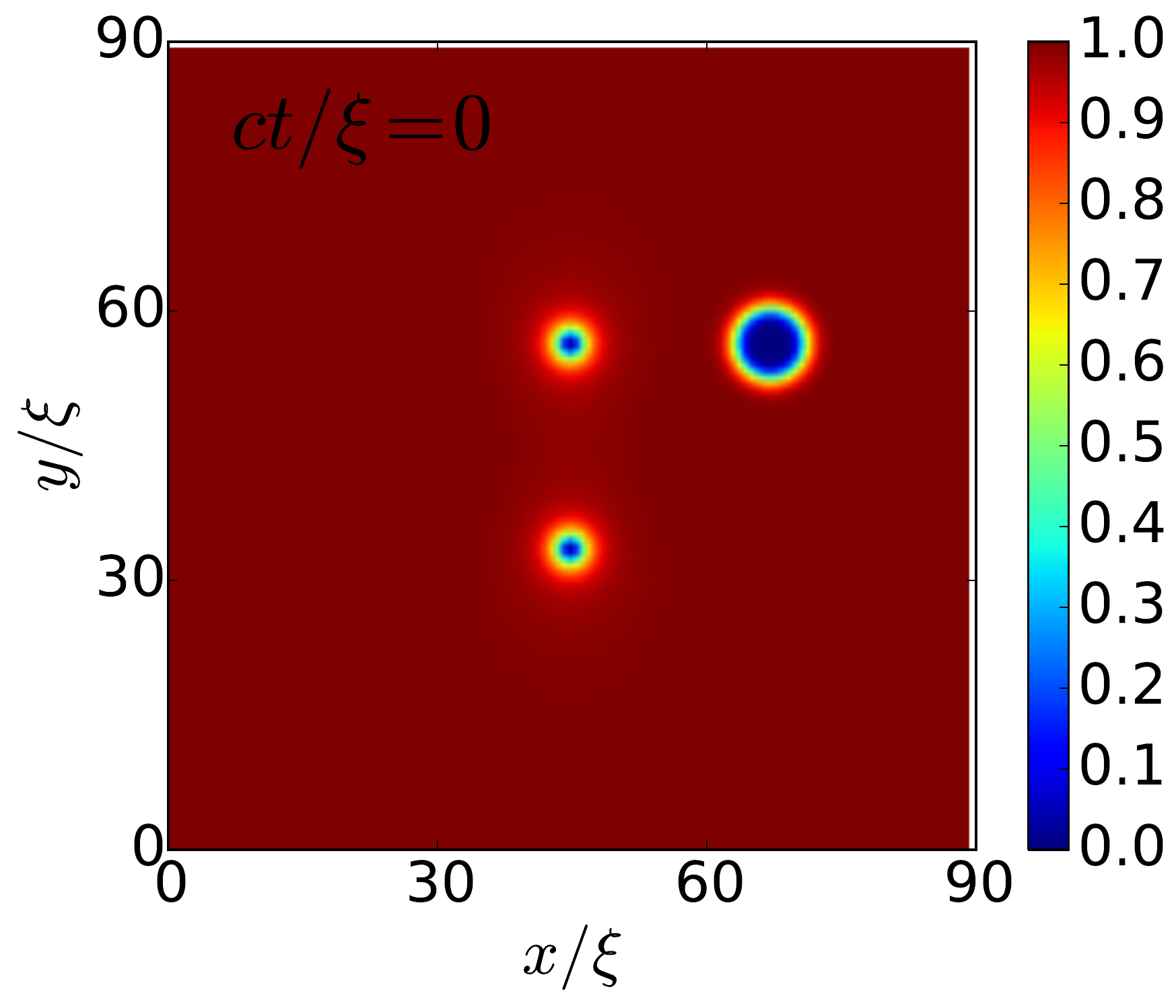}
\put(10.,10){\large{\bf (a)}}
\end{overpic}
\begin{overpic}
[height=4.5cm,unit=1mm]{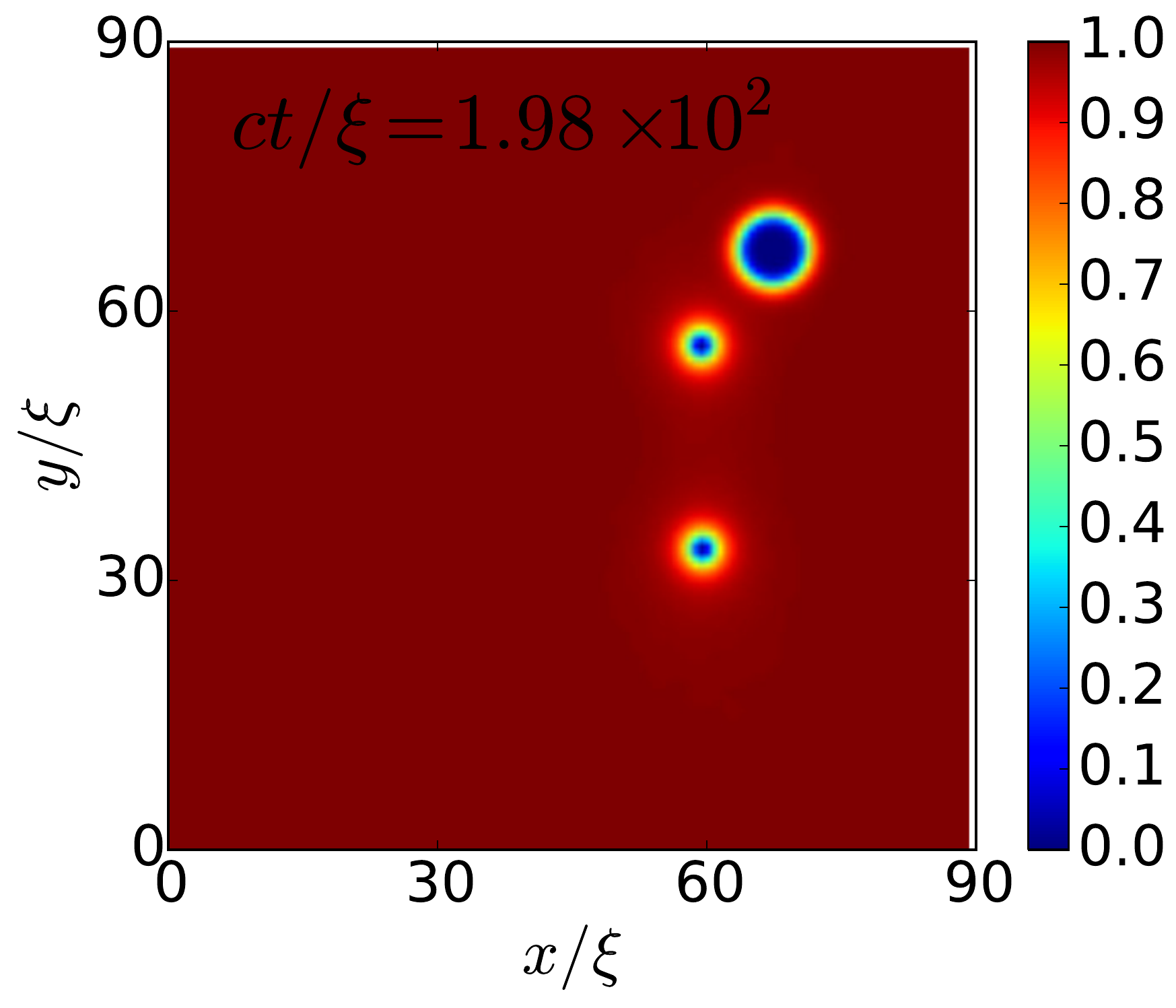}
\put(10,10){\large{\bf (b)}}
\end{overpic}
\begin{overpic}
[height=4.5cm,unit=1mm]{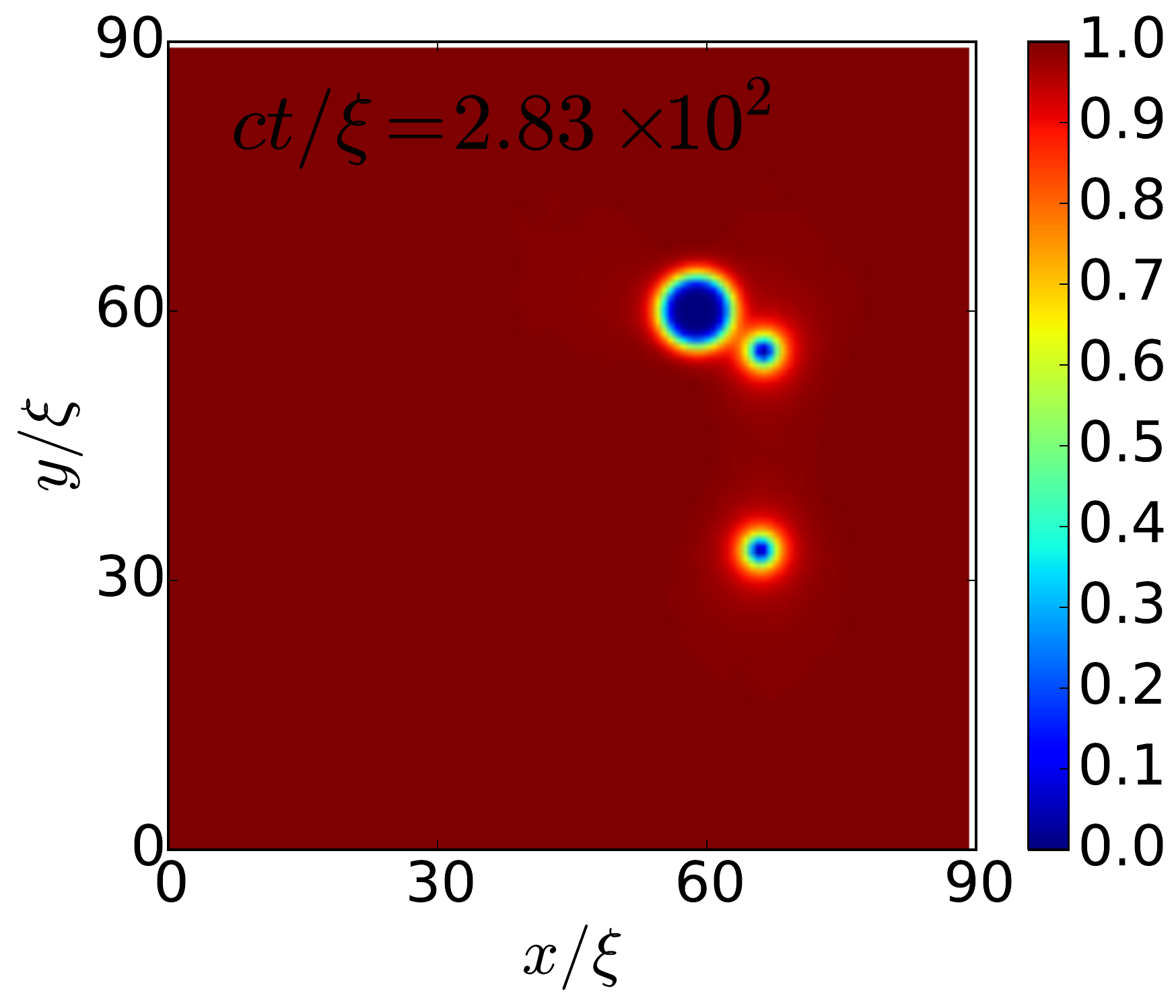}
\put(10,10){\large{\bf (c)}}
\end{overpic}
\\
\begin{overpic}
[height=4.5cm,unit=1mm]{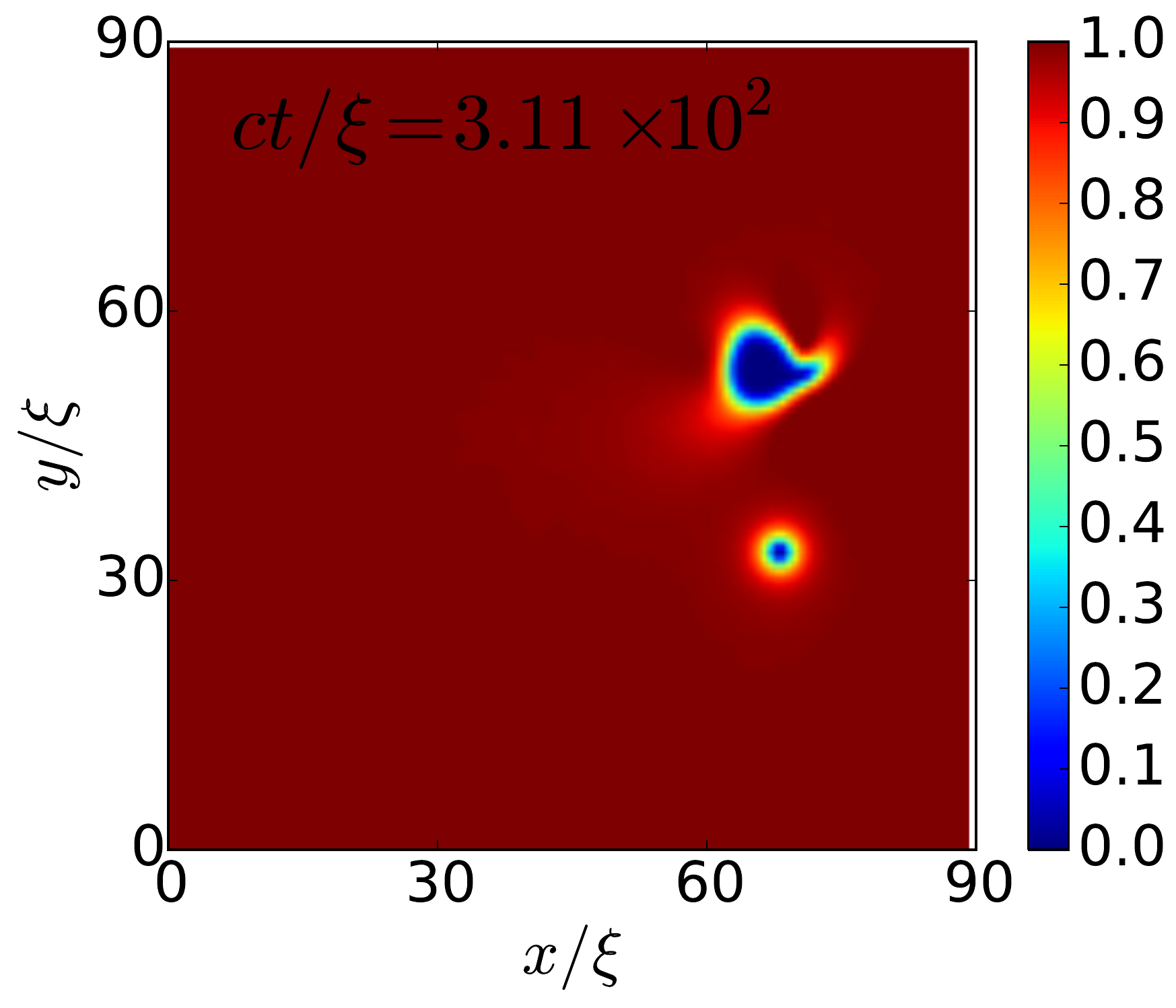}
\put(10.,10){\large{\bf (d)}}
\end{overpic}
\begin{overpic}
[height=4.5cm,unit=1mm]{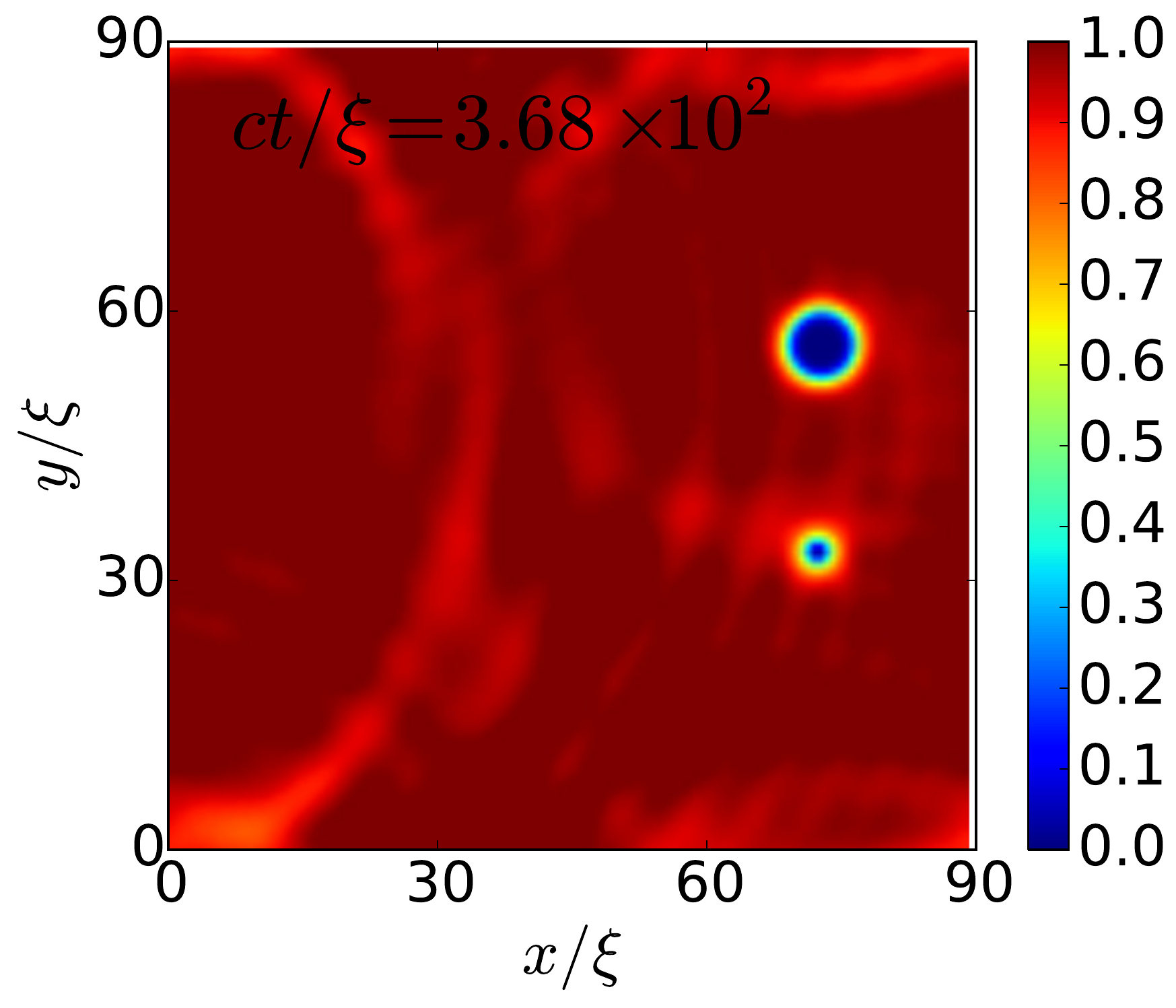}
\put(10,10){\large{\bf (e)}}
\end{overpic}
\begin{overpic}
[height=4.5cm,unit=1mm]{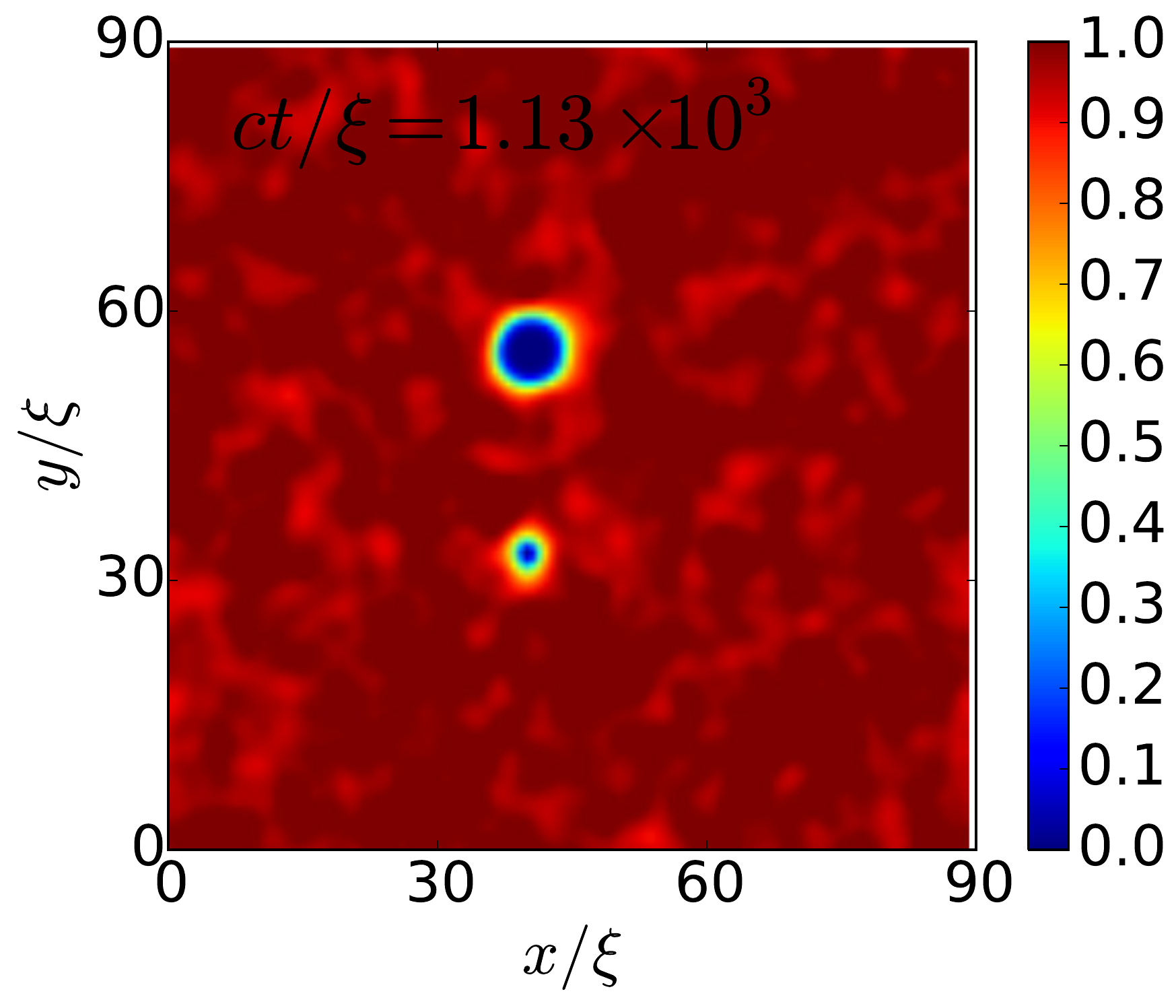}
\put(10,10){\large{\bf (f)}}
\end{overpic}
\caption{\small (Color online) \textbf{Light particle interacting with a translating
	vortex-antivortex pair}:
	Spatiotemporal evolution of the density field
$\rho(\mathbf{r},t)$ shown via pseudocolor plots, for a ligth particle
placed in the path of the positive (upper) vortex of a translating
vortex-antivortex pair (initial configuration $\tt ICP2A$).}
\label{fig:1partpairtranslpdL}
\end{figure*}

\begin{figure*}
\centering
\begin{overpic}
[height=4.5cm,unit=1mm]{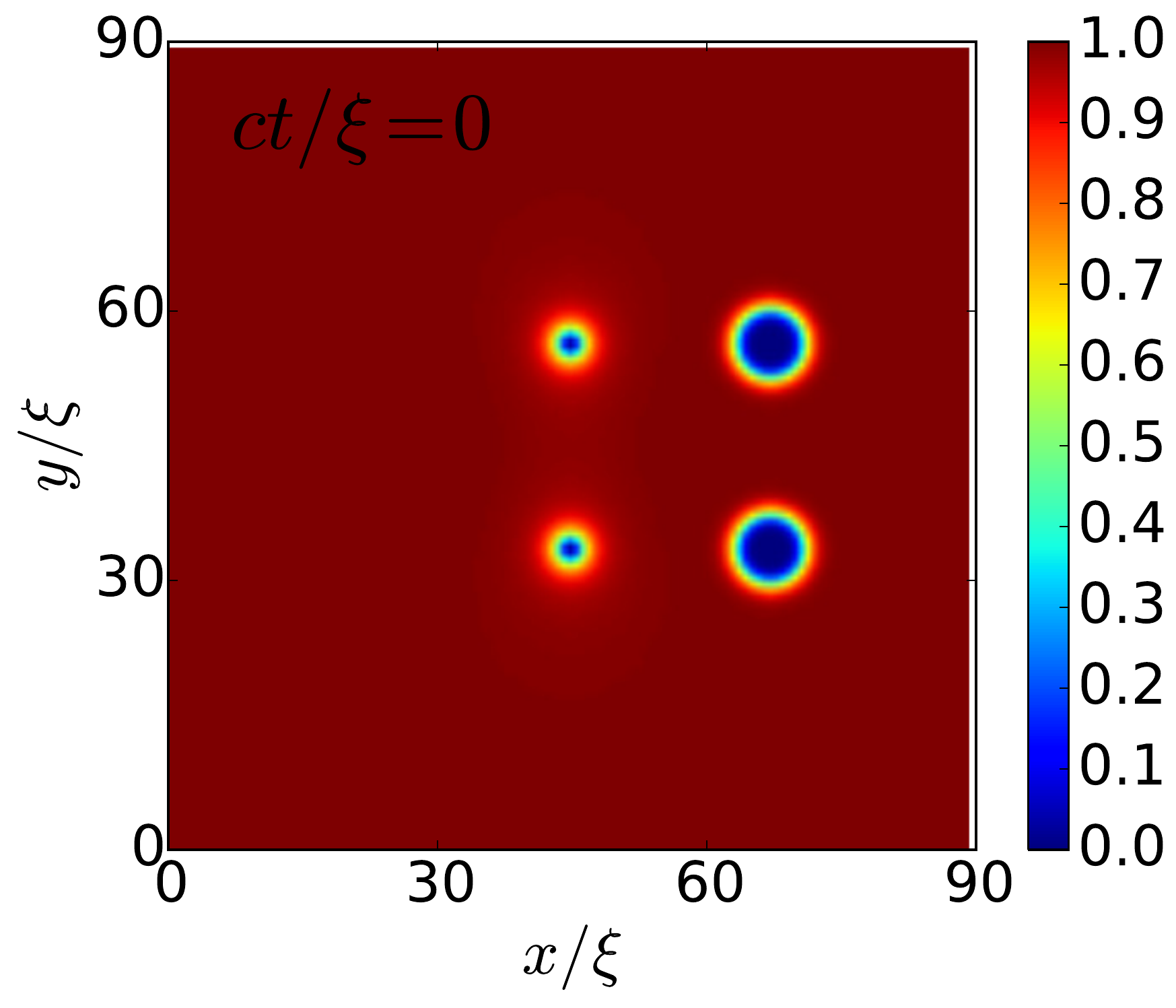}
\put(10.,10){\large{\bf (a)}}
\end{overpic}
\begin{overpic}
[height=4.5cm,unit=1mm]{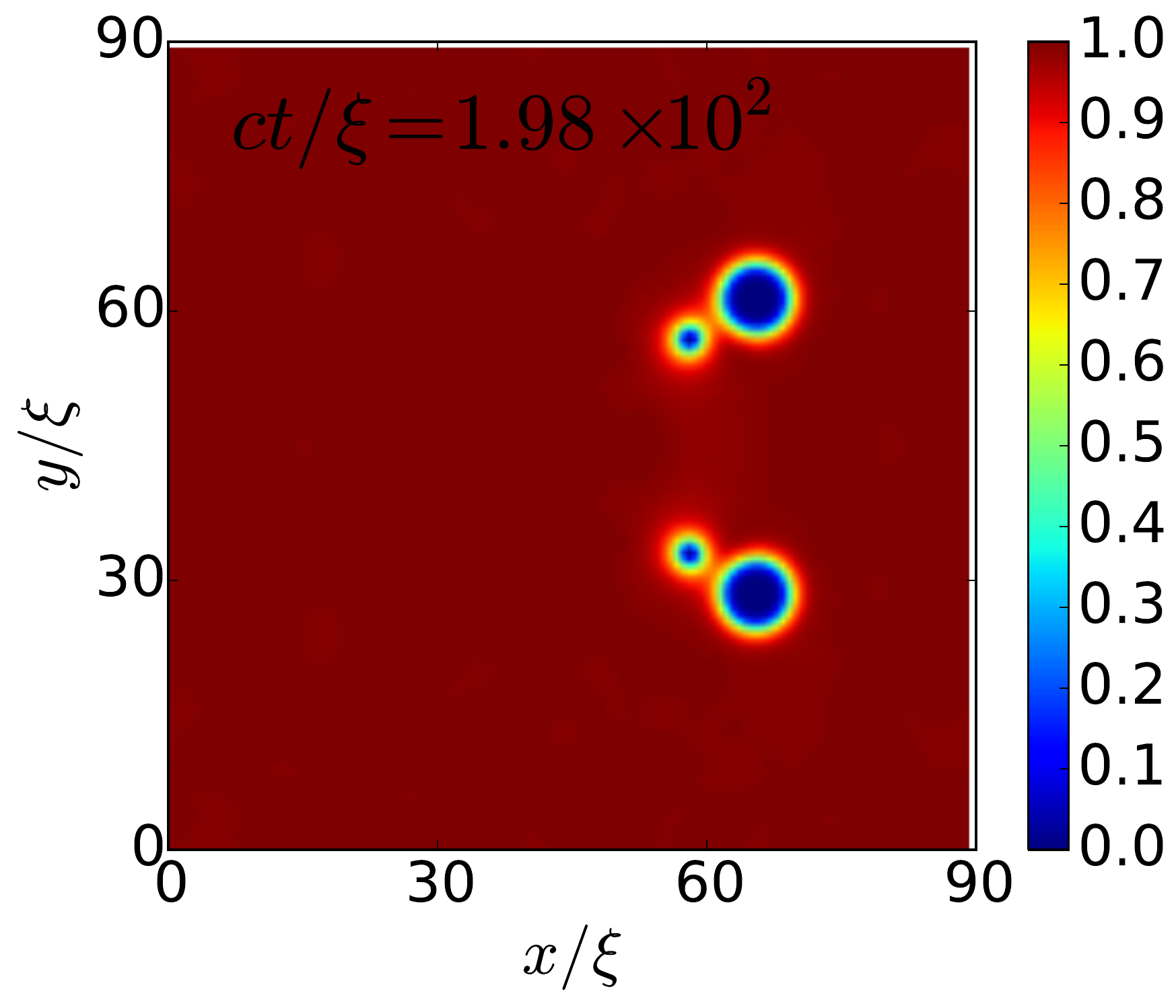}
\put(10,10){\large{\bf (b)}}
\end{overpic}
\begin{overpic}
[height=4.5cm,unit=1mm]{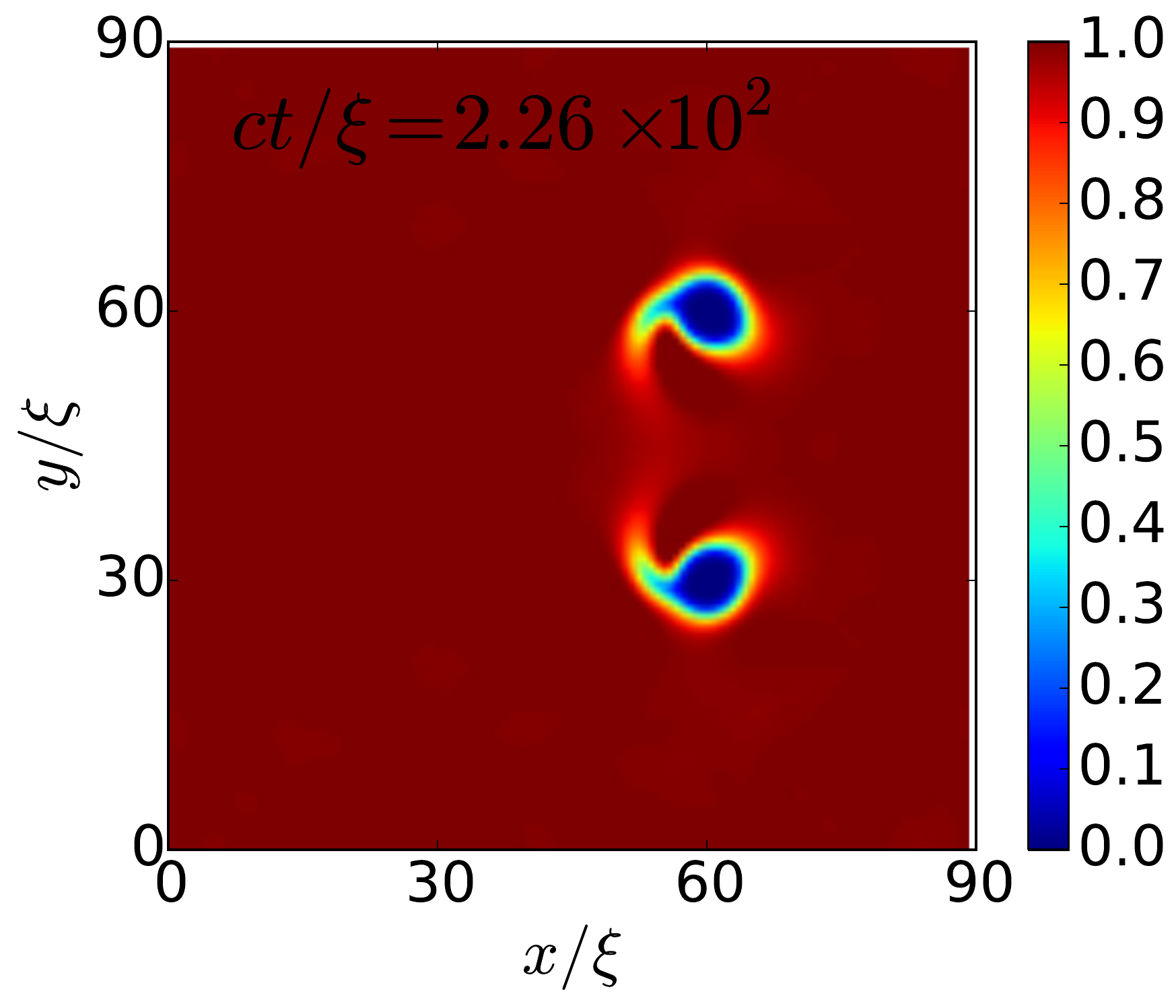}
\put(10,10){\large{\bf (c)}}
\end{overpic}
\\
\begin{overpic}
[height=4.5cm,unit=1mm]{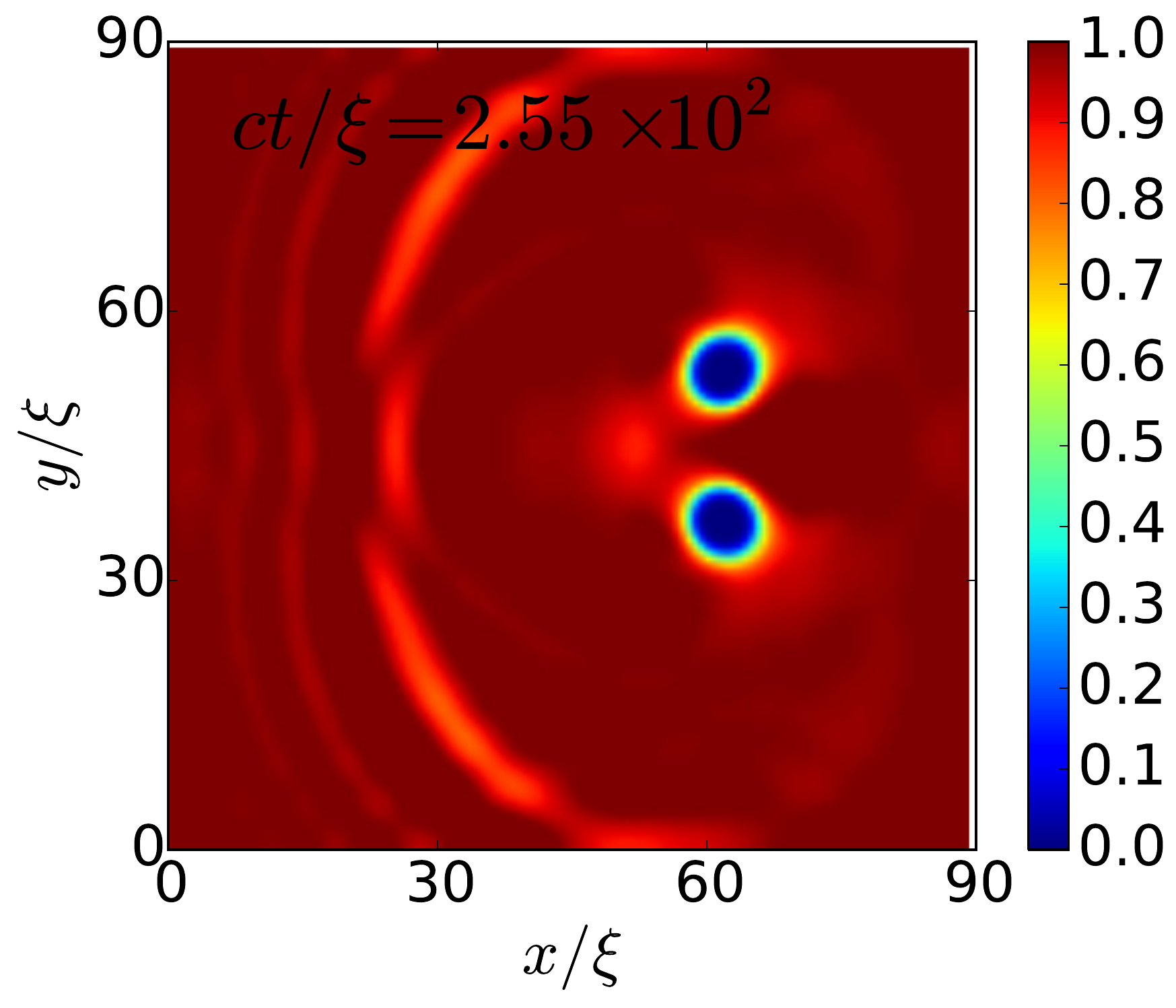}
\put(10.,10){\large{\bf (d)}}
\end{overpic}
\begin{overpic}
[height=4.5cm,unit=1mm]{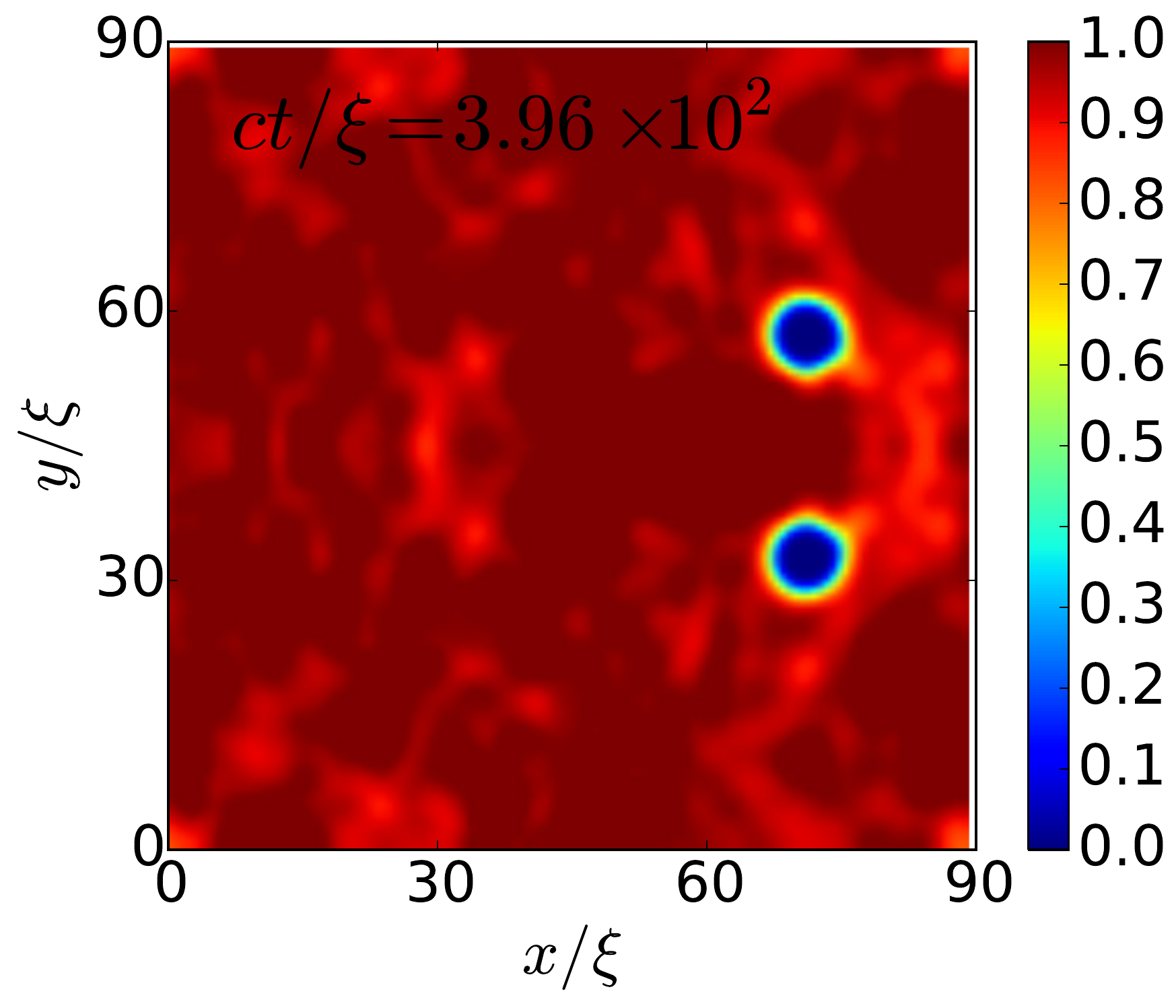}
\put(10,10){\large{\bf (e)}}
\end{overpic}
\begin{overpic}
[height=4.5cm,unit=1mm]{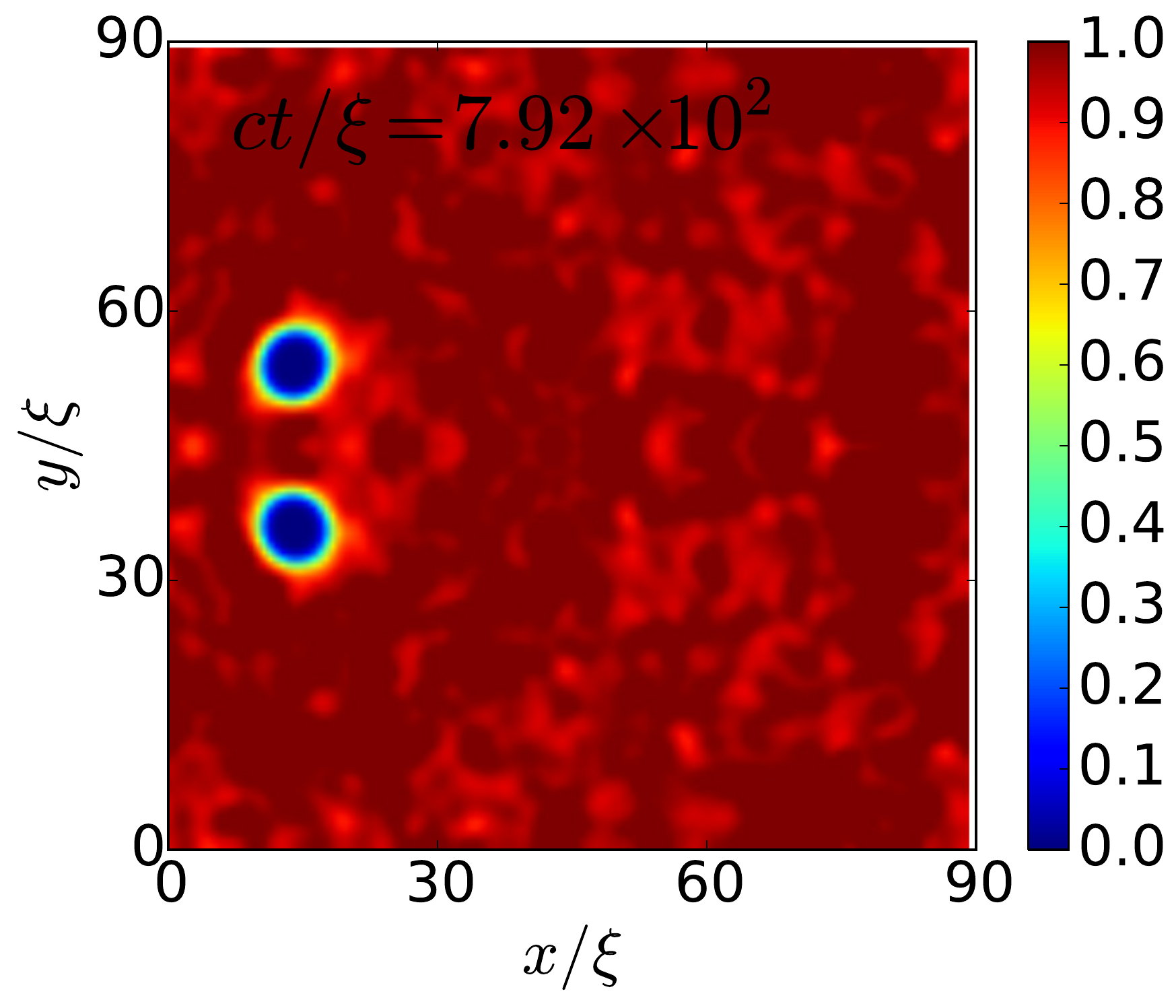}
\put(10,10){\large{\bf (f)}}
\end{overpic}
\caption{\small \textbf{Two neutral particles interacting with a translating
	vortex-antivortex pair}:
	Spatiotemporal evolution of the density field
$\rho(\mathbf{r},t)$ shown via
pseudocolor plots, for two neutral particle placed in the path of the positive (upper)
and negative (lower) vortices, respectively, of a translating vortex-antivortex pair 
(initial configuration $\tt ICP2B$).
}
\label{fig:2partpairtranslpdN}
\end{figure*}

\begin{figure*}
\centering
\begin{overpic}
[height=4.5cm,unit=1mm]{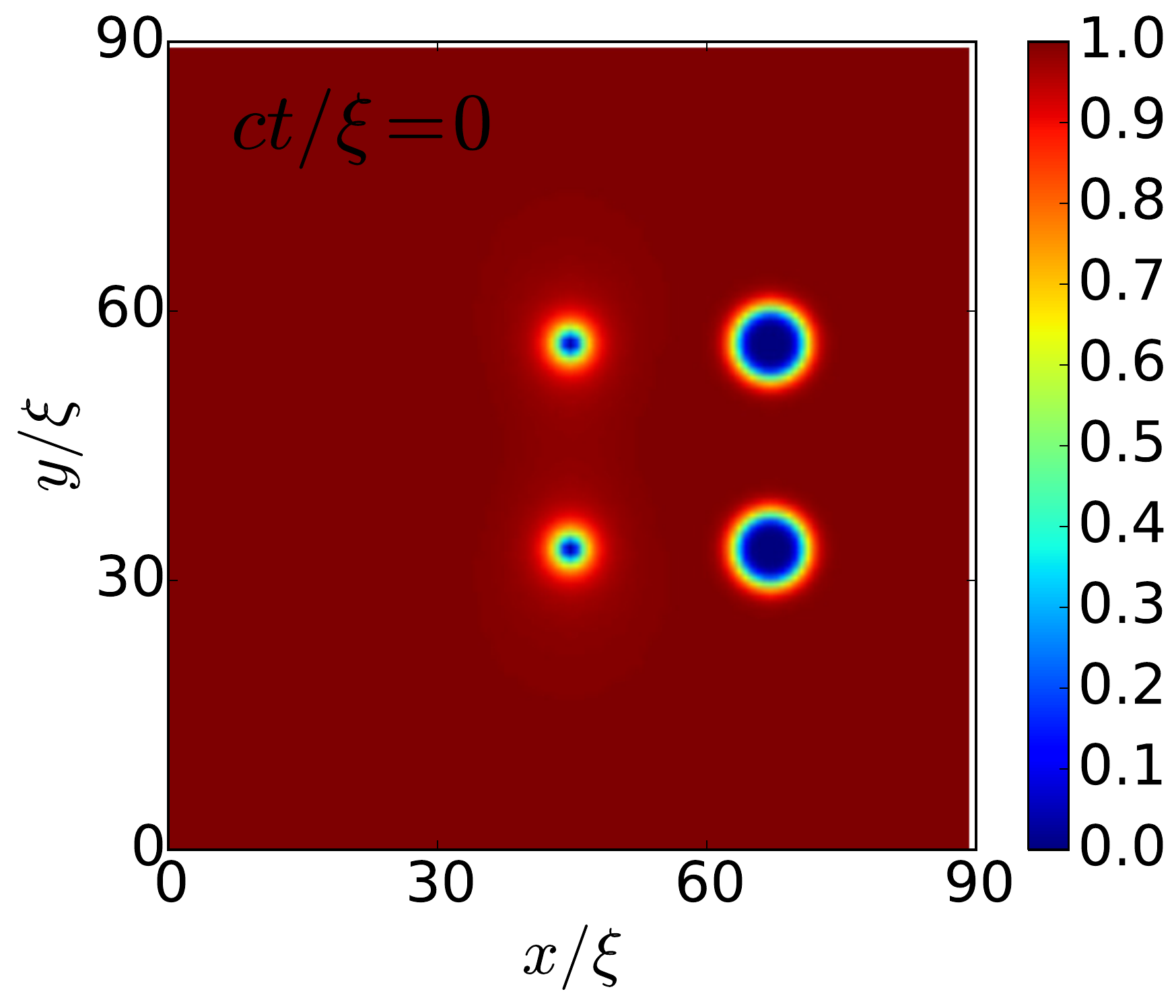}
\put(10.,10){\large{\bf (a)}}
\end{overpic}
\begin{overpic}
[height=4.5cm,unit=1mm]{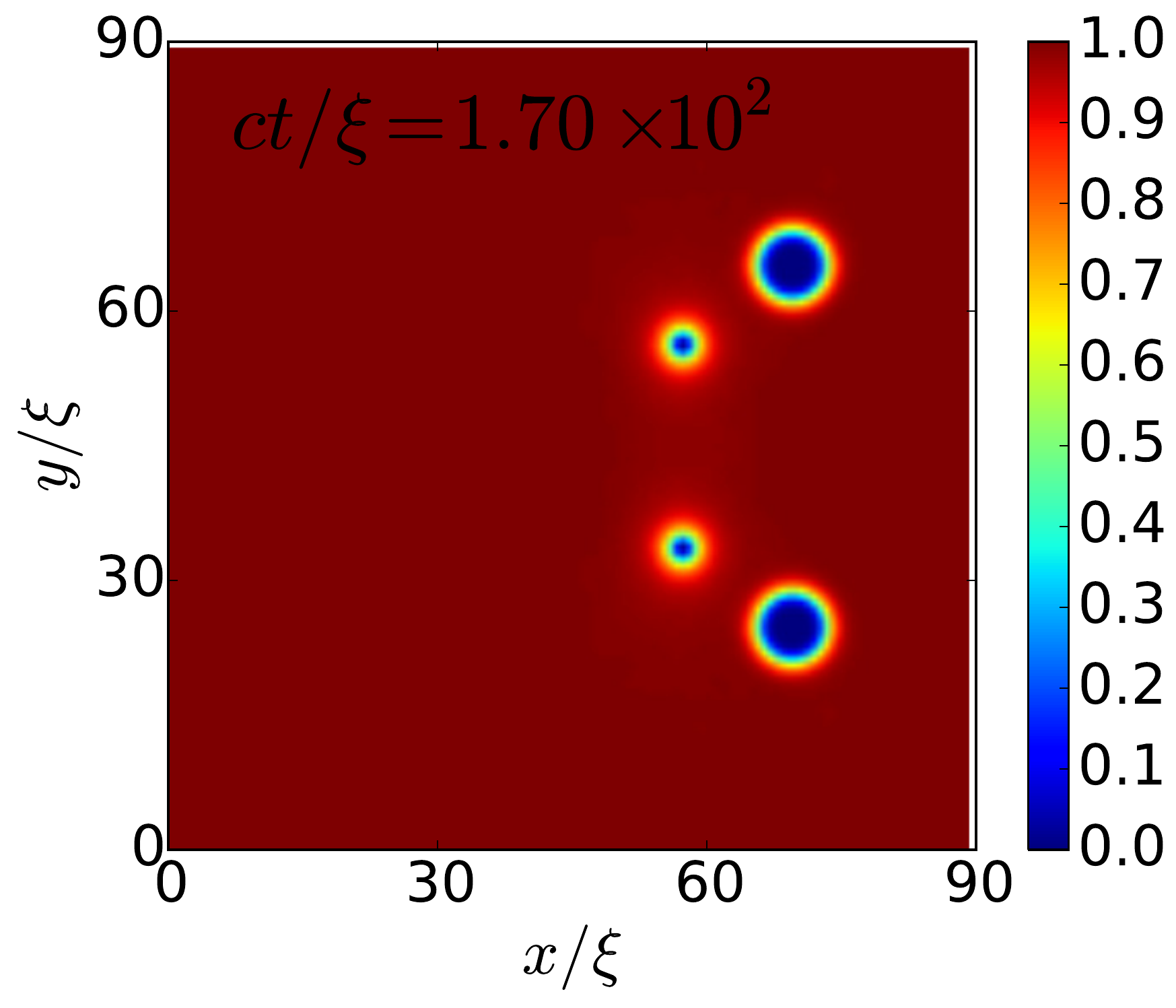}
\put(10,10){\large{\bf (b)}}
\end{overpic}
\begin{overpic}
[height=4.5cm,unit=1mm]{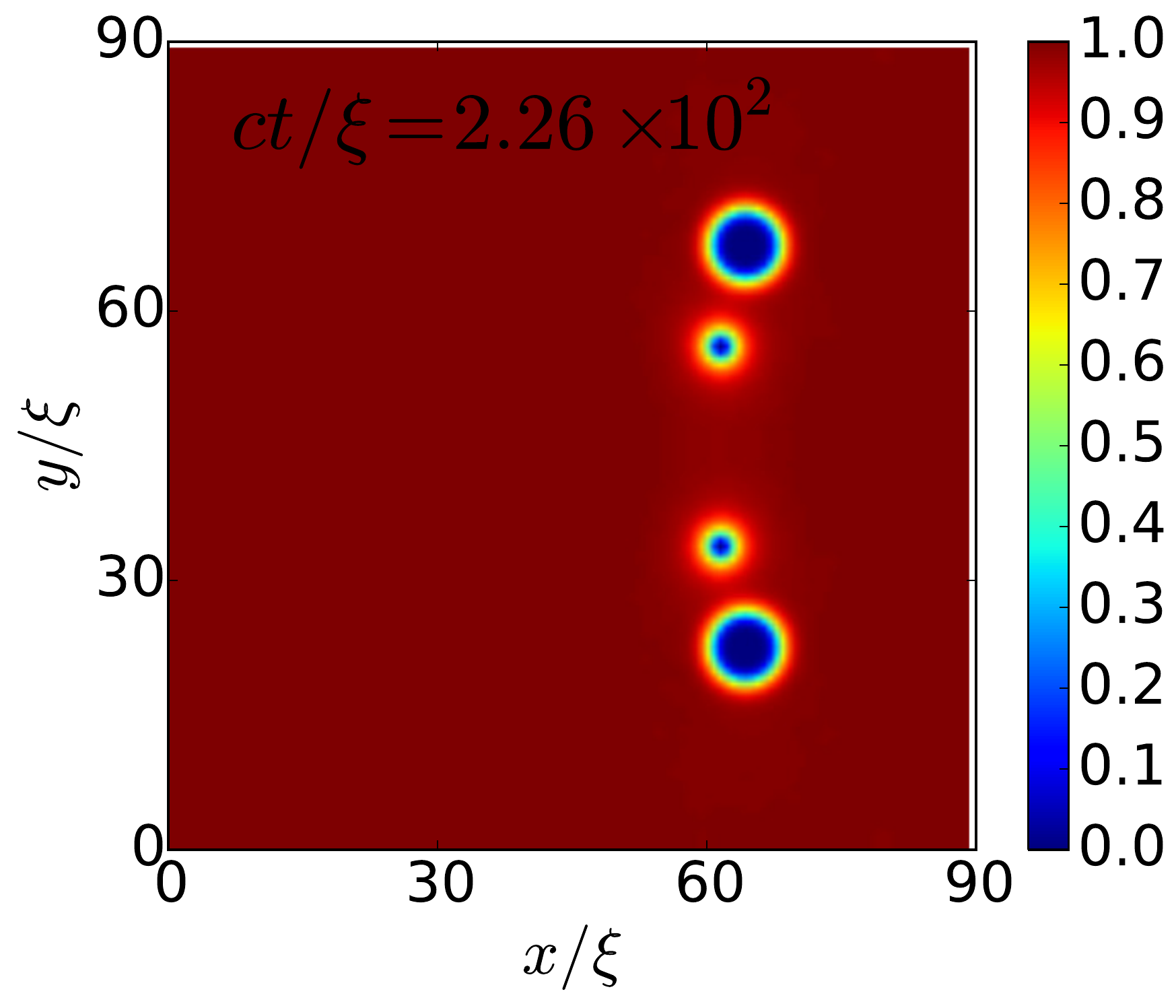}
\put(10,10){\large{\bf (c)}}
\end{overpic}
\\
\begin{overpic}
[height=4.5cm,unit=1mm]{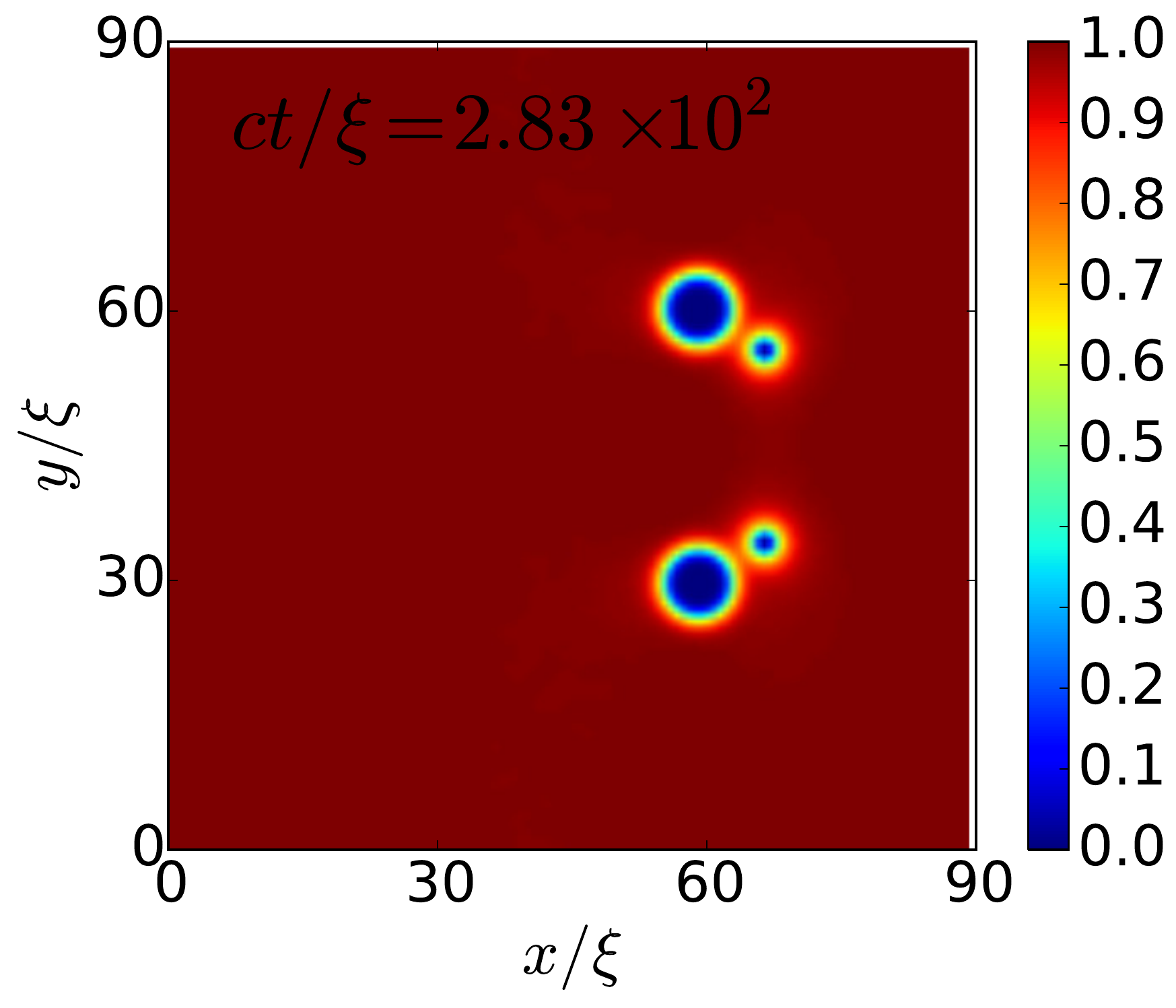}
\put(10.,10){\large{\bf (d)}}
\end{overpic}
\begin{overpic}
[height=4.5cm,unit=1mm]{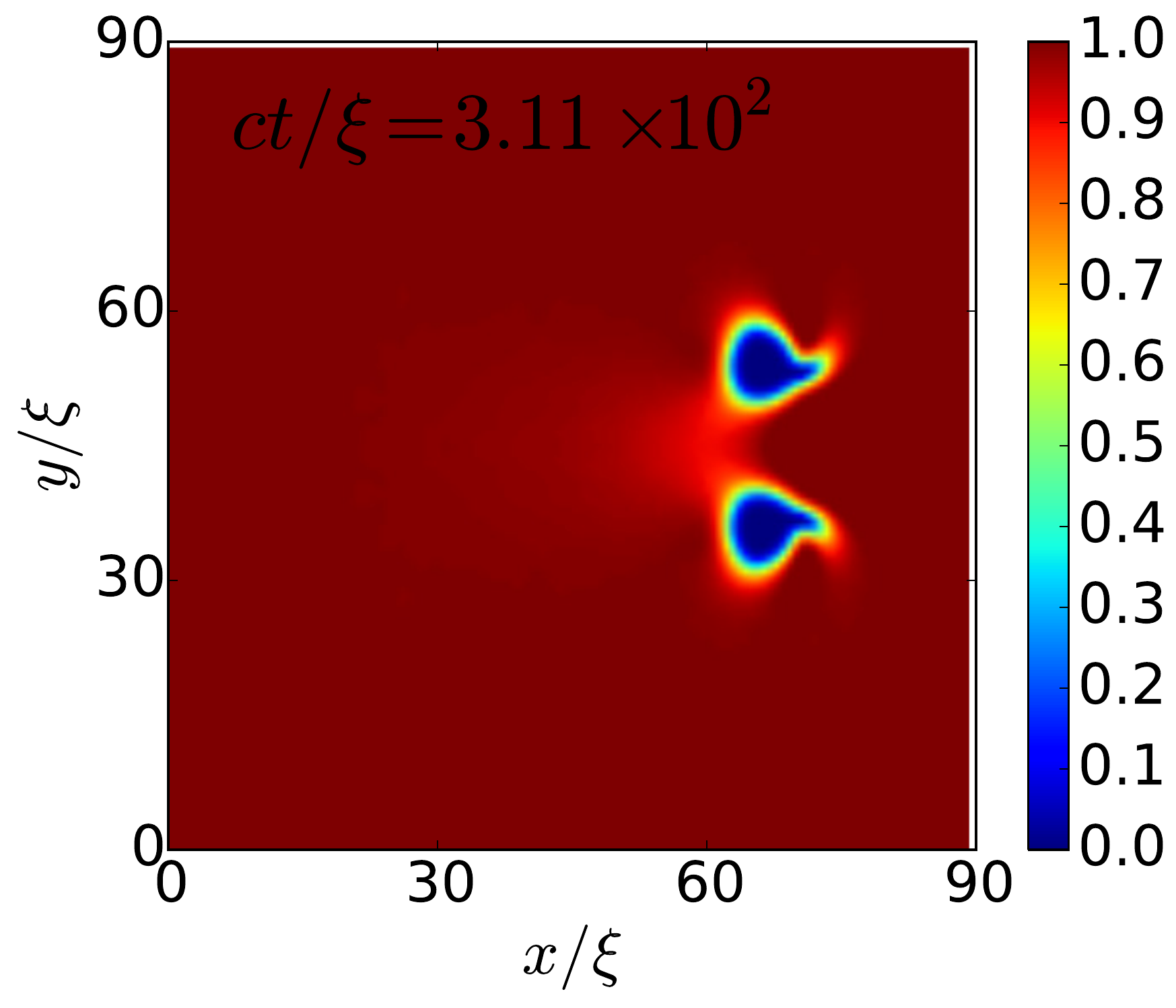}
\put(10,10){\large{\bf (e)}}
\end{overpic}
\begin{overpic}
[height=4.5cm,unit=1mm]{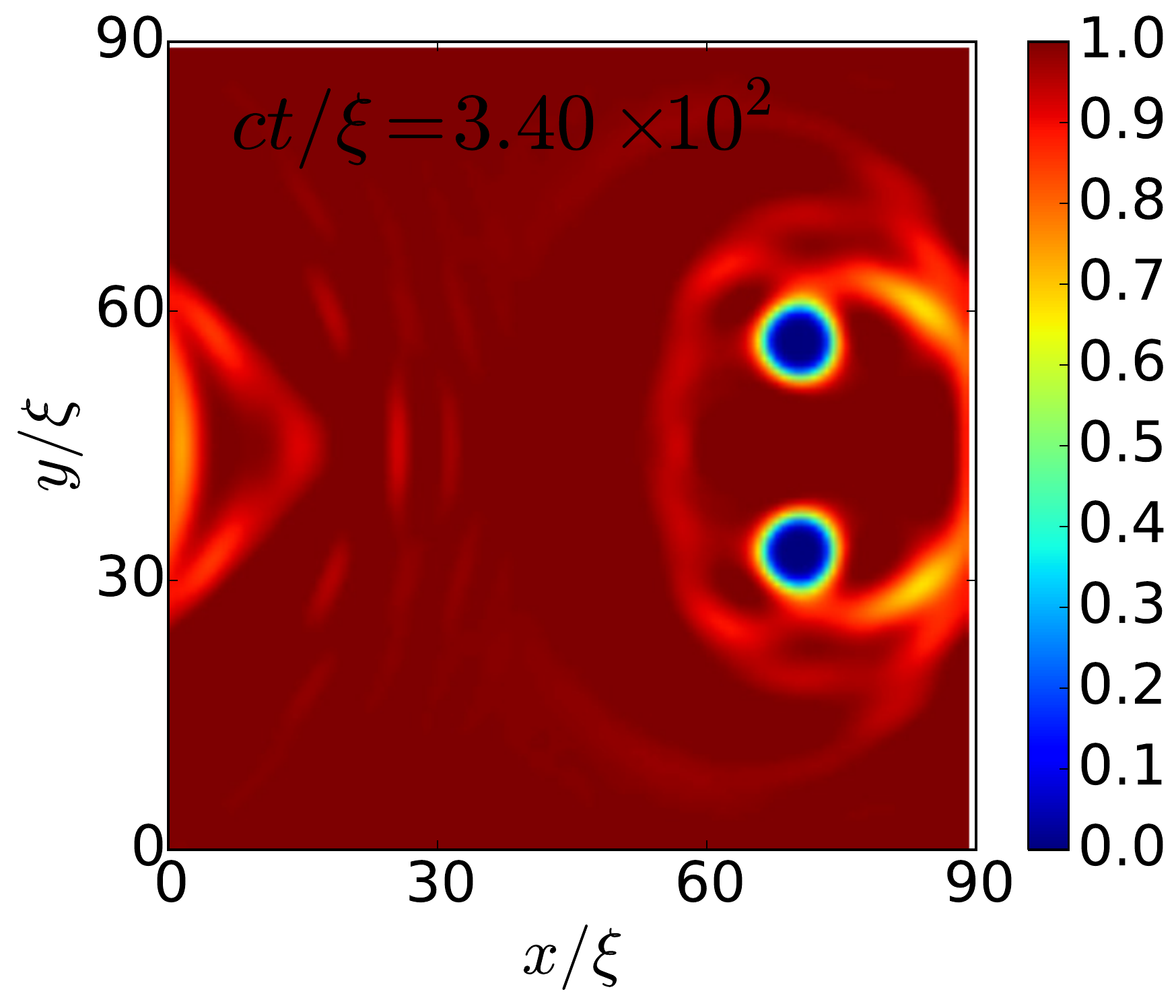}
\put(10,10){\large{\bf (f)}}
\end{overpic}
\caption{\small \textbf{Two Light particles interacting with a translating
	vortex-antivortex pair}:
	Spatiotemporal evolution of the density field
$\rho(\mathbf{r},t)$ shown via
pseudocolor plots, for two light particle placed in the path of the positive (upper)
and negative (lower) vortices, respectively, of a translating vortex-antivortex pair 
(initial configuration $\tt ICP2B$).
}
\label{fig:2partpairtranslpdL}
\end{figure*}

\begin{figure*}
\centering
\resizebox{\linewidth}{!}{
\includegraphics[height=4.cm,unit=1mm]{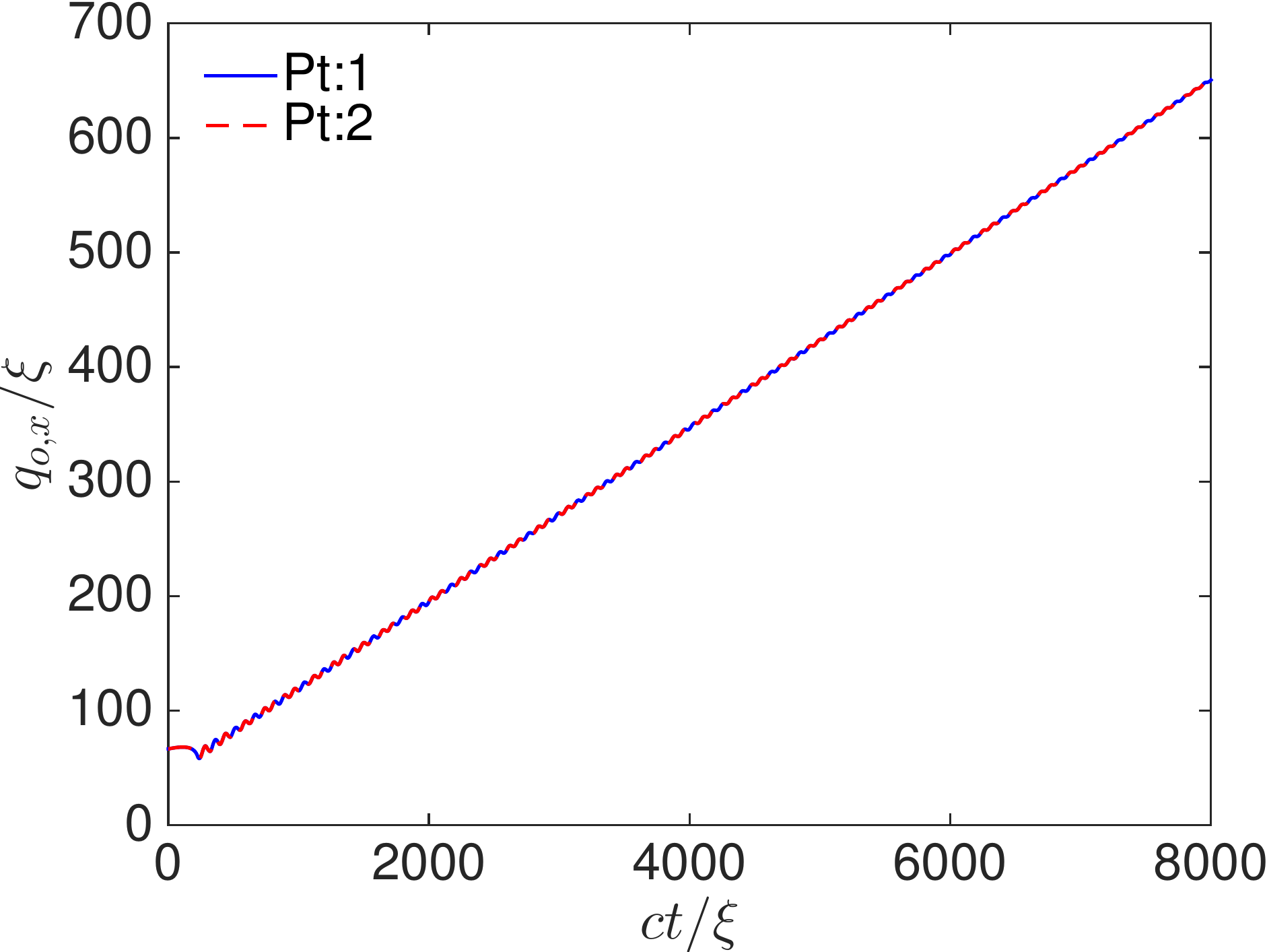}
\put(-75,20){\bf(a)}
\hspace{0.25cm}
\includegraphics[height=4.cm,unit=1mm]{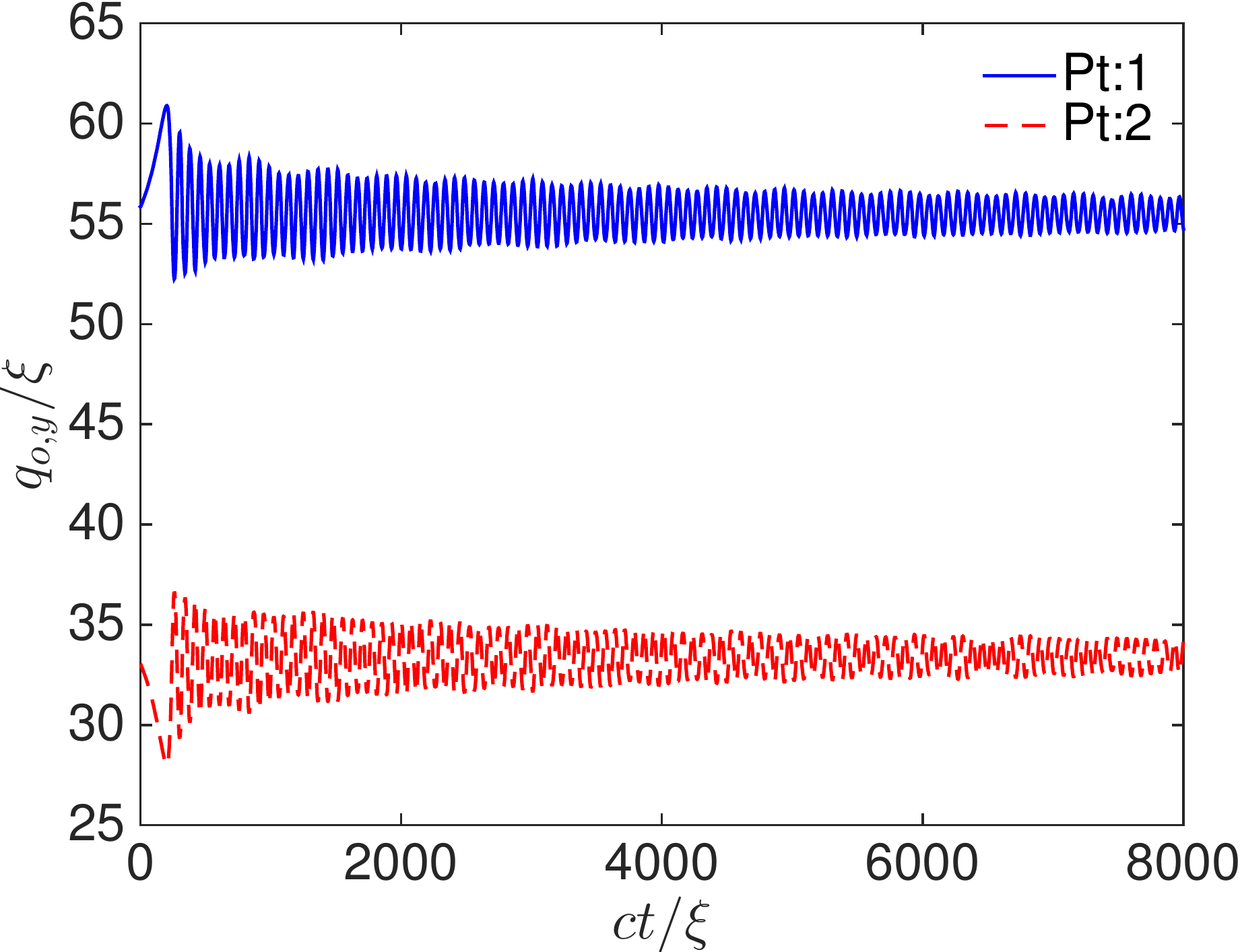}
\put(-75,20){\bf(b)}
}
\\
\vspace{0.30cm}
\resizebox{\linewidth}{!}{
\includegraphics[height=4.cm,unit=1mm]{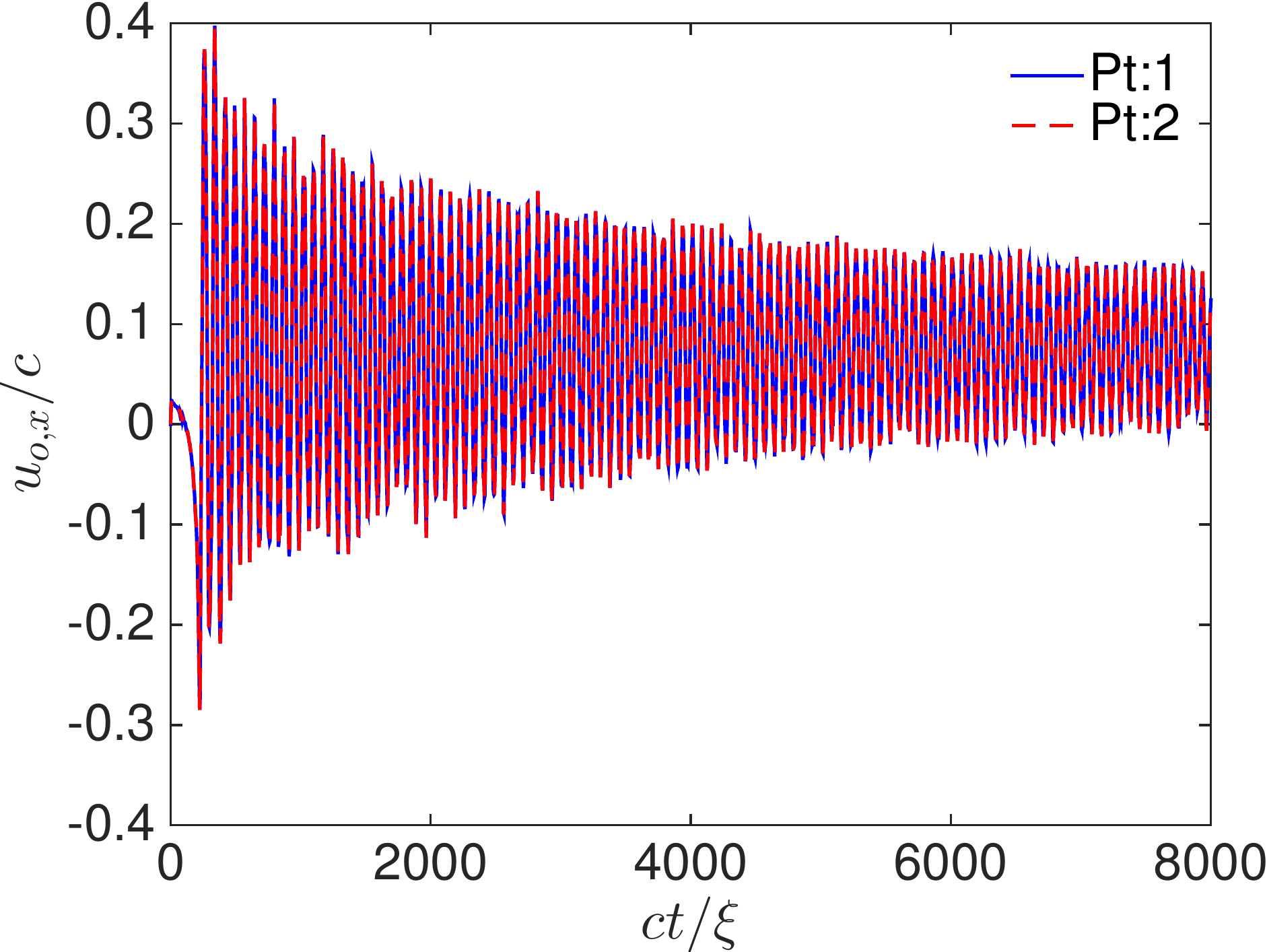}
\put(-75,20){\bf(c)}
\hspace{0.25cm}
\includegraphics[height=4.cm,unit=1mm]{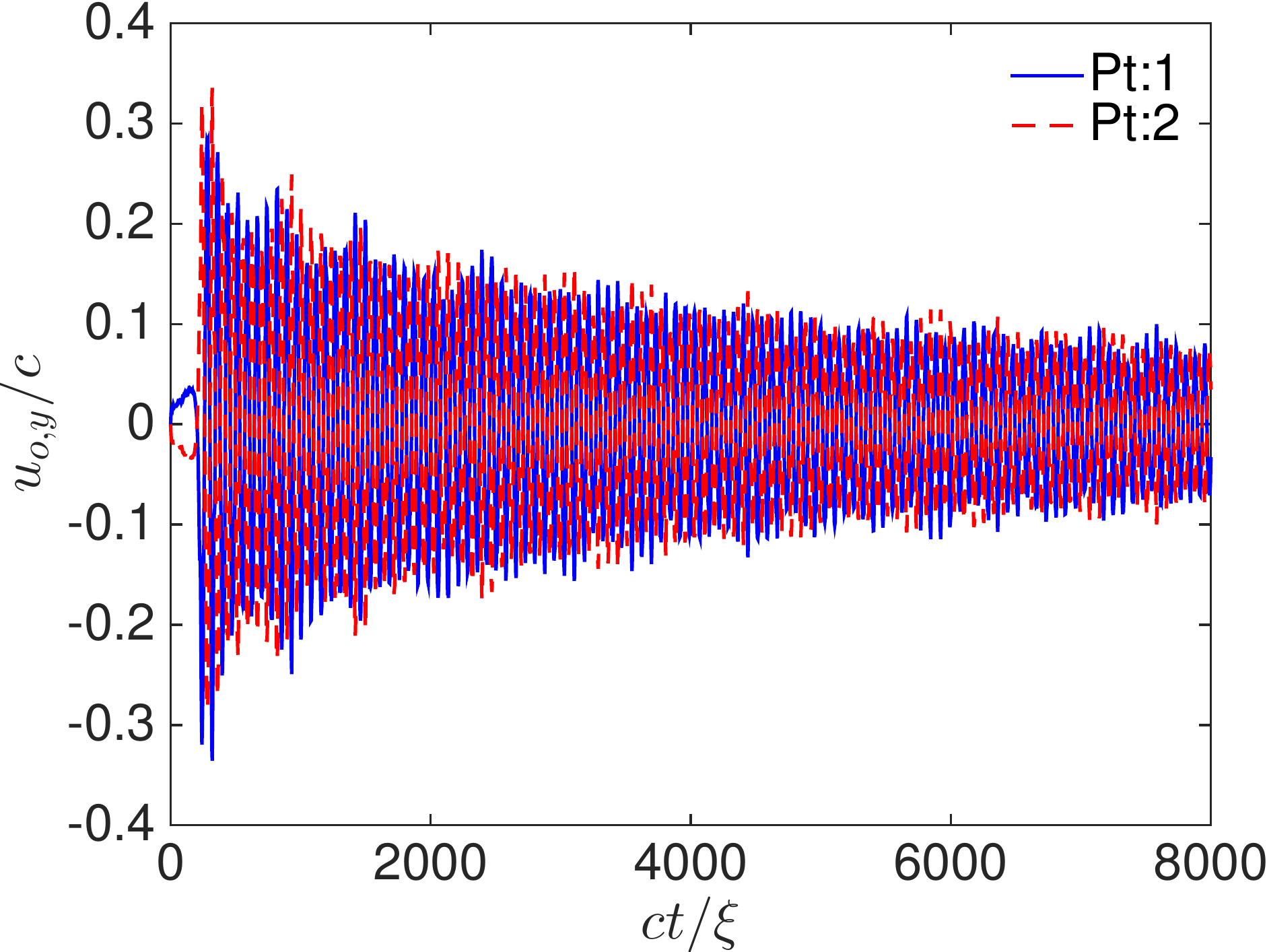}
\put(-75,20){\bf(d)}
}
\caption{\small Plots versus time $t$ of (a) $q_{\rm o,x}$, (b) $q_{\rm o,y}$, 
(c) $u_{\rm o,x}$, (d) $u_{\rm o,y}$ for two neutral particles
$Pt:1$ (blue solid curve) and $Pt:2$ (red dashed curve),
placed in the path of the positive (upper)
and negative (lower) vortices, respectively, of a translating vortex-antivortex pair 
(initial configuration $\tt ICP2B$).
In (a) and (b) the values of $q_{\rm o,x}$ and $q_{\rm o,y}$ are not modulo $2\pi$;
i.e., if particle goes around our periodic simulation domain once, say in the 
$\hat{\mathbf{x}}$ direction, then the values of $q_{\rm o,x}$ is its value in the
box plus $2\pi$.
}
\label{fig:2partpairtranslquN}
\end{figure*}

\begin{figure*}
\centering
\resizebox{\linewidth}{!}{
\includegraphics[height=4.cm,unit=1mm]{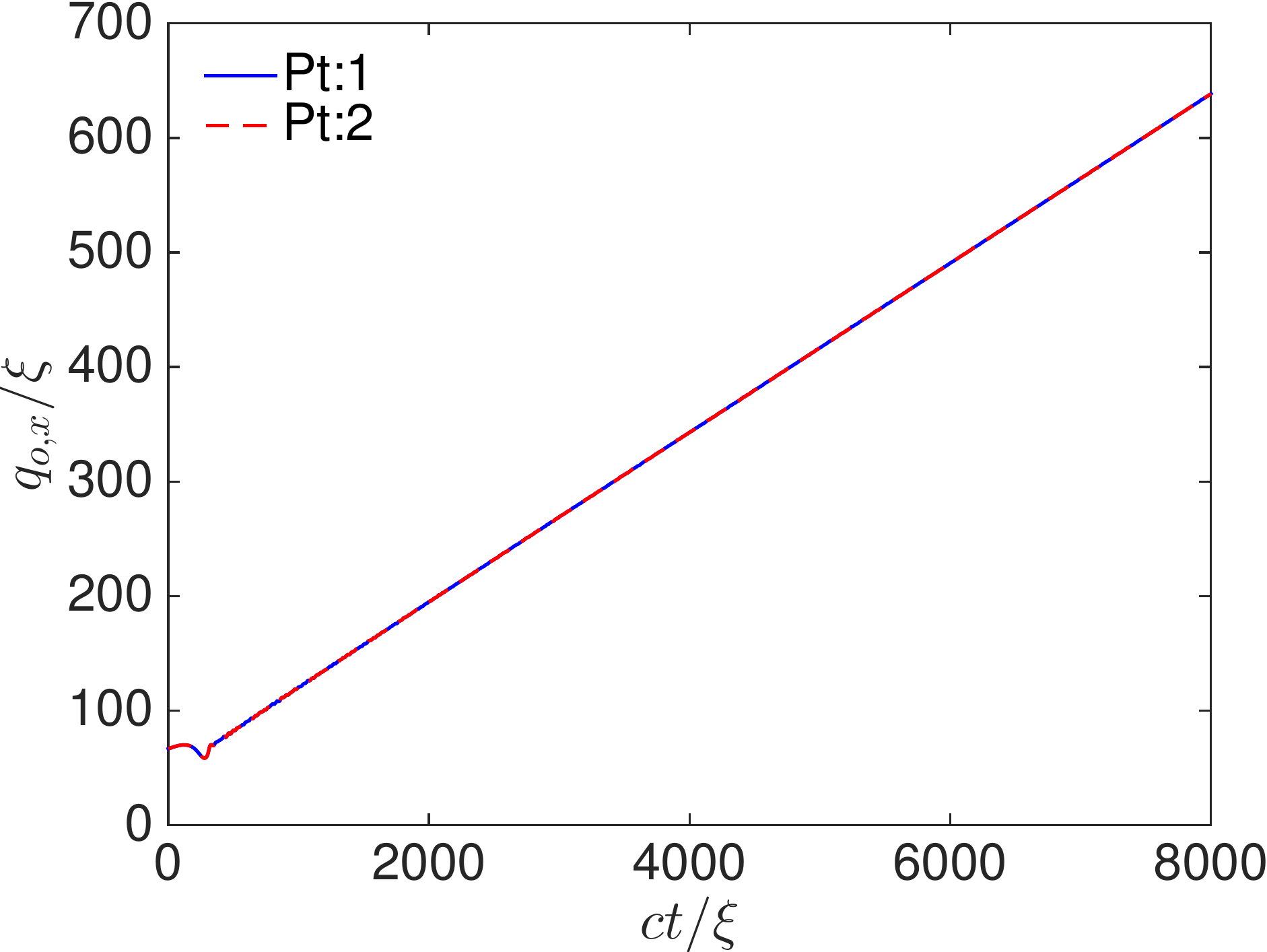}
\put(-75,20){\bf(a)}
\hspace{0.25cm}
\includegraphics[height=4.cm,unit=1mm]{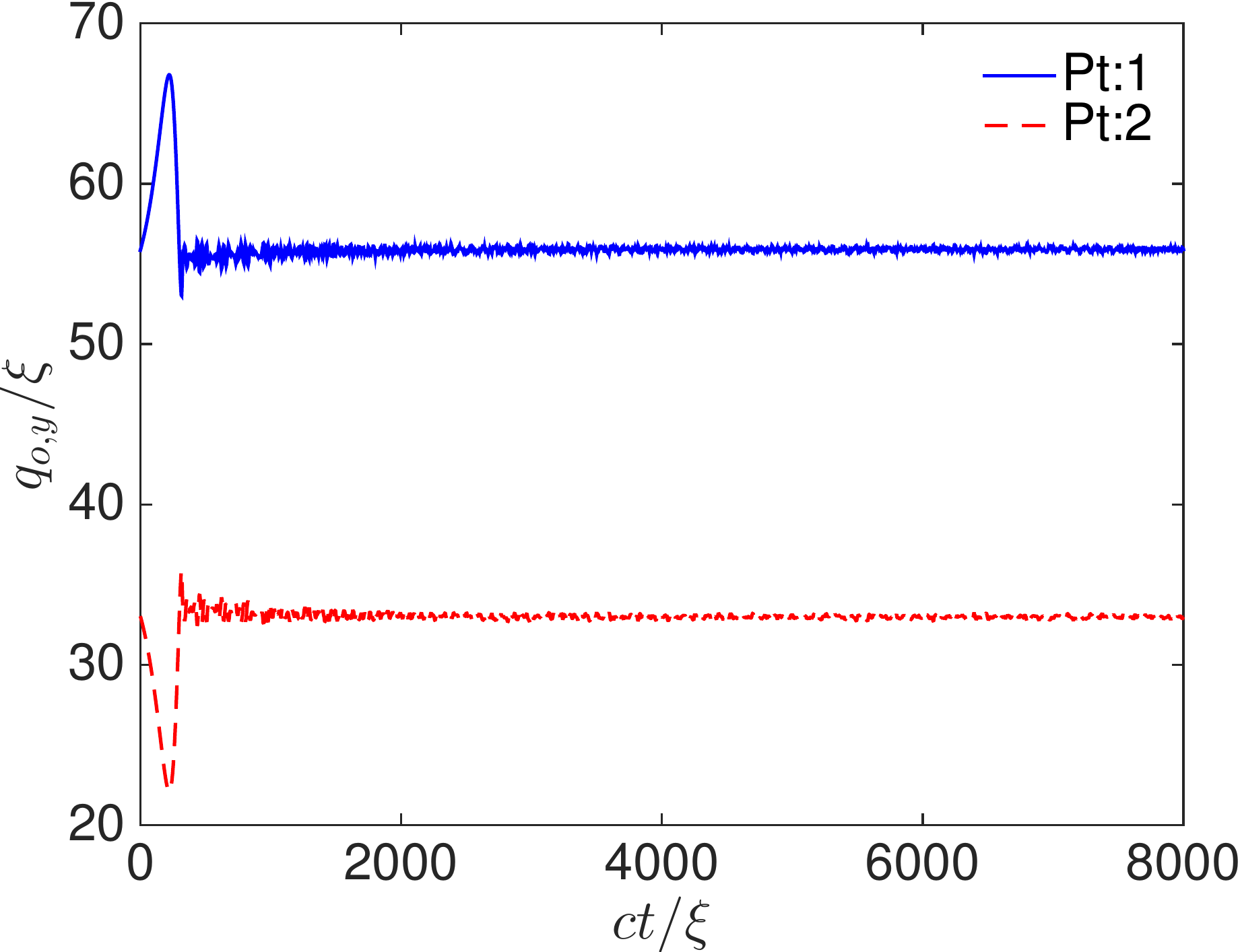}
\put(-75,20){\bf(b)}
}
\\
\vspace{0.30cm}
\resizebox{\linewidth}{!}{
\includegraphics[height=4.cm,unit=1mm]{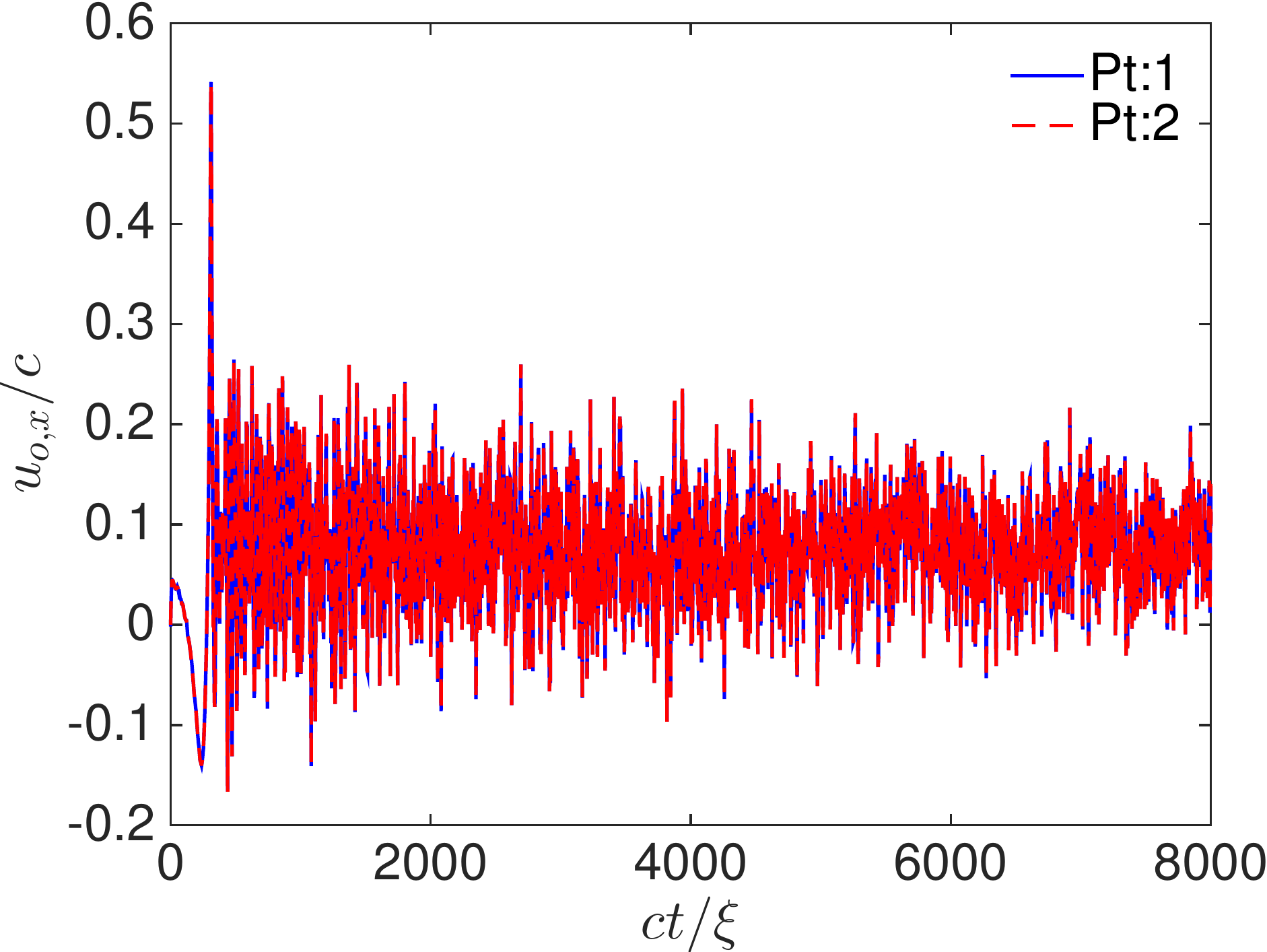}
\put(-75,20){\bf(c)}
\hspace{0.25cm}
\includegraphics[height=4.cm,unit=1mm]{24c.pdf}
\put(-75,20){\bf(d)}
}
\caption{\small Plots versus time $t$ of (a) $q_{\rm o,x}$, (b) $q_{\rm o,y}$, 
(c) $u_{\rm o,x}$, (d) $u_{\rm o,y}$ for two light particles
$Pt:1$ (blue solid curve) and $Pt:2$ (red dashed curve),
placed in the path of the positive (upper)
and negative (lower) vortices, respectively, of a translating vortex-antivortex pair 
(initial configuration $\tt ICP2B$).
In (a) and (b) the values of $q_{\rm o,x}$ and $q_{\rm o,y}$ are not modulo $2\pi$;
i.e., if particle goes around our periodic simulation domain once, say in the 
$\hat{\mathbf{x}}$ direction, then the values of $q_{\rm o,x}$ is its value in the
box plus $2\pi$.
}
\label{fig:2partpairtranslquL}
\end{figure*}

%

%

\vspace{0.5cm}
{\large\bf{SUPPLEMENTAL MATERIAL}}
\vspace{0.5cm}

{\textcolor{blue}{\bf Video M1}}(\url{https://youtu.be/pELKLOXaLc4}):
Spatiotemporal evolution of the density field
$\rho(\mathbf{r},t)$ shown via pseudocolor plots, illustrating the dynamics of a heavy particle, when a constant external force $\mathbf{F}_{\rm ext}=0.28\,c^2\xi\rho_0\,\hat{\mathbf{x}}$ acts on it (initial configuration $\tt ICP1$).  The
particle appears as a large blue patch and the vortices as blue dots.
\newline

{\textcolor{blue}{\bf Video M2}}(\url{https://youtu.be/Fi6IxpCiX0k}):
Spatiotemporal evolution of the density field
$\rho(\mathbf{r},t)$ shown via pseudocolor plots, illustrating the dynamics of a neutral particle, when a constant external force $\mathbf{F}_{\rm ext}=0.14\,c^2\xi\rho_0\,\hat{\mathbf{x}}$ acts on it (initial configuration $\tt ICP1$).  The
particle appears as a large blue patch and the vortices as blue dots.
\newline

{\textcolor{blue}{\bf Video M3}}(\url{https://youtu.be/t1ggxiei_Fw}):
Spatiotemporal evolution of the density field
$\rho(\mathbf{r},t)$ shown via pseudocolor plots, illustrating the dynamics of a light particle, when a constant external force $\mathbf{F}_{\rm ext}=0.14\,c^2\xi\rho_0\,\hat{\mathbf{x}}$ acts on it (initial configuration $\tt ICP1$).  The
particle appears as a large blue patch and the vortices as blue dots.
\newline

{\textcolor{blue}{\bf Video M4}}(\url{https://youtu.be/6Hk8d3fqRGE}):
Spatiotemporal evolution of the density field
$\rho(\mathbf{r},t)$ shown via pseudocolor plots, for a heavy particle
placed in the path of the positive (upper) vortex of a translating
vortex-antivortex pair (initial configuration $\tt ICP2A$).
\newline

{\textcolor{blue}{\bf Video M5}}(\url{https://youtu.be/KfMGJDgVhQM}):
Spatiotemporal evolution of the density field
$\rho(\mathbf{r},t)$ shown via pseudocolor plots, for a neutral particle
placed in the path of the positive (upper) vortex of a translating
vortex-antivortex pair (initial configuration $\tt ICP2A$).
\newline

{\textcolor{blue}{\bf Video M6}}(\url{https://youtu.be/hCs9XJpmtsM}):
Spatiotemporal evolution of the density field
$\rho(\mathbf{r},t)$ shown via pseudocolor plots, for a ligth particle
placed in the path of the positive (upper) vortex of a translating
vortex-antivortex pair (initial configuration $\tt ICP2A$).
\newline

{\textcolor{blue}{\bf Video M7}}(\url{https://youtu.be/m0WSbgLe1Go}):
Spatiotemporal evolution of the density field
$\rho(\mathbf{r},t)$ shown via pseudocolor plots, for two heavy particle
placed in the path of the positive (upper) and negative (lower) vortices,
respectively, of a translating vortex-antivortex pair (initial configuration
$\tt ICP2B$).
\newline

{\textcolor{blue}{\bf Video M8}}(\url{https://youtu.be/rn02wXt-pdA}):
Spatiotemporal evolution of the density field
$\rho(\mathbf{r},t)$ shown via
pseudocolor plots, for two neutral particle placed in the path of the positive (upper)
and negative (lower) vortices, respectively, of a translating vortex-antivortex pair 
(initial configuration $\tt ICP2B$).
\newline

{\textcolor{blue}{\bf Video M9}}(\url{https://youtu.be/XIYvhngJqaE}):
Spatiotemporal evolution of the density field
$\rho(\mathbf{r},t)$ shown via
pseudocolor plots, for two light particle placed in the path of the positive (upper)
and negative (lower) vortices, respectively, of a translating vortex-antivortex pair 
(initial configuration $\tt ICP2B$).
\newline

{\textcolor{blue}{\bf Video M10}}(\url{https://youtu.be/syzxk_upx30}):
Spatiotemporal evolution of the filtered
vorticity field (derived from the incompressible velocity field), for the
neutral particle initially in the presence of counter-rotating vortex clusters
(initial configuration $\tt ICP3A$). The instantaneous position of the particle
is shown by a black disk.
\newline

{\textcolor{blue}{\bf Video M11}}(\url{https://youtu.be/79gmHBd1pqE}):
Spatiotemporal evolution of the density field
$\rho(\mathbf{r},t)$ shown via pseudocolor plots, for the four neutral
particles (large blue patches), initially placed at the centers of the
counter-rotating vortex clusters (initial configuration $\tt ICP3B$).
\newline

\end{document}